\documentclass[titlepage,12pt]{book}
\usepackage{graphicx,rotating}

\def\textwidth{15.9cm}              

\begin{document}

\pagenumbering{roman}

\title{\Huge
\mbox{} \\
\mbox{} \\
Isomorphic Strategy Spaces in Game Theory  \\
\mbox{} \\
\mbox{} \\
\mbox{} \\}

\author{
$
   \begin{array}{l}
     \hspace{-0.6cm}\mbox{Michael J. Gagen}                   \\
     \hspace{-0.6cm}\mbox{Email: mjgagen at gmail.com}        \\
     \hspace{-0.6cm}\mbox{URL: http://www.millitangent.org/}  \\
   \end{array} \nonumber
$
 \mbox{} \\
 \mbox{} \\
 \mbox{} \\
 \mbox{} \\
 \mbox{} \\
 }

\date{10 April 2013}

\maketitle


\mbox{} \vspace{16cm}

Copyright \copyright Michael J. Gagen 2013.

\vspace{0.5cm}

All rights reserved.  No part of this book may be reproduced in
any form by any electronic or mechanical means (including
photocopying, recording, or information storage and retrieval)
without permission in writing from the author.

\vspace{0.5cm}

Michael J. Gagen assert his right to be identified as the author
of this work.

 \addcontentsline{toc}{section}{\bf Contents}
 \tableofcontents

 \addcontentsline{toc}{section}{\bf List of Figures}
 \listoffigures

 \mbox{} \newpage
 \addcontentsline{toc}{section}{\bf List of Tables}
 \listoftables

 \mbox{} \newpage
 \mbox{} \newpage
 \thispagestyle{myheadings}\markboth{PREFACE}{}
 \mbox{}\vspace{1.5cm}

\section*{\Huge Preface}
\addcontentsline{toc}{section}{\bf Preface}

\thispagestyle{myheadings}\markboth{PREFACE}{PREFACE}
\vspace{1cm}

This book summarizes ongoing research introducing probability
space isomorphic mappings into the strategy spaces of game
theory.

This approach is motivated by discrepancies between probability
theory and game theory when applied to the same strategic
situation.  In particular, probability theory and game theory
can disagree on calculated values of the Fisher information, the
log likelihood function, entropy gradients, the rank and
Jacobian of variable transforms, and even the dimensionality and
volume of the underlying probability parameter spaces. These
differences arise as probability theory employs structure
preserving isomorphic mappings when constructing strategy spaces
to analyze games.  In contrast, game theory uses weaker mappings
which change some of the properties of the underlying
probability distributions within the mixed strategy space. Here,
we explore how using strong isomorphic mappings to define game
strategy spaces can alter rational outcomes in simple games .

Specific example games considered are the chain store paradox,
the trust game, the ultimatum game, the public goods game, the
centipede game, and the iterated prisoner's dilemma. In general,
our approach provides rational outcomes which are consistent
with observed human play and might thereby resolve some of the
paradoxes of game theory.

\section{Acknowledgments}
The author gratefully acknowledges a fruitful collaboration with
Kae Nemoto.

 \mbox{} \newpage

 \pagenumbering{arabic}

\chapter{Strong isomorphisms in strategy spaces}
 \label{chap_strong_isomorphisms}

\section{Introduction}

\subsection{Irreducible complexity of strategic optimization}

The essential problem of economics and the rational for game
theory was first posed by von Neumann and Morgenstern
\cite{vonNeumann_44}. They described the fundamental economic
optimization problem by contrasting the non-strategic single
player case with the strategic multi-player situation. In
particular, they stated the non-strategic case is ``an economy
which is represented by the `Robinson Crusoe' model, that is an
economy of an isolated single person, or otherwise organized
under a single will." In this economy, ``Crusoe faces an
ordinary maximization problem, the difficulties of which are of
a purely technical---and not conceptual---nature".  This
non-strategic case was contrasted with a strategic ``social
exchange economy [where] the result for each one will depend in
general not merely upon his own actions but on those of the
others as well. \dots This kind of problem is nowhere dealt with
in classical mathematics. \dots this is no ordinary maximization
problem, no problem of the calculus of variations, of functional
analysis, etc" \cite{vonNeumann_44}.

Thus, von Neumann and Morgenstern essentially claimed that
strategic optimization problems were irreducibly more complex
than non-strategic optimization problems. And yet, after
learning a few new techniques, the solution of strategic games
turns out to be not significantly more complex than the solution
of non-strategic decision trees---larger and more difficult
certainly, but not irreducibly more complex. In this work, we
claim that the proposed solution to strategic analysis is
incomplete. We will argue that strategic optimization is indeed
irreducibly more complex than non-strategic optimization, and
this irreducible complexity is missing from current formulations
of strategic optimization.

We will look for this missing irreducible complexity by applying
probability theory and game theory to the same strategic
situation, and examining any differences that arise.  We will
show that when applied to the same strategic game, probability
theory and game theory can disagree on calculated values of the
Fisher information, the log likelihood and entropy gradients,
the rank and Jacobian of variable transforms, and even the
dimensionality and volume of the underlying probability
parameter spaces. These differences arise as probability theory
employs structure preserving, isomorphic mappings when
constructing a mixed strategy space to analyze games.  In
contrast, game theory uses weaker mappings which change some of
the properties of the underlying probability distributions
within the mixed strategy space.  We will explore how using
strong isomorphic mappings to define mixed strategy spaces can
alter rational outcomes in simple games, and might resolve some
of the paradoxes of game theory.

\subsection{Strategy spaces of game theory}

One possibly fruitful way to gain insight into the paradoxes of
game theory is to show that probability theory and game theory
analyze simple games differently.  It would be expected of
course that these two well developed fields should always
produce consistent results.  However, we will show in this paper
that probability theory and game theory can produce
contradictory results when applied to even simple games. These
differences arise as these two fields construct mixed strategy
spaces differently.

The mixed strategy space of game theory is constructed,
according to von Neumann and Morgenstern, by first making a
listing of every possible combination of moves that players
might make and of all possible information states that players
might possess. This complete embodiment of information then
allows every move combination to be mapped into a probability
simplex whereby each player's mixed strategy probability
parameters belong to ``disjoint but exhaustive alternatives,
\dots subject to the [usual normalization] conditions \dots and
to no others." \cite{vonNeumann_44}.  The resulting
unconstrained mixed strategy space is then a ``complete set" of
all possible probability distributions that might describe the
moves of a game
\cite{vonNeumann_44,Nash_50_48,Nash_51_28,Kuhn_1953,Hart_92_19}.
Further, the absence of non-normalization constraints ensures
``trembles" or ``fluctuations" are always present within the
mixed strategy space so every possible pure strategy probability
distribution is played with non-zero (but possibly
infinitesimal) probability \cite{Selton_1975}.  Together, these
properties of the mixed strategy space---a complete set of
``contained" probability distributions, no additional
constraints, and ever present trembles---lead to inconsistencies
with probability theory.

\subsection{Isomorphic probability spaces}

In constructing a mixed strategy space, probability theory first
examines how subsidiary probability distributions can be
``contained" within a mixed space and whether the properties of
the probability distributions are altered as a result.
Probability theory uses isomorphisms to implement mappings of
one probability space into another space. An isomorphism is a
structure preserving mapping from one space to another space. In
abstract algebra for instance, an isomorphism between vector
spaces is a bijective (one-to-one and onto) linear mapping
between the spaces with the implication that two vector spaces
are isomorphic if and only if their dimensionality is identical
\cite{Chatterjee_2005}.  When the preservation of structure is
exact, then calculations within either space must give identical
results. Conversely, if the degree of structure preservation is
less than exact, then differences can arise between calculations
performed in each space.  It is thus crucial to examine the
fidelity of the ``containment" mappings used to construct the
mixed spaces of game theory. Probability theory defines
isomorphic probability spaces as follows.  We give two
definitions for completeness, see Refs.
\cite{Ito_1984,Gray_2009,Walters_1982}.

{\bf Definition 1:} A probability space ${\cal
P}=\{\Omega,\sigma,P\}$ consists of a set of events $\Omega$, a
sigma-algebra of all subsets of those events $\sigma$, and a
probability measure defined over the events $P$. Two probability
spaces ${\cal P}=\{\Omega,\sigma,P\}$ and ${\cal
P}'=\{\Omega',\sigma',P'\}$ are said to be {\em strictly
isomorphic} if there is a bijective (1-to-1 and onto) map
$f:\Omega\rightarrow\Omega'$ which exactly preserves assigned
probabilities, so for all $e\in\Omega$ we have $P(e)=P'[f(e)]$.
A slight weakening of this definition defines an {\em
isomorphism} as a bijective mapping $f$ of some unit probability
subset of $\Omega$ onto a unit probability subset of $\Omega'$.
That is, the weakened mapping ignores null event subsets of zero
probability.

{\bf Definition 2:} Two probability spaces ${\cal
P}=\{\Omega,\sigma,P\}$ and ${\cal P}'=\{\Omega',\sigma',P'\}$
are isomorphic if there are null event sets $\Omega^0 \in
\Omega$ and $\Omega'^0 \in \Omega'$ and an isomorphism
$f:(\Omega-\Omega^0)\rightarrow (\Omega'-\Omega'^0)$ between the
two measurable spaces $(\Omega-\Omega^0,\sigma)$ and
$(\Omega'-\Omega'^0,\sigma')$ with the added properties that
$P'(F)=P[f^{-1}(F)]$ for $F\in\sigma'$ and $P(G)=P'[f(G)]$ for
$G\in\sigma$.  In other words, an isomorphism exists if there is
an invertible measure-preserving transformation between the unit
probability events in each space, $(\Omega-\Omega^0)\in\Omega$
and $(\Omega'-\Omega'^0)\in\Omega'$.  This also implies that the
null probability event sets of each space are mapped to each
other.

In particular, we note that strong isomorphisms between source
and target probability spaces require they have identical
dimensionality and tangent spaces \cite{Sernesi_1993}.

\subsection{Isomorphism choice alters optimization outcomes}

The mixed strategy space of game theory ``contains" different
probability distributions many possessing different
dimensionality (according to probability theory).  Their altered
dimensionality within the mixed space can alter those computed
outcomes dependent on dimensionality.  A simple illustration of
this process can make this clear.

A 1-dimensional function $f(x)$ can be embedded within a
2-dimensional function $g(x,y)$ in two ways: using constraints
$g(x,y_0)=f(x)$, or limits $\lim_{y\rightarrow y_0}
g(x,y)=f(x)$. In either case, many of the properties of the
source function $f(x)$ are preserved, but not necessarily all of
them.  In particular, these different methods alter gradient
optimization calculations. That is, the gradient is properly
calculated when constraints are used, $f'(x)=g'(x,y_0)$, but not
when a limit process is used, $f'(x)\neq \lim_{y\rightarrow y_0}
\nabla g(x,y)$ (where $\nabla$ indicates a gradient operator).

We note our use of gradient operators is unusual in game theory.
In lieu of gradient operators, the rational players of game
theory generally simply compare the values of expected payoff
functions at different points within a probability space.
However, we remind ourselves that every comparison of an
expected payoff function over a probability space is equivalent
to evaluating a gradient. Specifically, a function $\Pi(x,y)$
with expectation $\langle \Pi(a)\rangle$ compared at the points
$a_1$ and $a_2$ within a probability space employs the identity
\begin{equation}      \label{eq_expectation_gradient2}
 \langle\Pi(a_2)\rangle - \langle\Pi(a_1)\rangle =
    \nabla \langle\Pi(a)\rangle . d_{21},
\end{equation}
where the distance vector is $d_{21}=\hat{a} (a_2 - a_1)$. This
results as all expectations are poly-linear in each probability
parameter.

\subsection{Mismatch between probability and game theory}

In this paper, we will show that exactly the same discrepancies
arise when probability theory and game theory are applied to
simple probability spaces, and that these discrepancies can be
significant. It is useful to indicate the magnitude of these
discrepancies here to motivate the paper (with full details
given in later sections below).  We consider a simple card game
with two potentially correlated variables $x,y\in\{0,1\}$ with
joint probability distribution $P_{xy}$. In the case where $x$
and $y$ are perfectly correlated, probability theory (denoted by
P) and game theory (denoted by G) respectively assign different
dimensions to both the Fisher information matrix ($F$) and the
gradient of the log Likelihood function ($\nabla L$), and can
disagree on the value of the gradient of the joint entropy at
some points ($\nabla E_{xy}$):
\begin{equation}
  \begin{array}{c|cc}
                        & \hspace{1cm}{\rm P}\hspace{1cm}  &    \hspace{1cm}{\rm G}\hspace{1cm}    \\ \hline
  {\rm dim}(F)          &     1    &       3       \\
  {\rm dim}(\nabla L)   &     1    &       3       \\
   |\nabla E_{xy}|           &     0    &     \infty. \\
  \end{array}
\end{equation}
These fields also disagree on the probability space gradients of
both the normalization condition ($P_{00}+P_{11}=1$) and the
requirement that the joint entropy equates to the marginal
entropy ($E_{xy}-E_x=0$):
\begin{equation}
  \begin{array}{c|cc}
                        & \hspace{1cm}{\rm P}\hspace{1cm}  &    \hspace{1cm}{\rm G}\hspace{1cm}    \\ \hline
  \nabla \left(P_{00}+P_{11}\right) &     0    &  \neq 0  \\
  \nabla \left(E_{xy}-E_x\right)    &     0    &  \neq 0. \\
  \end{array}
\end{equation}
Should these fields model a change of variable within this game,
they further disagree on the rank of the transform matrix ($A$),
and on the invertibility of the Jacobian matrix ($J$):
\begin{equation}
  \begin{array}{c|cc}
                        & \hspace{1cm}{\rm P}\hspace{1cm}  &    \hspace{1cm}{\rm G}\hspace{1cm}    \\ \hline
  {\rm Rank}(A) &     1           &  2  \\
  J             &     {\rm Singular}    &  {\rm Invertible}. \\
  \end{array}
\end{equation}
These fields even disagree on the dimension ($d$) and volume
($V$) of the minimal probability space used to analyze the game:
\begin{equation}
  \begin{array}{c|cc}
                        & \hspace{1cm}{\rm P}\hspace{1cm}  &    \hspace{1cm}{\rm G}\hspace{1cm}    \\ \hline
  d &  1    &   3       \\
  V &  1    &   \frac{1}{6}.       \\
  \end{array}
\end{equation}
The differences between game theory and probability theory arise
due to the different use of isomorphic mappings to construct
mixed strategy spaces.

We now show the necessity for considering isomorphic probability
spaces using examples ranging from simple dice games to
bivariate normal distributions.

\begin{figure}[htb]
\centering
\includegraphics[width=0.8\columnwidth,clip]{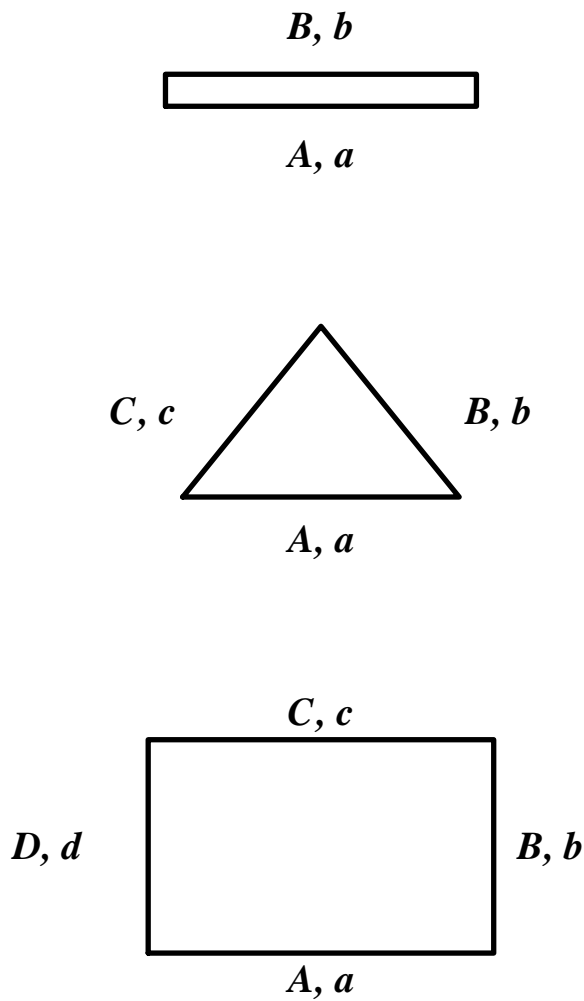}
\caption[Three alternate dice]{\em Three alternate dice with different numbers of
sides. A coin with sides $A$ and $B$ appearing with respective
probabilities $a$ and $b$, a triangle with faces $A, B$ and $C$
occurring with respective probabilities $a, b$ and $c$, and a
square die with faces $A, B, C$ and $D$ each occurring with
respective probabilities $a, b, c$ and $d$.
 \label{f_alternate_dice}}
\end{figure}

\section{Optimization and isomorphic probability spaces}
\label{sect_Optimization_and_isomorphic_probability_spaces}

In this section, we introduce the need to use isomorphic
mappings when embedding probability spaces within mixed spaces.

\subsection{Isomorphic dice}

Consider the three alternate dice shown in Fig.
\ref{f_alternate_dice} representing a 2-sided coin, a 3-sided
triangle, and a 4-sided square.  Faces are labeled with capital
letters and the probabilities of each face appearing are labeled
with the corresponding small letter.  The corresponding
probability spaces defined by these die are
\begin{eqnarray}              \label{eq_coin_tri_sq}
  {\cal P}_{\rm coin}
    &=& \big\{ x\in\{A,B\},\{a,b\} \big\}  \nonumber \\
  {\cal P}_{\rm triangle}
    &=& \big\{ x\in\{A,B,C\},\{a,b,c\} \big\}  \nonumber \\
  {\cal P}_{\rm square}
    &=& \big\{ x\in\{A,B,C,D\},\{a,b,c,d\} \big\}.
\end{eqnarray}
Here the required sigma-algebras are not listed, and each of
these spaces are subject to the usual normalization conditions.
For notational convenience we sometimes write
$(p_1,p_2,p_3,p_4)=(a,b,c,d)$ and denote the number of sides of
each respective die as $n\in\{2,3,4\}$.  In each respective die
space, the gradient operator is
\begin{equation}
   \nabla =   \sum_{i=1}^{n-1}  \hat{p}_i \frac{\partial}{\partial p_i}
\end{equation}
where a hatted variable $\hat{p}_i$ is a unit vector in the
indicated direction and we resolve the normalization constraint
via $p_n=1-\sum_{i=1}^{n-1}p_i$.

We now wish to optimize a nonlinear function over these spaces,
and we choose a function which cannot be optimized using
standard approaches in game theory. The chosen function is
\begin{equation}
  f = V^2 E_x,
\end{equation}
with
\begin{eqnarray}
  V  &=& \int_{\rm space} dv \nonumber \\
  E_x &=& - \sum_{i=1}^n p_i \log p_i,
\end{eqnarray}
where $V$ is the volume of each respective probability parameter
space and $E_x$ is the marginal entropy of each space
\cite{Georgii_2008}. We will complete this optimization in three
different ways, two of which will be consistent with each other
and inconsistent with the third.

As a first pass at optimizing the function $f$, we simply
maximize $f$ within each probability space and then compare the
optimal outcomes to determine the best achievable outcome.  As
is well understood, the entropy of a set of $n$ events is
maximized when those events are equiprobable giving a maximum
entropy of $E_{x, {\rm max}}=\log n$.  In addition, for the coin
we have
\begin{eqnarray}               \label{eq_coin_functions}
  V &=& \int_{0}^{1} da \int_{0}^{1} db \; \delta_{a+b=1} \nonumber \\
    &=& \int_{0}^{1}  da \nonumber \\
    &=& 1      \nonumber \\
  E_x &=&  -[ a \log(a) + (1-a) \log(1-a)] \nonumber \\
  \nabla E_x &=& - \hat{a} \log \frac{a}{1-a}.
\end{eqnarray}
For the triangle, the equivalent functions are
\begin{eqnarray}                \label{eq_triangle_functions}
  V &=& \int_{0}^{1} da \int_{0}^{1} db \int_{0}^{1} dc \; \delta_{a+b+c=1} \nonumber \\
    &=& \int_{0}^{1}  da \int_{0}^{1-a}  db \nonumber \\
    &=& \frac{1}{2}      \nonumber \\
  E_x &=&  -[ a \log(a) + b \log(b) + (1-a-b) \log(1-a-b)] \nonumber \\
  \nabla E_x &=& - \hat{a} \log \frac{a}{1-a-b} - \hat{b} \log \frac{b}{1-a-b}.
\end{eqnarray}
Finally, for the square, we have
\begin{eqnarray}               \label{eq_square_functions}
  V &=& \int_{0}^{1} da \int_{0}^{1} db \int_{0}^{1} dc  \int_{0}^{1} dd \; \delta_{a+b+c+d=1} \nonumber \\
    &=& \int_{0}^{1}  da \int_{0}^{1}  db \int_{0}^{1-a-b}  dc \nonumber \\
    &=& \frac{1}{6}      \nonumber \\
  E_x &=&  -[ a \log(a) + b \log(b) + c \log(c) + (1-a-b-c) \log(1-a-b-c)] \nonumber \\
  \nabla E_x &=& - \hat{a} \log \frac{a}{1-a-b-c} -
                   \hat{b} \log \frac{b}{1-a-b-c} - \hat{c} \log \frac{c}{1-a-b-c}.
\end{eqnarray}
Consequently, the function $f$ takes maximum values in the three
probability spaces of
\begin{eqnarray}               \label{eq_F_Opt}
  f_{\rm coin,\; max} &=& \log 2 \nonumber \\
  f_{\rm triangle,\; max} &=& \frac{\log 3}{4} \nonumber \\
  f_{\rm square,\; max} &=& \frac{\log 4}{36}.
\end{eqnarray}
Comparing these outcomes makes it clear that the best that can
be achieved is to use a coin with equiprobable faces.

\begin{figure}[htb]
\centering
\includegraphics[width=0.9\columnwidth,clip]{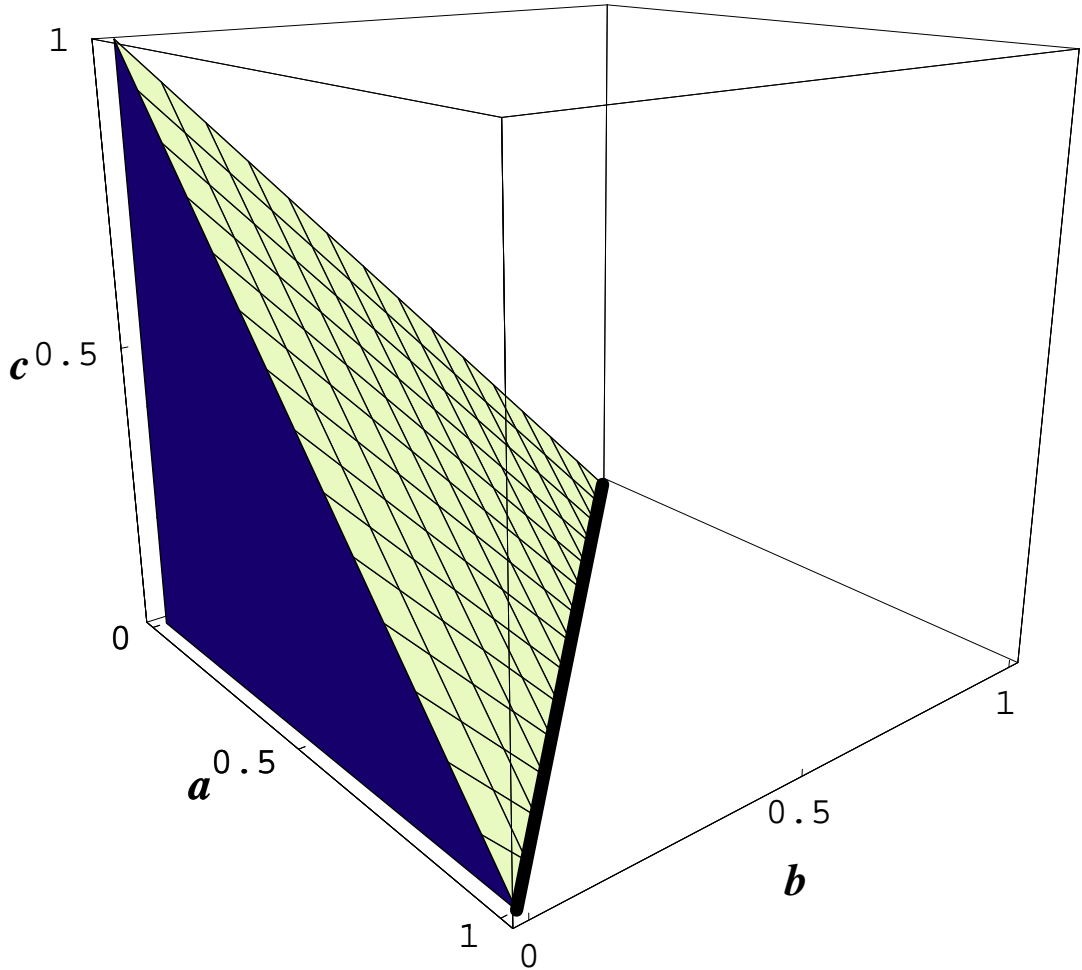}
\caption[The target strategy spaces for alternate dice]{\em The target space
containing points corresponding to the
probability spaces respectively
of the coin ${\cal P}'_{\rm coin}$ along the line $a+b=1$ with constraint
$(c,d)=(0,0)$ (heavy line),
of the triangle ${\cal P}'_{\rm triangle}$ along the surface $a+b+c=1$
with constraint $d=0$ (hashed surface), and
of the square ${\cal P}'_{\rm square}$ filling the volume $a+b+c+d=1$
(filled polygon).
Note that points such as $(a,b,c)=(0.5,0.5,0)$ correspond to all three
probability spaces and are only distinguished by which constraints are acting.
 \label{f_mixed_dice}}
\end{figure}

The second method uses isomorphisms to map all of the three
incommensurate source spaces into a single target space.  We
choose our mappings as follows:
\begin{eqnarray}
  {\cal P}'_{\rm coin}
    &=& \left. \left\{ x\in\{A,B,C,D\},\{a,b,c,d\} \right\}\right|_{(cd)=(00)}  \nonumber \\
  {\cal P}'_{\rm triangle}
    &=& \left. \left\{ x\in\{A,B,C,D\},\{a,b,c,d\} \right\}\right|_{d=0}  \nonumber \\
  {\cal P}'_{\rm square} &=& \left\{ x\in\{A,B,C,D\},\{a,b,c,d\} \right\}.
\end{eqnarray}
Here, while all probability spaces share a common event set and
probability distribution, the isomorphic mappings impose
constraints on the ${\cal P}'_{\rm coin}$ and ${\cal P}'_{\rm
triangle}$ spaces.  The constraints arise from mapping the null
sets of zero probability from each source space to the
corresponding events of the enlarged target space.  The target
probability space is shown in Fig. \ref{f_mixed_dice} where the
normalization condition $d=1-a-b-c$ is used. The points
corresponding to the probability spaces of the coin ${\cal
P}'_{\rm coin}$ are mapped along the line $a+b=1$ with
constraint $(c,d)=(0,0)$. Those points corresponding to the
probability spaces of the triangle ${\cal P}'_{\rm triangle}$
are mapped along the surface $a+b+c=1$ with constraint $d=0$.
Finally, the probability spaces corresponding to the square
${\cal P}'_{\rm square}$ fill the volume $a+b+c+d=1$ and are not
subject to any other constraint.

The interesting point about the target space is that many
points, e.g. $(a,b,c,d)=(\frac{1}{2},\frac{1}{2},0,0)$, lie in
all of the probability spaces of the coin, triangle, and square
die and are only distinguished by which constraints are acting.
That is, when this point is subject to the constraint
$(cd)=(00)$, then it corresponds to the probability space ${\cal
P}'_{\rm coin}$ (and not to any other). Conversely, when this
same point is subject to an imposed constraint $d=0$ then it
corresponds to the probability space ${\cal P}'_{\rm triangle}$.
Finally, when no constraints apply then, and only then does this
point correspond to the probability space of the square ${\cal
P}'_{\rm square}$. This means that it is not the probability
values possessed by a point which determines its corresponding
probability space but the probability values in combination with
the constraints acting at that point.

It is now straightforward to use the isomorphically constrained
target space to maximize the function $f$ over all embedded
probability spaces using standard constrained optimization
techniques. For instance, to optimize $f$ over points
corresponding to the coin and subject to the constraint
$(c,d)=(0,0)$ then either simply resolve the constraint via
setting $c=d=0$ before the optimization begins, or simply
evaluate the gradient of $f$ at all points $(a,b,0,0)$ in the
direction of the unit vector $\frac{1}{\sqrt{2}}(1,-1,0,0)$
lying along the line $a+b=1$. In more detail, the function
$f(a,b,c)$ has a directed gradient in the direction
$\frac{1}{\sqrt{2}}(1,-1,0)$ of
\begin{equation}
  \nabla f(a,b,c) . \frac{1}{\sqrt{2}}(1,-1,0)
   = V^2 \frac{1}{\sqrt{2}} \log \frac{b}{a}
\end{equation}
using Eq. \ref{eq_square_functions}. The rate of change of $f$
with respect to the only remaining variable $a$ is given by
\begin{equation}
  \frac{d f}{d a} = \sqrt{2} \nabla f . \frac{1}{\sqrt{2}}(1,-1,0).
\end{equation}
Altogether, at points where $(a,b,c)=(a,1-a,0)$ this gives a
directed gradient of
\begin{equation}       \label{eq_directed_grad}
  \frac{d f}{d a}
   = V^2  \log \frac{1-a}{a}
\end{equation}
which is optimized at $(a,b,c)=(\frac{1}{2},\frac{1}{2},0)$. An
optimization over all three isomorphic constraints leads to the
same outcomes as obtained previously in Eq. \ref{eq_F_Opt} with
the same result. This completes the second optimization analysis
and as promised, it is consistent with the results of the first.

The same is not true of the third optimization approach which
produces results inconsistent with the first two.  The reason we
present this method is that it is in common use in game theory.
The third optimization method commences by noting that the
probability space of the square is complete in that it already
``contains" all of probability spaces of the triangle and of the
coin. This allows a square probability space to mimic a coin
probability space by simply taking the limit
$(c,d)\rightarrow(0,0)$.  Similarly, the square mimics the
triangle through the limit $d\rightarrow 0$. In turn, this means
that an optimization over the space of the square is effectively
an optimization over every choice of space within the square.
Specifically, game theory discards constraints to model the
choice between contained probability spaces. This optimization
over the points of the square has already been completed above.
When optimizing the function $f$ over the unconstrained points
corresponding to the square, the maximum value is $f=\log(4)/36$
at
$(a,b,c,d)=(\frac{1}{4},\frac{1}{4},\frac{1}{4},\frac{1}{4})$,
and according to game theory, this is the best outcome when
players have a choice between the coin, the triangle, or the
square.

The optimum result obtained by the third optimization method,
that used by game theory, conflicts with those found by the
previous two methods as commonly used in probability theory. The
difference arises as game theory models a choice between
probability spaces by making players uncertain about the values
of their probability parameters within any probability space.
Consequently, their probability parameters are always subject to
infinitesimal fluctuations, i.e. $c>0^+$ or $d>0^+$ always.
These fluctuations alter the dimensions of the space which
impacts on the calculation of the volume $V$ and alters the
calculated gradient of the entropy. Game theory eschews the role
of isomorphism constraints within probability spaces on the
grounds that any such constraints restrict player uncertainty
and hence their ability to choose between different probability
spaces. The probability parameter fluctuations mean that players
have access to all possible probability dimensions at all times
so a single mixed space is the appropriate way to model the
choice between contained probability spaces. In contrast,
probability theory holds that the choice between probability
spaces introduces player uncertainty about which space to use,
but specifically does not introduce uncertainty into the
parameters within any individual probability space.  As a
result, probability theory employs isomorphic constraints to
ensure that the properties of each embedded probability space
within the mixed space are unchanged.

The upshot is that a game theorist cannot evaluate the Entropy
(or uncertainty) gradient of a coin toss while considering
alternate die because uncertainty about which dice is used
bleeds into the Entropy calculation. However, the probability
theorist will distinguish between their uncertainty about which
face of the coin will appear and their uncertainty about which
dice is being used.

\subsection{Alternate coin probability spaces}

The preceding section has shown the importance of using
isomorphism constraints to preserve the properties of the coin
probability space ${\cal P}_{\rm coin}$ when embedded within
larger spaces.  However, isomorphism constraints must also be
used in the very definition of a probability space.  If a
probability space is to be defined to match some physical
apparatus, then a structure preserving isomorphic mapping must
be established between the physical apparatus and the
probability space.  We illustrate this now by adopting several
different probability spaces for a coin.

In the preceding sections, we have the physical coin as shown in
Fig. \ref{f_alternate_dice} and its corresponding probability
space as defined in Eq. \ref{eq_coin_tri_sq}.  To reiterate,
\begin{equation}
   {\cal P}_{\rm coin} = \big\{ x\in\{A,B\},\{a,b\} \big\}.
\end{equation}
After taking account of the normalization constraint $b=1-a$,
the gradient operator in this space is
\begin{equation}
   \nabla = \hat{a} \frac{\partial}{\partial a}.
\end{equation}
If we define a payoff via the random variable $\Pi(A)=0$ and
$\Pi(B)=1$, then a gradient optimization gives
\begin{eqnarray}
  \nabla \langle\Pi\rangle &=& \nabla P(B) \nonumber \\
                           &=& -\hat{a}
\end{eqnarray}
indicating that expected payoffs are maximized by setting $a=0$
as expected.

There are many very different formulations possible for the
probability space of a simple two sided coin, and these are
considered to be functionally identical only after the
appropriate structure-preserving isomorphisms have been defined.
Every alternative introduces a different parameterization which
alters dimensionality and gradient operators and modifies the
optimization algorithm. We illustrate this now.

Our coin could be optimized using a probability measure space
${\cal P}_{\rm coin}^2$ involving two uncorrelated coins, namely
\begin{equation}               \label{eq_alternate_top}
   {\cal P}_{\rm coin}^2 = \big\{ (x,y) \in\{(0,0),(0,1),(1,0),(1,1)\},
     \{(1-p)(1-q),(1-p)q,p(1-q),pq\} \big\}.
\end{equation}
An isomorphism can be defined by mapping event $A$ onto the
event set $(x,y)\in\{(0,0),(1,1)\}$ and $B$ onto
$(x,y)\in\{(0,1),(1,0)\}$.  In this space, the gradient operator
is
\begin{equation}
   \nabla = \hat{p} \frac{\partial}{\partial p}
            + \hat{q} \frac{\partial}{\partial q}
\end{equation}
and a gradient optimization of the expected payoff gives
\begin{eqnarray}
  \nabla \langle\Pi\rangle &=& \nabla P(B) \nonumber \\
                           &=& \hat{p} (1-2q) + \hat{q} (1-2p).
\end{eqnarray}
This shows that when $q<\frac{1}{2}$ then payoffs are maximized
by setting $p=1$ and conversely, when $p<\frac{1}{2}$ then
payoffs are maximized by setting $q=1$.

Alternatively, the binary decision could be optimized using a
continuously parameterized probability measure space ${\cal
P}_{\rm coin}^3$. In this space, the choices $A$ and $B$ might
be determined using a continuously distributed variable
$u\in(-\infty,\infty)$ possessing a normally distributed
probability distribution
\begin{equation}
    P(u) = \frac{1}{\sqrt{2\pi}\sigma}
             e^{-\frac{1}{2}\frac{(u-\bar{u})^2}{\sigma^2}},
\end{equation}
with mean $\bar{u}$, standard deviation $\sigma$, and variance
$\sigma^2$.  We introduce a new parameter, $p$, so outcome $A$
occurs with probability
\begin{equation}
    P(A) = \frac{1}{\sqrt{2\pi}\sigma}
             \int_{-\infty}^{p} du\;
             e^{-\frac{1}{2}\frac{(u-\bar{u})^2}{\sigma^2}},
\end{equation}
while outcome $B$ occurs with probability
\begin{equation}
    P(B) = \frac{1}{\sqrt{2\pi}\sigma}
             \int_{p}^{\infty} du\;
             e^{-\frac{1}{2}\frac{(u-\bar{u})^2}{\sigma^2}}.
\end{equation}
This space has only one probability parameter $p$ so the
gradient operator is
\begin{equation}
   \nabla = \hat{p} \frac{\partial}{\partial p},
\end{equation}
and optimizing the expected payoff gives
\begin{eqnarray}
    \nabla \langle\Pi\rangle
     &=& \nabla  \frac{1}{\sqrt{2\pi}\sigma}
         \int_{p}^{\infty} du\;
             e^{-\frac{1}{2}\frac{(u-\bar{u})^2}{\sigma^2}}  \nonumber \\
     &=& -\nabla F(p),
\end{eqnarray}
where $F(p)$ is the cumulative normal distribution. As the
cumulative normal distribution is monotonically increasing,
$\nabla F(p)>0$, so the expected payoff is maximized by setting
$p\rightarrow -\infty$ giving $P(B)=1$ as expected.

For a more extreme alternative, consider a quantum probability
measure space ${\cal P}_{\rm coin}^4$ in which event $A$
corresponds to a measurement finding a two-state quantum system
in its ground state, and event $B$ occurs when the measurement
finds the system in its excited state. Writing the quantum
system state as
\begin{equation}
    |\Psi\rangle = \left[
     \begin{array}{c}
       a \\
        \\
       b \\
     \end{array}
    \right],
\end{equation}
where $a$ and $b$ are complex numbers satisfying
$|a|^2+|b|^2=1$, then we have $P(A)=|a|^2$ and $P(B)=|b|^2$. In
this space, the payoff is an operator
\begin{equation}
    \Pi = \left[
     \begin{array}{cc}
       0 &  0  \\
          &    \\
       0  & 1  \\
     \end{array}
    \right],
\end{equation}
giving the expected payoff as
\begin{eqnarray}
  \langle\Pi\rangle
     &=& \langle \Psi| \Pi |\Psi\rangle \nonumber \\
     &=& |b|^2  \nonumber \\
     &=& r^2,
\end{eqnarray}
where in the last line we write $b=re^{i\theta}$ with real
$0\leq r\leq 1$ and $0\leq\theta<2\pi$. Here, the expected
payoff depends only on the single real variable $r$ so
optimization is via the gradient operator
\begin{equation}
   \nabla = \hat{r} \frac{\partial }{\partial r}
\end{equation}
giving
\begin{equation}
    \nabla \langle\Pi\rangle = 2r.
\end{equation}
As required, maximization requires setting $r=1$, with $\theta$
arbitrary.

For a last example, consider a probability space ${\cal P}_{\rm
coin}^5$ which selects a number $u$ in the Cantor set ${\cal C}$
with uniform probability $P(u)$ such that when $u\leq p$ then
event $A$ occurs while when $p<u$ then event $B$ occurs. The
Cantor set ${\cal C}$ is interesting as it has an uncountably
infinite number of members and yet has measure zero
\cite{Burk_1998}. In this space, the expected payoff is
\begin{eqnarray}        \label{eq_alternate_bottom}
  \langle\Pi\rangle
     &=& \sum_{u\in{\cal C}} P(u) \Pi(u) \nonumber \\
     &=& \sum_{u> p \in{\cal C}} P(u)  \nonumber \\
     &=&  1-C(p),
\end{eqnarray}
where $C(p)$ is the cumulative probability distribution termed
the Cantor function.  Interestingly, the Cantor function is an
example of a ``Devil's staircase", a function which is
continuous but not absolutely continuous everywhere, and is
differentiable with derivative zero almost everywhere, and which
maps the measure zero Cantor set continuously onto the measure
one set $[0,1]$ \cite{Burk_1998}. As with the normal
distribution example above, the Cantor function is nondecreasing
allowing an intuitive maximization of the expected payoff via
the gradient operator
\begin{equation}
    \nabla = \frac{\partial }{\partial p}
\end{equation}
giving
\begin{equation}
    \nabla \langle\Pi\rangle = -\frac{dC(p)}{dp}.
\end{equation}
As the cumulative normal distribution is nondecreasing, we have
$\frac{dC(p)}{dp}\geq 0$ so the expected payoff is maximized by
setting $p=0$. This intuitive ansatz suffices for our purposes
here.

Lastly, the player is of course, not restricted to using only
simple probability measure spaces, and more complicated spaces
can be considered. In fact, players will most likely use a
pseudo-random number generator consisting of the correlated
dynamical interactions of some millions (or more) of electronic
components in a computer.  It is only the correlations of these
millions of variables that allows a dimensionality reduction to
the few variables required to model the player's chosen
probability space.  Isomorphisms underlie the dimensionality
reductions of random number generators.

To summarize, optimizing an expected payoff first requires the
adoption of a suitable probability measure space, and it is only
the adoption of such a space that permits the definition of
gradient operators and the expected payoff functions allowing
the optimization to be completed.  These steps involve
establishing an isomorphic mapping from the physically modeled
space to the probability space which is property conserving.  Of
course, should the probability space then be embedded within any
other probability space, these properties must still be
conserved, and this will require additional isomorphic
constraints.

\begin{figure}[htb]
\centering
\includegraphics[width=0.8\columnwidth,clip]{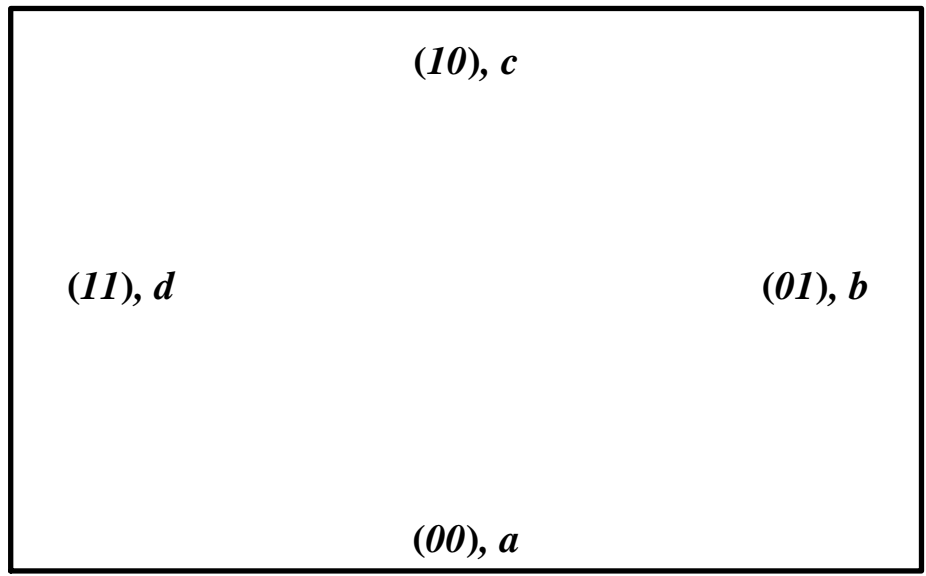}
\caption[A four-sided square probability space]{\em A four-sided square probability space where joint variables
$x$ and $y$ take values $(x,y)\in\{(0,0),(0,1),(1,0),(1,1)\}$ with respective
probabilities $(a,b,c,d)$.
 \label{f_square_joint_xy_var}}
\end{figure}

\subsection{Joint probability space optimization}

We will briefly now examine isomorphisms between the joint
probability spaces of two arbitrarily correlated random
variables.  In particular, we consider two random variables
$x,y$ as appear on the square dice of Fig.
\ref{f_square_joint_xy_var} with probability space
\begin{eqnarray}
  {\cal P}_{\rm square}
    &=& \big\{ (x,y)\in\{(0,0),(0,1),(1,0),(1,1)\}, \{a,b,c,d\} \big\}.
\end{eqnarray}
The correlation between the $x$ and $y$ variables is
\begin{eqnarray}
 \rho_{xy}
    &=& \frac{\langle xy\rangle-\langle x\rangle\langle y\rangle}{\sigma_x \sigma_y} \nonumber \\
    &=& \frac{ad-bc}{\sqrt{(c+d)(a+b)(b+d)(a+c)}}.
\end{eqnarray}
Here, $\sigma_x$ and $\sigma_y$ are the respective standard
deviations of the $x$ and $y$ variables.

The space ${\cal P}_{\rm square}$ of course contains many
embedded or contained spaces.  We will separately consider the
case where $x$ and $y$ are perfectly correlated, and where they
are independent. As noted previously, there are two distinct
ways for these spaces to be contained within ${\cal P}_{\rm
square}$, namely using isomorphism constraints or using limit
processes.  These two ways give the respective definitions for
the perfectly correlated case
\begin{eqnarray}
  {\cal P}_{\rm corr}
    &=& \left. \left\{ (x,y)\in\{(0,0),(0,1),(1,0),(1,1)\}, \{a,b,c,d\} \right\}\right|_{b=c=0}  \nonumber \\
  {\cal P}'_{\rm corr}
    &=& \lim_{(bc)\rightarrow(00)} \left\{ (x,y)\in\{(0,0),(0,1),(1,0),(1,1)\},  \{a,b,c,d\} \right\}
\end{eqnarray}
and for the independent case
\begin{eqnarray}
  {\cal P}_{\rm ind}
    &=& \left. \left\{ (x,y)\in\{(0,0),(0,1),(1,0),(1,1)\}, \{a,b,c,d\} \right\}\right|_{ad=bc} \nonumber \\
  {\cal P}'_{\rm ind}
    &=& \lim_{ad\rightarrow bc} \left\{ (x,y)\in\{(0,0),(0,1),(1,0),(1,1)\}, \{a,b,c,d\} \right\}.
\end{eqnarray}
Here, all spaces satisfy the normalization constraint
$a+b+c+d=1$, which we typically resolve using $d=1-a-b-c$. The
gradient operator in the probability space of the square dice
with probability parameters $(a,b,c)$ is
\begin{equation}
 \nabla = \hat{a} \frac{\partial }{\partial a} +
             \hat{b} \frac{\partial }{\partial b} +
             \hat{c} \frac{\partial }{\partial c},
\end{equation}
where a hat indicates a unit vector in the indicated direction.
Evaluating any function dependent on a gradient or completing an
optimization task using either isomorphic constraints or limit
processes can naturally result in different outcomes as we now
illustrate.

\subsubsection{Perfectly correlated probability spaces}

We first consider the case where the $x$ and $y$ variables are
perfectly correlated in the spaces ${\cal P}_{\rm corr}$ with
isomorphism constraints or ${\cal P}'_{\rm corr}$ using limit
processes.

The maximum achievable joint entropy \cite{Georgii_2008} for our
two perfectly correlated variables obviously occurs at the point
where they are equiprobable.  This can be found by evaluating
the gradient of the joint entropy function
\begin{eqnarray}
 E_{xy}(a,b,c)
   &=& -\sum_{xy} P_{xy} \log P_{xy} \\
   &=& - a \log a - b \log b - c \log c -  (1-a-b-c) \log(1-a-b-c) \nonumber
\end{eqnarray}
giving respective gradients in the ${\cal P}_{\rm corr}$ and
${\cal P}'_{\rm corr}$ spaces of
\begin{eqnarray}      \label{eq_entropy_max}
 \nabla E_{xy}|_{b=c=0}
   &=& - \hat{a} \log \left( \frac{a}{1-a} \right) \nonumber \\
 \nabla E_{xy}
   &=& - \hat{a} \log \left( \frac{a}{1-a-b-c} \right)
       - \hat{b} \log \left( \frac{b}{1-a-b-c} \right)
       - \hat{c} \log \left( \frac{c}{1-a-b-c} \right) \nonumber \\
 \lim_{(bc)\rightarrow (00)} \nabla E_{xy} &=& {\rm undefined}.
\end{eqnarray}
Equating these gradients to zero locates the maximum at
$(a,b,c)=(\frac{1}{2},0,0)$ in ${\cal P}_{\rm corr}$ and at
$(a,b,c)=(\frac{1}{4},\frac{1}{4},\frac{1}{4})$ in ${\cal
P}'_{\rm corr}$.

The Fisher Information is defined in terms of probability space
gradients as the amount of information obtained about a
probability parameter from observing any event
\cite{Georgii_2008}. Writing $(a,b,c)=(p_1,p_2,p_3)$, the Fisher
Information is a matrix with elements $i,j\in\{1,2,3\}$ with
\begin{eqnarray}            \label{eq_Fisher_3d}
  F_{ij} &=&  \sum_{xy} P_{xy}
     \left(\frac{\partial}{\partial p_i} \log P_{xy}\right)
      \left(\frac{\partial}{\partial p_j} \log P_{xy}\right).
\end{eqnarray}
When isomorphically constrained in the space ${\cal P}_{\rm
corr}$, the Fisher Information is $F_{ij}|_{b=c=0}$ with the
only nonzero term being
\begin{eqnarray}                \label{eq_Fisher_Inf}
 F_{11} &=& (1-a) \left[ \hat{a} \frac{\partial }{\partial a} \log (1-a) \right]^2
     + a \left[ \hat{a} \frac{\partial }{\partial a} \log a \right]^2 \nonumber \\
     &=& \frac{1}{a(1-a)}
\end{eqnarray}
This means that the smaller the Variance the more the
information obtained about $a$. In the unconstrained space
${\cal P}'_{\rm corr}$, the Fisher Information is a very
different, $3\times 3$ matrix.

Probability parameter gradients also allow estimation of
probability parameters by locating points where the Log
Likelihood function is maximized $\nabla \log L=0$
\cite{Georgii_2008}. This evaluation takes very different forms
in the isomorphically constrained space ${\cal P}_{\rm corr}$
and the unconstrained space ${\cal P}'_{\rm corr}$. The
likelihood function estimates probability parameters from the
observation of $n$ trials with $n_a$ appearances of event
$(x,y)=(0,0)$, $n_b$ appearances of event $(x,y)=(0,1)$, $n_c$
appearances of event $(x,y)=(1,0)$, and $n_d$ appearances of
event $(x,y)=(1,1)$.  We have $n_a+n_b+n_c+n_d=n$, giving the
Likelihood function
\begin{equation}
 L = f(n_a,n_b,n_c,n)  a^{n_a} b^{n_b} c^{n_c} (1-a-b-c)^{n-n_a-n_b-n_c}
\end{equation}
where $f(n_a,n_b,n_c,n)$ gives the number of combinations. The
optimization proceeds by evaluating the gradient of the Log
Likelihood function. When isomorphically constrained in the
space ${\cal P}_{\rm corr}$, the gradient of the Log Likelihood
function is
\begin{equation}
 \nabla \log L|_{b=c=0}
     =  \hat{a} \left[ \frac{n_a}{a} - \frac{n-n_a}{1-a}  \right],
\end{equation}
which equated to zero gives the optimal estimate at $a=n_a/n$
and $n_b=n_c=0$ as expected.  Conversely, when unconstrained in
the space ${\cal P}'_{\rm corr}$, the gradient of the Log
Likelihood function evaluates as
\begin{eqnarray}              \label{eq_max_likelihood_3d}
 \nabla \log L  &=&  \hat{a} \left[ \frac{n_a}{a} - \frac{n-n_a-n_b-n_c}{1-a-b-c}  \right]
      + \hat{b} \left[ \frac{n_b}{b} - \frac{n-n_a-n_b-n_c}{1-a-b-c}  \right] \nonumber \\
  && \hspace{2cm} + \hat{c} \left[ \frac{n_c}{c} - \frac{n-n_a-n_b-n_c}{1-a-b-c}  \right].
\end{eqnarray}
This is obviously a very different result.  However, in our case
the same estimated outcomes can be achieved in both spaces. For
example, if an observation of $n$ trials shows $n_a$ instances
of $(x,y)=(0,0)$ and $n-n_a$ instances of $(x,y)=(1,1)$ then
both constrained and unconstrained approaches give the best
estimates of the probability parameters of
$(a,b,c,d)=(\frac{n_a}{n},0,0,1-\frac{n_a}{n})$.

Finally, when $x$ and $y$ are perfectly correlated it is
necessarily the case that expectations satisfy $\langle
x\rangle-\langle y\rangle=0$, that variances satisfy
$V(x)-V(y)=0$, that the joint entropy is equal to the entropy of
each variable so $E_{xy}-E_x=0$, and that finally, the
correlation between these variables satisfies $\rho_{xy}-1=0$.
In the unconstrained probability space ${\cal P}'_{\rm corr}$,
the expectation, variance, and entropy relations of interest
evaluate as
\begin{eqnarray}        \label{eq_expectation_relations_3d}
 \langle x\rangle - \langle y\rangle  &=& c-b  \nonumber  \\
 V(x) - V(y)   &=& (c-b)(a-d)    \\
 E_x &=& -\left[ (a+b) \log(a+b) +  (1-a-b) \log(1-a-b) \right] \nonumber \\
 E_{xy}
   &=& -\left[ a \log a + b \log b + c \log c+  (1-a-b-c) \log(1-a-b-c)\right]. \nonumber
\end{eqnarray}
These functions lead to gradient relations in the ${\cal P}_{\rm
corr}$ and ${\cal P}'_{\rm corr}$ spaces of:
\begin{eqnarray}
 \nabla\left[\langle x\rangle-\langle y\rangle\right]|_{b=c=0} &=& 0 \nonumber \\
  \lim_{(bc)\rightarrow (00)}\nabla\left[\langle x\rangle-\langle y\rangle\right] &=&
       -\hat{b}+\hat{c}  \nonumber \\
 \nabla\left[V(x)-V(y)\right]|_{b=c=0} &=& 0 \nonumber \\
 \lim_{(bc)\rightarrow (00)}\nabla\left[V(x)-V(y)\right] &=&
      (1-2a)\hat{b}-(1-2a)\hat{c} \nonumber \\
 \nabla\left[E_{xy}-E_{x}\right]|_{b=c=0} &=& 0 \nonumber \\
 \lim_{(bc)\rightarrow (00)}\nabla\left[E_{xy}-E_{x}\right] &\neq &
      {\rm undefined} \nonumber \\
 \nabla \rho_{xy}|_{b=c=0} &=& 0 \nonumber \\
 \nabla \rho_{xy} &\neq & 0.
\end{eqnarray}
Obviously, taking the limit $(b,c)\rightarrow (0,0)$ does not
reduce the limit equations to the required relations.

\subsubsection{Independent probability spaces}

We next consider the case where the $x$ and $y$ variables are
independent using the spaces ${\cal P}_{\rm ind}$ with
isomorphism constraints or ${\cal P}'_{\rm ind}$ with limit
processes.

When random variables are independent, then their joint
probability distribution is separable for every allowable
probability parameter of ${\cal P}_{\rm ind}$ or ${\cal P}'_{\rm
ind}$.  This means the gradient of this separability property
must be invariant across these probability spaces. That is, we
must have $P_{xy}=P_xP_y$ and hence
$\nabla\left[P_{xy}-P_xP_y\right]=0$.  Similarly, separability
requires we also satisfy $\nabla\left[ \langle xy\rangle-\langle
x\rangle \langle y\rangle\right]=0$. Further, every independent
space must have conditional probabilities equal to marginal
probabilities and so satisfy $\nabla\left[P_{x|y}-P_x\right]=0$.
Finally, two independent variables have joint entropy equal to
the sum of the individual entropies so every independent space
must satisfy $\nabla\left[E_{xy}-E_{x}-E_{y}\right]=0$. These
relations evaluate differently in either ${\cal P}_{\rm ind}$
with isomorphism constraints or ${\cal P}'_{\rm ind}$ with limit
processes. For the square die under consideration, we have
probabilities and expectations of
\begin{eqnarray}      \label{eq_ind_space}
  P_{xy}(00)-P_x(0) &=& ad - bc \nonumber \\
  \langle xy \rangle - \langle x \rangle \langle y \rangle &=&
      ad - bc \nonumber \\
  P_{x|y}(0|0)-P_x(0) &=& \frac{ad-bc}{a+c},
\end{eqnarray}
and entropies of
\begin{eqnarray}
  E_x &=& -(a+b) \log(a+b) - (1-a-b)\log(1-a-b) \nonumber \\
  E_y &=& -(a+c) \log(a+c) - (1-a-c)\log(1-a-c) \nonumber \\
  E_{xy} &=& -a \log a - b \log b - c \log c - d \log d.
\end{eqnarray}

The resulting gradients are
\begin{eqnarray}
 \nabla \left[P_{xy}(00)- P_x(0)P_y(0)\right]|_{ad=bc} &=& 0  \nonumber \\
 \lim_{ad\rightarrow bc} \nabla \left[P_{xy}(00)- P_x(0)P_y(0)\right]
     &=& \lim_{ad\rightarrow bc} \nabla (ad-bc) \neq 0 \nonumber \\
 \nabla \left[\langle xy\rangle-\langle x\rangle\langle y\rangle\right]|_{ad=bc}
   &=& 0  \nonumber \\
 \lim_{ad\rightarrow bc} \nabla \left[\langle xy\rangle-\langle x\rangle\langle y\rangle\right]
     &=& \lim_{ad\rightarrow bc}\nabla (ad-bc) \;\neq\; 0 \nonumber \\
 \nabla \left[P_{x|y}(0|0)-P_x(0)\right]|_{ad=bc} &=& 0  \nonumber  \\
 \lim_{ad\rightarrow bc} \nabla \left[P_{x|y}(0|0)-P_x(0)\right]
     &=& \lim_{ad\rightarrow bc}\nabla \left[\frac{ad-bc}{a+c} \right] \;\neq\; 0 \nonumber \\
 \nabla \left[E_{xy}-E_x-E_y\right]|_{ad=bc} &=& 0 \nonumber  \\
 \lim_{ad\rightarrow bc} \nabla \left[E_{xy}-E_x-E_y\right]
     &=&  \\
 && \hspace{-4.5cm}\lim_{ad\rightarrow bc}  \nabla \left\{
    a \log \left[\frac{d}{a}\frac{a-ad+bc}{d-ad+bc}\right] +
    b \log \left[\frac{d}{b}\frac{b+ad-bc}{d-ad+bc}\right] + \right.   \nonumber \\
 && \hspace{-4cm} \left. c \log \left[\frac{d}{c}\frac{c+ad-bc}{d-ad+bc}\right] +
    \log \left[\frac{d-ad+bc}{d}\right] \right\} \;\neq\; 0.\nonumber
\end{eqnarray}

\subsection{Entropy maximization}

The joint entropy $E_{xy}$ reflects the uncertainty between the
$x$ and $y$ variables.  According to probability theory, this
uncertainty does not include any uncertainty about which
probability space is being chosen, while conversely, according
to game theory the uncertainty between these variables increases
when it includes additional uncertainty about which probability
space is being chosen.

\begin{figure}[htb]
\centering
\includegraphics[width=0.8\columnwidth,clip]{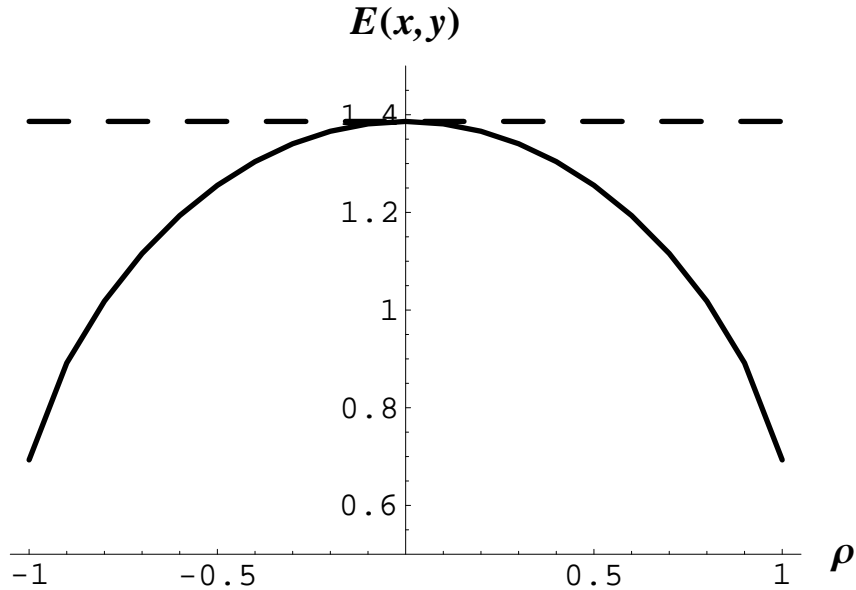}
\caption[Maximizing joint entropy]{\em  Maximizing the
joint entropy of two correlated random
variables $x,y\in\{0,1\}$.   Without isomorphism constraints,
the maximum entropy is equal to $2\log 2$ (dashed line). However,
when subject to
isomorphism constraints, the simplex will exactly reproduce the
different maximum entropy states of each of its embedded probability
spaces (solid line).
 \label{f_entropy_maximization}}
\end{figure}

We now present a numerical investigation of how to determine the
maximum joint entropy $E_{xy}$ of embedded probability states
featuring possibly correlated variables $x$ and $y$ as depicted
in Fig. \ref{f_square_joint_xy_var}. The joint entropy is
\begin{equation}
 E_{xy}(a,b,c) = -\sum_{xy} P_{xy} \log P_{xy}.
\end{equation}
Using isomorphism constraints, the maximization problem is
\begin{equation}
 \max \left.E_{xy}\right|_{\rho_{xy}=\bar{\rho}}
\end{equation}
for all $\bar{\rho}\in[-1,1]$. Here, the correlation function
between $x$ and $y$ is given by the later Eq.
\ref{eq_rho_correlation}.  This equation can be inverted to
solve for the variable $r$ as a function of $p$, $q$, and the
constant correlation $\bar{\rho}$, and the result
$r_+(p,q,\bar{\rho})$ is given in Eq. \ref{eq_r_plus_minus}. A
numerical optimization then generates the maximum entropy value
for every correlation state $\bar{\rho}$ with the results shown
in Fig. \ref{f_entropy_maximization}.  As expected, the presence
of isomorphism constraints ensures the entropy ranges from a
minimum of $\log 2$ up to a maximum of $2\log 2$.

In contrast, when the joint entropy is maximized over the entire
space using the techniques of game theory, then a single maximum
outcome is achieved giving the maximum entropy in the absence of
isomorphism constraints.  This line is also shown in Fig.
\ref{f_entropy_maximization} as the constant at $E_{xy,{\rm
max}}=2 \log 2$.

\subsection{Continuous bivariate Normal spaces}

The above results are general.  When source probability spaces
are embedded within target probability spaces, then the use of
isomorphic mapping constraints will preserve all properties of
the embedded spaces.  Conversely, when constraints are not used
then some of the properties of the embedded spaces will not be
preserved in general. We illustrate this now using normally
distributed continuous random variables.

Consider two normally distributed continuous independent random
variables $x$ and $y$ with $x,y\in(-\infty,\infty)$. When
independent, these variables have a joint probability
distribution $P_{xy}$ which is continuous and differentiable in
six variables, $P_{xy}(x,\mu_x,\sigma_x,y,\mu_y,\sigma_y)$ where
the respective means are $\mu_x$ and $\mu_y$ and the variances
are $\sigma_x^2$ and $\sigma_y^2$.  The marginal distributions
are $P_{x}(x,\mu_x,\sigma_x)$ and $P_{y}(y,\mu_y,\sigma_y)$. In
particular, we have
\begin{eqnarray}     \label{eq_bivariate_normal_ind}
  P_{xy} &=& \frac{1}{2\pi\sigma_x\sigma_y}
    e^{-\frac{1}{2}
    \left[ \frac{(x-\mu_x)^2}{\sigma_x^2} +
      \frac{(y-\mu_y)^2}{\sigma_y^2} \right] } \nonumber \\
  P_{x} &=& \frac{1}{\sqrt{2\pi}\sigma_x}
    e^{-\frac{1}{2}
    \frac{(x-\mu_x)^2}{\sigma_x^2} } \nonumber \\
  P_{y} &=& \frac{1}{\sqrt{2\pi}\sigma_y}
    e^{-\frac{1}{2}
    \frac{(y-\mu_y)^2}{\sigma_y^2} }.
\end{eqnarray}
The conditional distribution for $x$ given some value of $y$ is
\begin{equation}
  P_{x|y} = \frac{1}{\sqrt{2\pi}\sigma_x}
    e^{-\frac{1}{2}
     \frac{(x-\mu_x)^2}{\sigma_x^2} }.
\end{equation}

These independent joint distributions can now be embedded into
an enlarged distribution representing two potentially correlated
normally distributed variables $x$ and $y$.  This enlarged
distribution $P'_{xy}(x,\mu_x,\sigma_x,y,\mu_y,\sigma_y,\rho)$
differs from $P_{xy}$ in its dependence on the correlation
parameter $\rho_{xy}=\rho$ with $\rho\in(-1,1)$.  This
distribution is continuous and differentiable in seven
variables. The joint distribution is
\begin{eqnarray}            \label{eq_bivariate_normal_corr}
  P'_{xy} &=& \frac{1}{2\pi\sigma_x\sigma_y \sqrt{1-\rho^2}}
   e^{-\frac{1}{2(1-\rho^2)}
    \left[ \frac{(x-\mu_x)^2}{\sigma_x^2}
     - \frac{2 \rho (x-\mu_x) (y-\mu_y)}{\sigma_x \sigma_y} +
      \frac{(y-\mu_y)^2}{\sigma_y^2} \right] }.
\end{eqnarray}
The marginal distributions for the correlated case are identical
to those of the independent space so $P'_{x}=P_{x}$ and
$P'_{y}=P_{y}$. The conditional distribution for $x$ given some
value of $y$ is
\begin{equation}
  P'_{x|y} = \frac{1}{\sqrt{2\pi(1-\rho^2)}\sigma_x}
    e^{-\frac{1}{2(1-\rho^2)}
     \frac{(x-\bar{\mu}_x)^2}{\sigma_x^2} },
\end{equation}
where the new conditioned mean is
\begin{equation}
  \bar{\mu}_x=\mu_x+\rho \frac{\sigma_x}{\sigma_y} (y - \mu_y).
\end{equation}

An isomorphic embedding requires that the unit probability
subset of $P_{xy}$ be mapped onto the unit probability subset of
$P'_{xy}$ and this is achieved by imposing an external
constraint that $\rho=0$ in the enlarged space.  Hence, we
expect $\left.P'_{xy}\right|_{\rho=0}=P_{xy}$. It is readily
confirmed that when the isomorphism constraint is imposed on the
enlarged distribution all properties are preserved, while this
is not the case in the absence of the constraint.  The gradient
operator $\nabla$ is now a function of seven variables
\begin{eqnarray}
 \nabla &=& \frac{\partial}{\partial x} \hat{x}
           + \frac{\partial}{\partial y} \hat{y}
           + \frac{\partial}{\partial \mu_x} \hat{\mu}_x
           + \frac{\partial}{\partial \mu_y} \hat{\mu}_y
           + \frac{\partial}{\partial \sigma_x} \hat{\sigma}_x
           + \frac{\partial}{\partial \sigma_y} \hat{\sigma}_y
           + \frac{\partial}{\partial \rho} \hat{\rho}.
\end{eqnarray}
The probability distributions must satisfy a number of gradient
relations, but we have:
\begin{eqnarray}        \label{eq_bivariate_normal_corr_test}
   \left. \nabla \left[P'_{xy}-P'_xP'_y\right]\right|_{\rho=0}
         &=& 0 \nonumber \\
   \lim_{\rho\rightarrow 0} \nabla \left[P'_{xy}-P'_xP'_y\right]
     &=& \hat{\rho} \lim_{\rho\rightarrow 0} \frac{\partial}{\partial \rho} P'_{xy} \neq 0  \nonumber \\
   \left. \nabla\left[P'_{x|y}-P'_x\right]\right|_{\rho=0}  &=& 0 \nonumber \\
   \lim_{\rho\rightarrow 0} \nabla \left[P'_{x|y}-P'_x\right]
      &=& \hat{\rho} \lim_{\rho\rightarrow 0} \frac{\partial}{\partial \rho} P'_{x|y} \neq 0.
\end{eqnarray}
Similarly, the expectations of functions of the $x$ and $y$
variables must also satisfy a number of gradient relations. As
expectations integrate over the $x$ and $y$ variables, the
gradient operator is a function of only five variables now,
\begin{equation}
 \nabla = \frac{\partial}{\partial \mu_x} \hat{\mu}_x
           + \frac{\partial}{\partial \mu_y} \hat{\mu}_y +
            \frac{\partial}{\partial \sigma_x} \hat{\sigma}_x
           + \frac{\partial}{\partial \sigma_y} \hat{\sigma}_y
           + \frac{\partial}{\partial \rho} \hat{\rho}.
\end{equation}
We have
\begin{eqnarray}        \label{eq_bivariate_normal_corr_test2}
   \left. \nabla\left[ \langle xy\rangle'-\langle x\rangle' \langle y\rangle'\right]\right|_{\rho=0}
        &=& 0  \nonumber \\
   \lim_{\rho\rightarrow 0}  \nabla \left[ \langle xy\rangle'-\langle x\rangle' \langle y\rangle'\right]
       &=& \hat{\rho} \lim_{\rho\rightarrow 0} \frac{\partial}{\partial \rho} \langle xy\rangle' \neq 0.\nonumber
\end{eqnarray}

\subsection{Quantum probability spaces}

As noted above, the use of isomorphic mappings to preserve the
properties of probability spaces is general. As a last
illustration, we show the use of isomorphic mappings when
applied to quantum probability spaces.

Suppose a quantum probability space is to be embedded within
another enlarged quantum probability space. (See
\cite{Hayashi_2006} for an overview of quantum information
theory including quantum information geometry.) An $N$ level
quantum system has von Neumann entropy defined as
\begin{equation}
  E_N=-{\rm tr} \hat{R}_N \log \hat{R}_N
\end{equation}
where here $\hat{R}_N$ is the quantum density matrix and ${\rm
tr}$ indicates a trace operation applied to a matrix. Supposing
that matrix $D$ diagonalizes the density matrix so $D\hat{R}_N
D^{\dagger}$ is diagonal, and that its eigenvalues are
$\lambda_i$ for $1\leq i\leq N$, we have
\begin{equation}        \label{eq_quantum_entropy}
  E_N = - \sum_{i=1}^{N} \lambda_i \log \lambda_i.
\end{equation}
The eigenvalue $\lambda_i$ specifies the occupancy probability
of the $i^{\rm th}$ level. Hence, maximizing the $N$-level
system entropy requires that $\lambda_i=1/N$ for all $i$.
Consequently, a two level quantum system maximizes its entropy
$E_2$ when the density matrix is an equiprobable mixture equal
to half of the two level identity matrix, $\hat{R}_2=1/2 I_2$,
while a three level quantum system maximizes its entropy $E_3$
when the density matrix is an equiprobable mixture of
$\hat{R}_3=1/3 I_3$.

Now, if the two level system were isomorphically embedded within
a three level system, then the two level system entropy $E_2$ is
properly maximized only when isomorphism constraints are used to
decouple the third level so that it plays no part in the
optimization. This is achieved by using an isomorphism
constraint $\lambda_3=0$ to decouple and remove the third level
from the system.  That is, the optimization taking account of an
isomorphism constraint $\nabla_3 E_3|_{\lambda_3=0}=0$ will
determine the correct maximum value for $E_2$.  However, a
failure to use an isomorphism constraint will locate an
incorrect maximum point via $\lim_{\lambda_3\rightarrow
0}\nabla_3 E_3$.  We have
\begin{equation}
  \nabla_2 E_2 = \nabla_3 E_3 |_{\lambda_3=0}
               \neq \lim_{\lambda_3 \rightarrow 0} \nabla_3 E_3.
\end{equation}
Isomorphism constraints must be used to properly embed one
quantum probability space within another.

\begin{figure}[htb]
\centering
\includegraphics[width=0.6\columnwidth,clip]{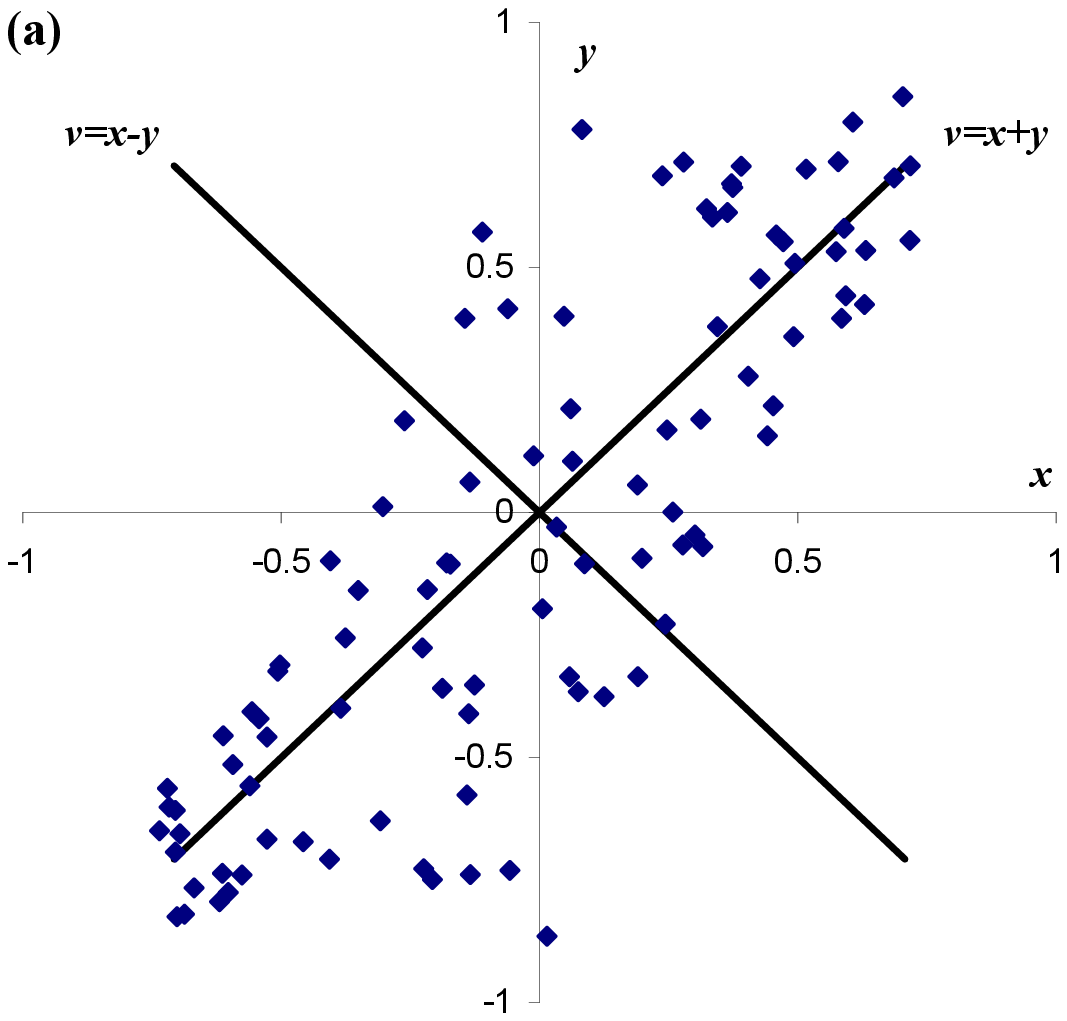}
\includegraphics[width=0.6\columnwidth,clip]{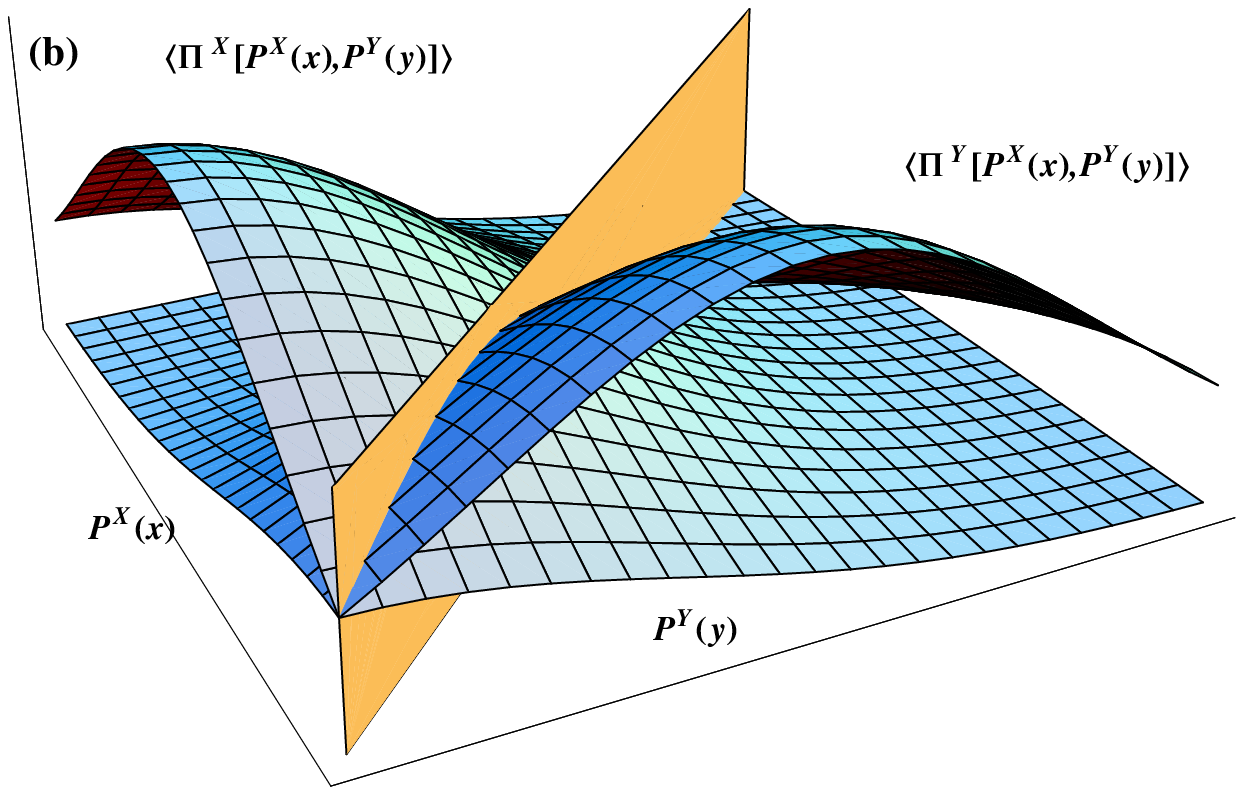}
\caption[Affine transforms of correlated variable]{\em (a)
An affine transformation of correlated variables $x$ and $y$
generates new orthogonal variables $u=x+y$
and $v=x-y$ which are uncorrelated. (b) When $x$ and $y$ are
perfectly correlated, $v=0$ and $u$ is the only free
variable and dimensionality is reduced.  Optimization solutions
must lie on the $u$-axis satisfying the constraint $x=y$.
\label{f_projections}}
\end{figure}

\subsection{Perfect correlation reduces dimensionality}

Standard probability theory holds that when two variables $x$
and $y$ are known to be perfectly correlated, then
$P(x,y)=P(x)P(y|x)=P(x)$.  That is, any optimization which
involves the joint distribution $P(x,y)$ does not involve two
dimensions but only one as $x=y$. Perfect correlation reduces
dimensionality which alters the gradient operators which in turn
can alter optima.

Probability theory takes account of this dimensionality
reduction when using Affine variable transforms. Typical
presentations of probability theory hold that ``any two
real-valued random variables $x$ and $y$ whose mean values and
variances exist may be represented as an Affine transformation
of a pair of uncorrelated random variables"
\cite{Pfeiffer_1990}. Such statements, carelessly interpreted,
would indeed suggest that perfect correlations involve no
reduction in the number of variables.  Writing the respective
mean values as $\langle x\rangle$ and $\langle y\rangle$, and
defining the translated variables
\begin{eqnarray}
   x^* &=& x-\langle x\rangle \nonumber \\
   y^* &=& y-\langle y\rangle,
\end{eqnarray}
then an affine transformation can always be used to define two
new variables
\begin{eqnarray}
   u &=& x^*+y^* \nonumber \\
   v &=& x^*-y^*.
\end{eqnarray}
These variables each have mean zero, $\langle u\rangle=\langle
v\rangle=0$, and are uncorrelated as
\begin{equation}
    {\rm cov}(u,v)=\langle uv\rangle=0.
\end{equation}
The zero covariance results from the orthogonality of the random
variables $u$ and $v$ in a suitable $L^2$ vector space, while
the possibly correlated original variables are generated from
the inverse affine transformation
\begin{eqnarray}
    x &=& \sigma_x x^* + \langle x\rangle \;=\;
               \frac{\sigma_x}{2} (u+v) + \langle x\rangle \nonumber \\
    y &=& \sigma_y y^* + \langle y\rangle \;=\;
               \frac{\sigma_y}{2} (u-v) + \langle y\rangle,
\end{eqnarray}
where here, $\sigma_z$ is the standard deviation of variable
$z\in\{x,y\}$.

If the $x$ and $y$ variables are perfectly correlated, then $v$
is identically zero and $u$ is the only surviving variable.
Perfect correlations reduce the dimensionality of the
optimization space and probability theory preserves the
dimensionality of perfectly correlated variables when using
Affine transforms.  (See Fig. \ref{f_projections}.)

A similar preservation of dimensionality occurs in the Hotelling
transform, a discrete version of the Karhunen-Lo\`{e}ve
transform \cite{Gersho_1991}.  This transform can also be used
to map the probability space of two uncorrelated centered
variables $(u,v)$ into the probability space of two correlated
centered variables $(x,y)$. If the state of correlation between
$x$ and $y$ is $\rho$, then the Hotelling transform is
implemented via
\begin{equation}      \label{eq_mixing_matrix}
   \left[
    \begin{array}{c}
      x   \\
      y
     \end{array}
   \right]
   =   \left[
          \begin{array}{cc}
            1  &   0 \\
            \rho  &   \sqrt{1-\rho^2}
          \end{array}
          \right]\;
   \left[
    \begin{array}{c}
      u   \\
      v
     \end{array}
   \right].
\end{equation}
Then, whenever the $x$ and $y$ variables are not perfectly
correlated both the $(u,v)$ and $(x,y)$ probability spaces are
two dimensional.  However, when $\rho=1$ and $x$ and $y$ are
perfectly correlated, then the mapping matrix becomes singular
and non-invertible ensuring that $x=y=u$ so that the $x$ and $y$
probability space is one dimensional even while the $u$ and $v$
probability space is two dimensional.  Probability theory again
acts to preserve the dimensionality of the joint probability
space of perfectly correlated variables.

\subsection{Example isomorphic functions}

There are different ways to embed a smaller source function
within an enlarged target function which can preserve different
amounts of the structure of the source function within the
target function. Consider for example, mapping a 1-dimensional
function $f(x)$ into a 2-dimensional function $g(x,y)$ along the
line $y=x$ so that $f(x)=g(x,x)$. One way to implement this
assignment is to use limit processes constraining most of the
neighbourhood of $g(x,y)$ in the vicinity of the line $y=x$ to
satisfy
\begin{equation}
     \lim_{y\rightarrow x} g(x,y) = f(x).
\end{equation}
Another way to do this is to ignore the values of $g(x,y)$ away
from the line $y=x$ and simply use externally imposed
constraints forcing the assignment on the line via
\begin{equation}
    g(x,y)|_{y=x} = f(x).
\end{equation}
This approach does not care about values $g(x,y)$ when $x\neq
y$.  The question then is, under what circumstances can
$\lim_{y\rightarrow x} g(x,y)$ or $g(x,y)|_{y=x}$ be used to
examine the properties of $f(x)$.

Hereinafter, for concreteness we will consider the simplified
example functions $f(x)=x^2$ and $g(x,y)=xy$. Each of the
implementations, $\lim_{y\rightarrow x} g(x,y)$ or
$g(x,y)|_{y=x}$, have different domains (dom) in each space, and
hence different integration volume elements ($dv$)
\begin{equation}
  \begin{array}{c|ccc}
                &     \hspace{1cm}f(x)\hspace{1cm}   &   \hspace{1cm}\lim_{y\rightarrow x} g(x,y)\hspace{1cm}     &   \hspace{1cm}g(x,y)|_{y=x}\hspace{1cm}   \\ \hline
  {\rm dom}     &     \Re    &  \Re\times\Re                     &   \Re               \\
     dv         &     dx     &    dx\; dy                        &   dx.                \\
  \end{array}
\end{equation}
The different dimensionalities of the domains impacts on any
attempt to change variables within each space.  The rank of the
change of variable transforms ($A$) and the dimensionality of
the Jacobian matrices ($J$) in each space are
\begin{equation}
  \begin{array}{c|ccc}
                &     \hspace{1cm}f(x)\hspace{1cm}   &   \hspace{1cm}\lim_{y\rightarrow x} g(x,y)\hspace{1cm}     &   \hspace{1cm}g(x,y)|_{y=x}\hspace{1cm}   \\ \hline
  {\rm rank}(A) &     1      &      2                            &    1                \\
  {\rm dim}(J)  &     1      &      2                            &    1.                \\
  \end{array}
\end{equation}
These differences impact on the evaluation of other properties
such as gradients, which should evaluate as
\begin{equation}
      \nabla f(x) = 2x\hat{x}
\end{equation}
where a hatted variable denotes a unit vector in the indicated
direction. In contrast, the gradient evaluated using a limit
assignment gives
\begin{equation}
 \nabla g(x,x) =  \lim_{y\rightarrow x} \nabla g(x,y)
     = x (\hat{x} + \hat{y}),
\end{equation}
which does not satisfy the required relation. Conversely, the
use of an externally imposed constraint ensures
\begin{equation}
 \nabla g(x,y)|_{y=x}  =  \nabla g(x,x) = 2x \hat{x}
\end{equation}
as required.

In summary, the definitions
\begin{equation}
 f(x) = g(x,y)|_{y=x} = \lim_{y\rightarrow x} g(x,y),
\end{equation}
do not generally carry over to the gradient relations, as
\begin{equation}
 \nabla f(x) = \nabla g(x,y)|_{y=x} \neq \lim_{y\rightarrow x} \nabla g(x,y),
\end{equation}
This results as the limit process $f(x) = \lim_{y\rightarrow x}
g(x,y)$ treats the $x$ and $y$ variables as being independent
and simply evaluates desired quantities at points $(x,y)$ lying
on the line $y=x$. In contrast, the constraint
$f(x)=g(x,y)|_{y=x}$ enforces a functional relation between the
$x$ and $y$ variables which preserves all the structures of
$f(x)$ within $g(x,x)$.  It is well understood that any
functional relation between the variables of a function will
impact on the properties of that function.  Such functional
relations must be preserved whenever that function is mapped
into a different space.  The need to take account of such
functional relations is a standard part of routine optimization
techniques such as differentiation via any of the chain rule,
Lagrangian multipliers, or directed vector gradients.

A number of standard techniques exist for evaluating the
gradient $f'(x)$ using the constrained function $g(x,y)|_{y=x}$.
For instance, the chain rule can be applied to the functions
$g(x,y)$ and $y(x)=x$ giving
\begin{eqnarray}                \label{eq_chain_rule}
 f'(x)  &=& \frac{\partial g}{\partial x} +
            \frac{\partial g}{\partial y}\frac{dy}{dx} \nonumber  \\
    & = & 2x \hat{x}.
\end{eqnarray}
Another common alternative is by using Lagrange multipliers in
which $f'(x)=L'(x)$ with
\begin{equation}                    \label{eq_langrange_mult}
  L(x,y,\lambda)= xy - \lambda (y-x)
\end{equation}
and
\begin{eqnarray}
  \frac{\partial L}{\partial x} &=& (y+\lambda) \hat{x} \nonumber \\
  \frac{\partial L}{\partial y} &=& (x-\lambda) \hat{y} \nonumber \\
  \frac{\partial L}{\partial \lambda} &=& (x-y) \hat{\lambda}.
\end{eqnarray}
Equating the last two lines to zero gives the required
constraints $y=x$ and $\lambda=x$ ensuring $f'(x)=L'(x)$. A
final way to perform this constrained optimization is to use
directed vector gradients where
\begin{equation}               \label{eq_directed_vector}
  f'(x)=\lim_{y\rightarrow x}\nabla g(x,y).v.\sqrt{2}
\end{equation}
with $v=(\hat{x}+\hat{y})/\sqrt{2}$.  Here, $v$ is normalized
and the extra factor of $\sqrt{2}$ properly calculates changes
in the $x$ direction.  This gives the magnitude of the gradient
as $f'(x)=2x$ as required.

There are two ways to embed the function $f(x)$ within the
surface $g(x,y)$ using either a limit process or an externally
imposed constraint.  The limit process fails to preserve many of
the properties of the source function within the target
function.  Conversely, the external constraint does ensure that
all source function structures are preserved within the target
function---dimensionality, gradient, and so on. In general, it
is not possible to embed a smaller space within a larger space
and preserve gradients and optimization outcomes without the use
of constraints.  These constraints reflect the use of isomorphic
mappings to preserve the properties of the source space with the
target space \cite{Insall_2009}.

\begin{figure}[htbp]
\centering
\includegraphics[width=0.8\columnwidth,clip]{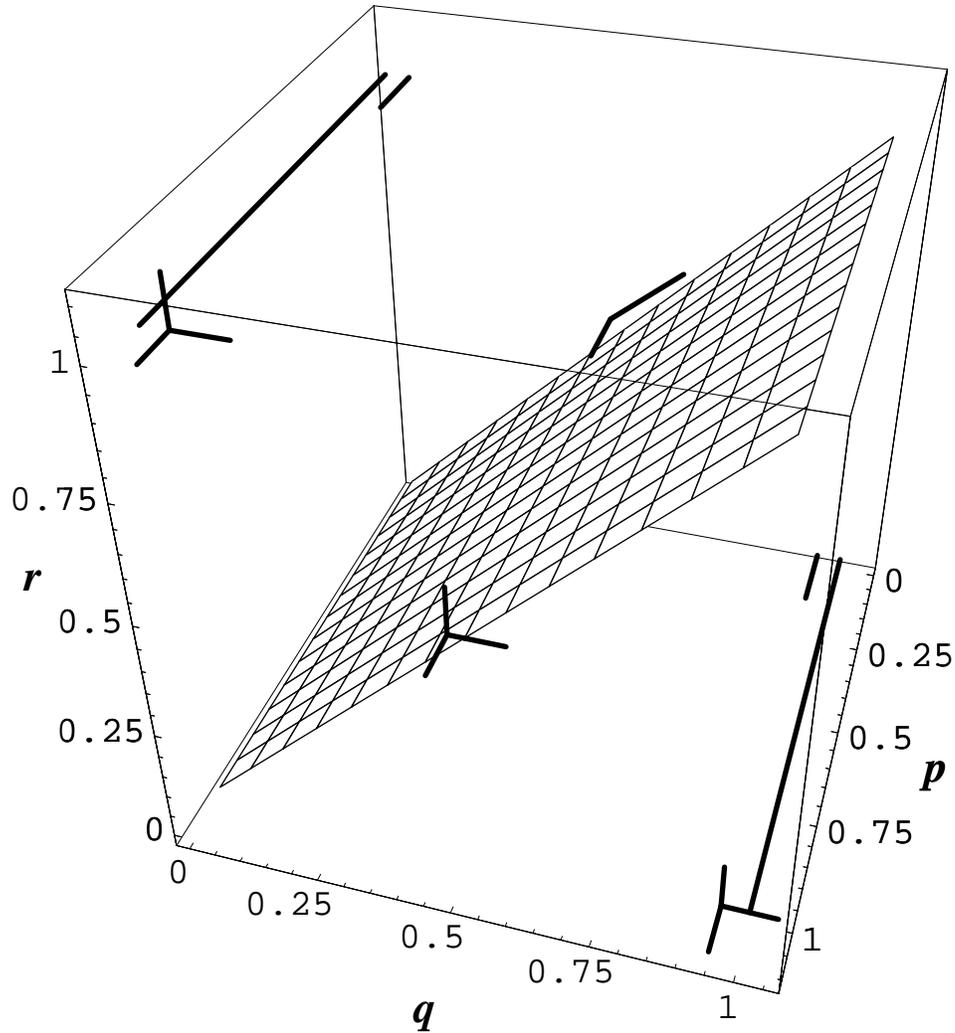}
\caption[Schematic representation of target strategy space]{\em  A schematic representation where a three dimensional
target probability strategy space $(p,q,r)$ embeds
respectively several one dimensional probability spaces associated
with perfectly correlated variables (lines, upper left and lower
right), and a two dimensional probability space associated with
independent variables (plane, middle). An exact isomorphism preserves
the respective original tangent spaces shown via one and two dimensional
axes offset in background.  A weak isomorphism fails to preserve the original
tangent spaces of the source probability distributions and assigns the
three dimensional tangent space of the target space to every embedded
distribution (as shown in foreground slightly
offset from each embedded space).
 \label{f_tangent_spaces}}
\end{figure}

\section{Isomorphisms and Optimization}

There are two approaches to optimization over probability spaces
presented here.  Probability theory uses isomorphic constraints
to exactly preserve the properties of embedded probability
spaces and then compares these exactly calculated values. Game
theory eschews the use of isomorphic constraints and in effect,
argues that any uncertainty about which probability space to
choose bleeds into many calculations within a given space and
alters the calculated outcomes.

When probability spaces are represented as geometries, then it
is expected that at least some of the properties of the
probability space will be rendered in geometric terms.  How
these geometrical properties are preserved when a probability
space is embedded within another is the question.  Probability
theory requires the exact preservation of all properties of
every source space and this is achieved by imposing different
constraints on different points within the target space. Game
theory in contrast, imposes a single target space geometry onto
every source probability space. One way to picture this is shown
in Fig. \ref{f_tangent_spaces}. This figure shows how
probability theory exactly preserves the dimensionality and
tangent spaces of embedded probability spaces, while game theory
overwrites these properties of the embedded spaces with the
corresponding properties of the mixed space.

\begin{figure}[htb]
\centering
\includegraphics[width=0.9\columnwidth,clip]{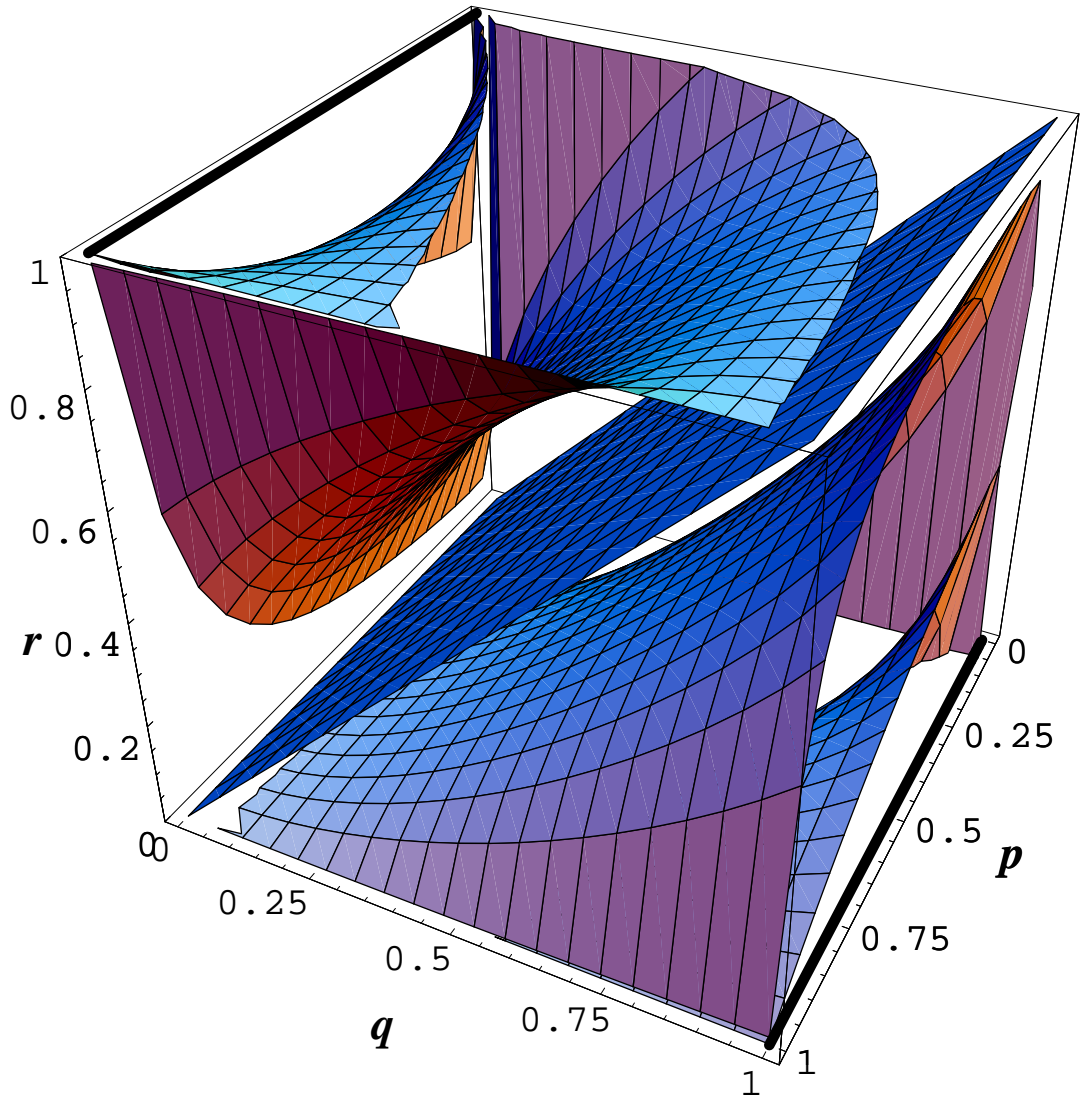}
\caption[Correlation constraints in the strategy space]{\em  Every point within the $(p,q,r)$
probability space shown specifies a particular state of
correlation $\rho_{xy}(p,q,r)$ between the $x$ and $y$ variables.
We show here several lines and surfaces of constant correlation
taking values from top left to bottom right of
$\rho_{xy}=+1,+0.75,+0.25,0,-0.25,-0.75,-1$.  The optimization of
expectations at any point $(p,q,r)$ must take account of
correlated changes between $x$ and $y$.
 \label{f_correlation_constraints}}
\end{figure}

In probability theory, the different isomorphism constraints and
tangent spaces acting at each point define non-intersecting
lines and surfaces within the target space. Some of these are
shown in Fig. \ref{f_correlation_constraints} representing the
$(p,q,r)$ simplex of the two potentially correlated $x$ and $y$
variables (this behavioural space is defined in the next
Chapter). Here, each state of correlation is a constant and
cannot vary during an optimization analysis so an optimization
procedure must sequentially take account of every possible
correlation state between these variables, setting
$\rho_{xy}=\rho$ for all $\rho\in[-1,1]$. These optimum points
can then be compared to determine which correlation state
between $x$ and $y$ returns the best value.

Unsurprisingly, these two distinct approaches can sometimes
generate conflicting results.

\subsection{Isomorphism constraints alter geometry}

In general, the imposition of any specific isomorphism
constraint can be expected to alter the geometry of optimization
space and alter optimization outcomes. We now illustrate this
briefly.

Consider a three dimensional volume in which Pythagoras's rule
specifies the distance $ds$ between points $(x,y,z)$ and
$(x+\Delta x,y+\Delta y,z+\Delta z)$ as
\begin{equation}
    ds^2 = dx^2 + dy^2 + dx^2.
\end{equation}
That Pythagoras's rule is satisfied indicates that the space is
flat. In contrast, when some constraint is adopted via
$z=f(x,y)$ then the shortest distance between two points no
longer satisfies Pythagoras's rule indicating that the
constraint has rendered the space curved.  Consider the example
relation
\begin{equation}
    z^2 = r^2 - x^2 - y^2,
\end{equation}
where $r$ denotes a radius of curvature.  The surface constraint
now requires
\begin{equation}
     z dz =  - x dx - y dy,
\end{equation}
so
\begin{equation}
     dz^2 =  \frac{(x dx + y dy)^2}{r^2 - x^2 - y^2}.
\end{equation}
In turn, this gives the shortest path distance between $(x,y)$
and $(x+\Delta x,y+\Delta y)$ as
\begin{eqnarray}
    ds^2
    &=& dx^2 + dy^2 + dz^2 \nonumber \\
    &=& dx^2 + dy^2 + \frac{(x dx + y dy)^2}{r^2 - x^2 - y^2} \\
    &=& \left[1+ \frac{x^2}{r^2 - x^2 - y^2}\right] dx^2 +
        \left[1+ \frac{y^2}{r^2 - x^2 - y^2}\right] dy^2 +
        \frac{2xy}{r^2 - x^2 - y^2} dxdy. \nonumber
\end{eqnarray}
Self-evidently, this shortest distance between the points
$(x,y)$ and $(x+\Delta x,y+\Delta y)$ does not satisfy
Pythagoras's rule reflecting the fact that the space is now
curved.

The adoption of a curvature imposing constraint ensures that
optimization problems (the shortest path distance) within the
plane are altered and so locate different optima. Further,
theorems valid in flat space are no longer applicable in the now
curved space.  When it is possible to impose curvature inducing
constraints on a space to alter optimization outcomes, then it
is necessary to examine every possibility to ensure a complete
optimization.

\section{Discussion}

A rational player must compare expected payoffs across the mixed
strategy space in order to locate equilibria.  As expectations
are polylinear, such comparisons are mathematically equivalent
to calculating gradients and the issues raised in this paper
apply. Further, it is perfectly possible that a rational player
might need to calculate the Fisher information defined in terms
of gradients of probability distributions in order to optimize
payoffs. It is perfectly possible that a rational player might
well need to optimize an Entropy gradient to maximize a payoff.
It is even perfectly possible to define games where payoffs
depend directly on the gradient of a probability
distribution---shine light through a sheet of glass painted by
players to alter transmission probabilities and make payoffs
dependent on the resulting light intensity gradients (call it
the interior decorating game). We have shown that rational
players working with the standard strategy spaces of game theory
will have difficulties with these games.

We have highlighted two alternate ways to optimize a
multivariate function $\Pi(x,y)$ where $x$ and $y$ might be
functionally related in different ways, $y=g_i(x)$ for different
$i$ say. The first approach, common to probability theory and
general optimization theory, considers each potential functional
relation as occupying a distinct space and approaches the
optimization as a choice between distinct spaces. Any
uncertainty about which space to choose does not leak into the
properties of any individual space.  If desired, isomorphic
constraints can be used to embed all these distinct spaces into
a single enlarged space for convenience, but if so, all the
properties of the optimization problem are exactly preserved.
The second approach, common to game theory, holds that the
uncertainty about which functional relation to choose should
appear in the same space as the variables $(x,y)$. This is
accomplished by expanding the size of the space to include both
the old variables $x$ and $y$ and sufficient new variables (not
explicitly shown here) to contain all the potential functional
relations and allow $\lim_{y\rightarrow
g_i(x)}\Pi(x,y)=\Pi[x,g_i(x)]$ for all $i$. This enlarged space
then allows gradient comparisons to be made at points
$\Pi[x,g_i(x)]-\Pi[x,g_j(x)]$ for all $i$ and $j$ to locate
optima. These two approaches can lead to conflicting
optimization outcomes as while these approaches generally assign
the same values to functions at all points,
\begin{equation}
  \left.\Pi(x,y)\right|_{y=g_i(x)} =
     \lim_{y\rightarrow g_i(x)} \Pi(x,y),
\end{equation}
they typically calculate different gradients at those same
points
\begin{equation}
  \left.\nabla \Pi(x,y)\right|_{y=g_i(x)} \neq
     \lim_{y\rightarrow g_i(x)} \nabla \Pi(x,y).
\end{equation}
These differences can be extreme when the function $\Pi(x,y)$
depends on global properties of the space---the dimension,
volume, gradient, information or entropy say.  In its approach,
game theory differs from many other fields in how it models
alternate functional dependencies including other fields of
economics.  For example, the Euler-Lagrange equations of
Ramsey-type models consider the functional variation of some
function $u$ while ensuring a consistent treatment of the
gradient of the function $u'$ \cite{Ramsey_1928_543}.  Gradients
are not taken in any limit in these fields.

Throughout this work, we have presumed that a rational player
should be able to use standard techniques from either
probability theory or optimization theory on the one hand, or
decision theory and game theory on the other, and expect all of
these methods to provide consistent results. We have shown that
when considering multiple, potentially correlated variables, and
functions of these variables dependent on the geometry of the
probability parameter space, then these methods can give rise to
contradictory optimization outcomes.  We have suggested decision
and game theory are incomplete when they require the adoption of
a single geometry for any decision or game tree, and that these
fields should consider applying the alternate geometries of
probability theory and optimization theory.  Recognizing that a
single multi-stage decision or game tree can encompass an
infinite number of incommensurate probability spaces might
resolve some of the paradoxes of game theory, and have broader
application.

The specification of a probability space determines which
variables exist and whether they are functionally constrained or
freely varying. Given the choice of a probability space,
optimization can only take place with respect to the freely
varying parameters within that adopted space. Should players
wish to explore a broader range of variation, then they must
seek to alter the functional assignments of some of their random
variables and functions, and so will alter their probability
spaces.  In other words, rational players of unbounded capacity
will search both among different probability spaces, which are
not always guaranteed to give the same outcomes, as well as
search within each space over all of the freely varying
parameters of each probability space. Rational players require a
decision procedure mediating this dual search of all possible
probability spaces and all possible variables within each space,
and that is what we seek to provide here.

Every probabilistic decision can be modeled by an infinite
number of different probability measure spaces.  For many
decisions, it is immediately obvious that every alternative
space leads to exactly the same optimized outcomes.  The
question is, is this true for every possible decision, for every
possible strategic interaction.  Before turning to answer this
question, we now turn to examine the probability spaces
typically encountered in game theory.  In particular, we focus
on mixed strategy probability measure spaces, behavioural
strategy probability measure spaces, and correlated equilibria
probability measure spaces.

\section{Appendix: Correlation and mutual information}

We employ probability space isomorphisms based on correlation.
However, it is  not clear that correlation is the appropriate
measure to use. It is well known that this measure of linear
correlation is insensitive to nonlinear correlations.  Because
of this, other measures might be more useful.  When two
variables are correlated, and if this correlation is ignored,
then information has been discarded. It might well be the case
that information based measures, in particular, mutual
information might provides a better way to take account of the
interrelatedness of random variables \cite{Pfeiffer_1990}.

\subsection{Nonlinear dependencies and correlation}

The correlation between arbitrary random variables $x$ and $y$
is
\begin{equation}
  \rho_{x,y} = \frac{{\rm cov}(x,y)}{\sigma_x\sigma_{y}}
   \; = \;  \frac{\langle xy\rangle - \langle x\rangle \langle y\rangle}{\sqrt{\langle x^2\rangle - \langle x\rangle^2}\sqrt{\langle y^2\rangle - \langle y\rangle^2}},
\end{equation}
defined in terms of the covariance ${\rm cov}(x,y)$, the
variance $\sigma_x^2={\rm cov}(x,x)$, and the mean $\langle
x\rangle$ \cite{Kelly_94}.

Consider two discrete random variables $x$ and $y$, with $x$
being any of $x\in\{-1,0,1\}$ with equal probability
$\frac{1}{3}$, and $y=x^2\in\{0,1\}$ so $P(y=0)=\frac{1}{3}$ and
$P(y=1)=\frac{2}{3}$. These variables would normally be
considered to be highly correlated as knowing $x$ immediately
specifies $y$, while knowing $y$ narrows the possible values of
$x$ to $x=\pm\sqrt{y}$. The respective probability distributions
are
\begin{eqnarray}
  P(x,y) &=& \frac{1}{3}
             \left( \delta_{x,-1}\delta_{y,1} +
                    \delta_{x,0}\delta_{y,0}  +
                    \delta_{x,1}\delta_{y,1}
              \right)  \nonumber \\
  P(x)   &=& \sum_{y=0}^1 P(x,y) \nonumber \\
         &=& \frac{1}{3}
             \left( \delta_{x,-1}+
                    \delta_{x,0} +
                    \delta_{x,1}
              \right)   \nonumber \\
  P(y)   &=& \sum_{x=-1}^1 P(x,y) \nonumber \\
         &=& \frac{1}{3}
             \left( \delta_{y,0} + 2 \delta_{y,1}
              \right)   \nonumber \\
  P(x|y) &=& \delta_{x,0}\delta_{y,0} +
                \frac{1}{2} \delta_{y,1}
                \left( \delta_{x,-1} +
                    \delta_{x,1}\right) \nonumber \\
  P(y|x) &=& \left( \delta_{x,-1}\delta_{y,1} +
                    \delta_{x,0}\delta_{y,0}  +
                    \delta_{x,1}\delta_{y,1}
              \right).
\end{eqnarray}
These distributions then give
\begin{eqnarray}
  {\rm cov}(x,y) &=& \langle xy\rangle -\langle x\rangle\langle y\rangle\nonumber \\
  & = & \sum_{x=-1}^{1} \sum_{y=0}^1 P(x,y) xy \nonumber \\
  & = & 0.
\end{eqnarray}
This zero covariance then specifies a zero coefficient of linear
correlation $\rho_{xy}=0$, but as noted above, this does not
mean these variables are uncorrelated.  Better measures of
correlation indicate this.

\subsection{Mutual Information}

A more general measure of the interrelatedness of discrete
variables is given by their mutual information \cite{Cover_91}.
This is defined in terms of their joint probability distribution
$P_{xy}$, the marginal distribution $P_x$ governing the $x$
variable, and the marginal distribution $P_y$ governing the $y$
variable. The information obtained from observing a single
instance of a discrete random variable $x$ is
\begin{equation}
    I(x)=- \log P(x).
\end{equation}
Consequently, the average information content of an entire
ensemble of observations of $x$ is obtained by averaging over
the entire distribution to give the entropy or uncertainty of
$x$,
\begin{equation}
    H(x)=-\sum_x P(x) \log P(x).
\end{equation}
Suppose now that a second discrete random variable $y$ is
observed.  In line with the above, the joint entropy or
uncertainty of $x$ and $y$ is
\begin{equation}
    H(x,y)=-\sum_{x,y}P(x,y)\log P(x,y).
\end{equation}
Consider now how much information we obtain about $x$ given
observations of $y$. The information obtained about $x$ given
knowledge of $y$ is $-\log P(x|y)$, which when averaged gives a
measure of the remaining uncertainty in $x$ given an observation
of $y$. This is the conditional entropy of $x$ given $y$ defined
as
\begin{equation}
    H(x|y)=-\sum_{x,y}P(x,y)\log P(x|y).
\end{equation}
Consequently, the average reduction in uncertainty in $x$ given
observations of $y$ is the mutual information content of the
joint probability distribution describing the two discrete
random variables $x$ and $y$, and is
\begin{equation}
    H(x;y)=H(x)-H(x|y).
\end{equation}
Then, when variables $x$ and $y$ are uncorrelated, we have
$P(x,y)=P(x)P(y)$ and $P(x|y)=P(x)$, so $H(x|y)=H(x)$, ensuring
their mutual information is minimized at $H(x;y)=0$, while their
joint entropy or uncertainty is maximized at $H(x,y)=H(x)+H(y)$.
Conversely, when these variables are perfectly correlated, then
$P(x,y)=P(x)P(y|x)=P(x)\delta_{yx}$ and $P(x|y)=1$, so
$H(x|y)=0$, ensuring their mutual information is maximized at
$H(x;y)=H(x)$, while their joint entropy or uncertainty is
minimized at $H(x,y)=H(x)$ \cite{Cover_91}.

For the example considered above, we have the entropies or
uncertainties in the respective $x$ and $y$ distributions of
\begin{eqnarray}
    H(x) &=& \log 3 \nonumber \\
    H(y) &=& \log 3 - \frac{2}{3} \log 2.
\end{eqnarray}
That is, there is less uncertainty in $y$ as there are only two
possible values taken by $y$ compared to the three possible
values taken by $x$.  Subsequently, the respective conditional
entropies are
\begin{eqnarray}
    H(x|y) &=& \frac{2}{3} \log 2 \nonumber \\
    H(y|x) &=& 0.
\end{eqnarray}
The difference between these conditional entropies results as
knowing $x$ uniquely specifies $y$ while knowing $y$ only
partially specifies $x$.  We can now calculate the mutual
information content $x$ and $y$ which is
\begin{equation}
    H(x;y) = H(y;x) = \log 3 - \frac{2}{3} \log 2.
\end{equation}
Lastly, the joint entropy or uncertainty of $x$ and $y$ is
\begin{equation}
    H(x,y) = H(y,x) =  \log 3.
\end{equation}

For the behavioural strategy distributions considered in this
paper, we have
\begin{equation}
    H_{x;y} = \log \left\{
             \frac{\left[(1-q)^{1-q} q^q \right]^{1-p}
                   \left[(1-r)^{1-r} r^r \right]^{p} }
                  {\left[1-q-p(r-q)\right]^{1-q-p(r-q)}
                   \left[q+p(r-q)\right]^{q+p(r-q)} }
                 \right\}.
\end{equation}
When $q=r$ indicating that $x$ and $y$ are uncorrelated, we have
a mutual information content of $H_{y;x}=0$. Conversely, when
$(q,r)=(0,1)$ and $x$ and $y$ are perfectly correlated, the
mutual information content is
\begin{eqnarray}
  H_{x;y} &=& H(x) \nonumber \\
         &=& - \left[ (1-p) \log(1-p) + p \log p \right].
\end{eqnarray}
Similarly, when $(q,r)=(1,0)$ and $x$ and $y$ are perfectly
anti-correlated, the mutual information content is
\begin{eqnarray}
  H_{x;y} &=& H(x) \nonumber \\
         &=& - \left[ (1-p) \log(1-p) + p \log p \right].
\end{eqnarray}
This duplicates the value for the perfect correlation case.

The case of continuous distributions is more complicated, where
for instance, the mutual information content evaluates as
\begin{equation}
    H(x;y)= \int dx \; \int dy \; P(x,y)
       \log \left( \frac{P(x,y)}{P(x)P(y)} \right).
\end{equation}

The upshot is that correlation corresponds to information. Every
different probability space that might be adopted by each player
corresponds to a physical randomization device, a ``roulette",
which defines certain correlations between random variables.
These correlations correspond to information, and should the
correlations be ignored, then this equates to the discarding of
information. In this paper, we assume that rational players will
make use of all available information including that implicit in
correlated joint probability measure spaces.

\subsubsection{Problem: Mutual information}

However that the mutual information is not a constant when $x$
and $y$ are perfectly correlated or anti-correlated. It is not
clear how mutual information might be used, but then again, it
is not clear why correlation should have the status desired for
it.  What is the connection between the functional dependencies
of our deterministic examples, and correlated variables?

 \chapter{Isomorphisms in Strategy Spaces}
 \label{chap_Isomorphisms_Strategy_Spaces}

\section{Introduction}

The preceding chapter has pointed out by example that there are
different ways to ``contain" one probability distribution within
another.  Probability theory uses strong isomorphic mappings,
while game theory uses weaker isomorphic mappings which preserve
fewer properties of the original distribution within the target
space. These differences arose (perhaps) as probability space
isomorphisms do not feature anywhere in the historical
definition of mixed strategy spaces. We briefly recap this
historical process below.

\subsection{Mixed strategy probability measure spaces}

{\bf Rationality, Utility:} Von Neumann and Morgenstern began
their formalization of game theory by defining the economic
problem as when ``rational players" seek to  ``obtain a maximum
of utility" using ``a complete set of rules of behavior in all
conceivable situations." \cite{vonNeumann_44}. Naturally, the
result ``is thus a combinatorial enumeration of enormous
complexity" \cite{vonNeumann_44}. Von Neumann and Morgenstern
aimed to formulate a complete plan, an analysis of every
possible move or variable or outcome" \cite{vonNeumann_44}.

{\bf Moves:} Each player makes moves in a game, where ``A move
is the occasion of a choice between various alternatives" at
each stage of the game \cite{vonNeumann_44}.

{\bf Pure Strategies:} The choices of moves combine into player
strategies: ``A {\em strategy} of the player $k$ is a function
\dots which is defined for every [personal move of that player],
and whose value [determines his choice at that move]"
\cite{vonNeumann_44}. A strategy is ``a complete plan: a plan
which specifies what choices [a player] will make in every
possible situation, for every possible actual information which
he may possess at that moment" \cite{vonNeumann_44}. Hence, for
von Neumann and Morgenstern, each different strategy for a given
player is a list of all the combinatorial play possibilities
available to that player throughout the game taking account of
every different possible history and information set in the
game. Each player chooses their strategy independently of all
the other players, as any dependencies and correlations are
already taken into account in the complete listing of
information sets and possibilities for every possible game that
might occur.  In particular, ``The player $k$ must choose his
strategy \dots without information concerning the choices of the
other players, or of the chance events (the umpire's choice).
This must be so since all the information he can at any time
possess is already embodied in his strategy"
\cite{vonNeumann_44}.  The choice of a strategy of play then
becomes the sole decision to be made by the player, and this is
made independently of any other choice.

{\bf Mixed Strategies:} Players can choose their pure strategies
according to some independent probability distributions, termed
a mixed strategy.  The probability parameters of each
distribution are subject to normalization constraints ``and to
no others" \cite{vonNeumann_44}.

{\bf Nash Equilibria:} Nash closely followed the von Neumann and
Morgenstern formalism \cite{Nash_50_48,Nash_51_28}.  Nash's
famous first paper commences ``One may define a concept of an
$n$-person game in which each player has a finite set of pure
strategies and in which a definite set of payments to the $n$
players corresponds to each $n$-tuple of pure strategies, one
strategy being taken for each player.  \dots For mixed
strategies, which are probability distributions over the pure
strategies, the pay-off functions are the expectations of the
players, thus becoming polylinear forms in the probabilities
with which the various players play their various pure
strategies." \cite{Nash_50_48}. In a second paper, Nash treated
the mixed strategy space as ``points in a simplex whose vertices
are the [pure strategies]. This simplex may be regarded as a
convex subset of a real vector space, giving us a natural
process of linear combination for the mixed strategies"
\cite{Nash_51_28}. Nash subsequently defined the set of all
mixed strategies for all players as ``a point in a vector space,
the product space of the vector spaces containing the mixed
strategies.  And the set of all such [points] forms, of course,
a convex polytope, the product of the simplices representing the
mixed strategies" \cite{Nash_51_28}. Because all the mixed
strategy probabilities are continuous, Nash was able to use
fixed point theorems to derive optimal points, referred to now
as Nash equilibria.

{\bf Behavioural strategy spaces:} Kuhn showed that the mixed
strategy spaces could be replaced by the more intuitively
accessible behavioural strategy space \cite{Kuhn_1953}. The
behavioural strategies are merely the player's choice
probabilities distributed over each branch of a game's decision
tree. These probabilities are `uncorrelated' or `locally
randomized' strategies wherein a local perspective decentralizes
the strategy decision of each player into a number of local
decisions \cite{Kuhn_1953,vanDamme_92_41}. In this, the
agent-normal game form, myopic agents at each history set
determine paths through the game tree using probability
distributions which are uncorrelated and independent. This
assumption allowed Kuhn to prove the equivalence of uncorrelated
behavioural strategies and the uncorrelated mixed strategies
introduced by von Neumann and Morgenstern \cite{vonNeumann_44}
and Nash \cite{Nash_51_28} in games of perfect recall
\cite{Kuhn_1953}.

{\bf Absent isomorphisms:} In the historical development painted
above, there is no room for isomorphic mappings and any
discussion of the properties of embedded probability
distributions.  A game definition provides a complete list of
moves and hence of strategies and hence of mixed strategies
which are independent and unconstrained (and complete).  Our
alternative approach posits that a game definition can be put
into a 1-1 correspondence with many alternate probability
spaces, with each choice of probability space altering the
complete list of moves and of strategies and hence of mixed
strategies.

In this chapter, we show that these two different approaches
lead to very different properties for mixed and behavioural
strategy spaces as defined by probability theory and game
theory.

\begin{figure}[htb]
\centering
\includegraphics[width=\columnwidth,clip]{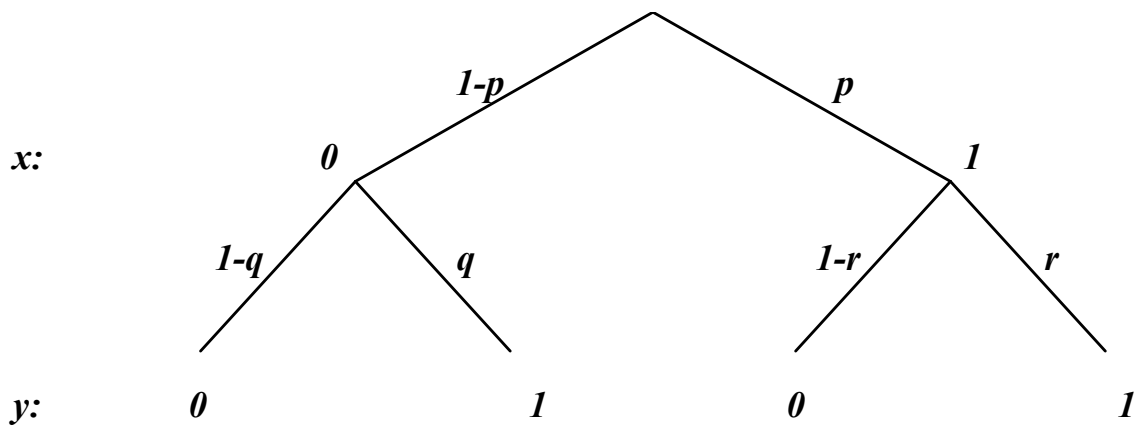}
\caption[A simple decision tree]{\em  A simple decision tree where potentially independent
or correlated variables $x$ and $y$ take values $\{0,1\}$ with the
probabilities shown.   This defines the $(p,q,r)$ behavioural
probability space.
 \label{f_xy_decision_tree}}
\end{figure}

\section{Mixed and behavioural strategy spaces}
\label{sect_Mixed_and_behavioural_strategy_spaces}

The different approaches of probability theory and game theory
to isomorphic embeddings impacts on the definitions of mixed and
behavioural strategy spaces. As previously, we will compare
these spaces both with and without isomorphism constraints. Our
focus will be on a simple decision problem involving two random
variables $x,y\in\{0,1\}$ where $y$ is potentially conditioned
on $x$ as shown in the behavioural strategy decision tree of
Fig. \ref{f_xy_decision_tree}.

\subsection{Mixed strategy space ${\cal P}_M$}

The mixed strategy space is denoted ${\cal P}_M$, and determines
the choice of $x$ via a probability distribution $\alpha$ while
the respective choices of $y$ on the left branch of the decision
tree $y_l$ and on the right branch $y_r$ are determined by an
independent probability distribution $\beta$ according to the
following table:
\begin{equation}                    \label{eq_mixed_strategies}
  \begin{array}{c|cccc}
 (y_l,y_r)=&   (0,0)    & (0,1)    &  (1,0)   &   (1,1)   \\\hline
     (x,y) &   \beta_0  & \beta_1  &  \beta_2 &   \beta_3 \\\hline
 \alpha_0  &    (0,0)   & (0,0)    &  (0,1)   &   (0,1)   \\
 \alpha_1  &    (1,0)   & (1,1)    &  (1,0)   &   (1,1).  \\
  \end{array}
\end{equation}
The mixed strategy simplex for each player is respectively
$S^X=\{(\alpha_0, \alpha_1)\in R_+^2: \sum_j \alpha_j=1\}$ and
$S^Y=\{(\beta_0, \beta_1, \beta_2, \beta_3)\in R_+^4: \sum_j
\beta_j=1\}$. The associated tangent spaces are $T^X=\{z\in R^2:
\sum_j z_j=0\}$ and $T^Y=\{z\in R^4: \sum_j z_j=0\}$, equivalent
to every possible positive or negative fluctuation in the
probabilities of the pure strategies of each player. The joint
probability distribution $P_{xy}(x,y)$ for $x$ and $y$ is
\begin{eqnarray}               \label{eq_mixed_prob}
  P_{xy}(0,0) &=& (1-\alpha_1) (1 - \beta_2 - \beta_3) \nonumber \\
  P_{xy}(0,1) &=& (1-\alpha_1) (\beta_2 + \beta_3) \nonumber \\
  P_{xy}(1,0) &=& \alpha_1 (1 - \beta_1 - \beta_3) \nonumber \\
  P_{xy}(1,1) &=& \alpha_1 (\beta_1 + \beta_3).
\end{eqnarray}
Here, we have used normalization constraints to eliminate
$\alpha_0$ and $\beta_0$.  The expectations of the $x$ and $y$
variables are given by
\begin{eqnarray}                \label{eq_mixed_expec}
 \langle x\rangle &=& \alpha_1 \nonumber \\
 \langle y\rangle &=& \beta_2+\beta_3 +
                      \alpha_1(\beta_1-\beta_2) \nonumber \\
 \langle x y\rangle &=& \alpha_1 (\beta_1+\beta_3),
\end{eqnarray}
while their variances are
\begin{eqnarray}               \label{eq_mixed_var}
 V(x) &=& \alpha_1(1-\alpha_1) \nonumber \\
 V(y) &=& \left[\beta_2+\beta_3 + \alpha_1(\beta_1-\beta_2)\right]
           \times \left[1-\beta_2-\beta_3 - \alpha_1(\beta_1-\beta_2)\right].
\end{eqnarray}
For completeness, we note the marginal and joint entropies are
\begin{eqnarray}
 E_{x}
   &=& - (1-\alpha_1)  \log(1-\alpha_1) - \alpha_1  \log\alpha_1  \nonumber \\
 E_{y}
   &=& - [1-\beta_2-\beta_3+\alpha_1(\beta_2-\beta_1)] \times
          \log[1-\beta_2-\beta_3+\alpha_1(\beta_2-\beta_1)] \nonumber \\
   &&  -[\beta_2+\beta_3-\alpha_1(\beta_2-\beta_1)] \times
            \log[\beta_2+\beta_3-\alpha_1(\beta_2-\beta_1)] \nonumber \\
 E_{xy}
   &=& - (1-\alpha_1) (1-\beta_2-\beta_3) \log[(1-\alpha_1) (1-\beta_2-\beta_3)] \nonumber \\
   &&    - (1-\alpha_1) (\beta_2+\beta_3) \log[(1-\alpha_1) (\beta_2+\beta_3)] \nonumber \\
   &&  - \alpha_1 (1-\beta_1-\beta_3) \log[\alpha_1 (1-\beta_1-\beta_3)] \nonumber \\
   &&    - \alpha_1 (\beta_1+\beta_3) \log[\alpha_1 (\beta_1+\beta_3)].
\end{eqnarray}
Naturally, the mixed strategy probability space can model any
state of correlation between $x$ and $y$ with the correlation
give by
\begin{equation}                    \label{eq_mixed_correlation}
    \rho_{xy}(\alpha_1,\beta_1,\beta_2,\beta_3)
    = \frac{\sqrt{\alpha_1(1-\alpha_1)}(\beta_1-\beta_2)}{
    \sqrt{\langle y\rangle
    \left[1-\langle y\rangle\right]}}.
\end{equation}
Then, when $x$ and $y$ are perfectly correlated we have
$\rho_{xy}=1$ requiring the constraints $\beta_1=1$ and
$\beta_0=\beta_2=\beta_3=0$.  When $x$ and $y$ are perfectly
anti-correlated we have $\rho_{xy}=-1$ requiring the constraints
$\beta_2=1$ and $\beta_0=\beta_1=\beta_3=0$. Finally, when $x$
and $y$ are independent we have $\rho_{xy}=0$ requiring the
constraint $\beta_1=\beta_2$.

\subsection{Behavioural strategy space  ${\cal P}_B$}

The behavioural strategy probability space \cite{Kuhn_1953} is
denoted ${\cal P}_B$ and is parameterized as shown in Fig.
\ref{f_xy_decision_tree}. The behavioural strategy space for the
players is $S^{XY}=\{(p,q,r)\in R_+^3: 0\leq p, q, r \leq 1\}$
after taking account of normalization. The associated tangent
space is $T^{XY}=\{z\in R^3\}$. The probability $P_{xy}(x,y)$
that $x$ and $y$ take on their respective values is
\begin{eqnarray}             \label{eq_behav_prob}
  P_{xy}(0,0) &=& (1-p)(1-q) \nonumber \\
  P_{xy}(0,1) &=& (1-p)q \nonumber \\
  P_{xy}(1,0) &=& p(1-r) \nonumber \\
  P_{xy}(1,1) &=& pr.
\end{eqnarray}
This distribution gives the following expected values:
\begin{eqnarray}             \label{eq_behav_expect}
 \langle x\rangle &=& p \nonumber \\
 \langle y\rangle &=& q + p (r - q) \nonumber \\
 \langle x y\rangle &=& p r,
\end{eqnarray}
while the variances of the $x$ and $y$ variables are
\begin{eqnarray}              \label{eq_behav_var}
 V(x) &=& p(1-p) \nonumber \\
 V(y) &=& \left[q + p (r - q)\right]\left[1-q - p (r - q)\right].
\end{eqnarray}
The marginal and joint entropies between the $x$ and $y$
variables are
\begin{eqnarray}
 E_{x}
   &=& - (1-p) \log(1-p) - p \log p \nonumber \\
 E_{y}
   &=& - [(1-p) (1-q)+p (1-r)] \times
              \log[(1-p) (1-q)+p (1-r)] \nonumber \\
   &&    - [(1-p) q+p r] \log[(1-p) q+p r] \nonumber \\
 E_{xy}
   &=& - (1-p) (1-q) \log[(1-p) (1-q)]
    - (1-p) q \log[(1-p) q] \nonumber \\
   &&  - p (1-r) \log[p (1-r)]
       - p r \log[pr].
\end{eqnarray}
The behavioural probability space also allows modeling any
arbitrary state of correlation between the $x$ and $y$ variables
where the correlation between $x$ and $y$ is
\begin{equation}           \label{eq_rho_correlation}
    \rho_{xy}
    = \frac{\sqrt{p(1-p)}(r-q)}{
        \sqrt{\left[q+p(r-q)\right]\left[1-q-p(r-q)\right]}}.
\end{equation}
Then, $x$ and $y$ are perfectly correlated at
$\rho_{xy}(p,0,1)=1$, perfectly anti-correlated at
$\rho_{xy}(p,1,0)=-1$, and uncorrelated if either $p=0$ or $p=1$
or $q=r$ giving $\rho_{xy}=0$. Hence, the decision tree of Fig.
\ref{f_xy_decision_tree} encompasses every possible state of
correlation between $x$ and $y$, and thus it can be used to
perform a complete analysis.

\begin{table*}
 \centering
 \scriptsize
\begin{tabular}{ccccc} \hline \hline
 $\rho_{xy}=1$ & ${\cal P}_M$  & ${\cal P}_B$  & $\left.{\cal P}_M\right|_{\beta_1=1}$ & $\left.{\cal P}_B\right|_{(q,r)=(0,1)}$ \\ \hline
 Parameters & $\alpha_1, \beta_1, \beta_2, \beta_3$ & $p, q, r$ & $\alpha_1$ & $p$ \\
 Dimensions & 4 & 3 & 1 & 1 \\
 $\nabla$ operator &
 $\frac{\partial}{\partial \alpha_1} \hat{\alpha}_1+\frac{\partial}{\partial \beta1} \hat{\beta}_1 + \frac{\partial}{\partial \beta_2} \hat{\beta}_2 + \frac{\partial}{\partial \beta_3} \hat{\beta}_3$ &
 $\frac{\partial}{\partial p} \hat{p}+\frac{\partial}{\partial q} \hat{q}+\frac{\partial}{\partial r} \hat{r}$ &
 $\frac{\partial}{\partial \alpha_1} \hat{\alpha}_1$ &
 $\frac{\partial}{\partial p} \hat{p}$ \\
  Gradient &
  $\lim_{\beta_1\rightarrow 1}\nabla (.)$ &
  $\lim_{(q,r)\rightarrow (0,1)}\nabla (.)$ &
  $\nabla$ &
  $\nabla$ \\
 \multicolumn{5}{l}{Probability Conservation}  \\
 $\nabla\left[P_{xy}(0,0)+P_{xy}(1,1)\right]$ &
 $ \alpha_1 \hat{\beta}_1 - (1-\alpha_1) \hat{\beta}_2+ (2\alpha_1 - 1) \hat{\beta}_3$ &
 $ - (1-p) \hat{q} + p \hat{r}$ &
 0 &
 0 \\
 $\nabla\left[P_{xy}(0,1)+P_{xy}(1,0)\right]$ &
 $ -\alpha_1 \hat{\beta}_1 + (1-\alpha_1) \hat{\beta}_2- (2\alpha_1 - 1) \hat{\beta}_3$ &
 $  (1-p) \hat{q} - p \hat{r}$ &
 0 &
 0 \\
 \multicolumn{5}{l}{Conditionals}  \\
 $\nabla P_{x|y}(0|0)$ &
 $\frac{\alpha_1}{1-\alpha_1} (\hat{\beta}_1 + \hat{\beta}_3)$ &
 $\frac{p}{1-p} \hat{r}$ &
 0 &
 0 \\
 $\nabla P_{x|y}(0|1)$ &
 $\frac{1-\alpha_1}{1\alpha_1} (\hat{\beta}_2+ \hat{\beta}_3)$ &
 $\frac{1-p}{p} \hat{q}$ &
 0 &
 0\\
 \multicolumn{5}{l}{Expectations}  \\
 $\nabla \langle x\rangle$ &
 $\hat{\alpha}_1$ &
 $\hat{p}$ &
 $\hat{\alpha}_1$ &
 $\hat{p}$ \\
 $\nabla \langle y\rangle$ &
 $\hat{\alpha}_1 + \alpha_1 \hat{\beta}_1 + (1-\alpha_1) \hat{\beta}_2 + \hat{\beta}_3$ &
 $\hat{p} + (1-p) \hat{q} + p \hat{r}$ &
 $\hat{\alpha}_1$ &
 $\hat{p}$\\
 $\nabla \langle x y\rangle$ &
 $\hat{\alpha}_1 + \alpha_1 \hat{\beta}_1 +\alpha_1 \hat{\beta}_3$ &
 $\hat{p} + p \hat{r}$ &
 $\hat{\alpha}_1$ &
 $\hat{p}$\\
 \multicolumn{5}{l}{Variance}  \\
 $\nabla \left[V(x)+V(y)-2\mbox{cov}(x,y)\right]$ &
 $-\alpha_1 \hat{\beta}_1+ (1-\alpha_1) \hat{\beta}_2+ (1-2\alpha_1)\hat{\beta}_3$ &
 $(1-p) \hat{q} - p \hat{r}$ &
 0 &
 0 \\
 \multicolumn{5}{l}{Entropy}  \\
  $\nabla\left[E_{xy}-E_{x}\right]$ & $\neq 0$ & $\neq 0$ &  0 & 0 \\
 \multicolumn{5}{l}{Correlation}  \\
  $\nabla \rho_{xy}$ & $\neq 0$ & $\neq 0$ &  0 & 0 \\
  &  &  &  &  \\
  &  &  &  & \\
  &  &  &  & \\   \hline \hline
 $\rho_{xy}=0$ & ${\cal P}_M$  & ${\cal P}_B$  & $\left.{\cal P}_M\right|_{\beta_1=\beta_2}$ & $\left.{\cal P}_B\right|_{r=q}$ \\  \hline
 Parameters & $\alpha_1, \beta_1, \beta_2, \beta_3$ & $p, q, r$ & $\alpha_1$, $\bar{\beta}=\beta_1+\beta_3$ & $p, q$ \\
 Dimensions & 4 & 3 & 2 & 2 \\
 $\nabla$ operator &
 $\frac{\partial}{\partial \alpha_1} \hat{\alpha}_1+\frac{\partial}{\partial \beta1} \hat{\beta}_1 + \frac{\partial}{\partial \beta_2} \hat{\beta}_2 + \frac{\partial}{\partial \beta_3} \hat{\beta}_3$ &
 $\frac{\partial}{\partial p} \hat{p}+\frac{\partial}{\partial q} \hat{q}+\frac{\partial}{\partial r} \hat{r}$ &
 $\frac{\partial}{\partial \alpha_1} \hat{\alpha}_1+\frac{\partial}{\partial \bar{\beta}} \hat{\bar{\beta}}$&
 $\frac{\partial}{\partial p} \hat{p}+\frac{\partial}{\partial q} \hat{q}$ \\
  Gradient &
  $\lim_{\beta_2\rightarrow \beta_1}\nabla  (.)$ &
  $\lim_{r\rightarrow q}\nabla (.)$ &
  $\nabla$ &
  $\nabla$ \\
 \multicolumn{5}{l}{Probability}  \\
 $\nabla \left[P_{xy}(0,0) - P_x(0)P_y(0)\right]$ &
 $\alpha_1 (1-\alpha_1) (\hat{\beta}_1 - \hat{\beta}_2)$ &
 $p (1-p) (\hat{r} - \hat{q})$ &
 0 &
 0\\
 $\nabla \left[P_{xy}(0,1) - P_x(0)P_y(1)\right]$ &
 $\alpha_1 (1-\alpha_1) (\hat{\beta}_2 - \hat{\beta}_1)$ &
 $p (1-p) (\hat{q} - \hat{r})$ &
 0 &
 0\\
 $\nabla \left[P_{xy}(1,0) - P_x(1)P_y(0)\right]$ &
 $\alpha_1 (1-\alpha_1) (\hat{\beta}_2 -  \hat{\beta}_1)$ &
 $p (1-p) (\hat{q} -  \hat{r})$ &
 0 &
 0\\
 $\nabla \left[P_{xy}(1,1) - P_x(1)P_y(1)\right]$ &
 $\alpha_1 (1-\alpha_1) (\hat{\beta}_1 - \hat{\beta}_2)$ &
 $p (1-p) (\hat{r} - \hat{q})$ &
 0 &
 0\\
 \multicolumn{5}{l}{Conditionals}  \\
 $\nabla \left[P_{x|y}(0|0)-P_x(0)\right]$ &
 $\frac{\alpha_1(1-\alpha_1)}{1-\beta_1-\beta_3} (\hat{\beta}_1-  \hat{\beta}_2)$ &
 $\frac{p(1-p)}{(1-q)} (\hat{r} - \hat{q})$ &
 0 &
 0\\
 $\nabla \left[P_{x|y}(0|1)-P_x(0)\right]$ &
 $\frac{\alpha_1(1-\alpha_1)}{\beta_1+\beta_3} (\hat{\beta}_2-\hat{\beta}_1)$ &
 $\frac{p(1-p)}{q} (\hat{q} -  \hat{r})$ &
 0 &
 0\\
 \multicolumn{5}{l}{Expectation}  \\
 $\nabla\left[\langle xy\rangle-\langle x\rangle \langle y\rangle\right]$ &
 $\alpha_1(1-\alpha_1) (\hat{\beta}_1 - \hat{\beta}_2)$ &
 $p(1-p) (\hat{r} -\hat{q})$ &
 0 &
 0\\
 \multicolumn{5}{l}{Entropy}  \\
  $\nabla\left[E_{xy}-E_{x}-E_{y}\right]$ & $\neq 0$ & $\neq 0$ &  0 & 0 \\
 \multicolumn{5}{l}{Correlation}  \\
 $\nabla \rho_{xy}$ & $\neq 0$ & $\neq 0$ &
 0 &
 0\\
  &  &  &  & \\
  &  &  &  & \\
\end{tabular}
 \caption[Mixed and Behavioural Spaces and Isomorphic Constraints]{\em A comparison of calculated results
 for mixed ${\cal P}_M$
 and behavioural ${\cal P}_B$ strategy spaces with those same spaces when
 subject to isomorphic constraints.  We examine points where respectively
 the $x$ and $y$ variables are first perfectly correlated with $\rho_{xy}=1$
 and then independent with $\rho_{xy}=1$. In the unconstrained behavioural spaces,
 all quantities are evaluated at points satisfying $\lim_{\beta_1 \rightarrow 1}$ or
 $\lim_{(q,r) \rightarrow (0,1)}$ when $\rho_{xy}=1$, and at points satisfying
 $\lim_{\beta_2 \rightarrow \beta_1}$ or  $\lim_{r\rightarrow q}$ when $\rho_{xy}=0$.
 The isomorphically constrained spaces are respectively
 indicated by $\left.{\cal P}_M\right|_{\beta_1=1}$ and $\left.{\cal P}_B\right|_{(q,r)=(0,1)}$
 for the perfectly correlated case, and
 $\left.{\cal P}_M\right|_{\beta_1=\beta_2}$ and $\left.{\cal
 P}_B\right|_{r=q}$ when the variables are independent.
 Game theory and probability theory assign
 different dimensionality and tangent spaces to these cases.   Many
 calculated results differ between these spaces. }
 \label{t_tangent_space_effects}
\end{table*}

\subsection{Isomorphic Mixed and Behavioural Spaces}

The mixed ${\cal P}_M$ and behavioural ${\cal P}_B$ strategy
spaces contain embedded probability spaces where $x$ and $y$ are
respectively perfectly correlated, independent, or partially
correlated. As previously, we will now perform a comparison of
probability spaces, both with and without isomorphic
constraints, for various correlation states between the $x$ and
$y$ variables.  That is, we will compare the mixed strategy
space ${\cal P}_M$ and behavioural strategy space ${\cal P}_B$
with isomorphically constrained mixed and behavioural strategy
spaces as indicated using the following notation.

The case of perfectly correlated $x$ and $y$ variables is
modeled by the spaces
\begin{equation}
\begin{array}{ll}
  \lim_{\beta_1\rightarrow 1} {\cal P}_M   &     {\rm mixed}   \\
  \left.{\cal P}_M\right|_{\beta_1=1}      &     {\rm constrained \; mixed}    \\
  \lim_{(q,r)\rightarrow (0,1)} {\cal P}_M &     {\rm behavioural}   \\
  \left.{\cal P}_B\right|_{(q,r)=(0,1)}    &     {\rm constrained \; behavioural}    \\
\end{array}
\end{equation}
In these spaces we expect all of the following to hold:
\begin{itemize}
\item $\nabla\left[P_{xy}(0,0)+P_{xy}(1,1)\right]=0$,
\item $\nabla\left[P_{xy}(0,1)+P_{xy}(1,0)\right]=0$,
\item $\nabla\left[P_{x|y}(0|0)\right]=0$,
\item $\nabla\left[P_{x|y}(0|1)\right]=0$,
\item $\nabla\left[\langle x\rangle - \langle
    y\rangle\right]=0$
\item $\nabla\left[\langle x\rangle - \langle
    xy\rangle\right]=0$
\item $\nabla\left[\langle y\rangle - \langle
    xy\rangle\right]=0$
\item
    $\nabla[V(x-y)]=\nabla\left[V(x)+V(y)-2\mbox{cov}(x,y)\right]=0$
\item $\nabla\left[E_{xy}-E_{x}\right]=0$.
\end{itemize}
Alternately, when $x$ and $y$ are independent, the relevant
spaces are
\begin{equation}
\begin{array}{ll}
  \lim_{\beta_1\rightarrow \beta_2} {\cal P}_M   &     {\rm mixed}   \\
  \left.{\cal P}_M\right|_{\beta_1=\beta_2}      &     {\rm constrained \; mixed}    \\
  \lim_{r\rightarrow q} {\cal P}_M               &     {\rm behavioural}   \\
  \left.{\cal P}_B\right|_{r=q}                  &     {\rm constrained \; behavioural}    \\
\end{array}
\end{equation}
In all these spaces, the probability distributions satisfy
\begin{itemize}
\item $\nabla\left[P_{xy}-P_xP_y\right]=0$
\item $\nabla\left[P_{x|y}-P_x\right]=0$
\item $\nabla\left[\langle xy\rangle-\langle x\rangle
    \langle y\rangle\right]=0$
\item $\nabla\left[E_{xy}-E_{x}-E_{y}\right]=0$.
\end{itemize}

Table \ref{t_tangent_space_effects} records whether each of the
expected relations is satisfied for each of the mixed and
behavioural spaces when they are either unconstrained, or
isomorphically constrained. As might be expected, the results
indicate that the weak isomorphisms used to construct the mixed
and behavioural spaces of game theory are not able to reproduce
necessarily true results from probability theory. Hence, the
rational player of game theory is unable to reliably reproduce
results from probability theory. These differences between game
theory and probability theory need to be resolved.

\section{Discussion}

The question posed in this chapter is whether a physical
situation involving variables $(x,y)$ defines a set of moves
$(x,y)\in\{(0,0),(0,1),(1,0),(1,1)\}$ which then defines a mixed
strategy space of three dimensions, or whether the variables
$(x,y)$ can be modeled by multiple distinct probability
distributions (perfectly correlated, independent,
anti-correlated, etc) each of which defines a set of possible
moves and corresponding mixed strategy space. These two
different approaches can each by modeled using a single mixed
strategy space with or without isomorphism constraints.  In this
case, the question is whether the simple physical decision or
game involving the variables $(x,y)$ is best modeled by a single
probability space which contains all others without using
isomorphic constraints and alters the properties of those
embedded spaces to reflect decision uncertainty, or by a single
probability space using isomorphic constraints to perfectly
preserve the properties of all embedded spaces.

 \chapter{A simple decision tree optimization}
 \label{chap_decision_tree}

\section{Optimizing simple decision trees}
\label{sect_Optimizing_simple_decision_trees}

We now turn to consider how the differences between probability
theory and game theory influence decision tree optimization. We
consider the usual two potentially correlated random variables
depicted in Fig. \ref{f_xy_decision_tree} and will use both the
unconstrained behavioural probability space ${\cal P}_B$ and the
isomorphically constrained behavioural spaces $\left.{\cal
P}_B\right|_{\rho_{xy}=\rho}$ for every value of the correlation
state $\rho\in[-1,1]$.  Our goal is to present an optimization
problem in which a rational player following the rules of game
theory cannot achieve the payoff outcomes of a player following
the rules of probability theory. We suppose that a player gains
a payoff by advising a referee of the parameters of the decision
tree probability space $(p,q,r)$ to optimize a given nonlinear
random function. The referee uses these parameters to determine
the value of the function and provides a payoff equivalent to
this value.  (If desired, the referee could estimate the
probability parameters by using indicator functions and
observing an ensemble average of decision tree outcomes.)

\subsection{Non-polylinear payoff functions}

There are many possible random functions which we could use, and
some are listed in Table \ref{t_tangent_space_effects}.  We
could choose any relations from this table of the form $f=0$
provided probability theory shows $\nabla f=0$ and game theory
has $\nabla f\neq 0$. When this is so, the function $\nabla f$
acts effectively as a discrepancy vector. We focus on the
squared magnitude of the length of the discrepancy vector and
examine functions of the form $F=1-|\nabla f|^2$. Immediately,
probability theory will optimize this function at the point
$F=1$ while game theory will locate an optimum at $F<1$.  In
particular, we choose
\begin{equation}
   f=P_{xy}(0,0)+P_{xy}(0,0)
\end{equation}
so
\begin{eqnarray}
   F &=& 1-\big|\nabla \left[P_{xy}(0,0)+P_{xy}(0,0)\right]\big|^2 \nonumber \\
     &=& 1-\big|\nabla \left[1-q+p(q+r-1)\right]\big|^2.
\end{eqnarray}

In the unconstrained behavioural space ${\cal P}_B$, a rational
player will evaluate this as
\begin{equation}
   F = 1-(1-q-r)^2-(1-p)^2-p^2.
\end{equation}
In turn, this will be maximized at points $p=\frac{1}{2}$ and
$q+r=1$ to give a maximum payoff of $F_{\rm max}=\frac{1}{2}$.

A contrasting result is obtained using the isomorphism
constraints of probability theory where our player faces the
optimization problem
\begin{eqnarray}
 \max F &=& 1-\big|\nabla \left[1-q+p(q+r-1)\right]\big|^2 \nonumber \\
     && \hspace{-1cm}
     \mbox{ subject to } \rho_{xy}=\rho, \;\;
                         \forall \rho\in[-1,1].
\end{eqnarray}
Our player might commence by adopting the constraint
$\rho_{xy}=1$ implemented by $(q,r)=(0,1)$ to give
\begin{eqnarray}
   \max F
      &=& \left. 1-\big|\nabla \left[1-q+p(q+r-1)\right]\big|^2 \right|_{(q,r)=(0,1)} \nonumber \\
      &=& 1.
\end{eqnarray}
This analysis leads to an optimum point at arbitrary $p$ and
$(q,r)=(0,1)$ and a maximum payoff of $F_{\rm max}=1$.
Self-evidently, the player would cease their optimization
analysis at this point as the achieved maximum can't be
improved.

\begin{figure}[htb]
\centering
\includegraphics[width=0.8\columnwidth,clip]{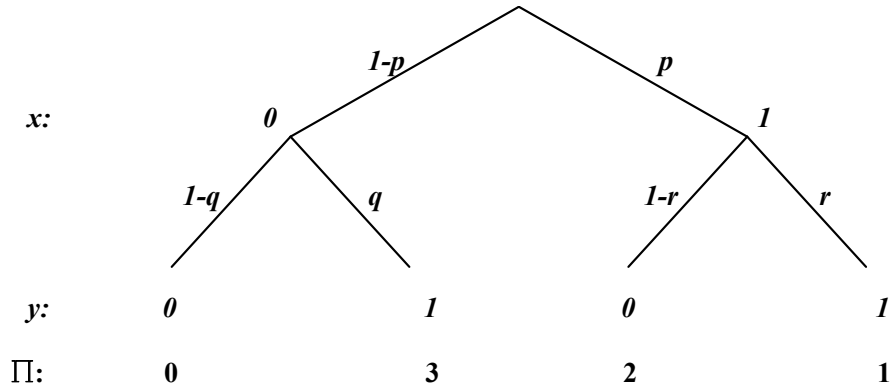}
\caption[A non-strategic decision tree]{\em A non-strategic decision tree over two stages
where a variable $x\in\{0,1\}$ is chosen in the first stage to condition the choice of a
second variable $y\in\{0,1\}$ in the second stage.  The attained payoffs $\Pi$ are as shown.
 \label{f_decision_game_full_tree}}
\end{figure}

\subsection{Polylinear payoff functions}

Of course, there are many random functions defined over decision
trees which produce identical results when using or not using
isomorphic constraints.  We now briefly illustrate this using
polylinear expected payoff functions, and consider optimizing
the function
\begin{eqnarray}
 \max \langle\Pi\rangle &=& 2 \langle x\rangle + 3 \langle y\rangle
                       - 4 \langle x y\rangle. \nonumber \\
     && \hspace{-1cm}
     \mbox{ subject to } \rho_{xy}=\rho, \;\;
                         \forall \rho\in[-1,1]
\end{eqnarray}
over the decision tree of Fig. \ref{f_decision_game_full_tree}.
Of course, simple inspection will locate the optimum at
$(\langle x\rangle,\langle y\rangle)=(0,1)$ giving an expected
payoff of $\langle\Pi\rangle=3$.  However, we step through the
process for later generalization to strategic games.

\begin{figure}[htb]
\centering
\includegraphics[width=0.8\columnwidth,clip]{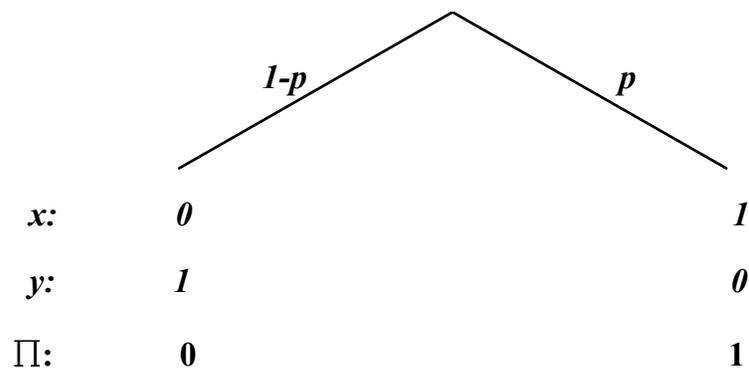}
\caption[A correlated decision tree]{\em The decision tree resulting when the variables
$x$ and $y$ are perfectly correlated.
 \label{f_decision_game_corr}}
\end{figure}

There are an infinite number of correlation constraints to be
examined, but several are straightforward.   As shown in Fig.
\ref{f_decision_game_corr}, when the variables are perfectly
correlated at $\rho_{xy}=1$ via the constraint $(q,r)=(0,1)$, we
have $\langle x\rangle=\langle y\rangle=\langle xy\rangle$
giving
\begin{equation}
 \langle\Pi\rangle = \langle x\rangle.
\end{equation}
This is optimized by setting $\langle x\rangle=1$ giving an
expected payoff of $\langle\Pi\rangle=1$.

\begin{figure}[htb]
\centering
\includegraphics[width=0.8\columnwidth,clip]{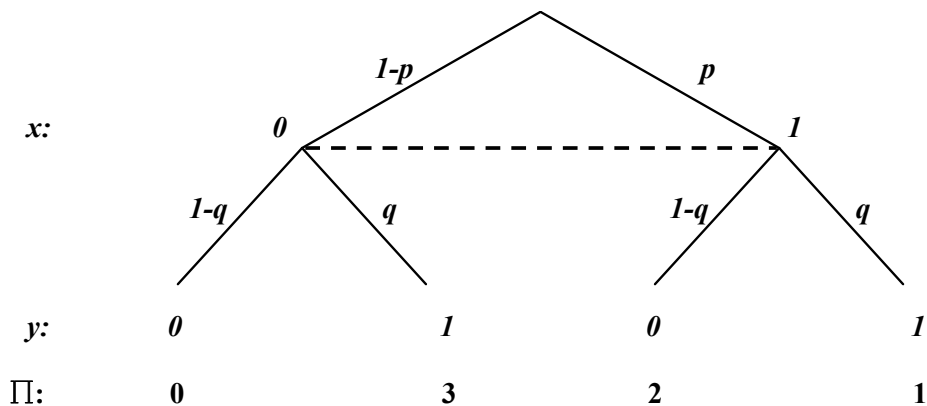}
\caption[An independent decision tree]{\em The decision tree resulting when the variables
$x$ and $y$ are independent.
 \label{f_decision_independent}}
\end{figure}

Fig. \ref{f_decision_independent} sets $\rho_{xy}=0$ so the $x$
and $y$ variables are independent by using the constraint $r=q$.
The expectations are now separable giving $\langle
xy\rangle=\langle x\rangle\langle y\rangle$ and
\begin{equation}
 \langle\Pi\rangle = 2 \langle x\rangle + 3 \langle y\rangle
                       - 4 \langle x\rangle \langle y\rangle.
\end{equation}
As the $\langle x\rangle$ and $\langle y\rangle$ variables are
independent, a check of internal stationary points and the
boundary leads to an optimal point at $(\langle x\rangle,\langle
y\rangle)=(0,1)$ and an expected payoff of
$\langle\Pi\rangle=3$.

\begin{figure}[htb]
\centering
\includegraphics[width=0.8\columnwidth,clip]{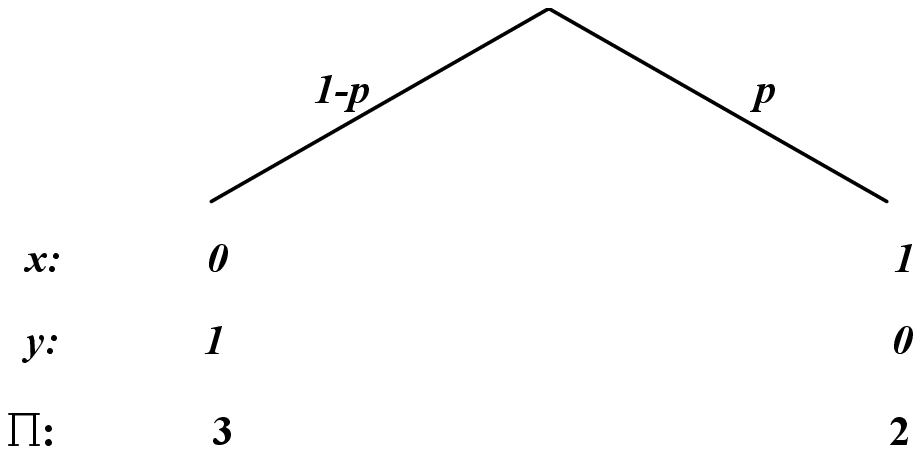}
\caption[An anti-correlated decision tree]{\em The decision tree resulting when the variables
$x$ and $y$ are perfectly anti-correlated.
 \label{f_decision_game_anticorr}}
\end{figure}

We lastly consider the case where the variables are perfectly
anti-correlated. As shown in Fig.
\ref{f_decision_game_anticorr}, when the variables are perfectly
correlated at $\rho_{xy}=-1$ via the constraint $(q,r)=(1,0)$,
we have $\langle y\rangle=(1-\langle x\rangle)$ and $\langle
xy\rangle=0$ giving
\begin{equation}
 \langle\Pi\rangle = 3-\langle x\rangle.
\end{equation}
This is optimized by setting $\langle x\rangle=0$ giving an
expected payoff of $\langle\Pi\rangle=3$.

More general correlation states require use of, for instance,
standard Lagrangian optimization procedures.

However, we here adopt a numerical optimization approach by
first using the correlation constraint to write the $r$ variable
as a function of $p$, $q$ and the correlation constant $\rho$,
giving a function $r=r_{+}(p,q,\rho)$.  In particular, when the
correlation (Eq. \ref{eq_rho_correlation}) between $x$ and $y$
is $\rho_{xy}=\rho$, and as long as both $p\neq 0$ and $p\neq
1$, then the correlation constraint defines two surfaces in the
$(p,q,r)$ simplex at height
\begin{equation}       \label{eq_r_plus_minus}
 r_{\pm}(p,q,\rho) =
     \frac{\rho^2 -2 q(1-p) (\rho^2 -1)
    \pm \rho  \sqrt{\rho^2+4q(1-q)\frac{(1-p)}{p}} }
    {2  \left[1 + p (\rho^2 -1) \right]}.
\end{equation}
The function $r_{+}(p,q,\rho)$ will give the correlation
surfaces we require within the simplex. That is, when $\rho=0$
we have $r_+(p,q,0)=q$ as required.  Similarly, when $\rho=1$ we
have $r_+(p,q,1)\geq 1$ across the entire $(p,q)$ plane with the
equality $r_+(p,q,1)=1$ only where $q=0$ or $q=1$. We require
$\rho=1$ at $(q,r)=(0,1)$. Finally, when $\rho=-1$ and $x$ and
$y$ are perfectly anti-correlated, we have $r_+(p,q,-1)\leq 0$
across the entire $(p,q)$ plane with the equality
$r_+(p,q,-1)=0$ only where $q=0$ or $q=1$. We require $\rho=-1$
at $(q,r)=(1,0)$.

The strict requirement that $0\leq r_+(p,q,\rho)\leq 1$
establishes permissible regions on the $(p,q)$ plane. For
$0<\rho<1$, the permissible region is bounded by the $q=0$ line
and the line
\begin{equation}    \label{eq_permissible_region1}
  q(p,\rho) = \frac{p}{p+ \frac{\rho^2}{1-\rho^2}}.
\end{equation}
Similarly, for $-1<\rho<0$, the $(p,q)$ region is bounded by the
$q=1$ line and the line
\begin{equation}   \label{eq_permissible_region2}
  q(p,\rho) = \frac{1}{1+p \frac{1-\rho^2}{\rho^2}}.
\end{equation}

The problem is then solved using a a typical Mathematica command
line of \cite{Pinter_2012}
{\bf%
\begin{eqnarray}
   && \mbox{NMaximize}[\{\mbox{inRange}[r_+(p,q,\rho)] \times
         \left[2p+3q-3pq-p r_+(p,q,\rho)\right], \nonumber \\
   && 0\leq p\leq 1 \mbox{ \&\& } 0\leq q\leq 1
       \},\{p,q\}].
\end{eqnarray}
}%
Here, a suitably defined ``inRange" function determines whether
$r_+$ is taking permissible values between zero and unity
allowing the payoff function to be examined over the entire
$(p,q)$ plane. The resulting optimal expected payoffs are
follows:
\begin{equation}
  \begin{array}{l|l|l|l}
  \rho&          (p,q,r)               & \langle\Pi\rangle \\ \hline
    +1      &  (1.,0.,1.)              &  1.               \\
    +0.75   &  (0.8138,0.3876,1.)      &  1.03032          \\
    +0.5    &  (0.4831,0.5917,1.)      &  1.40068          \\
    +0.25   &  (0.2590,0.7953,1.)      &  2.02693          \\
     0      &  (0.,1.,1.)              &  3.               \\
    -0.25   &  (0.,1.,0.9378)          &  3.               \\
    -0.5    &  (0.,1.,0.7506)          &  3.               \\
    -0.75   &  (0.,1.,0.4386)          &  3.               \\
    -1      &  (0.,1.,0.)              &  3.           \\
  \end{array}
\end{equation}
Some care must be taken to ensure convergence of the solution.
This analysis makes it evident that the player can maximize
expected payoffs by choosing a correlation constraint where $x$
and $y$ is independent (say) allowing the setting
$(p,q,r)=(0,1,1)$ to gain a payoff of $\langle\Pi\rangle=3$.
Other choices would also have been possible.

We now turn to applying isomorphism constraints to the strategic
analysis of game theory.

 \chapter{A simple two-player-two-stage optimization}
 \label{chap_simple_two_player}

\section{Optimizing a multistage game tree}
\label{sect_Optimizing_a_multistage_game_tree}

In this section, we show that the use of isomorphic constraints
can alter the outcomes of strategic games even when expected
payoff functions are being used.  We will consider either the
mixed strategy space ${\cal P}_M$ (Eq. \ref{eq_mixed_prob}) and
the behavioural strategy space ${\cal P}_B$ (Eq.
\ref{eq_behav_prob}) or the isomorphically constrained
behavioural spaces $\left.{\cal P}_B\right|_{\rho_{xy}=\rho}$
for every value of the correlation state $\rho\in[-1,1]$.

\begin{figure}[htb]
\centering
\includegraphics[width=0.8\columnwidth,clip]{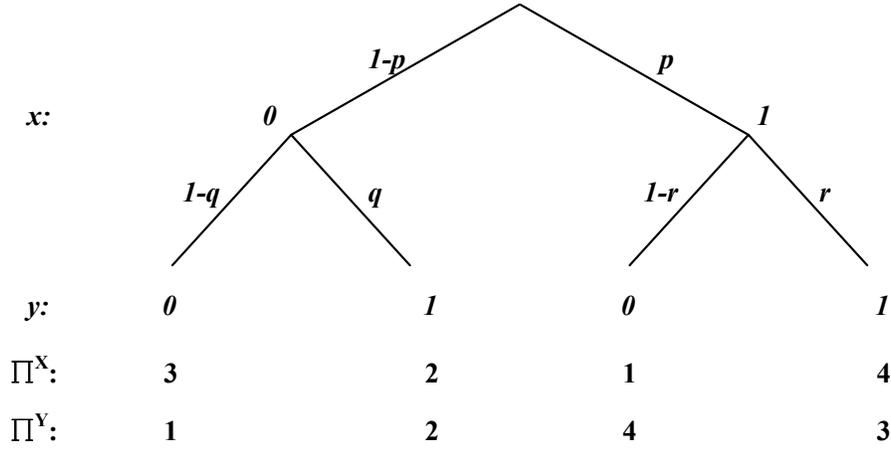}
\caption[A two-player strategic game]{\em Two players, $X$ and $Y$
conduct a two-stage sequential game where $X$ chooses the first variable
$x\in\{0,1\}$ and $Y$ chooses the second variable $y\in\{0,1\}$
conditioned on $x$.  The payoffs for players are $\Pi^X$ and $\Pi^Y$.
 \label{f_game_full_tree}}
\end{figure}

We consider a strategic interaction between two players over
multiple stages as depicted in Fig. \ref{f_game_full_tree}.
Here, two players denoted $X$ and $Y$ seek to optimize their
respective payoffs
\begin{eqnarray}
 X: \max \Pi^{X}(x,y) &=& 3 - 2 x - y + 4 x y \nonumber \\
 Y: \max \Pi^{Y}(x,y) &=& 1 + 3 x + y - 2 x y.
\end{eqnarray}
Again, we assume a domain $x,y\in\{0,1\}$ and that player $X$
chooses the value of $x$ and advises this to $Y$ before $Y$
determines the value of $y$. Players will either consider the
payoff functions above or their expectations
\begin{eqnarray}
 X: \max \langle\Pi^{X}\rangle &=& 3 - 2 \langle x\rangle - \langle y\rangle + 4 \langle x y\rangle \nonumber \\
 Y: \max \langle\Pi^{Y}\rangle &=& 1 + 3 \langle x\rangle + \langle y\rangle - 2 \langle x y\rangle.
\end{eqnarray}

\subsection{Unconstrained mixed space ${\cal P}_M$}

For the unconstrained mixed strategy space ${\cal P}_M$, the
expected payoffs for each player are
\begin{equation}                    \label{eq_mixed_payoffs}
  \begin{array}{c|cccc}
 (y_l,y_r)=&   (0,0)    & (0,1)    &  (1,0)   &   (1,1)   \\\hline
     (\langle\Pi^X\rangle,\langle\Pi^Y\rangle)
           &   \beta_0  & \beta_1  &  \beta_2 &   \beta_3 \\\hline
 \alpha_0  &    (3,1)   & (3,1)    &  (2,2)   &   (2,2)   \\
 \alpha_1  &    (1,4)   & (4,3)    &  (1,4)   &   (4,3).  \\
  \end{array}
\end{equation}
Using this table, the expected payoff functions take the form
\begin{eqnarray}
  \langle\Pi^X\rangle &=&
   3 - \beta_2 - \beta_3 + \alpha_1(-2 + 3 \beta_1 + \beta_2 + 4 \beta_3)   \nonumber \\
  \langle\Pi^Y\rangle &=&
   1 + \beta_2 + \beta_3 + \alpha_1(3 - \beta_1 - \beta_2 -2\beta_3)
\end{eqnarray}
while the unconstrained gradients evaluate as
\begin{eqnarray}
 \nabla  \langle\Pi^X\rangle &=&
       (-2 + 3 \beta_1 + \beta_2 + 4 \beta_3) \hat{\alpha}_1
              + 3 \alpha_1 \hat{\beta}_1
       + (\alpha_1 - 1) \hat{\beta}_2
       + (4 \alpha_1 - 1) \hat{\beta}_3   \nonumber \\
  \nabla  \langle\Pi^Y\rangle &=&
       (3 - \beta_1 - \beta_2 -2\beta_3) \hat{\alpha}_1
        - \alpha_1 \hat{\beta}_1
       + (1 - \alpha_1) \hat{\beta}_2
       +  (1 - 2 \alpha_1) \hat{\beta}_3.
\end{eqnarray}
The expected payoff can then optimized by either comparing
returns in the payoff table for each mixed strategy combination,
or by the equivalent strategy of comparing the simultaneous
rates of change of the payoff functions with the probability
parameters. (To illustrate the second approach, the rate of
change of $\langle\Pi^Y\rangle$ with $\beta_1$ is equal to
$-\alpha_1$ which is almost always negative indicating that
payoffs are maximized by setting $\beta_1=0$.) Either approach
then locates the optimal mixed strategy of
$(\alpha_1,\beta_1,\beta_2,\beta_3)=(0,0,1,0)$ leading to
expected payoffs of
$(\langle\Pi^X\rangle,\langle\Pi^Y\rangle)=(2,2)$.

\subsection{Unconstrained behavioural space ${\cal P}_B$}

The unconstrained behavioural strategy space ${\cal P}_B$ is
pictured in Fig. \ref{f_xy_decision_tree}. The unconstrained
optimization problem faced by each player is
\begin{eqnarray}          \label{eq_expected_Payoff_X_Y}
 X: \max_{p}   \langle\Pi^X\rangle &=& 3-2p-q+pq+3pr \nonumber \\
 Y: \max_{q,r} \langle\Pi^Y\rangle &=& 1+3p+q-pq-pr.
\end{eqnarray}
The unconstrained gradients of the expected payoffs evaluate as
\begin{eqnarray}
 \nabla \langle\Pi^X\rangle &=& (q+3r-2) \hat{p}
                                - (1-p) \hat{q}
                                + 3p \hat{r} \nonumber \\
 \nabla \langle\Pi^Y\rangle &=& (3-q-r) \hat{p}
                                + (1-p) \hat{q}
                                - p \hat{r}.
\end{eqnarray}
This perfect information game can then be optimized by
inspection, or by equating gradients to zero, or by using
backwards induction.  The resulting optimal pure strategy
choices are $(x,y)=(0,1)$ giving payoffs of
$(\Pi^X,\Pi^Y)=(2,2)$.

\subsection{Constrained behavioural space $\left.{\cal P}_B\right|_{\rho_{xy}=\rho}$}

We now consider the constrained behavioural spaces $\left.{\cal
P}_B\right|_{\rho_{xy}=\rho}, \forall \rho\in[-1,1]$. The two
players are non-communicating and it is generally not possible
to use a single value for the correlation $\rho$, and this
generally makes the analysis intractable. However, player $Y$
has total control over the setting of the correlation $\rho$ in
three cases---when $\rho=\pm 1$ and $\rho=0$.  We consider these
cases now.

First consider the space ${\cal P}_{B}|_{\rho_{xy}=1}$ in which
the variables are functionally equal so $y=x=xy$. (We can
consider the payoff functions directly rather than their
expected values.) In this space the players face the respective
optimization tasks
\begin{eqnarray}
 X: \max_x \Pi^{X}(x) &=& 3 +  x \nonumber \\
 Y:        \Pi^{Y}(x) &=& 1 +  2x.
\end{eqnarray}
As a result, player $X$ optimizes their payoff by setting $x=1$
giving the outcomes $(\Pi^{X},\Pi^{Y})=(4,3)$.

In contrast, in the space ${\cal P}_{B}|_{\rho_{xy}=-1}$, the
variables are functionally related by $y=1-x$ and $xy=0$.  These
constraints render the optimization tasks as
\begin{eqnarray}
 X: \max_x \Pi^{X}(x) &=& 2 -  x \nonumber \\
 Y:        \Pi^{Y}(x) &=& 2 + 2 x.
\end{eqnarray}
Here, player $X$ chooses $x=0$ to optimize their payoff leading
to the outcomes $(\Pi^{X},\Pi^{Y})=(2,2)$.

Finally, when player $Y$ chooses to discard all information
about the $x$ variable, then the variables $x$ and $y$ are
independent and the chosen space is ${\cal
P}_{B}|_{\rho_{xy}=0}$.  When the variables are independent,
there might not necessarily be a pure strategy solution and we
need to optimize expected payoffs.  In this space, we have
$\langle x\rangle=p$ and $\langle y\rangle=q$ and $\langle
xy\rangle=\langle x\rangle\langle y\rangle=pq$ giving the
optimization problem
\begin{eqnarray}
 X:\max_p \langle\Pi^X\rangle &=&
     3 - 2 p - q + 4 p q \nonumber \\
 Y:\max_q \langle\Pi^Y\rangle &=&
     1 + 3 p + q - 2 p q.
\end{eqnarray}
The best response functions or equivalent partial differentials
are
\begin{eqnarray}
 X:  \frac{\partial \langle\Pi^X\rangle}{\partial p}  &=&
     - 2 + 4  q \nonumber \\
 Y: \frac{\partial \langle\Pi^Y\rangle}{\partial q}  &=&
     1 - 2 p
\end{eqnarray}
locating the optimal point at $(p,q)=(\frac{1}{2},\frac{1}{2})$
with expected payoffs of
$(\langle\Pi^X\rangle,\langle\Pi^Y\rangle)=(\frac{5}{2},\frac{5}{2})$.

At this stage of the analysis, both players have separately
calculated an equilibrium point in three spaces ${\cal
P}_{B}|_{\rho_{xy}=\rho}$ for $\rho\in\{-1,0,1\}$, and the
selection of these correlation states is solely at the
discretion of player $Y$. The expected payoffs gained at each of
these ``local" equilibrium points can then be compared to obtain
a ``global" optimal expected payoff. For convenience, these are
summarized here:
\begin{equation}
 \begin{array}{cc}
   \rho &  (\langle\Pi^X\rangle,\langle\Pi^Y\rangle) \\ \hline
       -1       &     (2,2)                          \\
       0        &     (\frac{5}{2},\frac{5}{2})      \\
       +1       &     (4,3).                         \\
 \end{array}
\end{equation}
Based on these results, player $Y$ will then rationally optimize
their expected payoff by choosing to have their variables in a
state of perfect correlation with $\rho=1$ in the space ${\cal
P}_{B}|_{\rho_{xy}=1}$.  Player $X$, also being a rational
optimizer will play accordingly to give equilibrium payoffs of
$(\langle\Pi^X\rangle,\langle\Pi^Y\rangle)=(4,3)$.

\begin{figure*}[htb]
\centering
\includegraphics[width=\textwidth,clip]{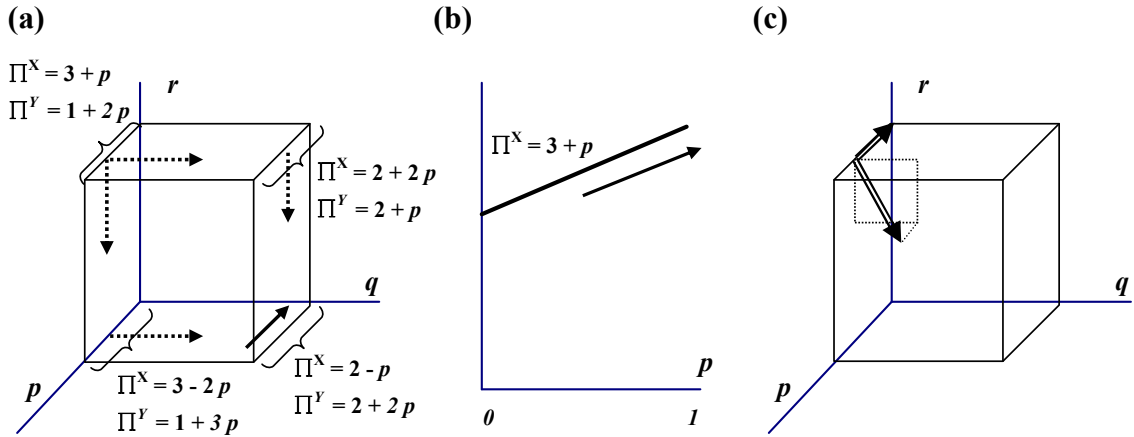}
\caption[Alternate probability space gradients]{\em (a) Game
theory adopts an unconstrained joint probability measure space
in which expected payoffs vary over three dimensions $(p,q,r)$
and where positive gradients with respect to $q$ and $r$ (dotted
arrows) and with respect to $p$ (solid arrow) ensure that
players maximize joint payoffs by choosing $(p,q,r)=(0,1,0)$.
(b) An alternate joint probability space where $x$ is perfectly
correlated to $y$ in which expected payoffs vary over a single
dimension $p$ with positive gradients with respect to $p$ (solid
arrow) ensuring that players optimize payoffs by choosing $p=1$.
(c) The choice of two alternate probability spaces (more are
possible) associates two different total gradients (double-lined
arrows) with any point along the perfect correlation line
$\rho_{xy}=1$ at $(q,r)=(0,1)$. In the absence of any effective
decision procedure privileging any one space over another,
players should examine all possible spaces, all possible
gradients, and all possible optimized outcomes.
 \label{f_multiple_spaces}}
\end{figure*}

It is useful again to reemphasize a geometric picture. As shown
in Fig. \ref{f_multiple_spaces}(a), an unconstrained behavioural
space has a three-dimensional gradient everywhere which is
non-zero even when $x$ and $y$ are perfectly correlated so
payoffs are not optimized at any such points. In contrast, the
use of isomorphic constraints when the $x$ and $y$ variables are
perfectly correlated gives the situation in Fig.
\ref{f_multiple_spaces}(b) where now a 1-dimensional gradient
points solely along the $\hat{p}$ axis.  A comparison in Fig.
\ref{f_multiple_spaces}(c) of the resulting outcomes can then be
made to determine which probability space should be chosen so as
to maximize outcomes.

\subsection{Strategic analysis difficulties}

The players might then seek to supplement the above solutions by
considering a wider range of correlation states.The optimization
task then becomes
\begin{eqnarray}              \label{eq_strategic_behavioural2}
 X: \max_{p}   \langle\Pi^X\rangle &=& 3-2p-q+pq+3pr \nonumber \\
 Y: \max_{q,r} \langle\Pi^Y\rangle &=& 1+3p+q-pq-pr \nonumber \\
     && \hspace{-2cm}
     \mbox{ subject to } \rho_{xy}=\rho, \;\;
                         \forall \rho\in[-1,1].
\end{eqnarray}
Unfortunately, there does not seem to be any straightforward way
to make progress with the general correlation case. Players are
non-communicating and hence cannot agree on a value of the
correlation state $\rho$.  If players adopt different values of
the correlation states they must model conflicting global
constraints and it is not clear how these can be resolved. An
attempt to model the use of a single correlation state generates
expected payoff functions which are not poly-linear in the
probability parameters and that are not generally quasi-concave.
This implies that existence theorems for Nash equilibria are
inapplicable in these cases so equilibrium points might not
exist for different correlation states. It is more than likely
that an acceptable solution methodology does not exist for
strategic interactions in the general correlation case, and it
is beyond the scope of this paper to consider this issue
further.  Here finally, we find the irreducible complexity of
strategic analysis expected by von Neumann and Morgenstern.

\subsection{More general constrained analysis}

The choice of variable $y$ is normally modeled as requiring two
separate and independent coin tosses---see the behavioural space
tree of Fig. \ref{f_game_full_tree}. When $x=0$ a coin is tossed
determining $y=u\in\{0,1\}$ with respective probabilities
$(1-q,q)$, while when $x=1$ another coin is tossed determining
$y=v\in\{0,1\}$ with respective probabilities $(1-r,r)$.  The
$u$ and $v$ coins are then simple, biassed, independent coins.

However, there is no need for this simplest possible treatment.
The $u$ and $v$ coins could themselves be modeled using any of
the alternate probability spaces of Eqs.
\ref{eq_alternate_top}---\ref{eq_alternate_bottom}.  These
alternate probability spaces would need to be checked by
rational players of unbounded capacity.

Another possible probability space might consider the $u$ and
$v$ variables themselves to be partially correlated.  That is,
the second stage player chooses to partially correlate their two
behavioural strategies by employing two sequential roulettes.
The first determines the variable $u\in\{0,1\}$ with
probabilities $(1-q,q)$ while the second gives $v\in\{0,1\}$
with respective probabilities $(1-r_1,r_1)$ if $u=0$ and
$(1-r_2,r_2)$ if $u=1$. The resulting correlation between the
variables $u$ and $v$ is then
\begin{equation}
    \rho_{uv}(q,r_1,r_2)
    = \frac{\sqrt{q(1-q)}(r_2-r_1)}{
    \sqrt{\left[r_1+q(r_2-r_1)\right]\left[1-r_1-q(r_2-r_1)\right]}}.
\end{equation}
When $r_1=r_2$ then these variables are uncorrelated as usual.
In turn, this correlation between the $u$ and $v$ variables
renders the correlation between the $x$ and $y$ variables as
\begin{equation}
    \rho_{xy}(p,q,r_1,r_2)
    = \frac{\sqrt{p(1-p)}[r_1-q(1+r_1-r_2)]}{
    \sqrt{\left[q+p(r_1-q)+pq(r_2-r_1)\right]\left[1-q+p(r_1-q)+pq(r_2-r_1)\right]}}.
\end{equation}
The second stage player might then choose to adopt a probability
space with a constant correlation between the $u$ and $v$
variables, say $\rho_{uv}(q,r_1,r_2)=\bar{\rho}_{uv}$ say. If
$\bar{\rho}_{uv}=0$ then we have the usual situation of
uncorrelated behavioural strategies normally considered by game
theory. Conversely, if $\bar{\rho}_{uv}=\pm 1$ we have
respectively either perfectly correlated or perfectly
anti-correlated behavioural strategies.  If such a correlation
constraint can be adopted, then both players should analyze this
possibility to determine whether it is optimal.

\begin{figure}[htbp]
\centering
\includegraphics[width=0.6\columnwidth,clip]{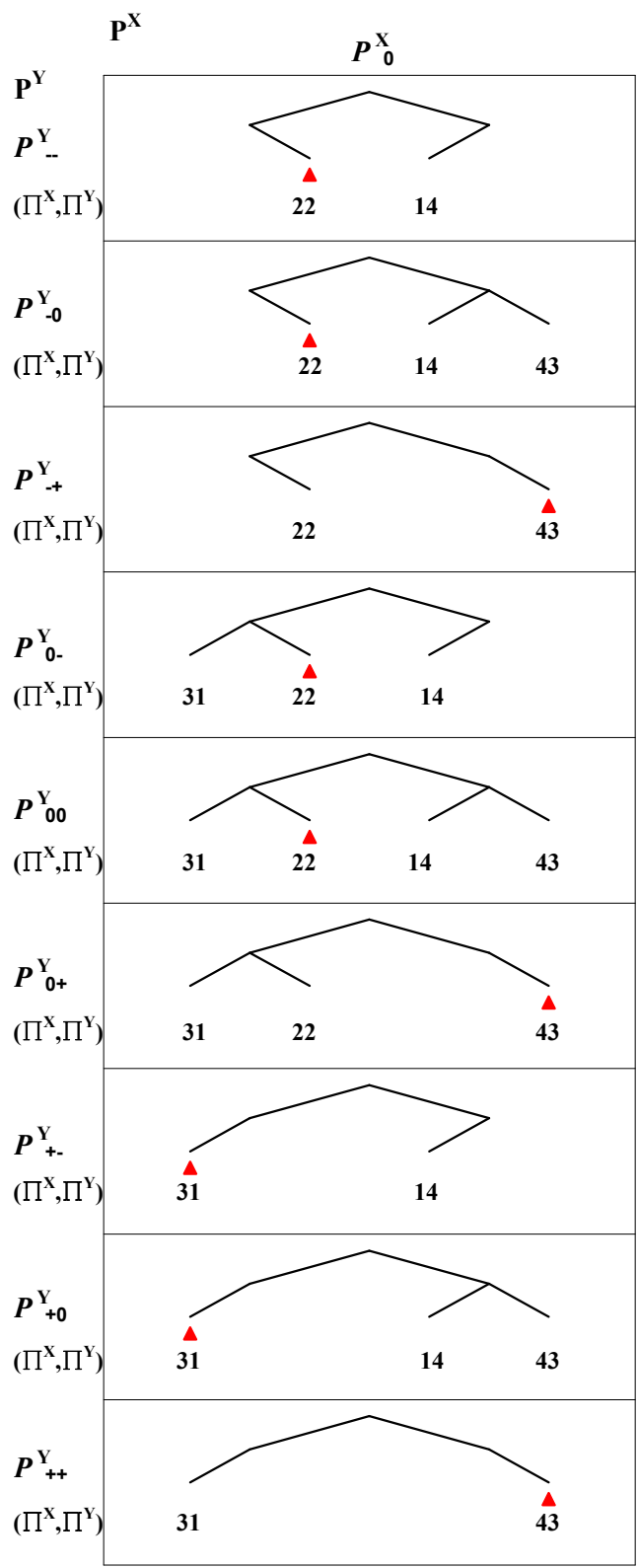}
\caption[Decision trees, payoffs and equilibria for
a simple game]{\em The nine distinct trees, payoffs and equilibria
(indicated by triangles) given that players $X$ and $Y$ adopt the
indicated joint probability space.  The two subscript symbols here respectively
indicate whether each of player $Y$'s second stage choices are
perfectly anti-correlated (``$-$"), uncorrelated (``$0$"), or
perfectly correlated (``$+$") to the previously observed random
variable $x$.
 \label{f_simple_extended}}
\end{figure}

Even more strangely, the $u$ and $v$ coin tosses could
themselves be partially correlated to the previous choice of
$x$.  That is, the $u$ and $v$ variables can be correlated with
$x$, and only after they have been chosen is the value for $y$
assigned.  For example, we might have $u$ perfectly
anti-correlated with $x$ so $u=1-x$ and  $v$ perfectly
correlated with $x$ so $v=x$, and then we assign $y=u$ if $x=0$
and $y=v$ if $x=1$. There are many possible choices that might
be considered. In particular, we might consider the 9 possible
cases which arise when firstly the $u$ variable is either
perfectly anti-correlated to $x$ (denoted ${\cal P}^Y_{-.}$),
independent of $x$ (${\cal P}^Y_{0.}$) or perfectly correlated
to $x$ (${\cal P}^Y_{+.}$), and the $v$ variable is either
perfectly anti-correlated to $x$ (denoted ${\cal P}^Y_{.-}$),
independent of $x$ (${\cal P}^Y_{.0}$) or perfectly correlated
to $x$ (${\cal P}^Y_{.+}$). We have introduced subscript symbols
indicating these possibilities. That is, we separately have
\begin{eqnarray}
    P^Y(u) &=& \left\{
      \begin{array}{cc}
        \delta_{u(1-x))} & {\cal P}^Y_{-.} \\
         &  \\
        (1-q,q) & {\cal P}^Y_{0.} \\
         &  \\
        \delta_{ux} & {\cal P}^Y_{+.} \\
      \end{array} \right.  \nonumber \\
    P^Y(v) &=& \left\{
      \begin{array}{cc}
        \delta_{v(1-x)} & {\cal P}^Y_{.-} \\
         &  \\
        (1-r,r) & {\cal P}^Y_{.0} \\
         &  \\
        \delta_{vx} & {\cal P}^Y_{.+} \\
      \end{array} \right. .
\end{eqnarray}
The right hand column here lists the shorthand notation for each
adopted strategy.  This notation shows that if $u$ is
independent of $x$ while $v$ is perfectly correlated to $x$, the
second stage probability distribution adopted by player $Y$ is
${\cal P}^Y_{0+}$. Similarly, when both $u$ and $v$ are
perfectly correlated to $x$ we have the probability distribution
${\cal P}^Y_{++}$. Each of these choices of a different
probability space generates a different optima within that
space, and these optima must be compared so that players can
decide which space they can rationally choose. Without showing
the details, the generated outcomes in these possible spaces are
\begin{equation}
    \begin{array}{c|c}
                                                & (\langle\Pi^X\rangle,\langle\Pi^Y\rangle) \\ \hline
       {\cal P}^Y_{--}                          &      (2,2)    \\
       {\cal P}^Y_{-0}                          &      (2,2)    \\
       {\cal P}^Y_{-+}                          &      (4,3)    \\
       {\cal P}^Y_{0-}                          &      (2,2)    \\
       {\cal P}^Y_{00}                          &      (2,2)    \\
       {\cal P}^Y_{0+}                          &      (4,3)    \\
       {\cal P}^Y_{+-}                          &      (3,1)    \\
       {\cal P}^Y_{+0}                          &      (3,1)    \\
       {\cal P}^Y_{++}                          &      (4,3).   \\
    \end{array}
\end{equation}
These outcomes can easily be verified by drawing the different
trees generated by each choice of joint probability space as
shown in Fig. \ref{f_simple_extended}. This extended table of
distinct trees makes evident that again, within this range of
considered joint probability spaces, player $Y$ optimizes their
outcomes by choosing, for instance, the space ${\cal P}^Y_{++}$
ensuring that their choice is perfectly correlated with that of
their opponent.

We argue that optimizing multiple-player-multiple-stage games is
more complicated than envisaged in conventional game analysis.
As noted earlier, the strategic optimization of expected payoffs
first requires the adoption of a suitable joint probability
measure space, and it is only the adoption of such a space that
permits the functional definition of both the expected payoff
and suitable gradient operators allowing the optimization to be
completed. For the above simple two player game, the expected
payoffs and gradient operators have been respectively defined
variously as
\begin{equation}
    \left( \langle\Pi^X\rangle, \langle\Pi^Y\rangle\right)=
    \left\{
 \begin{array}{cc}
  \left( 2-p,2+2p\right) & {\cal P}^Y_{--} \\
  &  \\
  \left( 2-p+3pr,2+2p-pr\right) & {\cal P}^Y_{-0} \\
  &  \\
  \left( 2+2p, 2+p\right) & {\cal P}^Y_{-+}  \\
  &  \\
  \left( 3-2p-q+pq, 1+3p+q-pq\right) & {\cal P}^Y_{0-} \\
  &  \\
  \left( 3-2p-q+pq+3pr, 1+3p+q-pq-pr \right) & {\cal P}^Y_{00} \\
  &  \\
  \left( 3+p-q+pq, 1+2p+q-pq \right) & {\cal P}^Y_{0+} \\
  &  \\
  \left( 3-2p, 1+3p\right)   & {\cal P}^Y_{+-}  \\
  &  \\
  \left( 3-2p+3pr, 1+3p-pr\right)   & {\cal P}^Y_{+0} \\
  &  \\
  \left( 3+p, 1+2p \right)  & {\cal P}^Y_{++}  \\
 \end{array}
    \right.
\end{equation}
and
\begin{equation}
    \left[ \nabla^X,\nabla^Y\right] =
    \left\{
 \begin{array}{cc}
  \left[ \left( \frac{\partial}{\partial p} \right), . \right] & {\cal P}^Y_{--} \\
  &  \\
  \left[ \left( \frac{\partial}{\partial p} \right),\left( \frac{\partial}{\partial r} \right)\right] & {\cal P}^Y_{-0} \\
  &  \\
  \left[ \left( \frac{\partial}{\partial p} \right), . \right]  & {\cal P}^Y_{-+} \\
  &  \\
  \left[ \left( \frac{\partial}{\partial p} \right),\left( \frac{\partial}{\partial q} \right)\right] & {\cal P}^Y_{0-} \\
  &  \\
  \left[ \left( \frac{\partial}{\partial p} \right),\left( \frac{\partial}{\partial q},\frac{\partial}{\partial r} \right)\right] & {\cal P}^Y_{00} \\
  &  \\
  \left[ \left( \frac{\partial}{\partial p} \right),\left( \frac{\partial}{\partial q}  \right) \right] & {\cal P}^Y_{0+}  \\
  &  \\
  \left[ \left( \frac{\partial}{\partial p} \right), . \right]  & {\cal P}^Y_{+-} \\
  &  \\
  \left[ \left( \frac{\partial}{\partial p} \right), \left( \frac{\partial}{\partial r} \right)\right] & {\cal P}^Y_{+0} \\
  &  \\
  \left[ \left( \frac{\partial}{\partial p} \right), . \right]  & {\cal P}^Y_{++}. \\
 \end{array}
    \right.
\end{equation}
That is, the expected payoff is defined here as a joint
functional mapping from the various probability measure spaces
to the reals via
\begin{eqnarray}
  \left( \langle\Pi^X\rangle, \langle\Pi^Y\rangle\right)&:&
      \left\{
      {\cal P}^X_0\times{\cal P}^Y_{--},
      {\cal P}^X_0\times{\cal P}^Y_{-0},
      {\cal P}^X_0\times{\cal P}^Y_{-+},
      {\cal P}^X_0\times{\cal P}^Y_{0-},
      {\cal P}^X_0\times{\cal P}^Y_{00},  \right. \nonumber \\
  && \left.
      {\cal P}^X_0\times{\cal P}^Y_{0+},
      {\cal P}^X_0\times{\cal P}^Y_{+-},
      {\cal P}^X_0\times{\cal P}^Y_{+0},
      {\cal P}^X_0\times{\cal P}^Y_{++}
      \right\} \rightarrow{\rm I\mkern-3mu R}\times{\rm I\mkern-3mu R}.
\end{eqnarray}
Again, this is in sharp contrast to the usual definition of game
theory that it is sufficient for optimization to consider that
the expected payoff is defined as the joint function mapping
\begin{equation}
    \left( \langle\Pi^X\rangle, \langle\Pi^Y\rangle\right):
      {\cal P}^X_0\times{\cal P}^Y_{00}
       \rightarrow{\rm I\mkern-3mu R}\times{\rm I\mkern-3mu R}.
\end{equation}

\section{Backwards induction and isomorphism constraints}

We have mentioned above that backwards induction can be used to
solve the unconstrained optimization problem.  This approach is
often presented as a `proof' that no alternative procedure could
possibly be considered by a rational player.  It is worth taking
a closer look at what is involved in the backwards induction
algorithm, and how it interacts with isomorphic constraints.

Backwards induction first constrains the values of first stage
probability parameters and then evaluates the gradients of the
expected payoff function $\langle\Pi^Y\rangle$ at different
nodes in the last stage of the game.  These last stage gradients
are then used to set the optimal values of the $(q,r)$
probability variables. These values are then applied as
constraints to the evaluation of the gradient of the expected
payoff function $\langle\Pi^X\rangle$ in the first stage of the
game---the first stage probability parameters are now treated as
variables. To illustrate these steps, we choose to begin our
analysis at a point in the behavioural strategy space where the
variables are perfectly correlated at $(q,r)=(0,1)$. The steps
involved are:
\begin{eqnarray}
  \lim_{(q,r)\rightarrow (0,1)} \frac{\partial \langle\Pi^Y\rangle|_{p=0}}{\partial q}
     &=& 1 > 0, \mbox{ so } q\rightarrow 1 \nonumber \\
  \lim_{(q,r)\rightarrow (0,1)} \frac{\partial \langle\Pi^Y\rangle|_{p=1}}{\partial r}
     &=& -1 < 0, \mbox{ so } r\rightarrow 0 \nonumber \\
  \left.\frac{\partial \langle\Pi^X\rangle}{\partial p}\right|_{(q,r)=(1,0)}
     &=& -p < 0, \mbox{ so } p\rightarrow 0.
\end{eqnarray}
The optimal point is then at $(p,q,r)=(0,1,0)$ giving payoffs of
$(\langle\Pi^X\rangle,\langle\Pi^Y\rangle)=(2,2)$.

It is very easy and straightforward to apply the backwards
induction algorithm to an isomorphically constrained space,
provided that the global isomorphic constraints and the altered
geometry is taken into account.  If the variables $x$ and $y$
are perfectly correlated then the game tree reduces to a single
stage and backwards induction is properly applied to that single
stage.  However, problems arise when as is common, it is argued
that backwards induction must be applied to both stages even
when the $x$ and $y$ variables are perfectly correlated.  This
argument presupposes that backwards induction overrides
isomorphic constraints and the altered game geometry.

To see how this is done, let us imagine trying to apply the
backwards induction algorithm to an isomorphically constrained
perfectly correlated space $\left.{\cal
P}_B\right|_{(q,r)=(0,1)}$ with $\rho=1$. The above evaluations
then try to combine limit processes, gradient evaluations, and
isomorphic constraints with global scope.  That is:
\begin{eqnarray}
  \left. \lim_{(q,r)\rightarrow (0,1)} \frac{\partial \langle\Pi^Y\rangle|_{p=0}}{\partial q}\right|_{(q,r)=(0,1)}
     &=& ? \nonumber \\
  \left. \lim_{(q,r)\rightarrow (0,1)} \frac{\partial \langle\Pi^Y\rangle|_{p=1}}{\partial r}\right|_{(q,r)=(0,1)}
     &=& ? \nonumber \\
  \left.\left.\frac{\partial \langle\Pi^X\rangle}{\partial p}\right|_{(q,r)=(0,1)}\right|_{(q,r)=(1,0)}
     &=& ?.
\end{eqnarray}
Mathematically and logically, these statements make little
sense.  An isomorphic constraint of global scope sets the values
$(q,r)=(0,1)$ and then backwards induction seeks to treat these
parameters as variables and evaluate a gradient with respect to
these variables. In actuality, these variables no longer exist
in this constrained probability space as there is no second
stage in this probability space.  The altered probability space
geometry has altered the game try to include only one stage and
one probability parameter.

Let us try a slightly more general treatment. Consider briefly
the optimization by player $Z\in\{X,Y\}$ of an example two stage
game where $x$ is known before $y$ is decided giving
\begin{equation}
    \langle\Pi^Z\rangle = \sum_{x,y=0}^{1}
       P^X(x) P^Y(y|x) \Pi^Z(x,y).
\end{equation}
The conventional analysis begins by drawing a single game tree
capturing every possible move that might be made along every
history, and assigning independent distributions to each
decision point which can then be optimized via backwards
induction. Then, backwards induction begins by optimizing the
last stage first via, for instance, evaluations like
\begin{eqnarray}
  \frac{\partial \langle\Pi^Z\rangle}{\partial P^Y(y'|x')}
   &=& \frac{\partial }{\partial P^Y(y'|x')}
         \sum_{x,y=0}^{1}
       P^X(x) P^Y(y|x) \Pi^Z(x,y) \nonumber \\
   &=& P^X(x')  \frac{\partial }{\partial P^Y(y'|x')}
         \left[
          \left(1-P^Y(y'|x')\right) \Pi^Z(x',1-y') +
          P^Y(y'|x') \Pi^Z(x',y')
          \right] \nonumber \\
   &=&     P^X(x') \left(\Pi^Z(x',y')- \Pi^Z(x',1-y')\right).
\end{eqnarray}
Implicit in this evaluation, is the assumption that the gradient
operator $\frac{\partial }{\partial P^Y(y'|x')}$ commutes with
the distribution $P^X(x')$ via
\begin{equation}
  \frac{\partial }{\partial P^Y(y'|x')} P^X(x') =
   P^X(x') \frac{\partial }{\partial P^Y(y'|x')}.
\end{equation}
This is only true under the assumption that the distributions
$P^Y(y|x)$ and $P^X(x)$ are not functionally dependent.  When
this is not the case, then obviously, the above commutation
relation cannot be used. Speaking figuratively, for longer $N$
stage games, backwards induction relies on similar independence
assumptions allowing gradients with respect to $i^{\rm th}$
stage distributions $P_i$ to commute with all earlier stage
distributions, giving (loosely)
\begin{equation}
 \max_{P_1,P_2,\dots,P_{N-1},P_N} \langle\Pi^Z\rangle
 = \max_{P_1} \; \left[ \sum \dots
     \max_{P_2} \; \left[ \sum \dots
     \max_{P_{N-1}} \; \left[ \sum \dots
     \max_{P_N} \; \left[ \sum \dots
    \right]
   \right]
  \right]
 \right]
\end{equation}
Again, commuting latter stage gradient operators through all
preceding earlier stage distributions is only valid under the
assumption that these distributions are not functionally
dependent. These assumptions are not necessarily true, and we
suggest that rational players will consider the case where they
are not warranted.

In our approach in contrast, we hold that the functionals
$\langle\Pi^Z\rangle$ cannot be represented by a single game
tree of finite size, and that they possess neither
dimensionality nor continuity properties. While they are a
mapping into a range of reals, their domain sets are essentially
unspecified.  In fact, and crudely put, if $S$ is the set of all
possible feasible spaces for this game, say $S=\{{\rm
I\mkern-3mu R}^1,{\rm I\mkern-3mu R}^2,\dots\}$, then the
functional is a mapping from the set of all possible feasible
spaces to the reals, $\langle\Pi^Z\rangle:S\rightarrow{\rm
I\mkern-3mu R}$.  Just as a topological space possesses
dimensionality but lacks any measure of distance and only gains
such measures with the adoption of a metric, these expected
payoff functionals do not even possess dimensionality prior to
the adoption of a suitable probability measure space.  In fact,
the mapping $\langle\Pi^Z\rangle$ must be defined over every
possible probability measure space.  For all these possible
space, within any adopted probability measure space,
$\langle\Pi^Z\rangle$ becomes a function of fixed dimensionality
and specified continuity and differentiability properties which
can be described by a suitable decision tree. Such a tree then
supports the backwards induction and subgame decomposition
operations which can then be used to optimize pathways through
this particular tree, one instance among many of the trees
definable using the entire mapping $\langle\Pi^Z\rangle$.

The adoption of a probability measure space inducing
correlations between any game variables alters the structure of
the decision tree to create an irreducible whole entity which
must be optimized as a single unit.  Backwards induction and
subgame decompositions cannot be improperly used to break these
indivisible units as any such attempt is simply mathematically
invalid.  This has profound implications canvassed later for the
evolution of hierarchical complexity.

When player $Y$ chooses an alternate probability space such as
${\cal P}^Y_{++}$ in which all of the second stage choices are
perfectly correlated with their opponent's previous move, then
they possess no free parameters and so have nothing to vary to
optimize their payoff. This restriction of their ability to vary
their second stage choice has been implicitly considered to be a
reason for not using the correlated probability space ${\cal
P}^Y_{++}$ in favour of the conventional space ${\cal
P}^Y_{00}$. This latter probability space allows players to
consider all possibilities in the second stage, thus justifying
the use of this probability space. However, this is a misleading
argument. No reasons have ever been provided for why a player
should restrict their analysis to a single space. Lifting this
restriction requires them in turn to choose which space offers
them the greatest range of choice. Rather, the player can
perform their optimization by first choosing among the infinite
number of available probability spaces, and then optimizing over
every parameter defined within each space. In some spaces they
consider, they will possess a certain number of parameters to
vary, and in other spaces they will possess a different number
of parameters to vary. Certainly, some spaces will offer no free
parameters to vary, but nothing is lost by having a player
consider this as one option among many. It is the conventional
analysis which restricts player searches by forcing them to
consider only a single type of probability space.

It has also been argued that, even when player $Y$ intends to
adopt correlated second stage play, their observation that
player $X$ chooses $x=0$ in the first stage will require player
$Y$ to rethink their desire to adopt a correlated strategy so
they should then seek to optimize their outcomes given that the
choice $x=0$ has been made. In effect, this argument presupposes
that player $Y$ has adopted the conventional probability space
which allows this player to have a further choice in the second
stage. As emphasized above, one of the firmest results of
probability measure theory is that joint probability
distributions are separable if and only if all the variables are
independent. That is, different variables can be separately
optimized if and only if they are are described by separable
joint probability distributions and this occurs if and only if
they are independent. This means that it is only when variables
are independent that a subgame decomposition be performed
allowing players to separately optimize decisions in each
subgame. It is a nonsense to argue that non-independent and
non-separable variables are actually separable and hence
separately optimized. When player $Y$ has made a prior choice to
adopt the probability space ${\cal P}^Y_{++}$, then they have
freely chosen not to have a choice in the second stage, and they
will compare the payoffs stemming from this choice with those
available from alternate choices.

To reiterate previous points, a coin consists of many components
possessing correlated dynamics, and these correlations permit
the construction of a coin decision tree with only two branches
indicating Heads or Tails.  A pseudo-random number generator
consists of millions of components all possessing correlated
dynamics so again, the total decision tree might possess only
two branches.  Correlation between variables reduces the size of
decision trees, and alters the dimensionality of expected payoff
functional spaces.

\section{Optimizing over multiple joint probability spaces}

We now have multiple possible joint probability spaces.  In
these alternate spaces, the expected payoff functions possess
exactly the same value when $x$ and $y$ are perfectly correlated
but possess entirely different gradients at this point.
Variational optimization principles insist that every possible
functional form and gradient must be taken into account in any
complete optimization.  These principles permit players to
infinitely vary the ``immutable" functional assignments defining
any space (i.e. $y=\delta_{x0}u+\delta_{x1}v$ and $y=x$ above),
providing access to a vastly larger decision space than usually
analyzed in game theory. It is not a question of which space is
best, rather, it is a question of either restricting the
analysis to a single space or allowing players to analyze all
possible spaces.

Game theory adopts expected payoff ``functions" allowing
examination of every possible combination of payoff values and
assumes that this is sufficient for optimization.  However,
while these functions can duplicate every possible payoff value,
they cannot duplicate every possible functional dependency or
gradient---and optimization depends on these dependencies and
gradients.

More generally, in our approach, rational players are able to
perform an entirely unconstrained search of every possible joint
probability space to optimize their payoffs via
\begin{eqnarray}
 X\!\!: \; \max_{{\cal P}^X}\;\; \langle \Pi^X \rangle \!\!
    &=& \!\! \int_{\Omega^X\!\!\times\Omega^Y}
         dP^{XY}_{xy} \;\;
         \Pi^X(x,y) \nonumber \\
    & & \\
 Y\!\!: \; \max_{{\cal P}^Y} \;\; \langle \Pi^Y \rangle \!\!
    &=& \!\! \int_{\Omega^X\!\!\times\Omega^Y}
         dP^{XY}_{xy} \;\;
         \Pi^Y(x,y). \nonumber
\end{eqnarray}
Here, each player's optimization is over every possible
probability space that might be applied to their problem. Game
analysis then requires players to jointly define a product
probability space ${\cal P}^X\times{\cal P}^Y$ where player $X$
is responsible for ${\cal P}^X$ and player $Y$ is responsible
for ${\cal P}^Y$.  As noted above, each player $Z$ can use any
of an infinite number of alternate probability spaces which we
here enumerate ${\cal P}^Z_i$ for $i=0,1,2,\dots$.  (The number
of probability spaces is non-denumerable.)  Because each player
must optimize their choices given the choices made by their
opponent, then both players must analyze every possible joint
probability space ${\cal P}^X_i\times{\cal P}^Y_j$ for
$i,j=0,1,2,\dots$. Each player is then faced with the task of
sequentially analyzing what happens given the adoption of every
possible joint probability space, and then optimizing their own
payoffs within each adopted probability space, and then
comparing the payoffs attainable from each joint probability
space to determine which space both they and their opponents
will adopt.

In contrast, conventional analysis mandates that players must
necessarily adopt a single probability space (whether mixed or
behavioural) leading to what is effectively a heavily
constrained optimization
\begin{eqnarray}
 X\!\!: \; \max_{{\cal P}^X}\;\; \langle \Pi^X \rangle \!\!
    &=& \!\! \int_{\Omega^X\!\!\times\Omega^Y}
         dP^{XY}_{xy} \;\;
         \Pi^X(x,y) \nonumber \\
    & &  \nonumber \\
 Y\!\!: \; \max_{{\cal P}^Y} \;\; \langle \Pi^Y \rangle \!\!
    &=& \!\! \int_{\Omega^X\!\!\times\Omega^Y}
         dP^{XY}_{xy} \;\;
         \Pi^Y(x,y)  \nonumber \\
        & &  \nonumber  \\
  \mbox{subject to} &&
       {\cal P}^X = {\cal P}^X_0, \;\;\;
       {\cal P}^Y = {\cal P}^Y_0.
\end{eqnarray}
That is, of all the possible joint probability spaces that might
be adopted, game theory restricts its rational players to a
single mandated choice.  And this without ever proving that this
single choice is somehow optimal.

We argue that optimization theory and probability theory are
entirely consistent with the fact that a known correlation state
between random variables will influence the dimensionality and
gradients of an optimization problem.  In view of this, these
fields offer no reasons whatsoever for the necessity of the
constraint shown in the last line above.

\begin{figure}[htbp]
\centering
\includegraphics[width=0.7\columnwidth,clip]{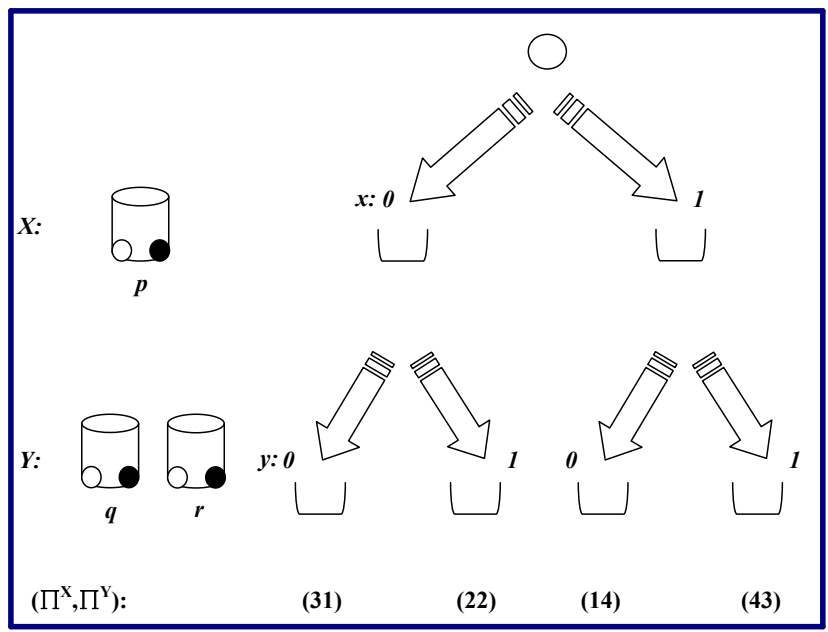}
\caption[Rational play using the conventional probability
space]{\em The conventional play of the two stage game features a
closed room containing players $X$ and $Y$, their respective
randomizing urns used to implement mixed strategies, and a large
metallic apparatus featuring a ball, and different channels and
cups to act as a decision recording device. Player $X$ implements
their ``$p$" randomization by draw either a white or black marble
from their urn, and correspondingly drops the ball down the $x=0$
or $x=1$ channel. Player $Y$ picks up the ball, selects the
relevant urn implementing either their ``$q$" or ``$r$"
randomizations, draws either a white or black marble, and
correspondingly drops the ball down the appropriate $y=0$ or $y=1$
channel into the waiting cups. Payoffs are assigned as shown.
 \label{f_story_conventional}}
\end{figure}

\subsection{Rational game play: A story}

Let us make the mathematics more concrete by telling a story in
an attempt to assist conceptualization of the new methods
presented here.

Suppose that you are the first player, player $X$, in the
example two stage game. As shown in Fig.
\ref{f_story_conventional}, you are in a room with your
opponent, player $Y$, and together, you are looking at the game
playing equipment. As player $X$, you play first and have to
drop a large ball down one of two channels marked $x=0$ or
$x=1$. To assist your decision, you have an urn containing a
prepared number of white or black marbles allowing you to
implement a randomized mixed strategy by selecting $x=0$ with
probability $1-p$ or $x=1$ with probability $p$.  You have
chosen $p$ so as to maximize your payoff.  You are also aware
that after your ball has landed in the appropriate cup, your
opponent, player $Y$, will choose one of their two randomizing
urns which each contain appropriate numbers of white and black
marbles. The first urn allows player $Y$ to choose $y=0$ with
probability $1-q$ and $y=1$ with probability $q$, while the
second urn allows them to choose $y=0$ with probability $1-r$
and $y=1$ with probability $r$. Player $Y$ has chosen $q$ and
$r$ so as to maximize their payoff. After determining their
choice of $y=0$ or $y=1$, player $Y$ will drop the ball down the
appropriate channel so that it lands in the waiting cup to
provide a permanent record of each players decisions.  The
players then divide a payoff accordingly as shown in Fig.
\ref{f_story_conventional}.  As shown in previous sections, a
conventional analysis results in the play combinations
$(x,y)=(0,1)$ and respective payoffs of
$\left(\langle\Pi^X\rangle,\langle\Pi^Y\rangle\right)=(2,2)$.
The above situation captures the conventionally mandated
procedure for payoff maximization in this particular strategic
interaction. It is presumed that the specified use of the
respective urns by each player along with the conventional
analysis specifying the values of $p$, $q$ and $r$ suffices to
optimize player payoffs. What could be simpler?

Notice however that game theory has never provided a proof that
the above procedure is complete, necessary, or sufficient.  In
particular, von Neumann and Morgenstern explicitly used a method
of ``indirect proof" subject to later falsification and so did
not prove the completeness, the necessity, or the sufficiency of
their methods.  Nash simply adopted a mixed strategy probability
space as the simplest way to provide an existence proof for what
are now called Nash equilibria. Kuhn established only that mixed
and behavioral probability spaces were equivalent in games of
perfect recall, but did not establish that they were complete,
necessary, or sufficient. In fact, no-one has ever provided a
mathematical proof of the completeness, the necessity, or the
sufficiency of preferring one probability space over all others.
Absent such proof, we suggest that rational players will explore
every feasible probability space describing any given game.  In
the absence of any confirmed decision procedure mandating the
use of one probability space over all others, we suppose that
players have the capacity to examine alternate probability
spaces, and choose between them so as to maximize their payoffs.

\begin{figure}[htbp]
\centering
\includegraphics[width=0.7\columnwidth,clip]{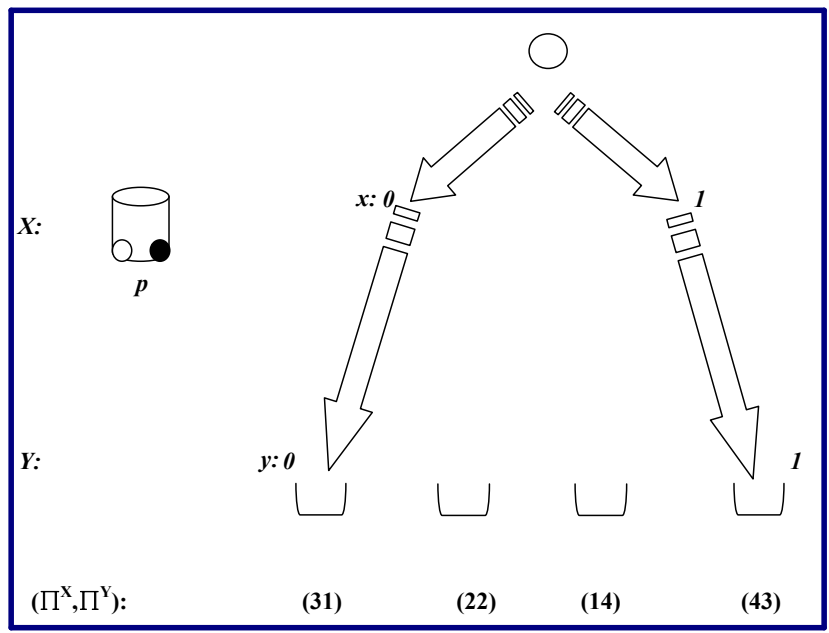}
\caption[Rational play using a correlated probability space]{\em
The correlated play of the two stage game features players
$X$ and $Y$ and an altered decision recording apparatus. Player
$X$ implements their ``$p$" randomization as usual and drops the
ball down either the $x=0$ or $x=1$ channel. Player $Y$ has used
their toolkit to alter the device so they no longer have any
decision to make as the ball simply continues falling down an
extended channel to the waiting cups. Payoffs are assigned as
shown.
 \label{f_story_correlated}}
\end{figure}

Accordingly, suppose now that player $Y$ adopts a different
procedure to that conventionally mandated.  Suppose in fact that
player $Y$ walks into the game room equipped with a toolkit
containing hacksaws, hammers, and welding equipment, and suppose
that before the game commences they set to work to reconfigure
the decision recording device.  As player $X$, you gaze in
appalled fascination as $Y$ hammers, cuts, and welds away until
the result is as shown in Fig. \ref{f_story_correlated}.  As the
time to start the game approaches, you have a decision to make.
Your eyes provide you with evidence that the decision making
device has been altered.  Your previous analysis was based on
the conventionally mandated device structure, but its alteration
makes the previous analysis irrelevant and in all likelihood,
wrong. As player $X$, you might seek to remonstrate with your
opponent by saying that they cannot alter the definition of the
game and that it is mandatory that they use the conventionally
mandated space. In response, player $Y$ simply responds that
they have not altered the game structure in any way, but have
merely adopted a probability space which correlates their
decision to the previous choice by $X$.  Every single move of
the game is still present but some have zero probability
assigned.  This is always possible. Conventional analysis allows
such assignments of zero probability but then insists that these
assignments can be altered by gradient optimization operations.
In contrast, $Y$ asserts that they have assigned zero
probabilities to certain moves which cannot be altered by
gradient optimization operations as is specifically allowed by
probability measure theory. Further, $Y$ knows of no proof
proving the conventional mandate, and as they are solely
motivated by a desire to maximize their payoff, they will take
any steps appropriate to that goal. Your decision is whether to
close your eyes to the altered nature of the decision making
device and continue to argue that any such alteration is
irrational and non-payoff maximizing, or to take the evidence of
your eyes into account and to alter your analysis. What decision
will you make? Self-evidently, as player $X$, after the game has
commenced, you will now choose to drop your ball down the $x=1$
channel as that maximizes your payoff.  Any other choice will
minimize your payoff, and as a payoff maximizing rational
player, you will not make such a choice.  The result, as shown
in previous sections, is that a correlated analysis results in
the play combinations $(x,y)=(1,1)$ and respective payoffs of
$\left(\langle\Pi^X\rangle,\langle\Pi^Y\rangle\right)=(4,3)$.
This provides an increased payoff for player $Y$ justifying
their rebuilding of the decision recording device.

\begin{figure}[htbp]
\centering
\includegraphics[width=0.7\columnwidth,clip]{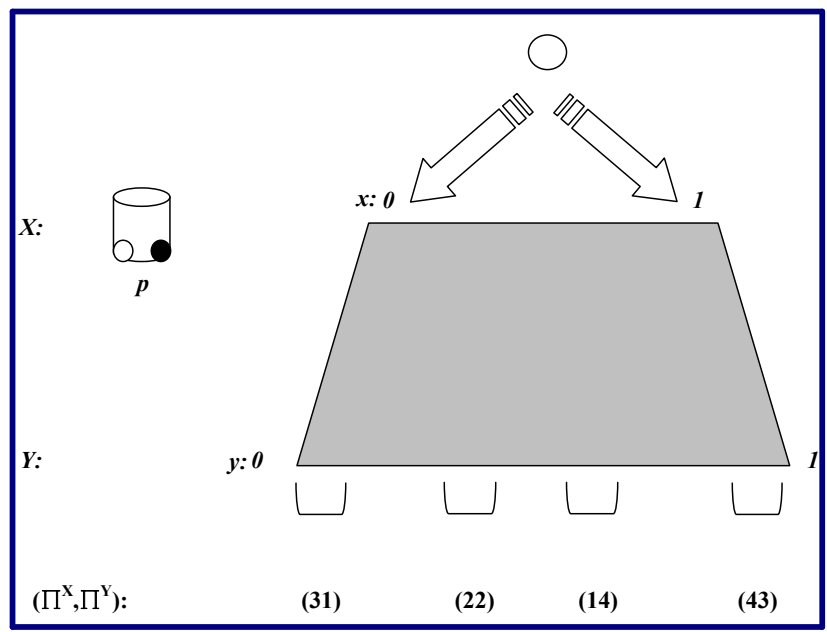}
\caption[Rational play using a hidden probability space]{\em The
play of the two stage game when player $X$
is unsure how player $Y$ has reconstructed the decision recording
apparatus. Player $X$ implements their ``$p$" randomization as
usual and drops the ball down either the $x=0$ or $x=1$ channel.
Player $Y$ might be using the conventional apparatus of Fig.
\protect\ref{f_story_conventional} or the correlated apparatus of
Fig. \protect\ref{f_story_correlated}.  Payoffs are assigned as
shown.
 \label{f_story_hidden}}
 \end{figure}

But that doesn't end the story as it is entirely unreasonable
that player $X$ perfectly knows how $Y$ is making their
decisions.  We now suppose that you, as player $X$, have watched
your opponent walk into the game room with their toolkit and a
large rectangular metal shield.  Player $Y$ erects their shield
to entirely hide their part of the decision making device from
your gaze, and behind this shield, they proceed to saw, hammer
and weld away. You, as player $X$, are however entirely unsure
what $Y$ is doing behind their shield.  Perhaps $Y$ is
reconstructing the original channel arrangements of the
conventionally mandated device of Fig.
\ref{f_story_conventional}.  Perhaps on the other hand, player
$Y$ is leaving the channels exactly as configured in the
correlated decision device of Fig. \ref{f_story_correlated} and
the welding is required to reconstruct the required ``$q$" and
``$r$" urns. The resulting situation, as perceived by yourself,
is as shown in Fig. \ref{f_story_hidden}.  Here, both you and
player $Y$ are depicted as being certain about how player $X$
will optimize their payoff.  Namely, $X$ will use an urn to
implement some mixed strategy ``$p$" to optimize their payoff.
However, you, as player $X$ have no information about how player
$Y$ will make their decision.  Again, you have a decision to
make.  A conventional analysis mandates that player $Y$ should
use a conventionally configured decision device and you should
play accordingly.  In this case, $Y$ will gain a payoff of
$\langle\Pi^Y\rangle=2$. However, $Y$ could alternatively choose
to adopt a correlated probability space in which case they will
gain a payoff of $\langle\Pi^Y\rangle=3$. Being rational, $Y$
can be expected to seek to maximize their expected payoff.  What
will you do? Will you assume that $Y$ has adopted a
conventionally mandated space and drop the ball down the $x=0$
channel in the hope that it stops half way requiring $Y$ to walk
over to the device to place it in the $y=1$ channel.  What a
disappointment then if the ball drops all the way down both the
$x=0$ and $y=0$ channels into the leftmost cup. Or
alternatively, will you assume that $Y$ is indeed a payoff
maximizer able to alter their choice of decision device leading
to the conclusion that $Y$ will have chosen to reconfigure the
channels to implement correlated play. In this case, you should
drop the ball into the $x=1$ channel in the hope that the ball
will drop all the way through both the $x=1$ and $y=1$ channels
into the rightmost cup. What a disappointment then if you see
the ball stop half way requiring $Y$ to walk over to place the
ball into the $y=0$ channel.  What is your choice?

We suggest that if you know (by observing) that $Y$ has
perfectly correlated their choice of $y$ to your choice of $x$,
then you must take this information into account.  Similarly,
even without direct observation, if you can deduce that $Y$ will
perfectly correlate their choice of $y$ to your choice of $x$,
then likewise, you must take this information into account.

\begin{figure}[htbp]
\centering
\includegraphics[width=0.7\columnwidth,clip]{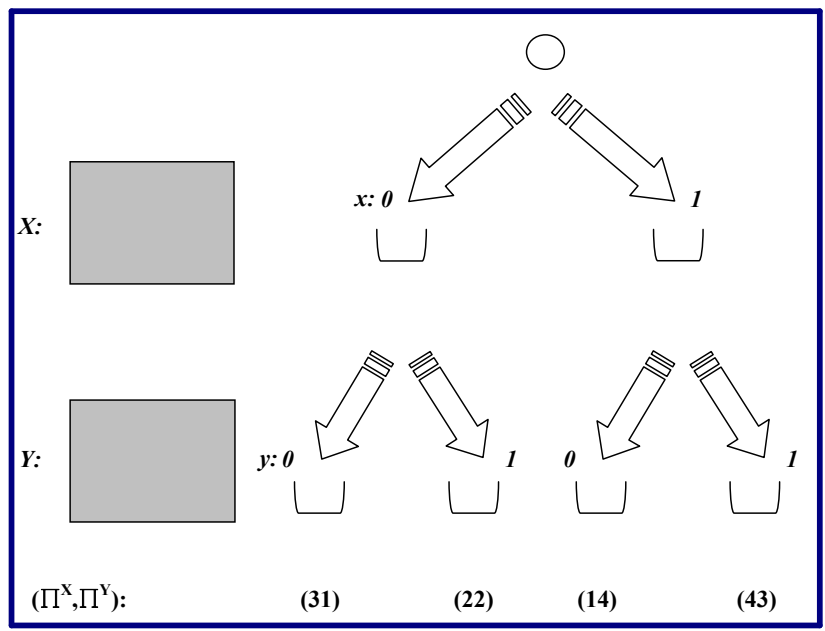}
\caption[Rational play using alternate probability spaces]{\em The
play of the two stage game when both player
$X$ and player $Y$ are unsure about which probability spaces and
randomization devices have been adopted by their opponents. In
this case, each player perfectly shields their decision making
apparatus from their opponent (shaded devices), and so might be
adopting the conventionally mandated analysis or any of an
infinite number of alternate possible probability spaces. Rational
players will analyze all these possibilities in order to maximize
their payoffs.  Payoffs are assigned as shown.
 \label{f_story_actual}}
 \end{figure}

In reality of course, the situation in a real strategic exchange
is more akin to that shown in Fig. \ref{f_story_actual}.  Here,
each player knows precisely the rules of the game including all
possible moves in their specified sequences.  What they don't
know is the choice of probability space made by their opponent.
This ignorance is represented by the coloured shields shown in
the figure. In fact, prior to their completing their own
analysis, they do not know which probability space they will
adopt, or whether they will choose a single space or randomize
over a number of spaces.  This is in sharp contrast to the
presumption of conventional game theory which mandates that each
player must use a particular probability space (or one of their
equivalents).  As noted above, there has never been a proof of
the completeness, necessity or sufficiency of this mandated type
of space. In view of this, we suggest that rational players will
simply optimize their choice of probability space to maximize
their expected payoff.  In Fig. \ref{f_story_actual}, you, as
player $X$, must deduce which space player $Y$ will use to
maximize their payoff. In the situation depicted here, $Y$ has
not physically reconstructed the decision recording device
before your eyes, but they have likely chosen to adopt a
particular probability space and physical randomization device.
Their roulette might involve their preprogramming one or more
random number generators, or might involve their providing
instructions to an agent who will act autonomously once the game
has begun allowing $Y$ to leave the room and take no further
part in the game.  As player $X$, you have absolutely no
information whatsoever about which roulette will be adopted by
$Y$.  The only fact you are sure of is that $Y$ will act so as
to maximize their payoff.

The question is, as always, is it possible for $Y$ to vary their
choice of probability space, of their roulette, or is this
impossible?  If it is impossible, provide a proof of this
conjecture, and then optimize accordingly.  If it is possible,
determine your optimal choices taking into account your
opponent's optimal choices.

\section{Discussion}

We propose that rational players will optimize their expected
payoff functionals (not functions) in strategic situations using
generalized calculus of variations approaches. These generalized
variational functional optimization methods examine every
possible value of a functional at every point as well as every
possible gradient through that point. A rational player, seeking
to perform a complete optimization, must examine every one of
these possibilities against all of the equivalent range of
possibilities of their opponents. These generalized methods give
access to an infinity of non-independent and functionally
constrained probability measure spaces defining non-continuous
expected payoff functionals defined over discontinuous domains
possessing, perhaps, a gradient nowhere.

The resulting generalized optimization approach corresponds to
optimizing an infinite number of alternate game decision trees
exhibiting altered optimal pathways and equilibria.

In this work, we follow the same methodology used by von Neumann
and Morgenstern \cite{vonNeumann_44}.   These authors initially
focussed on single players, typified by Robinson Crusoe, who
tried to optimize their payoff by choosing their actual moves or
pure strategies in a consumption game. They then showed that
this optimization method (focussed solely on pure strategies)
did not generalize to all multiple player games leading to the
introduction of probability distributions over pure strategies,
defining mixed strategies. That is, it was established by these
and later authors that while certain games (single player or
multiple-player-perfect-information games) had solutions in pure
strategies, this was not always true of more general games, and
as a mixed strategy analysis entirely subsumes a pure strategy
analysis, it was always advisable for a rational player to
perform a complete mixed strategy analysis for general games.
Here, we suggest similar results.  It seems to be sufficient to
employ conventional analysis for single-player or
multiple-player-single-stage games.  However, we suggest that
the complete analysis of multiple-player-multiple-stage games
requires more than a conventional analysis. Again, as the
conventional analysis is entirely subsumed within our augmented
optimization approach, it seems advisable for rational players
to perform an augmented analysis in general.

In earlier chapters, we have alluded to the possibility that our
expanded optimization analysis would produce results which
differ from standard results in game theory.  This does not mean
that game theory is wrong. Just as a theorem valid in a flat
geometry---the interior angles of all triangles sum to 180
degrees---can be invalid in a curved geometry, then so can
results validly derived in game theory be invalid in our
extended analysis.  Game theory is incomplete, rather than
wrong.

For instance, Kuhn established that games of perfect recall
could always be decomposed into discreet subgames, and that the
equilibrium pathway of the entire game consisted of concatenated
portions of the equilibrium pathways of all the relevant
subgames \cite{Kuhn_1953}. Crucial to the proof of this result,
is the separability of the joint probability distributions of
the entire game, and such separability exists only for the
independent behavioural probability spaces developed by Kuhn. In
our approach, behavioural strategies are not necessarily
independent so their governing probability spaces are not
necessarily separable.  A theorem derived assuming that
probability distributions is separable, is not applicable when
distributions are inseparable.

Similarly, in the same paper, Kuhn established that games of
perfect information always have pure strategy equilibria
\cite{Kuhn_1953}.  In our approach, even in perfect information
games, players are uncertain about which probability space might
be adopted by their opponents, and this allows equilibria to be
probabilistic.  Again, there is no contradiction with existing
results, as theorems derived assuming separable probability
distributions are inapplicable when distributions are
inseparable.

All of the results and theorems of game theory are derived under
certain assumptions about the joint probability spaces governing
game analysis. When players can adopt alternate probability
spaces invalidating these assumptions, then naturally, they can
derive results which differ from those of game theory.  Such
differences reflect limitations in the optimization analysis of
game theory, rather than errors in our more general optimization
approach.

Finally, we again remind ourselves that conventional analysis
routinely predicts outcomes at odds with observation.  As we
later show, the extended analysis that we argue must be
available to players of unbounded rationality, will produce
outcomes entirely consistent with observation.

Obviously, there are immediate applications of our new methods
to sequential games such as the chain store paradox, the trust
game, the ultimatum game, the public goods game, the centipede
game, and the iterated prisoner's dilemma.  We turn to this now.

 \chapter{Correlated Equilibria}
 \label{chap_correlated_eq}

\section{Introduction}

We are introducing isomorphism constraints into the strategy
spaces of game theory.  These constraints alter strategy space
geometries to allow the location of new equilibria. It is useful
to contrast out approach with Aumann's ``correlated equilibria".

\section{Correlated equilibria}

In 1974, Aumann modeled a nominally competitive game in which
players coopt public roulettes and share information to improve
their payoffs. This possibility arises as the Nash equilibria
for non-communicating players has them locating the best payoff
regardless of their opponent's choices so correlated changes of
strategy are impossible. Given the ability to communicate
however, correlated strategies become possible allowing novel
equilibria. Following Aumann's terminology, these are now termed
``correlated equilibria".

Our work here differs from Aumann's approach.  We allow players
to alter their chosen private randomization devices but do not
permit communication between players. We show that even without
additional communication channels, if players use different
physical randomization devices with different numbers of
independent coordinates and functionally constrained
coordinates, then these possible probability spaces must be
taken into account. To clarify the difference and similarities
between our entirely non-communicating analysis and Aumann's
correlated equilibria, we here go through one of the examples
used by Aumman in detail.

To model correlated equilibria, Aumman introduced probability
measures into his definitions of needed
\begin{quote}
equipment for randomizing strategies, and for defining utilities
and subjective probability for the players.  Thus to the
description of the game we append the following:

(5) A set $\Omega$ (the states of the world), together with a
$\sigma$-field ${\cal B}$ of subsets of $\Omega$ (the events);

(6) For each player $i$, a sub-$\sigma$-field ${\cal I}_i$ of
${\cal B}$ (the events in ${\cal I}_i$ are those regarding which
$i$ is informed).

(7) For each player $i$, a relation $\succeq_i$ (the preference
order of $i$) on the space of lotteries on the outcome space
$X$, where a lottery on $X$ is a ${\cal B}$-measurable function
from $\Omega$ to $X$ \cite{Aumann_74_67}.
\end{quote}
This welter of definitions was made understandable by use of a
series of worked examples, and we here follow the same route by
examining in detail Aumann's example (2.7).

\begin{figure}[htb]
\centering
\includegraphics[width=\columnwidth,clip]{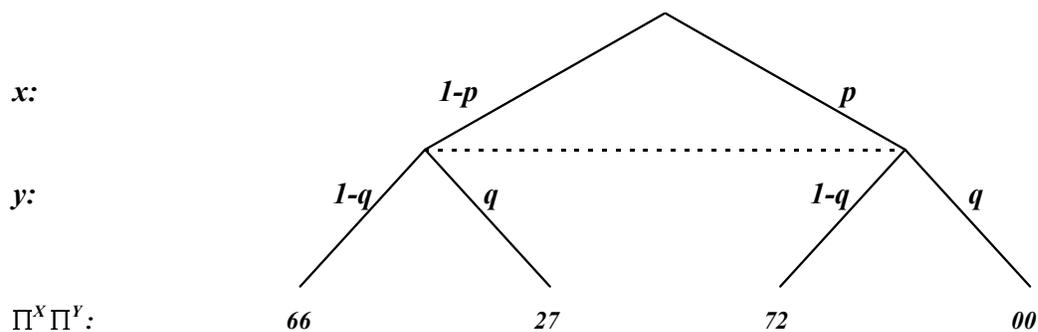}
\caption[Correlated equilibria: An example]{\em The
game tree for the two player non-zero-sum game considered by
Aumann in his example (2.7) \protect\cite{Aumann_74_67}.  Here,
two players $X$ and $Y$ simultaneously and independently choose
one of two options $x,y\in\{0,1\}$ to gain the payoff combinations
shown. \label{f_correlated_eq_27a}}
\end{figure}

\begin{figure}[htb]
\centering
\includegraphics[width=0.8\columnwidth,clip]{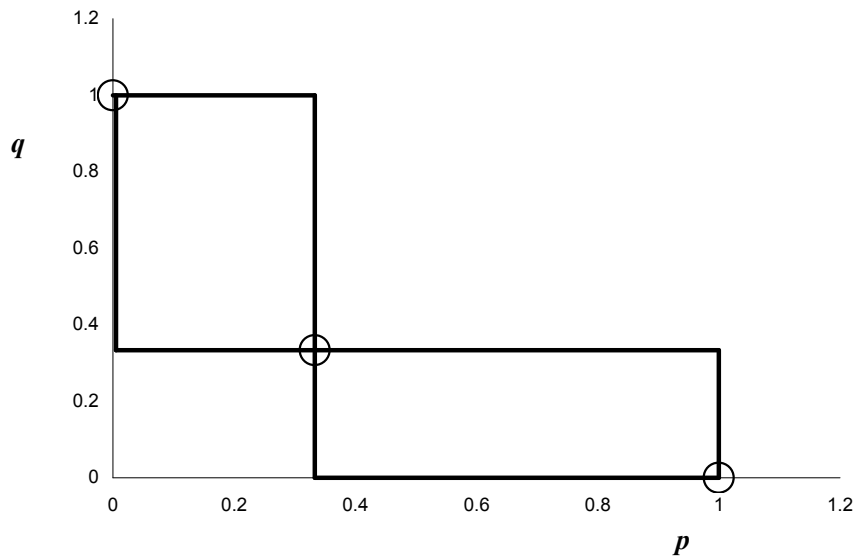}
\caption[Correlated equilibria: Nash equilibria without
communication]{\em The intersection of the gradient conditions
specifying Nash equilibria for the two player non-zero-sum game
considered by Aumann in his example (2.7)
\protect\cite{Aumann_74_67}. The three Nash equilibria points are
circled.  \label{f_correlated_eq_27b}}
\end{figure}

In Aumman's example (2.7), the two-person payoff matrix is
\begin{equation}
  \begin{array}{cc}
      & P_y \\
    P_x &
    \begin{array}{c|cc}
      (\Pi^X,\Pi^Y)  & 0 & 1 \\   \hline
      0 & (6,6) & (2,7) \\
      1 & (7,2) & (0,0) \\
    \end{array}
  \end{array}.
\end{equation}
In terms of the behavioural probability space defined in Fig.
\ref{f_correlated_eq_27a}, the expected payoff optimization
problems are
\begin{eqnarray}
    X: \max_{p}\;\; \langle\Pi^X\rangle
       &=& 6+p-4q-3pq \nonumber \\
    Y: \max_{qr}\;\; \langle\Pi^Y\rangle
       &=& 6-4p+q-3pq.
\end{eqnarray}
These expected payoffs are continuous multivariate functions
dependent only on the freely varying parameters $(p,q)$ so the
relevant gradient operator used by both players to analyze this
particular probability space is
\begin{equation}
    \nabla =
    \left[
    \frac{\partial}{\partial p},\frac{\partial}{\partial q}
    \right].
\end{equation}
Optimization then proceeds as usual via
\begin{eqnarray}
    \frac{\partial\langle\Pi^X\rangle}{\partial p} &=&
      1-3q \nonumber  \\
    \frac{\partial\langle\Pi^Y\rangle}{\partial q} &=&
      1-3p
\end{eqnarray}
so equilibria appear at the intersections shown in Fig.
\ref{f_correlated_eq_27b}. As noted by Aumann, there are three
Nash equilibria for this game at choices $(p,q)=(0,1)$, $(1,0)$,
and $(\frac{1}{3},\frac{1}{3})$ generating respective payoffs
$(\langle\Pi^X\rangle,\langle\Pi^Y\rangle)=(2,7)$, $(7,2)$, and
$(\frac{14}{3},\frac{14}{3})$.

Aumann now supposes that the players share a public 3-sided fair
dice allowing events ``A", ``B", and ``C" to be selected with
probability $\frac{1}{3}$, and that $X$ is informed whether or
not event ``A" appeared, while $Y$ is told whether or not ``C"
appeared.  Aumann then asks, given this altered environment with
additional communications, how will players now optimize their
expected payoffs. As a first step, the players must alter their
probability spaces to reflect the changed physical randomization
devices being used.

\begin{figure}[htb]
\centering
\includegraphics[width=\columnwidth,clip]{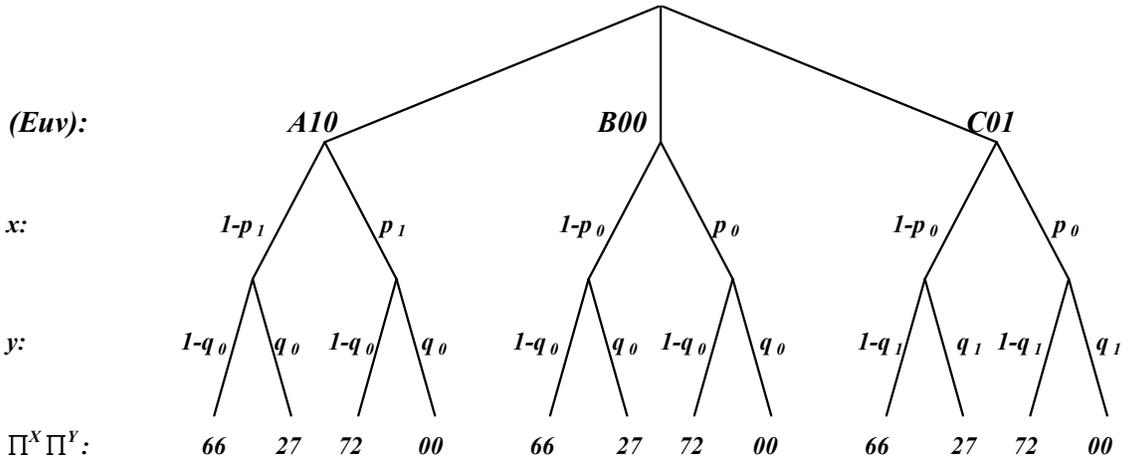}
\caption[Correlated equilibria: Communication alters decision tree]{\em The modified game tree corresponding to
the players sharing a three-sided dice selecting an event $E=A$,
$B$, or $C$ with equal probability $\frac{1}{3}$ with player $X$
advised whether event $A$ occurs or not (specified by the
indicator variable $u$) while player $Y$ is advised whether event
$C$ occurs or not (indicated by the indicator variable $v$). The
players can then appropriately condition their decisions on their
available information sets, as indicated.  The respective
information sets are not adequately represented on this figure.
\label{f_correlated_eq_27c}}
\end{figure}

One possibility is depicted Fig. \ref{f_correlated_eq_27c}.
Here, event $E\in\{A,B,C\}$ occurs each with probability of
$1/3$ and conditions two additional variables $u,v\in\{0,1\}$.
Player X knows the value of the variable $u$ while player $Y$
knows the value of $v$.  The variable $u$ is set to $u=1$ when
$E=A$ and $u=0$ otherwise, while $v=1$ when $E=C$ and $v=0$
otherwise.  The players can condition their subsequent choices
on the $u$ and $v$ variables.

The altered expected payoff functions are then

\begin{eqnarray}
    X: \max_{P^X}\;\; \langle\Pi^X\rangle
       &=& \sum_{Euv,x,y} P(Euv,x,y) \Pi^X(x,y) \nonumber \\
       &=& \sum_{Euv,x,y} P(Euv) P^X(x|Euv) P^Y(y|Euv) \Pi^X(x,y) \nonumber \\
       &=& \frac{1}{3} \left[ 18 + 2 p_0 + p_1 - 8 q_0 - 4 q_1 -
          3 \left[ p_1 q_0 + p_0 q_0 + p_0 q_1 \right] \right] \nonumber \\
    Y: \max_{P^Y}\;\; \langle\Pi^Y\rangle
       &=& \sum_{Euv,x,y} P(Euv,x,y) \Pi^Y(x,y) \nonumber \\
       &=& \sum_{Euv,x,y} P(Euv) P^X(x|Euv) P^Y(y|Euv) \Pi^Y(x,y) \nonumber \\
       &=& \frac{1}{3} \left[ 18 - 8 p_0 - 4 p_1 + 2 q_0 + q_1 -
         3 \left[ p_1 q_0 + p_0q_0 + p_0 q_1 \right] \right] .
\end{eqnarray}
written in terms of the joint probability distribution
$P(Euv,x,y)$ spanning the probability space, and where we
recognize that the payoff functions $\Pi^Z(x,y)$ depend only on
the choices $x$ and $y$, and we also take account of the various
conditioning possibilities of the variables.

\begin{figure}[htb]
\centering
\includegraphics[width=0.6\columnwidth,clip]{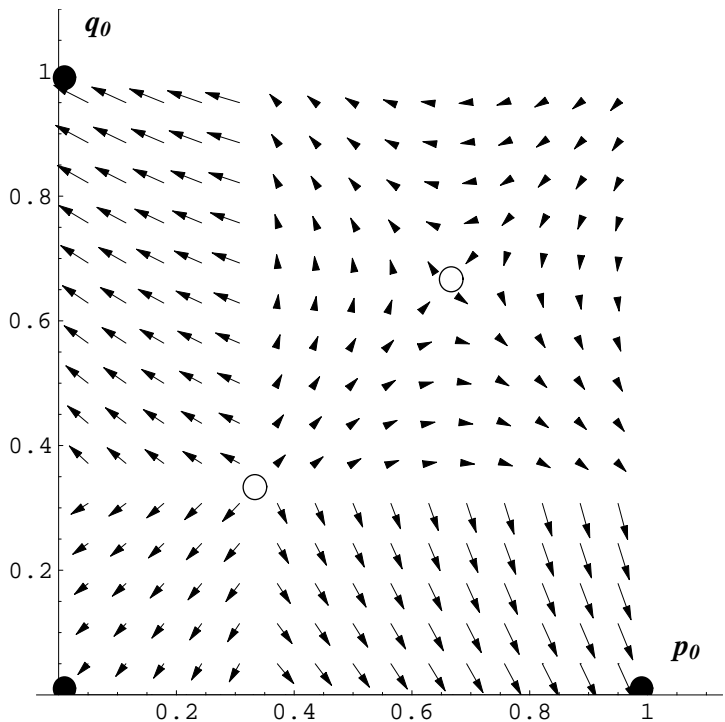}
\caption[Correlated equilibria: Flow diagram]{\em The flow diagram showing the direction of the
gradient of the respective expected payoffs
$[\frac{\partial\langle\Pi^X\rangle}{\partial
p_0},\frac{\partial\langle\Pi^Y\rangle}{\partial q_0}]$
identifying two unstable stationary points at
$(p_0,q_0)=(\frac{1}{3},\frac{1}{3})$ and
$(\frac{2}{3},\frac{2}{3})$ (open circles), as well as three
stable stationary points locating correlated equilibria at
$(p_0,q_0)=(0,0)$, $(0,1)$, and $(1,0)$ (closed disks). The
respective payoffs at the correlated equilibria are
$(\langle\Pi^X\rangle,\langle\Pi^Y\rangle)=(5,5)$, $(2,7)$, and
$(7,2)$.  \label{f_correlated_phase_diagram}}
\end{figure}

Consequently, in this expanded probability space the relevant
gradient operator is
\begin{equation}
    \nabla =
    \left(
    \frac{\partial}{\partial p_0},\frac{\partial}{\partial p_1},
    \frac{\partial}{\partial q_0},\frac{\partial}{\partial q_1}
    \right)
\end{equation}
in terms of which the players evaluate
\begin{eqnarray}
    \frac{\partial\langle\Pi^X\rangle}{\partial p_0} &=&
      \frac{1}{3} (2-3q_0-3q_1) \nonumber  \\
    \frac{\partial\langle\Pi^X\rangle}{\partial p_1} &=&
      \frac{1}{3} (1-3q_0) \nonumber  \\
    \frac{\partial\langle\Pi^Y\rangle}{\partial q_0} &=&
      \frac{1}{3} (2-3p_0-3p_1) \nonumber  \\
    \frac{\partial\langle\Pi^Y\rangle}{\partial q_1} &=&
      \frac{1}{3} (1-3p_0).
\end{eqnarray}
The second and fourth lines here specify that
\begin{eqnarray}
    p_1 &=& \left\{
    \begin{array}{ll}
      1                & \mbox{if } q_0<\frac{1}{3} \\
                       &                             \\
      \mbox{arbitrary} & \mbox{if } q_0=\frac{1}{3} \\
                       &                             \\
      0                & \mbox{if } q_0>\frac{1}{3} \\
    \end{array}
    \right. \nonumber  \\
    &&  \nonumber \\
    q_1 &=& \left\{
    \begin{array}{ll}
      1                & \mbox{if } p_0<\frac{1}{3} \\
                       &                             \\
      \mbox{arbitrary} & \mbox{if } p_0=\frac{1}{3} \\
                       &                             \\
      0                & \mbox{if } p_0>\frac{1}{3} \\
    \end{array}
    \right. ,
\end{eqnarray}
which in turn allows calculating the flow diagram for the
remaining gradients in terms of the variables $p_0$ and $q_0$ as
shown in Fig. \ref{f_correlated_phase_diagram}. This locates two
unstable stationary points at
$(p_0,q_0)=(\frac{1}{3},\frac{1}{3})$ and
$(\frac{2}{3},\frac{2}{3})$ and three stable stationary points
defining correlated equilibria at $(p_0,q_0)=(0,0)$, $(0,1)$,
and $(1,0)$.  The respective payoffs for each player at these
correlated equilibria points are
$(\langle\Pi^X\rangle,\langle\Pi^Y\rangle)=(5,5)$, $(2,7)$, and
$(7,2)$.  There is then an additional correlated equilibria
giving an increased expected payoff for each player motivating
them to use the additional available information to correlate
their strategy choices to their opponent's moves.

The location of a correlated equilibrium point with improved
payoffs to both players,
$(\langle\Pi^X\rangle,\langle\Pi^Y\rangle)=(5,5)$, lying
strictly outside the convex hull of the Nash equilibrium payoffs
concludes Aumann's example. To reiterate, every change of the
physical randomization device adopted by players, whether secret
or public, must be modelled by altered probability spaces.
Aumann introduced these tools to model correlated equilibria
generated by players sharing a public randomization device and
shared communication.  This communication means that novel
correlated equilibria can be located even in two-player single
stage games.

In contrast, our work with isomorphic constraints based on
correlations eschews any additional communication between the
players.  Rather, players can adopt different secret
randomization devices modelled by altered probability spaces
possessing different dimensionality, continuity properties,
differentiability conditions, and gradients, all of which allow
the location of novel equilibria. The continued absence of
communication between the players means that, as far as we can
tell, novel constrained equilibria appear only in
multiple-player-multiple-stage games.

 \chapter{The chain store paradox}
 \label{chap_chain_store}

\section{Introduction}

The chain store paradox examines predatory pricing to maintain
monopoly profits.  It gains its ``paradoxical" moniker as (so it
has been argued \cite{Milgrom_1982_280}) a substantial
proportion of the economics profession finds itself disagreeing
with the clear predictions of game theory in this game.  That
is, many economists would hold that it is irrational for any
firm to engage in predatory pricing to drive rivals out of
business and so gain a monopolist position as predation is
costly to the predator while potential new entrants well
understand that any price cutting is temporary. It is also
generally held that any attempt to extract monopoly pricing
benefits in some industry would quickly attract new entrants so
any monopoly gains will be short lived.  An extensive literature
has demonstrated the implausibility of these claims, with Ref.
\cite{Milgrom_1982_280} examining predatory pricing in the
shipping industry, IBM pricing strategies against competitors,
and coffee price wars, for instance.

Selton first proposed the chain store paradox as a complement to
the finite iterated prisoner's dilemma \cite{Selten_78_12} in
order to highlight inadequacies in game theory.  These lacks
would then justify the necessity of bounding rationality in game
theory. Terming the conventional game theoretic analysis and
predicted outcome as the ``induction" argument, and contrasting
this with an alternate ``deterrence" theory, Selton noted
\begin{quote}
``\dots only the induction theory is game theoretically correct.
Logically, the induction argument cannot be restricted to the last
periods of the game.  There is no way to avoid the conclusion that
it applies to all periods of the game.

Nevertheless the deterrence theory is much more convincing.  If I
had to play the game in the role of [the monopolist], I would
follow the deterrence theory.  I would be very surprised if it
failed to work.  From my discussions with friends and colleagues,
I get the impression that most people share this inclination.  In
fact, up to now I met nobody who said that he would behave
according to the induction theory.  My experience suggests that
mathematically trained persons recognize the logical validity of
the induction argument, but they refuse to accept it as a guide to
practical behavior.

It seems safe to conjecture that even in a situation where all
players know that all players understand the induction argument
very well, [the monopolist] will adopt a deterrence policy and the
other players will expect him to do so.

The fact that the logical inescapability of the induction theory
fails to destroy the plausibility of the deterrence theory is a
serious phenomenon which merits the name of a paradox.  We call it
the `chain store paradox'" \cite{Selten_78_12}.
\end{quote}

Efforts to resolve the paradox include recognizing that players
might not be sure that their opponents are rational payoff
maximizers due to the impact of mistakes or trembles,
rationality bounds, incomplete information, or altered
definitions of rationality, all of which necessitate use of
subjective probabilities \cite{Rosenthal_1981_92}.  In addition,
introducing asymmetric information whereby entrants are
uncertain whether monopolists are governed by behavioural rules
which eliminate common knowledge of rationality and provide a
rationale for entrants to base their expectations of the
monopolist's future behaviour on its past actions
\cite{Milgrom_1982_280}, while the use of imperfect information
or uncertainty about monopolist payoffs allows the replication
of observed behaviours \cite{Kreps_1982_253}. Other approaches
include dropping common knowledge of rationality
\cite{Davis_85_13}, or by introducing incomplete and imperfect
information \cite{Trockel_86_16}.  For a good review of how this
paradoxical game contributes to economic understanding appears,
see \cite{Wilson_1992_305}.

Selton's construction of the paradox hinges on the use of
``deterrence" theory in a multiple stage game (involving repeated
choices by the monopolist), whereby the monopolist can adopt a
non-rational strategy in early stages of the game to build a
reputation for implementing that strategy which induces their
opponent's to alter their own choices in latter stages. All
subsequent treatments have followed Selton in modelling such
multiple stage games and have then introduced some mechanism to
justify ``reputational" effects.

In contrast, in our treatment here, by introducing isomorphic
constraints into our strategy spaces, we can establish that it
is rational for the monopolist to adopt the seemingly irrational
choice even in a minimal game (where the monopolist makes a
single response to a single entrance) where it is commonly
thought that reputation or deterrence effects cannot make an
appearance. The conventional analysis of this minimal game is
immediately solved via backwards induction dependent on the
assumptions of a common knowledge of rationality (CKR),
independent behavioural strategies defining separable joint
probability distributions and allowing subgame decompositions.
In our extended analysis, the adoption of isomorphically
constrained joint probability spaces allows non-independent
behavioural strategies described by non-separable joint
probability distributions all of which invalidate subgame
decompositions and alter the optima located via backwards
induction. We demonstrate this now.

\begin{figure}[htb]
\centering
\includegraphics[width=0.8\columnwidth,clip]{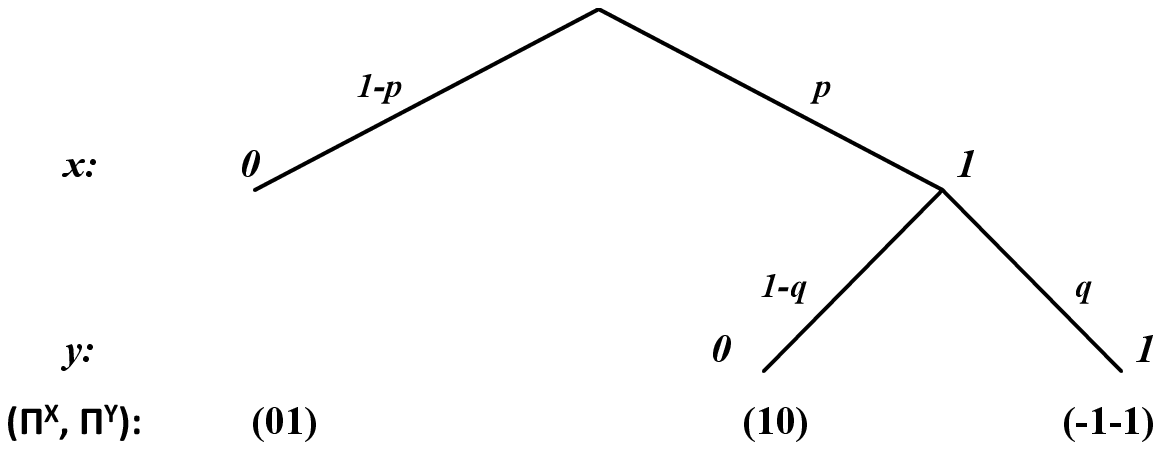}
\caption[The chain store paradox]{\em A minimal chain store game
decision tree in an unconstrained behavioural space
where a potential new market entrant $X$ must decide
to either stay out of a new market $x=0$ with probability $1-p$
or enter the market $x=1$ with probability $p$, in which case
the monopolist $Y$ chooses to either acquiesce $y=0$ with
probability $1-q$ or fight $y=1$ with probability $q$
their entry, with the corresponding payoffs shown.
\label{f_chain_store}}
\end{figure}

\section{The chain store paradox}

The minimal chain store paradox, conventionally pictured in Fig.
\ref{f_chain_store}, is defined over two sequential stages where
first, a potential market entrant $X$ must decide to either stay
out of a new market $x=0$ or enter that market $x=1$ where their
opponent, the monopolist $Y$, observes this choice. Should $X$
stay out of the market, they neither gain nor lose any payoff
while $Y$ gains monopolist profits so $(\Pi^X,\Pi^Y)=(0,1)$. In
contrast, should $X$ enter the market, $Y$ must then decide
whether to acquiesce to their opponent's entry $y=0$ by leaving
prices unchanged and losing profits so $(\Pi^X,\Pi^Y)=(1,0)$ or
by driving $X$ out of business by price cutting so payoffs are
$(\Pi^X,\Pi^Y)=(-1,-1)$.

\subsection{Unconstrained behaviour strategy spaces}

\begin{figure}[htb]
\centering
\includegraphics[width=0.8\columnwidth,clip]{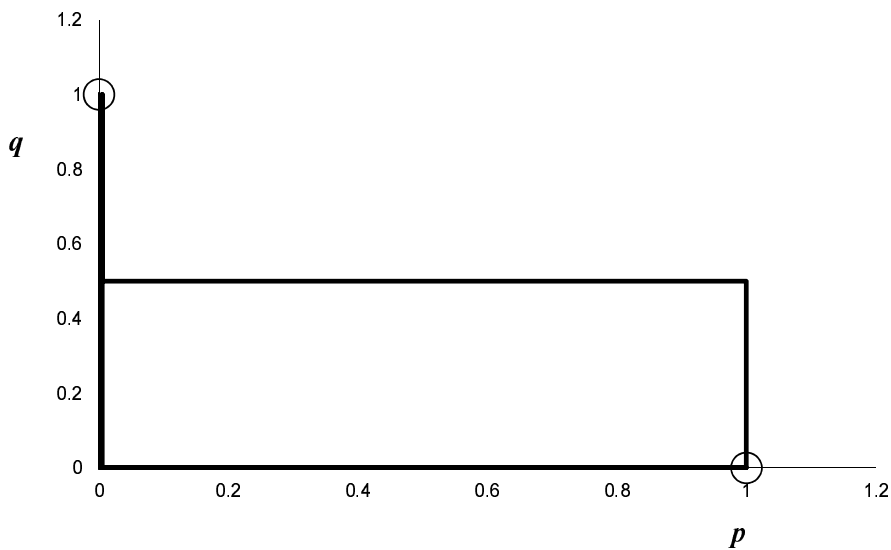}
\caption[Nash equilibria for the chain store paradox]{\em The
intersection of the gradient conditions specifying Nash equilibria
for the minimal chain store paradox. The two Nash equilibria
points are circled. \label{f_chain_store_eq}}
\end{figure}

A standard analysis frames the behaviour strategy spaces of each
player as being
\begin{eqnarray}
  {\cal P}^X_B &=& \left\{x\in\{0,1\},\{1-p,p\}\right\} \nonumber \\
  {\cal P}^Y_B &=& \left\{y\in\{0,1\},\{1-q,q\}|x=1\right\}.
\end{eqnarray}
Here, player $Y$ chooses their value of $y$ only when advised
that $x=1$.  In the joint behaviour space ${\cal
P}^X_B\times{\cal P}^Y_B$, the respective optimization problems
for the players are
\begin{eqnarray}
    X: \max_{p}\;\; \langle\Pi^X\rangle &=& p - 2 p q \nonumber \\
    Y: \max_{q}\;\; \langle\Pi^Y\rangle &=& 1 - p - p q,
\end{eqnarray}
so the only independent parameters are $p$ and $q$. In this
joint space, the gradient operator used by each player in their
analysis is
\begin{equation}
    \nabla =
    \left[
    \frac{\partial}{\partial p}, \frac{\partial}{\partial q} \right],
\end{equation}
so optimal solutions are obtained via
\begin{eqnarray}
    \frac{\partial\langle\Pi^X\rangle}{\partial p} &=& 1 - 2 q  \nonumber  \\
    \frac{\partial\langle\Pi^Y\rangle}{\partial r} &=& - p.
\end{eqnarray}
The solutions to these conditions are graphed in Fig.
\ref{f_chain_store_eq}.  Here, the gradient of the payoff for
the monopolist $Y$ is essentially always negative so $Y$ sets
$q=0$ and so always acquiesces to new market entrants.  In turn,
realizing this, $X$ determines that the gradient of their payoff
is always positive and so always sets $p=1$ and decides to enter
the market. There is also an equilibria at the point $p=0$ and
$q=1$, termed imperfect as it requires $Y$ to adopt an
irrational strategy (to fight) when $X$ stays out of the market
even though this intention cannot be sustained if indeed it
turns out that $X$ enters the market. The resulting expected
payoffs given that players adopt the sole perfect Nash
equilibria of $p=1$ and $q=0$ are
$\left(\langle\Pi^X\rangle,\langle\Pi^Y\rangle \right)=(1,0)$.

It is useful to again remind ourselves how this conventional
analysis without isomorphism constraints models perfect
correlations between $x$ and $y$ to show that the monopolist
cannot rationally sustain a perfectly correlated strategy.
Suppose that $Y$ seeks to perfectly correlate $y$ with $x$ via
$q=1$. As usual, both players are perfectly capable of
evaluating the expected payoff gradients in the appropriate
limit to obtain
\begin{eqnarray}
 \lim_{q\rightarrow 1} \frac{\partial\langle\Pi^X\rangle}{\partial p}
    &=& \lim_{q\rightarrow 1} (1 - 2 q) \; = \; -1 \nonumber  \\
 \lim_{q\rightarrow 1} \frac{\partial\langle\Pi^Y\rangle}{\partial q}
    &=& \lim_{q\rightarrow 1} - p  \; = \; -p.
\end{eqnarray}
That is, even when the monopolist seeks to perfectly correlate
their choice $y$ with $x$, the non-zero gradients present at
these points ensure they must rationally alter their intention
so as to maximize their payoff. This conclusion is of course
valid only when isomorphism constraints are absent so that
behavioural strategy probability distributions are separable
allowing subgame decompositions and optimization via backwards
induction. Conversely, this result does not pertain when
isomorphism contraints are in use.

Rational players of unbounded capacity are able to alter their
choice of probability space, and will optimize this choice so as
to maximize their expected payoffs.  In each alternate space,
the generated joint probability distributions might well involve
non-independent variables so the joint probability distributions
are nonseparable preventing conventional subgame decompositions
and ensuring that novel equilibria can be located. We now
complete a partial search of the possible joint probability
spaces.

\subsection{Isomorphically correlated space ${\cal P}^X_B\times{\cal P}^Y_B|_{q=1}$}

Suppose that player $Y$ employs an isomorphism constraint $q=1$
ensuring that variable $y$ is perfectly correlated to $x$ via
$y=x$ and $y^2=x^2=xy=x$. We denote this space ${\cal
P}^Y_B|_{q=1}$. In this space, the optimization tasks facing the
players are
\begin{eqnarray}
    X: \max_{x}\;\; \Pi^X &=& -x \nonumber \\
    Y: \;\; \Pi^Y &=& 1 - 2 x.
\end{eqnarray}
It is immediately evident that player $X$ maximizes their payoff
in this space by setting $x=0$.  The same result arises when
expected payoffs are used where we have the relations $\langle
y\rangle=\langle x\rangle$ and $\langle y^2\rangle=\langle
x^2\rangle=\langle xy\rangle=\langle x\rangle$ giving
\begin{eqnarray}
    X: \max_{p}\;\; \langle\Pi^X\rangle &=& -p \nonumber \\
    Y: \;\; \langle\Pi^Y\rangle &=& 1 - 2p.
\end{eqnarray}
As usual, the decision by $Y$ to adopt the ${\cal P}^Y_B|_{q=1}$
probability space leaves them with no further decisions to
optimize. The relevant gradient operator used by both players to
analyze this particular probability space is
\begin{eqnarray}
    \nabla &=& \frac{\partial}{\partial p}
\end{eqnarray}
so optimization proceeds as usual via
\begin{equation}
    \frac{\partial\langle\Pi^X\rangle}{\partial p} = -1
\end{equation}
ensuring that player $X$ chooses not to enter the market via
$p=0$ giving $x=0$.  Consequently, this means that $Y$ chooses
$y=0$ but this setting does not influence payoffs. That is, when
players $(X,Y)$ adopt the ${\cal P}^X_B\times{\cal
P}^Y_B|_{q=1}$ joint probability space, they maximize their
payoffs via the combination $(x,y)=(0,0)$ to garner payoffs
$\left(\langle\Pi^X\rangle,\langle\Pi^Y\rangle\right)= (0,1)$.
In short, the monopolist has deterred any new entry into the
market so they retain their profit.  The threat they made to
retaliate was not empty and indeed, was sufficient to modify
rational outcomes.

\subsection{The functionally anti-correlated space: ${\cal P}^X_B\times{\cal P}^Y_B|_{q=0}$}

Alternatively, player $Y$ might choose the alternate probability
space ${\cal P}^Y_B|_{q=0}$ in which player $Y$ chooses to
functionally anti-correlate their $y$ variable to the previous
choice of $x$ via $y=1-x$ and $xy=0$. In the joint probability
space ${\cal P}^X_B\times{\cal P}^Y_B|_{q=0}$, the expected
payoff optimization problem becomes
\begin{eqnarray}
    X: \max_{x}\;\; \Pi^X &=& x \nonumber \\
    Y: \;\; \Pi^Y &=& 1 - x.
\end{eqnarray}
It is immediately evident that player $X$ maximizes their payoff
in this space by setting $x=1$.  The use of expected payoffs
will lead to the same result as we have the relations $\langle
y\rangle=1-\langle x\rangle$ and $\langle xy\rangle=0$ giving
\begin{eqnarray}
    X: \max_{p}\;\; \langle\Pi^X\rangle &=& p \nonumber \\
    Y: \;\; \langle\Pi^Y\rangle &=& 1 - p.
\end{eqnarray}
Again, the adoption of the ${\cal P}^Y_B|_{q=0}$ probability
space leaves $Y$ with no decisions to optimize. As a result, the
gradient operator is again
\begin{eqnarray}
    \nabla &=& \frac{\partial}{\partial p},
\end{eqnarray}
with optimization giving
\begin{equation}
    \frac{\partial\langle\Pi^X_{0-}\rangle}{\partial p} = 1,
\end{equation}
ensuring that player $X$ chooses to enter the market via $p=1$
with $x=1$.  Consequently, this means that $Y$ chooses $y=0$ but
this setting does not influence payoffs.  The result is that
when players $(X,Y)$ adopt the ${\cal P}^X_B\times{\cal
P}^Y_B|_{q=0}$ joint probability space, they maximize their
payoffs via the combination $(x,y)=\{(1,0)\}$ to garner payoffs
$\left(\langle\Pi^X\rangle,\langle\Pi^Y\rangle\right)= (1,0)$.
In this space, $X$ is undeterred and enters the market to garner
the profits

\subsection{Expected payoff comparison across multiple probability spaces}

Altogether, the various joint probability spaces which might be
adopted by the players lead to a table of expected payoff
outcomes of
\begin{equation}
    \begin{array}{c|c}
      (\langle\Pi^X\rangle,\langle\Pi^Y\rangle) & {\cal P}^X_B  \\ \hline
                                                &               \\
       {\cal P}^Y_B|_{q=0}                      &      (1,0)  \\
                                                &               \\
       {\cal P}^Y_B                             &      (1,0)  \\
                                                &               \\
       {\cal P}^Y_B|_{q=1}                      &      (0,1)  \\
    \end{array}
\end{equation}
making it evident that to maximize their payoff, player $Y$ must
rationally elect to use probability space ${\cal P}^Y_B|_{q=1}$
in preference to either ${\cal P}^Y_B$ or ${\cal P}^Y_B|_{q=0}$.
That is, $Y$ will undertake to functionally correlate their
choice to the previous choice of the potential market entrant,
and thereby deny themselves a choice about the setting of $y$
once the game has commenced.  They do this knowing it to be the
payoff maximizing choice of probability space (among the few
examined here). Knowing this, player $X$ will not enter the
market even in this minimal chain store game. Similar results
apply for extended games with multiple markets and potential
entrants. The clear prediction of our analysis is that players
of unbounded rationality will always fight entrants in the chain
store game even though this strategy appears to be non-rational
when examined using conventional analysis.  That is, in the
chain store game, a monopolist does not need to build a
reputation for aggression over initial stages to try to
discourage potential entrants in later stages.  A monopolist, of
unbounded rationality, is well aware that making a choice to
adopt a probability space in which their choices are
functionally assigned to be correlated to their opponent's is
both payoff maximizing and rational.

It is of course possible to consider a broader range of joint
probability spaces for both players $X$ and $Y$, but these do not
alter the conclusion here that it can be rational for a monopolist
to punish market entrants to resolve the chain store paradox.

 \chapter{The trust game}
 \label{chap_trust_game}

\section{Introduction}

The previous chapter considered what conventional analysis holds
to be anomalous aggression, anomalous as it decreases the
payoffs of the aggressive player.  In this chapter, we consider
trusting behaviour where players transfer their own payoffs to
their opponent in the hope that their opponent will return the
favour and transfer an enlarged pool of funds back to them.
Needless to say, the conventional analysis holds that each of
these trusting actions is anomalous.  In this chapter, we
consider the single shot trust game.

In earlier formulations, the trust game took place over repeated
stages \cite{Kreps_1990_1} allowing reputation and punishment
theories to explain why players can exhibit trust and increase
their payoffs over those predicted by game theory. Such results
motivated investigations of single shot trust games (initially
termed the investment game) where the minimal number of stages
ensures that reputation and punishment effects are absent.
Despite this, players continue to exhibit trust to increase
their payoff \cite{Berg_1995_122}. More recently, players
involved in the trust game have undergone functional magnetic
resonance imaging of their brains during play
\cite{King-Casas_2005_78}. Other minimal games eliminating
reputation and punishment effects are the ultimatum and the
dictator game among others.

\begin{figure}[htb]
\centering
\includegraphics[width=0.8\columnwidth,clip]{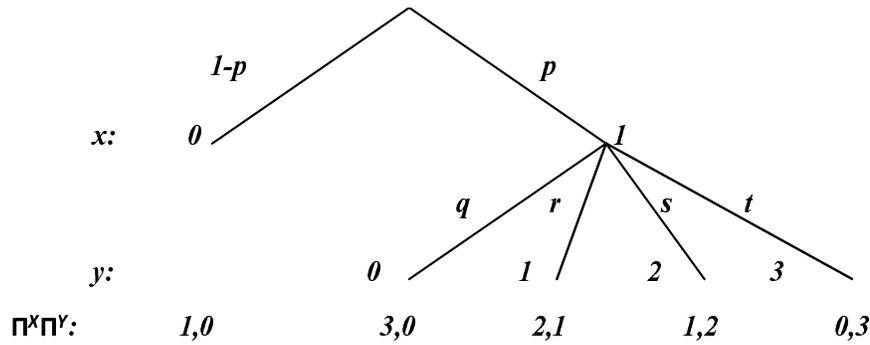}
\caption[The trust game]{\em A minimal trust game wherein player
$X$ possesses funds of one unit and must choose to either retain
these funds $x=0$ generating payoffs of
$\left(\Pi^X,\Pi^Y\right)=(1,0)$ or
trust their opponent by investing their funds with $Y$ via $x=1$.
Should this investment occur, both players are aware that $Y$
receives three units, and must then decide how much of this total
to keep and how much to return to $X$. That is, $Y$ decides to
retain an amount $y\in\{0,1,2,3\}$ while returning an amount $3-y$
to $X$ generating payoffs of
$\left(\Pi^X,\Pi^Y\right)=(3-y,y)$.
\label{f_trust}}
\end{figure}

\section{A simplified trust game}

In this section, we simplify the trust game as far as possible
without losing any of its character.

The minimal trust game, as conventionally pictured in Fig.
\ref{f_trust}, is defined over two sequential stages where
first, player $X$ possess a single unit of funds and must choose
to either retain these funds $x=0$ generating payoffs of
$\left(\Pi^X,\Pi^Y\right)=(1,0)$, or trust their opponent by
investing their funds with $Y$ via $x=1$. Should this investment
occur, both players are aware that $Y$ receives three units and
must then decide how much of this total to keep and how much to
return to $X$. That is, $Y$ decides to retain an amount
$y\in\{0,1,2,3\}$ while returning an amount $3-y$ to $X$
generating payoffs of $\left(\Pi^X,\Pi^Y\right)=(3-y,y)$.
Altogether, the payoffs to the players are
\begin{eqnarray}
    \Pi^X &=& 1 - x + x(3-y)   \nonumber \\
    \Pi^Y &=& xy.
\end{eqnarray}

\subsection{Unconstrained behaviour strategy spaces}

Conventional game analysis commences with the assumption that
players $X$ and $Y$ each adopt a probability space lacking
isomorphism constraints. Possible spaces include
\begin{eqnarray}
  {\cal P}^X_B &=& \left\{x\in\{0,1\},\{1-p,p\}\right\} \nonumber \\
  {\cal P}^Y_B &=& \left\{y\in\{0,1,2,3\},\{q,r,s,t\}|x=1\right\}.
\end{eqnarray}
Here, player $Y$ chooses their value of $y$ only when advised
that $x=1$ and we have the normalization condition $q+r+s+t=1$.
In the joint behaviour space ${\cal P}^X_B\times{\cal P}^Y_B$,
the respective optimization problems for the players are
\begin{eqnarray}
    X: \max_{p}\;\; \langle\Pi^X\rangle  &=& 1 - p + p (3q+2r+s) \nonumber \\
    Y: \max_{q,r,s}\;\; \langle\Pi^Y\rangle &=& p (3-3q-2r-s ).
\end{eqnarray}
The only independent variables here are $p, q, r$ and $s$ (subject
to normalization constraints) so the relevant gradient operator is
\begin{equation}
    \nabla =
    \left[
    \frac{\partial}{\partial p},
    \frac{\partial}{\partial q},
    \frac{\partial}{\partial r},
    \frac{\partial}{\partial s}, \right].
\end{equation}
Consequently, optimal solutions are obtained via
\begin{eqnarray}
    \frac{\partial\langle\Pi^X\rangle}{\partial p}
         &=& - 1 +  3q+2r+s \nonumber  \\
    \frac{\partial\langle\Pi^Y\rangle}{\partial q}
         &=& -3p \nonumber \\
    \frac{\partial\langle\Pi^Y\rangle}{\partial r}
         &=& -2p \nonumber \\
    \frac{\partial\langle\Pi^Y\rangle}{\partial s}
         &=& -p.
\end{eqnarray}
The last three equations here straightforwardly show that $Y$
maximizes their expected payoff by setting $q=r=s=0$ ensuring
$t=1$ to give $y=3$. In turn, this result simplifies the
optimization condition for $X$ establishing that $X$ maximizes
their payoff by setting $p=0$ to give $x=0$. The Nash equilibria
for this simplified trust game is then $(x,y)=(0,3)$ so both $X$
and $Y$ selfishly retain all the funds they can generating
expected payoffs of
$\left(\langle\Pi^X\rangle,\langle\Pi^Y\rangle \right)=(1,0)$.

As noted previously, these payoffs are not optimal as they could
be improved by both players adopting different choices, as is
commonly observed in human play.

\begin{figure}[htb]
\centering
\includegraphics[width=0.8\columnwidth,clip]{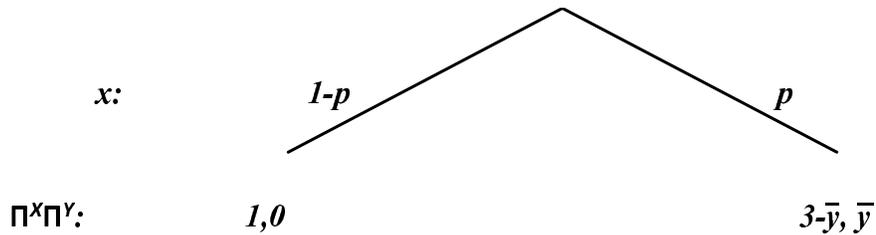}
\caption[Isomorphically correlated choices in the trust game]{\em
The case where players $(X,Y)$ adopt the ${\cal P}^X_B\times{\cal
P}^Y_B|_{y=\bar{y}}$ joint probability space where player $Y$ functionally
correlates their second stage choice to their opponent's first
stage choice.  In this case, a decision by $X$ to invest funds
with $Y$ automatically invokes a partial return of funds.
\label{f_trust_cor}}
\end{figure}

\subsection{The isomorphically correlated space ${\cal P}^X_B\times{\cal P}^Y_B|_{y=\bar{y}}$}

Rational players are able to alter their choice of probability
space, and will optimize this choice so as to maximize their
expected payoffs.  Suppose that player $Y$ considers an
alternate probability space denoted ${\cal P}^Y_B|_{y=\bar{y}}$
in which the choice of the variable $y$ is determined by the
preceding choice of $x$ via
\begin{equation}
  y = 3 (1-x) + x \bar{y}.
\end{equation}
This means that when $x=0$ we have $y=3$ while the choice $x=1$
enforces the setting $y=\bar{y}$ for $\bar{y}\in\{0,1,2,3\}$.
This possibility is shown in Fig. \ref{f_trust_cor}. Noting we
still have $x^2=x$ and $x(1-x)=0$, the payoffs to each player
are
\begin{eqnarray}
    X: \max_{x}\;\; \Pi^X  &=& 1 + x (2-\bar{y}) \nonumber \\
    Y: \;\; \Pi^Y &=& x \bar{y}.
\end{eqnarray}
It is evident that player $X$ will set $x=1$ provided
$\bar{y}<2$ and $x=0$ when $\bar{y}>2$.  They are indifferent
when $\bar{y}=2$ and so will play safe with $x=0$.  The same
results appear when the expected payoffs are maximized via
\begin{eqnarray}
    X: \max_{p}\;\; \langle\Pi^X\rangle &=& 1 + 2p - p \bar{y} \nonumber \\
    Y:  \langle\Pi^Y\rangle &=& p \bar{y}.
\end{eqnarray}
The relevant gradient operator is
\begin{equation}
    \nabla =
    \left[
    \frac{\partial}{\partial p} \right],
\end{equation}
and optimization proceeds via
\begin{equation}
 \frac{\partial\langle\Pi^X\rangle}{\partial p} = (2-\bar{y}).
\end{equation}
As a result, $X$ maximizes their payoff by setting $p=1$
whenever $\bar{y}<2$, and $p=0$ otherwise. Subsequently, because
$Y$ has left themselves no free choices during the game, the
outcomes $(\bar{y},x,y,\langle\Pi^X\rangle,\langle\Pi^Y\rangle)$
are respectively $(0,1,0,3,0)$, $(1,1,1,2,1)$, $(2,0,3,1,0)$,
and $(3,0,3,1,0)$.

\subsection{Expected payoff comparison across multiple probability spaces}

The optimal payoffs in the various joint probability spaces
considered here which might be adopted by the players are
\begin{equation}
    \begin{array}{c|c}
      (\langle\Pi^X\rangle,\langle\Pi^Y\rangle) & {\cal P}^X_B   \\ \hline
                                                &                \\
       {\cal P}^Y_B                             &    (1,0)       \\
                                                &                \\
       {\cal P}^Y_B|_{y=0}                      &    (3,0)       \\
                                                &                \\
       {\cal P}^Y_B|_{y=1}                      &    (2,1)       \\
                                                &                \\
       {\cal P}^Y_B|_{y=2}                      &    (1,0)       \\
                                                &                \\
       {\cal P}^Y_B|_{y=3}                      &    (1,0)       \\
    \end{array}.
\end{equation}
This makes it evident that to maximize their payoff, $Y$ must
rationally elect to use the joint probability space ${\cal
P}^Y_B|_{y=1}$ in preference to any alternate probability space
considered here. That is, player $Y$ will undertake to
functionally correlate their second stage decision to the
previous choice of their opponent, and thereby deny themselves a
second stage choice during the game knowing this to be the
payoff maximizing choice. Knowing this, $X$ is confident enough
to send all of their funds to $Y$ with the clear expectation of
making a profit. This prediction of our extended analysis is in
accord with observation.

 \chapter{The ultimatum game}
 \label{chap_ultimatum}

\section{Introduction}

The prevalence and importance of bargaining in society justifies
the examination of simple bargaining models such as the ultimatum
game, particularly in view of the discrepancy between observed
player strategies and rational equilibrium solutions
\cite{Guth_82_36}. In the ultimatum game, two players must divide
an item of equal utility to both (generally money).  One player,
the proposer, offers a proportional division to the other, the
responder, who must either accept it in which case the division
proceeds as suggested, or reject it in which case neither player
receives any money. The assumption that players are rational and
payoff maximizing allows derivation of the subgame perfect
equilibrium where in each stage the proposer offers the smallest
positive amount of money possible which the responder accepts as
receiving some amount of money, however small, is always better
than receiving none.  This solution is seldom observed in
experiments making the ultimatum game an ideal vehicle for testing
the assumptions of game theory.

This role as a game theory test-bed has long been explored
\cite{Stahl_72,Rubinstein_82_97,Guth_82_36,Binmore_85_11,Ochs_89_35,Bolton_91_10,Oosterbeek_2004_171},
and tested by many experiments including examination of the
influence of variable stake sizes
\cite{Hoffman_96_28,Slonim_98_56,Cameron_99_47} and of culture
\cite{Roth_91_10,Henrich_00_97}.  See experimental surveys in
\cite{Thaler_88_19,Roth_95_12,Camerer_95_20}. Experimental
results typically demonstrate offers closer to a fair split
(50\%), and frequent rejections of offers even substantially
above 0\% (approximately the predicted equilibrium offer).
Further, more detailed analysis shows that players, while
failing to locate the subgame perfect equilibrium, are
performing a sophisticated matching of offers to acceptance
probabilities so as to maximize payoffs \cite{Zamir_00}, while
the ability to track a changing game environment demonstrates
that proposers can be induced to vary their offer ranges and
that responders can expand their acceptance sets---in effect
offers and acceptances are contingent on the possibly changing
game environment \cite{Winter_97}.

Proposed modifications to game theory to generate the observed
payoff maximizing behaviour have focused on introducing mechanisms
to complement player self-interest. In the main, these proposed
additions either exploit modified utility functions interdependent
on both player's payoffs by taking account of psychological
factors (so player utility increases with player equity or player
intentionality say), or by embedding the ultimatum game within a
larger, perhaps societal game (taking account of player reputation
and self image for instance).  These differing approaches include
fairness
\cite{Ochs_89_35,Ruffle_1998_247,Bolton_91_10,Roth_91_10,Prasnikar_92_86,%
Rabin_93_12,Loewenstein_93_13,Forsythe_94_34,Blount_95_13,%
Bolton_00_16}, though with equity definitions generally
self-serving and modified by player information and payoff
asymmetries \cite{Kagel_96_10}, rivalry \cite{Burnell_99_22},
reciprocity \cite{Dufwenberg_98_0,Falk_00_0}, envy
\cite{Kirchsteiger_94_37}, punishment and revenge
\cite{Bolton_95_95}, competition and cooperation
\cite{Fehr_99_81}, altruism and spitefulness \cite{Levine_98_59},
and reputation \cite{Hoffman_94_34}. In these approaches, player
strategies effectively become contingent on both player's payoffs
generating novel equilibria allowing more equitable play.

Player learning can be modelled via algorithms modifying current
strategy selections (offers and acceptance probabilities) in the
light of prior game events \cite{Slonim_98_56,Roth_95_16} which
again makes player strategies contingent on those of their
opponents to generate novel equilibria.  See also
\cite{Gale_95_56,Vriend_97_9,Duffy_99_13}.  Essentially the same
algorithm can be implemented at the population level using
evolutionary games theory in which players observe and learn about
previous acceptances and rejections of other players and modify
their strategies accordingly \cite{Nowak_00_17}, or simply learn
which payoff splits maximize payoffs \cite{Huck_99_13}.  See also
\cite{Guth_92_23,Peters_00_31}.  Again, these approaches
effectively make current strategy choice contingent on prior game
events to generate novel equilibria.

\begin{figure}[htb]
\centering
\includegraphics[width=\columnwidth,clip]{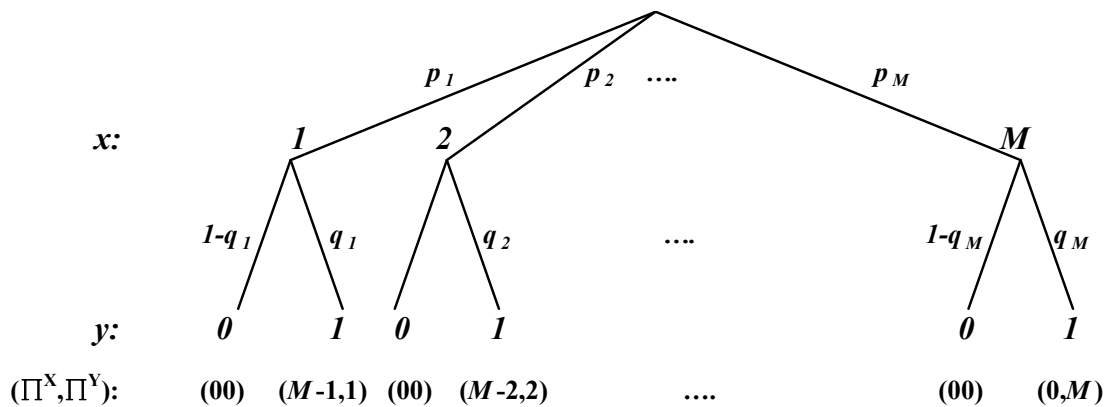}
\caption[The Ultimatum game: Conventional analysis]{\em A
conventional tree of the two stage ultimatum game.  In this
decision tree, $X$ makes an integral offer $1\leq x\leq M$ with
probability $p_x$ to $Y$ who either accepts the offer by choosing $y=1$ with
probability $q_x$ or who rejects the offer by setting $y=0$ with probability
$1-q_x$. If the offer is accepted, the player payoffs are
$(\Pi^X,\Pi^Y)=(M-x,x)$ while if the offer is rejected,
player payoffs are $(\Pi^X,\Pi^Y)=(0,0)$.
\label{f_ultimatum_game}}
\end{figure}

\section{The Ultimatum game}

As shown in Fig. \ref{f_ultimatum_game}, the ultimatum game is
defined here over two sequential stages where first $X$
communicates an integral offer $1\leq x\leq M$ to $Y$. Player
$Y$ must then decide whether to accept the offer by choosing
$y=1$ in which case $Y$ keeps the offer amount $x$ and $X$
receives an amount $M-x$.  Alternately, $Y$ rejects the offer by
choosing $y=0$ in which case neither player receives any payoff.
That is, the payoffs are
\begin{eqnarray}
    \Pi^X &=& (M-x) y \nonumber \\
    \Pi^Y &=& x y.
\end{eqnarray}

A quick optimization analysis (achieved by straightforwardly
embedding the discrete payoffs in the corresponding continuous
functions) has
\begin{eqnarray}
    \frac{\partial \Pi^X}{\partial x}
         &=& - y <0 \nonumber  \\
    \frac{\partial \Pi^Y }{\partial y}
         &=& x >0,
\end{eqnarray}
indicating that player $X$ can increase their payoff by setting
$x$ as small as possible, so $x=1$, while player $Y$ increases
their payoff by making $y$ as large as possible, so $y=1$.  This
gives the equilibrium point $(x,y)=(1,1)$ generating payoffs of
$(\Pi^x,\Pi^Y)=(M-1,1)$.  However, few human players adopt this
equilibrium point.

A more detailed analysis has players seeking to alter their
choices of probability spaces ${\cal P}^X$ and ${\cal P}^Y$ so
as to maximize their respective payoffs. As previously, players
must determine which joint probability space defining the joint
probability distributions will optimize payoff outcomes.

\subsection{The isomorphically unconstrained space: ${\cal P}^X_B\times{\cal P}^Y_B$}

The conventional analysis of the ultimatum game commences with
players $X$ and $Y$ each adopting a probability space lacking
isomorphism constraints. Possible spaces include
\begin{eqnarray}
  {\cal P}^X_B &=& \left\{x\in\{1,2,\dots,M\},\{p_1,p_2,\dots,p_M\}\right\} \nonumber \\
  {\cal P}^Y_B &=& \left\{y\in\{0,1\},\{P^Y(y=0|x=i)=(1-q_i),P^Y(y=1|x=i)=q_i,\forall i\}\right\}.
\end{eqnarray}
Here, we have the normalization condition $\sum_i p_i=1$.

In the joint behaviour space ${\cal P}^X_B\times{\cal P}^Y_B$,
the respective optimization problems for the players are
\begin{eqnarray}
    X: \max_{p_2,\dots,p_M}\;\; \langle\Pi^X\rangle
    &=& q_1 (M-1) - \sum_{i=2}^{M}
          p_i \left[ q_1 (M-1) - q_i (M-i) \right]  \nonumber \\
    Y: \max_{q_1,\dots,q_M}\;\; \langle\Pi^Y\rangle
    &=& q_1 + \sum_{i=2}^{M}
          p_i ( q_i i - q_1).
\end{eqnarray}
We have here resolved the normalization condition via
$p_1=1-\sum_{i=2}^{M}p_i$.  Consequently, the expected payoffs
are continuous multivariate functions dependent on the
probability parameters $(p_2,\dots,p_M,q_1,\dots,q_M)$, so the
relevant gradient operator used by both players to analyze this
particular probability space is
\begin{equation}
    \nabla =
    \left[
    \frac{\partial}{\partial p_2},\dots,
    \frac{\partial}{\partial p_M},
    \frac{\partial}{\partial q_1},\dots,
    \frac{\partial}{\partial q_M} \right].
\end{equation}
Immediately then, the optimization conditions evaluated by each
player are
\begin{eqnarray}
    \frac{\partial\langle\Pi^X\rangle}{\partial p_i}
         &=& - [ (M-1) q_1 - (M-i) q_i ], \;\;\; \forall i\in[2,M]   \nonumber  \\
    \frac{\partial\langle\Pi^Y\rangle}{\partial q_i}
         &=& i  p_i  \;\;\; \forall i\in[1,M].
\end{eqnarray}
The conditions for rates of change of $Y$'s payoff with respect
to $q_1,\dots,q_M$ here are all non-negative ensuring that $Y$
sets $q_1=\dots=q_M=1$ and thus accepts any offer from $X$
greater than or equal to $x=1$.  In turn, these determinations
simplify the optimization conditions for $X$ wherein the rates
of change for $X$'s payoff with respect to all of
$p_2,\dots,p_M$ are negative so $X$ sets $p_2=\dots=p_M=0$ and
$p_1=1$. The resulting choices by each player are $(x,y)=(1,1)$
generating expected payoffs of
$\left(\langle\Pi^X\rangle,\langle\Pi^Y\rangle \right)=(M-1,1)$.
This is the unique Nash equilibrium point for this ultimatum
game, given the adoption of the joint probability space ${\cal
P}^X_B\times{\cal P}^Y_B$.  Unfortunately, it is not an
equilibrium adopted by many human players.

Rational players will be very aware that both they and their
opponent can alter their choice of probability space, and will
optimize this choice so as to maximize their expected payoffs.
In these alternate spaces, the random probability variables used
in player optimizations might well be non-independent so joint
probability distributions are nonseparable preventing
conventional subgame decompositions and ensuring that novel
equilibria can be located. We illustrate this now accomplishing,
as usual, only a partial search of the available infinity of
probability spaces.

\subsection{An isomorphically constrained space: ${\cal P}^X_B\times{\cal P}^Y_B|_{y=\bar{y}}$}

Suppose that player $Y$ adopts one of a possible $M-1$ alternate
probability spaces ${\cal P}^Y_B|_{y=\bar{y}}$ for integral
$2\leq\bar{y}\leq M$ in which they correlate their $y$ variable
with the previous value $x$. In particular, suppose that $Y$
undertakes to reject any offer $x$ less than $\bar{y}$ and to
accept any offer $x$ equal to or greater than $\bar{y}$. That is
$Y$ adopts the functional assignment
\begin{equation}
   y  = \left\{
        \begin{array}{ll}
          0 & \mbox{if } x<\bar{y} \\
            &    \\
          1 & \mbox{if } x\geq\bar{y}. \\
        \end{array}
       \right.
\end{equation}
In other words, we have $y=\delta_{x\geq\bar{y}}$ giving the
payoff functions
\begin{eqnarray}
    X: \max_{x}\;\; \Pi^X  &=& (M-x)\delta_{x\geq\bar{y}}  \nonumber \\
    Y: \;\; \Pi^Y &=& x \delta_{x\geq\bar{y}}.
\end{eqnarray}
It is then evident that player $X$ will set $x=\bar{y}$ to
maximize their payoff at $\Pi^X=(M-\bar{y})$ giving player $Y$ a
payoff of $\Pi^Y=\bar{y}$.  Similar results are obtained from
optimizing the expected payoff functions obtained using the
probability distribution
\begin{equation}
   P^Y(y|x)  = \left\{
        \begin{array}{l}
          P^Y(y=0|x) = 1 - \sum_{j=\bar{y}}^M\delta_{jx} \\
                \\
          P^Y(y=1|x) = \sum_{j=\bar{y}}^M\delta_{jx}. \\
        \end{array}
       \right.
\end{equation}
Players of unbounded rationality must then sequentially assume
that players $X$ and $Y$ have adopted the joint probability
space ${\cal P}^X_B\times{\cal P}^Y_B|_{y=\bar{y}}$ for
$2\leq\bar{y}\leq M$, and within each space, locate the
constrained equilibria optimizing outcomes, all of which can be
subsequently compared in a later comparison table. We complete
this process now.

\begin{figure}[htb]
\centering
\includegraphics[width=\columnwidth,clip]{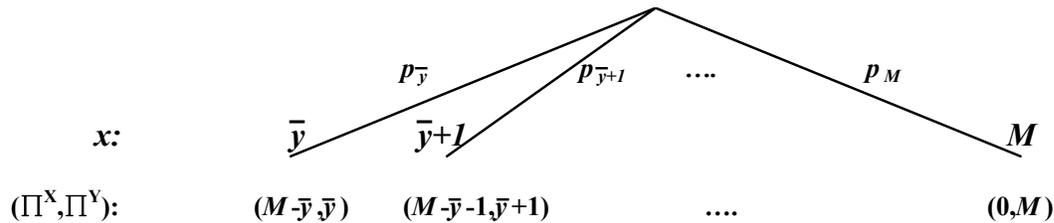}
\caption[The Ultimatum game: Isomorphically constrained strategies]{\em The case where players $(X,Y)$ adopt the ${\cal
P}^X_B\times{\cal P}^Y_B|_{y=\bar{y}}$ joint probability space where
player $Y$ is functionally constrained to reject any offer
$x<\bar{y}$ and to accept any offer $x\geq\bar{y}$. As a result
offers of a lesser amount appear neither in the expected payoff
functions nor in the corresponding game tree.
\label{f_ultimatum_bar_y}}
\end{figure}

With the adoption of the joint probability space ${\cal
P}^X_B\times{\cal P}^Y_B|_{y=\bar{y}}$, and taking account of
the the normalization condition
$p_{\bar{y}}=1-\sum_{i=\bar{y}+1}^M p_i$, the expected payoff
optimization problems for the players becomes
\begin{eqnarray}
    X: \max_{p_{\bar{y}+1},\dots,p_M}\;\; \langle\Pi^X\rangle
               &=& (M-\bar{y}) +
                       \sum_{i=\bar{y}+1}^M p_i (\bar{y}-i) \nonumber \\
    Y: \langle\Pi^Y\rangle
               &=& \sum_{i=\bar{y}}^M p_i i,
\end{eqnarray}
which are now dependent only on the freely varying parameters
$(p_{\bar{y}+1},\dots,p_M)$.  That is, given their previous
choice of probability space, player $Y$ has no further
independent parameters, while player $X$ is indifferent to any
choice with $1\leq i<\bar{y}$ because these variables have
disappeared from the problem specification.  The resulting game
tree is as shown in Fig. \ref{f_ultimatum_bar_y}.  The relevant
gradient operator used by both players to analyze this
particular probability space is
\begin{eqnarray}
    \nabla &=&
    \left[
    \frac{\partial}{\partial p_{\bar{y}+1}},\dots,
    \frac{\partial}{\partial p_M}\right].
\end{eqnarray}
Optimization then proceeds as usual via
\begin{eqnarray}
    \frac{\partial\langle\Pi^X\rangle}{\partial p_i}
     & = &  \bar{y}-i,  \;\;\; \forall i \in [\bar{y}+1,M].
\end{eqnarray}
All of the terms on the right hand side are negative ensuring
that player $X$ sets $p_{\bar{y}+1}=\dots=p_M=0$.  In turn, this
means that $X$ sets $p_{\bar{y}}=1$ and only ever offers
$x=\bar{y}$. (When $\bar{y}=M$, player $X$ gains zero payoff
regardless of their offer and so is indifferent.) Consequently,
in the joint probability space ${\cal P}^X_B\times{\cal
P}^Y_B|_{y=\bar{y}}$, players $(X,Y)$ choose the combination
$(x,y)=\{(\bar{y},1)\}$ to garner payoffs
$\left(\langle\Pi^X\rangle,\langle\Pi^Y\rangle\right)=
(M-\bar{y},\bar{y})$.

\subsection{Payoff comparison across probability spaces}

The above analysis has considered a total of one conventional
joint probability space ${\cal P}^X_B\times{\cal P}^Y_B$ and
$M-1$ alternate probability spaces ${\cal P}^X_B\times{\cal
P}^Y_B|_{y=\bar{y}}$ for $2\leq\bar{y}\leq M$. Altogether, the
various joint probability spaces adopted by the players lead to
a table of expected payoff outcomes of
\begin{equation}
    \begin{array}{c|c}
                                                &               \\
      (\langle\Pi^X\rangle,\langle\Pi^Y\rangle) & {\cal P}^X_B  \\ \hline
       {\cal P}^Y_B                             &      (M-1,1)  \\
       {\cal P}^Y_B|_{y=2}                      &      (M-2,2)  \\
         \vdots                                 &      \vdots   \\
         {\cal P}^Y_B|_{y=M-2}                  &      (2,M-2)  \\
         {\cal P}^Y_B|_{y=M-1}                  &      (1,M-1)  \\
    \end{array},
\end{equation}
making it evident that to maximize their payoff, player $Y$ must
rationally elect to use probability space ${\cal
P}^Y_B|_{y=M-1}$ in preference to ${\cal P}^Y_B$.  Knowing this,
player $X$ will offer $x=(M-1)$ to $Y$ to ensure that they gain
a payoff greater than zero.

\subsection{An indicative solution reflecting symmetries}

Obviously, in normal play of the ultimatum game, $X$ does not
normally expect that they need to offer all of the available
funds to avoid rejection, and $Y$ seldom elects to reject every
offer less than all of the funds. This might result as the game
is now highly symmetric.

A conventional analysis shows that player $X$ can garner a
payoff of $M-1$ and force $Y$ to accept a payoff of $1$.  The
isomorphic constrained analysis here shows that $Y$ can force a
payoff of $M-1$ for themselves leaving $X$ with a minimal payoff
of $1$.  Player $X$, facing a minimal payoff of $1$ could then
seek to modify their own probability space and undertake to not
even consider offers greater than $\bar{x}$ say.  It is possible
that an extended analysis taking account of the ability of both
$X$ and $Y$ to veto offers will settle in a choice around
$\bar{x}=\bar{y}=M/2$ or thereabouts.

The analysis presented here is indicative only and we do not
attempt to resolve the ultimatum game.  It suffices for our
purposes to show that including isomorphic constraints within
the strategy spaces of the ultimatum game allows a broader range
of equilibria outcomes than considered by conventional game
theory.

\section{Discussion}

This paper presents an analysis of isomorphically constrained
play in the finitely iterated Ultimatum game. The use of
isomorphic constraints reduces the dimensionality of the game
strategy spaces and can modify game properties and equilibrium
points. We suggest that these constraints are routinely
exploited in human play to maximize player outcomes.  We crudely
suggested that fair play might be one possible outcome of our
extended analysis.

Experiments across a wide range of cultures show human players
as commonly adopting fair play. This carries the implication
that human game players in a diversity of cultures have a
natural ability to exploit isomorphic constraints to their own
ends. Further, we suggest that use of isomorphic constraints are
common in bargaining situations and in economics in general, and
it is necessary that games theory be able to properly model
these isomorphic constraints in strategic interactions. Further,
our analytical approach is likely to be more broadly applicable
to the wider economic sphere as modeled by game theory.

 \chapter{The public goods game}
 \label{chap_public_goods}

\section{Introduction}

There are many situations in which a number of players must
jointly participate in creating some common resource but where no
player can be prevented from exploiting that resource. This
creates a ``free-rider" or ``tragedy of the commons" style problem
as while all players benefit if the public good is provided, any
individual player can increase their benefits if they avoid paying
their share of the costs \cite{Hardin_1968_1243}.  As a result,
players do not cooperate and the public good is not provided.
These results are altered if players are able to punish free
riders, even when punishment carries significant costs to the
initiator \cite{Fehr_2000_980}. The public goods game allows
experimental examination of how norms of cooperative behaviour are
established and enforced using a wide range of theoretical
approaches
\cite{Nowak_98_573,Wedekind_2000_850,Fehr_04_185,Nowak_2005_1291},
including a proposed quantum solution \cite{Chen_03_0301013}.

\begin{figure}[htb]
\centering
\includegraphics[width=\columnwidth,clip]{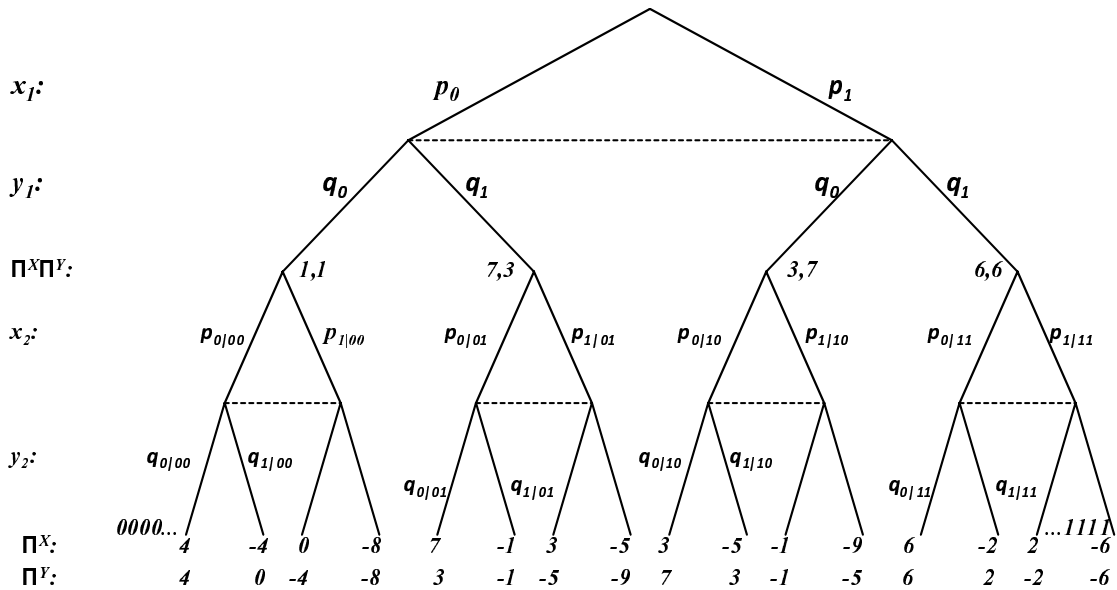}
\caption[The public goods game]{\em A minimal public goods game
involving two players $X$ and $Y$ who simultaneously choose to
make an investment of some amount $x_1,y_1\in\{0,1\}$ in stage one.
The return to each player of their own investment is negative
whilst the return to them from their opponent's investment is
positive.  Thus, investment is a public good which creates a free rider
problem. In the second stage, each player can choose to either
punish their opponent for their first stage actions $x_2,y_2=1$ at
some cost to themselves, or not $x_2,y_2=0$, with the corresponding
payoffs shown.
\label{f_public_goods}}
\end{figure}

\section{A simplified public goods game}

Here as usual, we simplify the public goods game as far as
possible without losing any of its character.  In particular, we
restrict the number of players to two, designated as usual $X$
and $Y$, and also restrict both the amounts that can be
exchanged and the amounts used to punish opponents.

The minimal public goods game, as pictured in Fig.
\ref{f_public_goods}, is defined over two sequential stages.  In
stage one, players $X$ and $Y$ both choose whether four units of
payoff is either retained $x_1,y_1=0$ or invested $x_1,y_1=1$.
The return to each player from their own investment is negative
whilst the return to them from their opponent's investment is
positive. The payoffs to the players from their joint actions in
stage one are
\begin{eqnarray}
    \Pi^X_1  &=& 4 -  x_1 + 3 y_1  \nonumber \\
    \Pi^Y_1  &=& 4 + 3 x_1 - y_1.
\end{eqnarray}
Thus, should both $X$ and $Y$ make no investments via
$x_1=y_1=0$ then their payoffs are $\Pi^X_1=\Pi^Y_1=4$ while if
both invest all their funds via $x_1=y_1=1$ then their payoffs
are improved to $\Pi^X_1=\Pi^Y_1=6$.  Unfortunately however, it
pays for each player to free ride on their opponent's
investment: should $X$ invest their funds $x_1=1$ while $Y$
retains all of their funds $y_1=0$, the joint payoffs are
$\left(\Pi^X_1,\Pi^Y_1\right)=\left(3,7\right)$, making it
tempting for $Y$ to free ride.  Conversely, should $X$ retain
their funds while $Y$ invests, the payoffs are
$\left(\Pi^X_1,\Pi^Y_1\right)=\left(7,3\right)$.  The net result
is that game theory predicts that both players attempt to free
ride on the investment of their opponent resulting in non-Pareto
optimal payoffs.

The willingness of players to incur costs to punish their free
riding opponents can then be studied by adding a second stage as
shown in Fig. \ref{f_public_goods}.  Here, each player can
choose to either not punish their opponent $x_2,y_2=0$ leaving
all payoffs unchanged, or can choose to punish their opponent
$x_2,y_2=1$ at some cost to themselves.  That is, should a
player choose to punish their opponent, they decrease their
payoff by four units while at the same time decreasing their
opponent's payoff by eight units. Consequently, by the end of
stage two, the joint payoffs are
\begin{eqnarray}
    \Pi^X  &=& 4 - x_1 + 3 y_1 - 4 x_2 - 8 y_2 \nonumber \\
    \Pi^Y  &=& 4 + 3 x_1 - y_1 - 8 x_2 - 4 y_2.
\end{eqnarray}
It is this two stage form of the game that generates significant
discrepancies between game theoretic predictions and observed
human play.  In particular, because punishment is costly then game
theory makes the firm prediction that rational players will never
choose to punish their opponents.  However, precisely the opposite
tends to occur in practise. People exhibit a strong tendency to
punish their free riding opponents even when this reduces their
own payoffs.  Herein lies the interest in the public goods game.

\subsection{Unconstrained behavioural strategy spaces: ${\cal P}^X_B\times{\cal P}^Y_B$}

Conventional game analysis commences with the assumption that
both players $X$ and $Y$ together adopt a joint probability
space ${\cal P}^X_B\times{\cal P}^Y_B$ in which every
behavioural strategy on every history set is independent. One
possibility for the joint behavioural strategy space is shown in
Fig. \ref{f_public_goods}.  We have chosen a terminology
allowing the expected payoff function for player $Z\in\{X,Y\}$
to be written as
\begin{eqnarray}
    Z: \max\;\; \langle\Pi^Z\rangle
    &=& \sum_{x_1,y_1,x_2,y_2=0}^{1}
          P^{XY}(x_1,y_1,x_2,y_2) \Pi^Z(x_1,y_1,x_2,y_2)  \nonumber \\
    &=& \sum_{x_1,y_1,x_2,y_2=0}^{1}
          P^{X}(x_1)P^Y(y_1)
          P^X(x_2|x_1y_1)P^Y(y_2|x_1y_1) \Pi^Z(x_1,y_1,x_2,y_2)  \nonumber \\
    &=& \sum_{x_1,y_1,x_2,y_2=0}^{1}
          p_{x_1} q_{y_1} p_{x_2|x_1y_1} q_{y_2|x_1y_1} \Pi^Z(x_1,y_1,x_2,y_2).
\end{eqnarray}
We also have implicit normalization conditions such as
$p_0+p_1=1$ and $p_{0|x_1y_1}+p_{1|x_1y_1}=1$, and so on.  The
expected payoff functions for each player are then
\begin{eqnarray}                \label{eq_public_goods1}
    X: \max_{p_1,p_{1|x_1y_1}}\;\; \langle\Pi^X\rangle
    &=&  4 - \langle x_1\rangle + 3 \langle y_1\rangle - 4 \langle x_2\rangle - 8 \langle y_2\rangle \nonumber \\
    &=&  4 - p_1 + 3 q_1
         - 4 \sum_{x_1y_1x_2=0}^{1} p_{x_1} q_{y_1} p_{x_2|x_1y_1} x_2
         - 8 \sum_{x_1y_1y_2=0}^{1} p_{x_1} q_{y_1} q_{y_2|x_1y_1} y_2 \nonumber \\
    &=&  4 - p_1 + 3 q_1
         - 4 \sum_{x_1y_1=0}^{1} p_{x_1} q_{y_1} p_{1|x_1y_1}
         - 8 \sum_{x_1y_1=0}^{1} p_{x_1} q_{y_1} q_{1|x_1y_1} \nonumber \\
    Y: \max_{q_1,q_{1|x_1y_1}}\;\; \langle\Pi^Y\rangle
    &=& 4 + 3 \langle x_1\rangle - \langle y_1\rangle - 8 \langle x_2\rangle - 4 \langle y_2\rangle    \nonumber \\
    &=& 4 + 3 p_1 - q_1
        - 8 \sum_{x_1y_1x_2=0}^{1} p_{x_1} q_{y_1} p_{x_2|x_1y_1} x_2
        - 4 \sum_{x_1y_1y_2=0}^{1} p_{x_1} q_{y_1} q_{y_2|x_1y_1} y_2   \nonumber \\
    &=& 4 + 3 p_1 - q_1
        - 8 \sum_{x_1y_1=0}^{1} p_{x_1} q_{y_1} p_{1|x_1y_1}
        - 4 \sum_{x_1y_1=0}^{1} p_{x_1} q_{y_1} q_{1|x_1y_1}.
\end{eqnarray}
Here, the expected payoff functions are continuous multivariate
functions dependent on the probability parameters
$\left[p_1,p_{1|00},p_{1|01},p_{1|10},p_{1|11}\right]$ and
$\left[q_1,q_{1|00},q_{1|01},q_{1|10},q_{1|11}\right]$, so the
relevant gradient operator  is
\begin{eqnarray}
    \nabla &=&
    \left[
    \frac{\partial}{\partial p_1},
    \frac{\partial}{\partial p_{1|00}},
    \frac{\partial}{\partial p_{1|01}},
    \frac{\partial}{\partial p_{1|10}},
    \frac{\partial}{\partial p_{1|11}},
    \frac{\partial}{\partial q_1},
    \frac{\partial}{\partial q_{1|00}},
    \frac{\partial}{\partial q_{1|01}},
    \frac{\partial}{\partial q_{1|10}},
    \frac{\partial}{\partial q_{1|11}} \right].
\end{eqnarray}
Normalization conditions mean that any term dependent on $p_0$
or $p_{0|x_1y_1}$ contributes a negative term to any gradient
with respect to $p_1$ or $p_{1|x_1y_1}$ respectively.  Similar
considerations apply to the $q$ parameters.

Taking account of normalization, the optimization conditions
evaluated by each player are
\begin{eqnarray}
    \frac{\partial\langle\Pi^X\rangle}{\partial p_1}
    &=& -1 + 4 \sum_{y_1=0}^{1} q_{y_1} \left( p_{1|0y_1} -  p_{1|1y_1} \right)
           + 8 \sum_{y_1=0}^{1} q_{y_1} \left( q_{1|0y_1} - q_{1|1y_1} \right)  \nonumber  \\
    \frac{\partial\langle\Pi^X\rangle}{\partial p_{1|00}}
         &=& - 4 p_0 q_0  \nonumber  \\
    \frac{\partial\langle\Pi^X\rangle}{\partial p_{1|01}}
         &=& - 4 p_0 q_1  \nonumber  \\
    \frac{\partial\langle\Pi^X\rangle}{\partial p_{1|10}}
         &=& - 4 p_1 q_0  \nonumber  \\
    \frac{\partial\langle\Pi^X\rangle}{\partial p_{1|11}}
         &=& - 4 p_1 q_1  \nonumber  \\
    \frac{\partial\langle\Pi^Y\rangle}{\partial q_1}
        &=& -1 + 8 \sum_{x_1=0}^{1} p_{x_1} \left( p_{1|x_10} -  p_{1|x_11} \right)
           + 4 \sum_{x_1=0}^{1} p_{x_1} \left( q_{1|x_10} - q_{1|x_11} \right)  \nonumber  \\
    \frac{\partial\langle\Pi^Y\rangle}{\partial q_{1|00}}
        &=& - 4 p_0 q_0 \nonumber \\
    \frac{\partial\langle\Pi^Y\rangle}{\partial q_{1|01}}
        &=& - 4 p_0 q_1 \nonumber \\
    \frac{\partial\langle\Pi^Y\rangle}{\partial q_{1|10}}
        &=& - 4 p_1 q_0 \nonumber \\
    \frac{\partial\langle\Pi^Y\rangle}{\partial q_{1|11}}
        &=& - 4 p_1 q_1,
\end{eqnarray}
Thus, player $X$ finds the rate of change of their payoff with
respect to $p_{1|ij}$ is always negative so they set
$p_{1|ij}=0$ for all $i$ and $j$.  Similarly, player $Y$ sets
$q_{1|ij}=0$ as the rate of change of their payoff with respect
to $q_{1|ij}$ is also always negative for all $i$ and $j$. That
is, there are no histories in which it is payoff maximizing for
either player to punish their opponent. In turn, these results
simplify the remaining two conditions for first stage moves
giving
\begin{eqnarray}
    \frac{\partial\langle\Pi^X\rangle}{\partial p_1}
    &=& -1  \nonumber  \\
    \frac{\partial\langle\Pi^Y\rangle}{\partial q_1}
        &=& -1.
\end{eqnarray}
This establishes that both players maximize their expected
payoffs by setting $p_1=0$ and $q_1=0$ in the first stage. Thus,
both players make no investment in the first round confident in
the knowledge that their opponent will not punish them for this.
The Nash equilibria for this simplified public goods game is
then $(x_1,y_1,x_2,y_2)=(0,0,0,0)$ generating expected payoffs
of $\left(\langle\Pi^X\rangle,\langle\Pi^Y\rangle
\right)=(4,4)$. As noted previously, these payoffs are not
Pareto optimal as they could be improved by both players
adopting different choices, as is commonly observed in human
play.

Rational players are able to alter their choice of probability
space, and will optimize this choice so as to maximize their
expected payoffs.  We here suppose that players might each
consider a total of two alternate probability spaces.

\begin{figure}[htb]
\centering
\includegraphics[width=\columnwidth,clip]{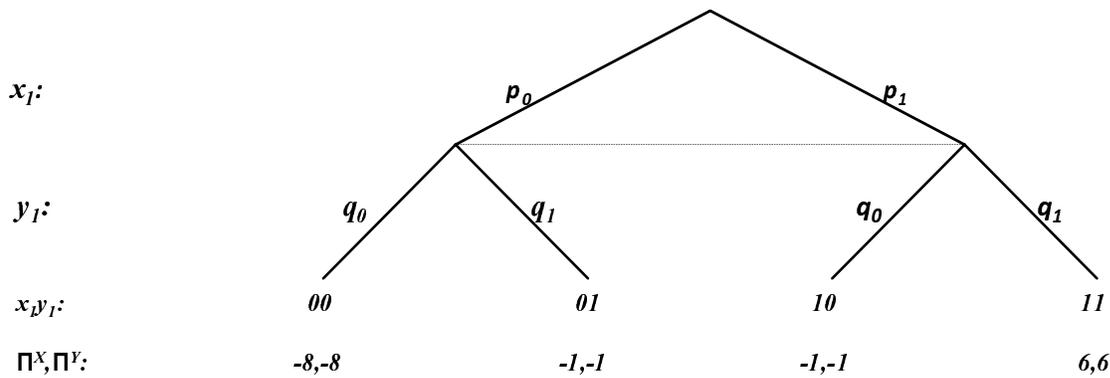}
\caption[The public goods game: Anti-correlated choices]{\em The case where players $(X,Y)$ adopt the
${\cal P}^X_B|_{x_2=1-y_1}\times{\cal P}^Y_B|_{y_2=1-x_1}$ joint probability space where both
players functionally anti-correlate their second stage choices to
their opponent's first stage choices.  Then a failure to invest
automatically invokes punishment while investment invokes no
punishment. \label{f_public_goods_cor_cor}}
\end{figure}

\subsection{Isomorphically anti-correlated space ${\cal P}^X_B|_{x_2=1-y_1}\times{\cal P}^Y_B|_{y_2=1-x_1}$}

Suppose first that both players $X$ and $Y$ choose to adopt a
joint probability space ${\cal P}^X_B|_{x_2=1-y_1}\times{\cal
P}^Y_B|_{y_2=1-x_1}$ as shown in Fig.
\ref{f_public_goods_cor_cor}, in which they each functionally
anti-correlate their second stage choices to the previous
choices of their opponents. This is implemented via
\begin{eqnarray}
   x_2            &=& 1-y_1   \nonumber \\
   p_{1|x_1y_1}   &=& \delta_{1,(1-y_1)}  \nonumber \\
   y_2            &=& 1-x_1   \nonumber \\
   q_{1|x_1y_1}   &=& \delta_{1,(1-x_1)}.
\end{eqnarray}

This choice of probability space alters the dimensions of the
game space, the game trees, and the payoff functions to be
\begin{eqnarray}
    \Pi^X  &=& 4 - x_1 + 3 y_1 - 4 x_2 - 8 y_2 \nonumber \\
           &=& 4 - x_1 + 3 y_1 - 4 (1-y_1) - 8 (1-x_1) \nonumber \\
           &=& -8 +7 x_1 + 7 y_1  \nonumber \\
    \Pi^Y  &=& 4 + 3 x_1 - y_1 - 8 x_2 - 4 y_2 \nonumber \\
           &=& 4 + 3 x_1 - y_1 - 8 (1-y_1) - 4 (1-x_1) \nonumber \\
           &=& -8 +7 x_1 + 7 y_1.
\end{eqnarray}
It is then immediately evident that players maximize their own
payoffs by choosing to invest $(x_1,y_1)=(1,1)$ which invokes a
subsequent lack of punishment in stage two giving
$(x_2,y_2)=(0,0)$.  The final payoffs are then
$(\Pi^X,\Pi^Y)=(6,6)$.

Optimization of the expected payoffs must reproduce this result.
The isomorphically constrained expected payoff functions can
simply be read from the tree in Fig.
\ref{f_public_goods_cor_cor} and are
\begin{eqnarray}
    X: \max_{p_1}\;\; \langle\Pi^X\rangle &=& -8 + 7 p_1 + 7 q_1  \nonumber \\
    Y: \max_{q_1}\;\; \langle\Pi^Y\rangle &=& -8 + 7 p_1 + 7 q_1.
\end{eqnarray}
These expected payoffs are continuous multivariate functions
dependent only on the freely varying parameters $p_1$ and $q_1$,
so the relevant gradient operator used by both players is
\begin{equation}
    \nabla =  \left[
    \frac{\partial}{\partial p_1},
    \frac{\partial}{\partial q_1} \right].
\end{equation}
Immediately then, the optimization conditions evaluated by each
player are
\begin{eqnarray}
 \frac{\partial\langle\Pi^X\rangle}{\partial p_1} &=& 7  \nonumber \\
 \frac{\partial\langle\Pi^Y\rangle}{\partial q_1} &=& 7,
\end{eqnarray}
ensuring that both players $X$ and player $Y$ maximize their
expected payoffs by investing their funds by setting $p_1=1$
giving $x_1=1$ and $q_1=1$ giving $y_1=1$. The functionally
assigned punishment choices then ensure that neither player
punishes the other so the equilibria choice of play is
$(x_1,y_1,x_2,y_2)=(1,1,0,0)$ generating expected payoffs of
$\left(\langle\Pi^X\rangle,\langle\Pi^Y\rangle
\right)=\left(6,6\right)$.

\begin{figure}[htb]
\centering
\includegraphics[width=\columnwidth,clip]{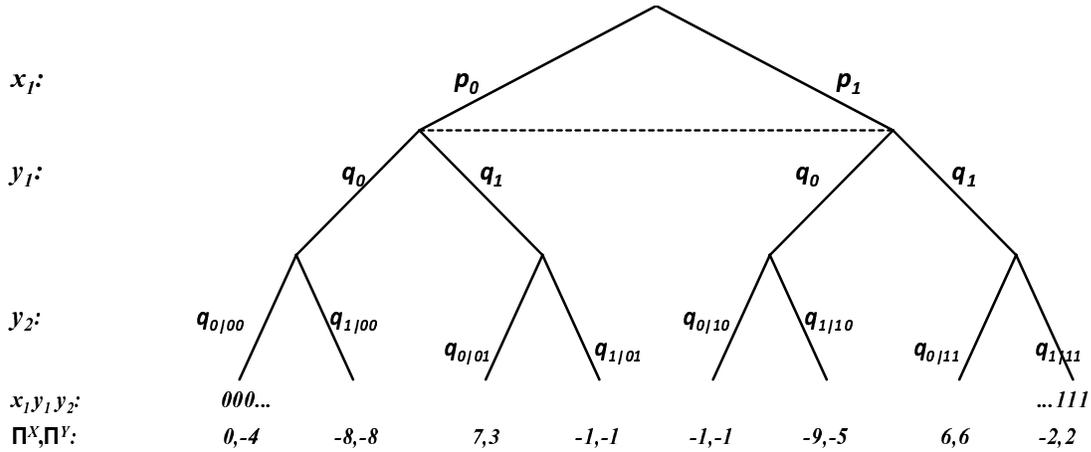}
\caption[The public goods game: Correlated and
independent strategies]{\em The case where players $(X,Y)$ adopt the
${\cal P}^X_B|_{x_1=1-y_1}\times{\cal P}^Y_B$ joint probability space where $X$
functionally anti-correlates their second stage choices to their
opponent's first stage choice and so automatically punishes a
failure to invest, while $Y$ adopts everywhere independent
behavioural strategies in both their stages. \label{f_public_goods_cor_ind}}
\end{figure}

\subsection{Anti-correlated and independent space: ${\cal P}^X_B|_{x_2=1-y_1}\times{\cal P}^Y_B$}

To complete this simplified analysis of the reduced public goods
game considered here, both players might also examine the
possible joint probability space ${\cal
P}^X_B|_{x_2=1-y_1}\times{\cal P}^Y_B$ in which $X$
anti-correlates their second stage choice to their opponent's
first stage choice while $Y$ does not employ any isomorphic
constraints---see Fig. \ref{f_public_goods_cor_ind}. (Symmetry
allows these results to be used for the space ${\cal
P}^X_B\times{\cal P}^Y_B|_{y_2=1-x_1}$ after an appropriate
reflection.) The required functional anti-correlations are
implemented via
\begin{eqnarray}
   x_2  &=& 1-y_1   \nonumber \\
   p_{1|x_1y_1} &=& \delta_{1,(1-y_1)}.
\end{eqnarray}

In the adopted probability space, the payoff functions for the
players are then
\begin{eqnarray}
    \Pi^X  &=& 4 - x_1 + 3 y_1 - 4 x_2 - 8 y_2 \nonumber \\
           &=& 4 - x_1 + 3 y_1 - 4 (1-y_1) - 8 y_2 \nonumber \\
           &=& - x_1 + 7 y_1 - 8 y_2 \nonumber \\
    \Pi^Y  &=& 4 + 3 x_1 - y_1 - 8 x_2 - 4 y_2 \nonumber \\
           &=& 4 + 3 x_1 - y_1 - 8 (1-y_1) - 4 y_2 \nonumber \\
           &=& -4 + 3 x_1 + 7 y_1 - 4 y_2.
\end{eqnarray}
Here, player $X$ sets $x_1=0$ to maximize their payoff while $Y$
sets $y_1=1$ and $y_2=0$ to maximize their payoff.  The final
outcome is $(\Pi^X,\Pi^Y)=(7,3)$.

A similar result is obtained from optimizing the expected payoff
functions.  The isomorphically constrained joint probability
space ${\cal P}^X_B|_{x_2=1-y_1}\times{\cal P}^Y_B$ specifies
the expected payoff optimization problem after the resolution of
the imposed functional constraints as
\begin{eqnarray}
    X: \max_{p_1}\;\; \langle\Pi^X\rangle
    &=&  4 - p_1 + 3 q_1
         - 4 \sum_{x_1y_1=0}^{1} p_{x_1} q_{y_1} p_{1|x_1y_1}
         - 8 \sum_{x_1y_1=0}^{1} p_{x_1} q_{y_1} q_{1|x_1y_1} \nonumber \\
    &=&  4 - p_1 + 3 q_1
         - 4 \sum_{x_1y_1=0}^{1} p_{x_1} q_{y_1} \delta_{1,(1-y_1)}
         - 8 \sum_{x_1y_1=0}^{1} p_{x_1} q_{y_1} q_{1|x_1y_1} \nonumber \\
    &=&  4 - p_1 + 3 q_1
         - 4 \sum_{x_1=0}^{1} p_{x_1} q_{0}
         - 8 \sum_{x_1y_1=0}^{1} p_{x_1} q_{y_1} q_{1|x_1y_1} \nonumber \\
    &=&  4 - p_1 + 3 q_1
         - 4 q_{0}
         - 8 \sum_{x_1y_1=0}^{1} p_{x_1} q_{y_1} q_{1|x_1y_1} \nonumber \\
    Y: \max_{q_1,q_{1|x_1y_1}}\;\; \langle\Pi^Y\rangle
    &=& 4 + 3 p_1 - q_1
        - 8 \sum_{x_1y_1=0}^{1} p_{x_1} q_{y_1} p_{1|x_1y_1}
        - 4 \sum_{x_1y_1=0}^{1} p_{x_1} q_{y_1} q_{1|x_1y_1}  \nonumber \\
    &=& 4 + 3 p_1 - q_1
        - 8 \sum_{x_1y_1=0}^{1} p_{x_1} q_{y_1} \delta_{1,(1-y_1)}
        - 4 \sum_{x_1y_1=0}^{1} p_{x_1} q_{y_1} q_{1|x_1y_1}  \nonumber \\
    &=& 4 + 3 p_1 - q_1
        - 8 \sum_{x_1=0}^{1} p_{x_1} q_{0}
        - 4 \sum_{x_1y_1=0}^{1} p_{x_1} q_{y_1} q_{1|x_1y_1}  \nonumber \\
    &=& 4 + 3 p_1 - q_1
        - 8 q_{0}
        - 4 \sum_{x_1y_1=0}^{1} p_{x_1} q_{y_1} q_{1|x_1y_1}.
\end{eqnarray}
These expected payoffs are continuous multivariate functions
dependent only on the first stage freely varying parameters
$p_1$ and $q_1$ and the second stage independent parameters
$\left[q_{1|00},q_{1|01},q_{1|10},q_{1|11}\right]$, so the
relevant gradient operator used by both players to analyze this
particular probability space is
\begin{equation}
    \nabla =
    \left[
    \frac{\partial}{\partial p_1},
    \frac{\partial}{\partial q_1},
    \frac{\partial}{\partial q_{1|00}},
    \frac{\partial}{\partial q_{1|01}},
    \frac{\partial}{\partial q_{1|10}},
    \frac{\partial}{\partial q_{1|11}} \right].
\end{equation}
The resulting optimization conditions evaluated by each player are
\begin{eqnarray}
 \frac{\partial\langle\Pi^X\rangle}{\partial p_1}
    &=& -1  + 8 \sum_{y_1=0}^{1} q_{y_1} \left( q_{1|0y_1}- q_{1|1y_1} \right) \nonumber \\
 \frac{\partial\langle\Pi^Y\rangle}{\partial q_1}
    &=&  7 + 4 \sum_{x_1=0}^{1} p_{x_1} \left( q_{1|x_10}-q_{1|x_11} \right) \nonumber \\
 \frac{\partial\langle\Pi^Y\rangle}{\partial q_{1|00}}
     &=& - 4 p_0 q_0 \nonumber \\
 \frac{\partial\langle\Pi^Y\rangle}{\partial q_{1|01}}
     &=& - 4 p_0 q_1 \nonumber \\
 \frac{\partial\langle\Pi^Y\rangle}{\partial q_{1|10}}
     &=& - 4 p_1 q_0 \nonumber \\
 \frac{\partial\langle\Pi^Y\rangle}{\partial q_{1|11}}
     &=& - 4 p_1 q_1.
\end{eqnarray}
The last four conditions here ensure that $Y$ maximizes their
expected payoff by setting $q_{1|x_1y_1}=0$ on any history
$x_1y_1$.  That is, $Y$ chooses the second stage choice $y_2=0$
and never punishes $X$ irrespective of $X$'s first stage move.
In turn, substituting these results into the second condition
establishes that $Y$ maximizes their expected payoff by setting
$q_1=1$ giving $y_1=1$.  That is, $Y$ always invests their funds
in stage one. Consequently, these results substituted into the
first condition shows that $X$ maximizes their payoff by setting
$p_1=0$ giving $x_1=0$ and so free rides on their opponent's
inability to punish them. The resulting equilibria choice of
play is $(x_1,y_1,x_2,y_2)=(0,1,0,0)$ generating expected
payoffs of
$\left(\langle\Pi^X\rangle,\langle\Pi^Y\rangle\right)=\left(7,3\right)$.

\subsection{Expected payoff comparison}

Altogether, the various joint probability spaces as considered
here which might be adopted by the players gives a table of
expected payoff outcomes of
\begin{equation}
    \begin{array}{c|cc}
      (\langle\Pi^X\rangle,\langle\Pi^Y\rangle) &     {\cal P}^Y_B                      &  {\cal P}^Y_B|_{y_2=1-x_1}               \\ \hline
                                                &                                       &                                          \\
       {\cal P}^X_B                             &         (4,4)                         &     (3,7)                                \\
                                                &                                       &                                          \\
       {\cal P}^X_B|_{x_2=1-y_1}                &         (7,3)                         &     (6,6)                                \\
    \end{array}
\end{equation}
making it evident that to maximize their payoff, both players
must rationally elect to use joint probability space ${\cal
P}^X_B|_{x_2=1-y_1}\times{\cal P}^Y_B|_{y_2=1-x_1}$ in
preference to any of the alternate probability space considered
here.  That is, players $X$ and $Y$ will undertake to
functionally anti-correlate their second stage decision to the
previous choice of their opponent, and thereby deny themselves a
second stage choice during the game. Again, they do this knowing
it to be the payoff maximizing choice of probability space
(among the few examined here).

The clear predictions of our analysis is that players of
unbounded rationality will choose to not free ride on their
neighbours and will punish free riders even at considerable cost
to themselves.

 \chapter{The centipede game}
 \label{chap_centipede}

\section{Introduction}

The centipede game was introduced by Rosenthal
\cite{Rosenthal_1981_92}.  A readily accessible treatment can be
found in \cite{Kreps_1990}.  The centipede game is of interest
due to the extreme discrepancy between experimentally observed
play and the predictions of game theory---see the experimental
investigations in \cite{McKelvey_1992_803} with discrepancies
explained by allowing players to altruistically consider their
opponent's payoffs, or by using learning approaches to explain
observed discrepancies in a normal form centipede game
\cite{Nagel_1998_356}.  More generally, the centipede game has
had a prime role in arguments over the definitions of
rationality, common knowledge of rationality, and backwards
induction
\cite{Binmore_87_17,Binmore_88_9,Binmore_1994_150,Aumann_1995_6,Binmore_1996_135,Aumann_1996_138,Aumann_1998_97}.
In part, this ongoing debate has led to the wider impugning of
backwards induction
\cite{Binmore_87_17,Binmore_88_9,Pettit_99_17,Broome_1999_237},
but see the defence of backwards induction in
\cite{Sobel_1993_114}. For an indication of the role of this
game in the wider economics and social sciences, see
\cite{Sigmund_2000_949}.

\begin{figure}[htb]
\centering
\includegraphics[width=0.9\columnwidth,clip]{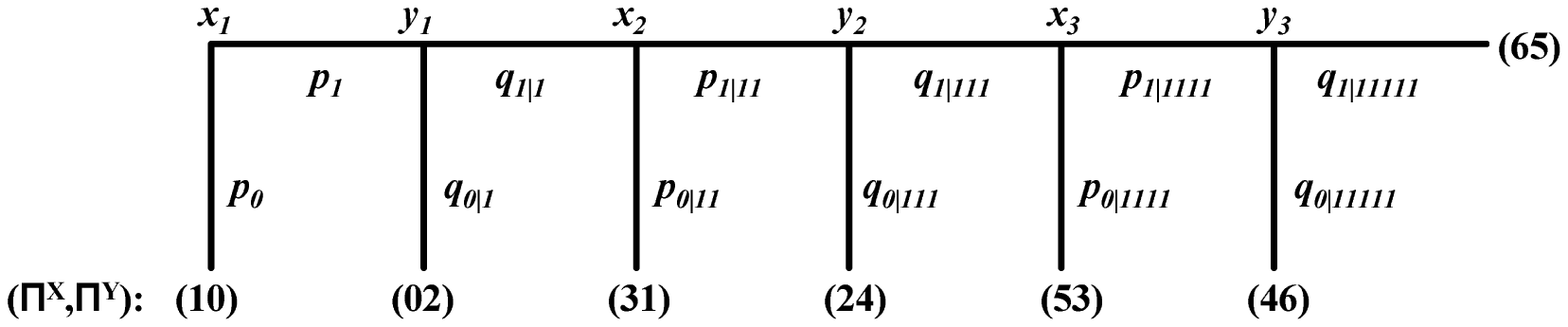}
\caption[The centipede game]{\em A truncated centipede game
decision tree over 6 stages where two players $X$ and $Y$
alternately choose to either play down ($x_i,y_i=0$ for $1\leq
i\leq 3$) in which case the game stops, or play across
($x_i,y_i=1$ for $1\leq i\leq 3$) so that either their opponent
faces a similar choice or the game terminates in stage 6.
\label{f_centipede}}
\end{figure}

\section{The centipede game}

The centipede game gains its peculiar name as it normally
features two players playing over 100 turns so that, when drawn
horizontally as in Fig. \ref{f_centipede}, the game tree takes
the appearance of a centipede.  Here, we truncate the game
without loss of generality at only 6 stages allowing a tractable
analysis. In this truncated centipede game, each player $X$ or
$Y$ must alternately elect to either play down ($x_i,y_i=0$ for
$1\leq i\leq 3$) in which case the game immediately terminates
and players gain the respective payoffs shown, or play across
($x_i,y_i=1$ for $1\leq i\leq 3$) in which case either their
opponent plays or the game terminates with the payoffs shown.
When either player hands play to their opponent, they suffer a
short term loss of potential payoff with the prospect of a long
term gain.  The interest in this game comes from the
countervailing effects of these short term losses and long term
gains which combine together to ensure that human players
typically fail to follow the recommendations of game theory and
yet significantly improve their payoffs by doing so.

In fact, the centipede game has a unique subgame perfect
equilibrium solution, which can be readily located by simply
inspecting Fig. \ref{f_centipede} and applying backwards
induction.  In the last (far right) stage, $Y$ can choose
$y_3=0$ to obtain a payoff of $\Pi^Y=6$, or can choose $y_3=1$
to obtain a payoff of $\Pi^Y=5$.  Obviously, $Y$ will prefer to
play down with $y_3=0$ in this final stage to maximize their
payoff.  Player $X$ is well able to deduce this to conclude that
if they choose $x_3=1$ to play across in the second last stage
then they will obtain a payoff of $\Pi^X=4$ when $Y$
subsequently plays down.  In contrast, should $X$ play down
themselves by choosing $x_3=0$, they will gain the improved
payoff of $\Pi^X=5$. Obviously, $X$ will choose $x_3=0$ to
preempt $Y$'s choice of $y_3=0$.  Exactly the same argument
applies to $Y$'s choice in the fourth stage, to $X$'s choice in
the third stage, to $Y$'s choice in the second stage, and
finally to $X$'s choice in the first stage.  That is, being able
to deduce that $Y$ will play down in the second stage by
choosing $y_1=0$ to give $X$ a payoff of $\Pi^X=0$, then player
$X$ will choose to maximize their payoff by preempting $Y$ and
playing down in the first stage through the choice $x_1=0$ to
gain an improved payoff of $\Pi^X=1$.  The associated payoff for
$Y$ is $\Pi^Y=0$.

And here lies the conundrum.  The sole conventionally mandated
choice of play lies in the first player $X$ choosing down at the
first opportunity to gain a mere fraction of the potential
payoff should they and their opponent play across a few times.
Interestingly, most people playing this game will indeed ignore
the conventionally sanctioned choice with both players typically
playing across repeatedly to drastically improve their payoffs.
Just as in the other games under consideration here, it seems
intuitively obvious to human players that adopting
``non-rational" play will improve payoffs.  However,
conventional analysis has had trouble explaining these
propensities.  Here, we show that lifting implicit conventional
bounds on rationality to allow players to take into account
alternate probability spaces easily produces game theoretic
predictions in agreement with observation.

Altogether, the payoffs to the players in the centipede game
considered here are
\begin{eqnarray}
    \Pi^X &=& (1 - x_1) + x_1
       \left( y_1
       \left[ 3 (1-x_2) + x_2
       \left\{ 2 (1-y_2) + y_2
       \left( 5(1-x_3) + x_3
       \left[ 4(1-y_3) + 6 y_3 \right] \right) \right\}\right]\right)   \nonumber \\
    \Pi^Y &=&  x_1
       \left( 2(1-y_1)+y_1
       \left[ 1 (1-x_2) + x_2
       \left\{ 4 (1-y_2) + y_2
       \left( 3(1-x_3) + x_3
       \left[ 6(1-y_3) + 5 y_3 \right] \right) \right\}\right]\right). \nonumber \\
       &&
\end{eqnarray}
As usual, players must then choose amongst their possible
probability spaces ${\cal P}^X$ and ${\cal P}^Y$ to optimize
their payoffs.  A first choice will be the examination of the
conventionally mandated probability space, which we turn to now.

\subsection{The unconstrained space ${\cal P}^X_B\times{\cal P}^Y_B$}

To replicate the standard conventional analysis (the backwards
induction analysis above), both players $X$ and $Y$ together
adopt a joint probability space ${\cal P}^X_B\times{\cal P}^Y_B$
in which every behavioural strategy on every history set is
independent---see Figs. \ref{f_centipede}. The expected payoff
optimization problem for each player $Z\in\{X,Y\}$ can be
written
\begin{eqnarray}
    Z: \max \;\; \langle\Pi^Z\rangle
    &=& \sum_{x_1,y_1,x_2,y_2,x_3,y_3=0}^{1}
          P^{XY}(x_1,y_1,x_2,y_2,x_3,y_3) \Pi^Z(x_1,y_1,x_2,y_2,x_3,y_3)  \nonumber \\
    &=& \sum_{x_1,y_1,x_2,y_2,x_3,y_3=0}^{1}
          P^{X}(x_1) P^Y(y_1|x_1)
          P^X(x_2|x_1y_1) P^{Y}(y_2|x_1y_1x_2) \times  \\
        &&  \hspace{1.5cm} \times P^X(x_3|x_1y_1x_2y_2) P^{Y}(y_3|x_1y_1x_2y_2x_3)
          \Pi^X(x_1,y_1,x_2,y_2,x_3,y_3).  \nonumber
\end{eqnarray}
To simplify notation, we write $P^X(x_2|x_1y_1)\rightarrow
p_{x_2|x_1y_1}$, $P^Y(y_2|x_1y_1)\rightarrow q_{y_2|x_1y_1}$ and
so on, and we take account of normalization conditions
$p_{0|x_1y_1}+p_{1|x_1y_1}=1$ and $q_{0|x_1y_1}+q_{1|x_1y_1}=1$
on all histories.

Consequently, the expected payoff optimization problem becomes
\begin{eqnarray}   \label{eq_centipede_payoffs}
    X: \max_{p_1,p_{1|11},p_{1|1111}}\;\; \langle\Pi^X\rangle
    &=&   \left[1-p_1\right] + \nonumber \\
       &&   \hspace{0.5cm} p_1 \left\{ 0 +  \right.  \nonumber \\
       &&   \hspace{1cm} q_{1|1} \left( 3 \left[1-p_{1|11}\right] + \right.  \nonumber \\
       &&   \hspace{1.5cm} p_{1|11} \left\{ 2 \left[1-q_{1|111}\right] + \right.  \nonumber \\
       &&   \hspace{2cm} q_{1|111} \left( 5 \left[1-p_{1|1111}\right] + \right.  \nonumber \\
       &&   \hspace{2.5cm} \left. \left. \left. \left.
            p_{1|1111} \left[ 4 \left[1-q_{1|11111}\right] + 6 q_{1|11111} \right]
            \right) \right\} \right) \right\} \nonumber \\
    Y: \max_{q_{1|1},q_{1|111},q_{1|11111}}\;\; \langle\Pi^Y\rangle
    &=&  p_1 \left[ 2 \left[1-q_{1|1}\right] +  \right.   \nonumber \\
       && \hspace{0.5cm} q_{1|1} \left(  \left[1-p_{1|11}\right] + \right.  \nonumber \\
       &&   \hspace{1cm} p_{1|11} \left\{ 4 \left[1-q_{1|111}\right] + \right.  \nonumber \\
       &&   \hspace{1.5cm} q_{1|111} \left( 3 \left[1-p_{1|1111}\right] + \right.  \nonumber \\
       &&   \hspace{2cm} \left. \left. \left. \left.
            p_{1|1111} \left[ 6 \left[1-q_{1|11111}\right] + 5 q_{1|11111} \right]
            \right) \right\} \right) \right].
\end{eqnarray}
In these optimization problems, the players $X$ and $Y$ have
respective independent probability parameters of
$p_1,p_{1|11},p_{1|1111}$ and $q_{1|1},q_{1|111},q_{1|11111}$
all of which can vary freely over $[0,1]$. Consequently, in the
joint space ${\cal P}^X_B\times{\cal P}^Y_B$, each player
optimizes using the gradient operator
\begin{equation}
    \nabla =
    \left[
    \frac{\partial}{\partial p_1},
    \frac{\partial}{\partial q_{1|1}},
    \frac{\partial}{\partial p_{1|11}},
    \frac{\partial}{\partial q_{1|111}},
    \frac{\partial}{\partial p_{1|1111}},
    \frac{\partial}{\partial q_{1|11111}} \right],
\end{equation}
as all other parameters disappear.  The easiest way to complete
the optimization is via backwards induction, so both players first
evaluate the last stage choice of player $Y$ via
\begin{equation}
    \frac{\partial\langle\Pi^Y\rangle}{\partial q_{1|11111}}
    = -p_1q_{1|1}p_{1|11}q_{1|111}p_{1|1111} \leq 0,
\end{equation}
which is either zero should any player have played down in any
preceding stage in which case $Y$ is indifferent to any choice
in this final stage, or always negative so essentially $Y$ plays
down via $q_{1|11111}=0$ and $y_3=0$.  This result allows player
$X$ to optimize their choice in the second last stage via
\begin{equation}
    \frac{\partial\langle\Pi^X\rangle}{\partial p_{1|1111}}
    = - p_1q_{1|1}p_{1|11}q_{1|111} \leq 0,
\end{equation}
which again, leads to the setting $p_{1|1111}=0$ and $x_3=0$.  A
similar analysis proceeds backwards through all the stages to
give the final solution, deducible by both players, of
$(x_1,y_1,x_2,y_2,x_3,y_3)=(0,0,0,0,0,0)$.  This choice garners
players the conventionally mandated payoffs of
$\left(\langle\Pi^X\rangle,\langle\Pi^Y\rangle \right)=(1,0)$.

\subsection{Isomorphically constrained spaces}

Naturally, players of unbounded rationality will not be content
to merely examine the conventionally mandated joint probability
space ${\cal P}^X_B\times{\cal P}^Y_B$ and will turn to consider
alternative joint probability spaces. In each alternative space,
isomorphic constraints alter game spaces and trees and thereby
alter the subgame decompositions used in the conventional
analysis to locate novel equilibria. We consider such
alternatives now.

As usual, there are an infinity of possible probability spaces
that might be adopted by the players in the sequential centipede
game, and we can here consider only a partial search of these
possible spaces.  We first suppose that the players restrict
their attention to ``Markovian" strategies in which the variable
of a given stage is only conditioned on the outcome of the
immediately preceding stage.  The alternative---correlating
variables in the given stage to the outcomes in every preceding
stage---simply generates to many options without adding
significantly to the analysis.  Given this restriction, a
moments reflection will make it obvious that there is little
point in a player choosing to anti-correlate their choice in a
given stage to their opponent's previous choice.  There opponent
must have played across so an anti-correlation would simply
force a move down and this merely duplicate the outcomes of the
conventional analysis above.  The same considerations make it
immediately attractive to have players consider perfect
correlations between the opponent's choices in the preceding
stage and the current choices in the present stage as a previous
choice of across then implies a current choice of across.  We
therefore suppose that players, in each stage after the first,
can make their choices either independently or by correlation to
their opponent's previous choice.

These consideration leave four possible probability spaces to be
enacted by player $X$, namely
\begin{equation}
    \begin{array}{l}
   {\cal P}^X_B              \\
   {\cal P}^X_B|_{x_2=y_1}   \\
   {\cal P}^X_B|_{x_3=y_2}   \\
   {\cal P}^X_B|_{x_2=y_1,x_3=y_2}.
    \end{array}
\end{equation}
Similarly, there are eight possible spaces to be enacted by
player $Y$, namely
\begin{equation}
    \begin{array}{l}
   {\cal P}^Y_B                       \\
   {\cal P}^Y_B|_{y_1=x_1}            \\
   {\cal P}^Y_B|_{y_2=x_2}            \\
   {\cal P}^Y_B|_{y_3=x_3}            \\
   {\cal P}^Y_B|_{y_1=x_1,y_2=x_2}    \\
   {\cal P}^Y_B|_{y_1=x_1,y_3=x_3}    \\
   {\cal P}^Y_B|_{y_2=x_2,y_3=x_3}    \\
   {\cal P}^Y_B|_{y_1=x_1,y_2=x_2,y_3=x_3}.
    \end{array}
\end{equation}
Altogether, this makes 32 joint probability spaces that need be
considered. We now turn to follow the players in their analysis
of the outcomes from their joint adoption of all of these
combinations of spaces.

\subsection{The space ${\cal P}^X_B|_{x_2=y_1,x_3=y_2}\times{\cal P}^Y_B|_{y_1=x_1,y_2=x_2,y_3=x_3}$}

Given the joint probability space ${\cal
P}^X_B|_{x_2=y_1,x_3=y_2}\times{\cal
P}^Y_B|_{y_1=x_1,y_2=x_2,y_3=x_3}$ in which every variable after
the first stage is isomorphically constrained to be perfectly
correlated to the preceding choice by their opponent, we have
the variable assignment reduces to $y_3=x_3=y_2=x_2=y_1=x_1$.
Subsequently, the payoff functions for both players become
\begin{eqnarray}
    \Pi^X &=& (1 - x_1) + x_1
       \left( y_1
       \left[ 3 (1-x_2) + x_2
       \left\{ 2 (1-y_2) + y_2
       \left( 5(1-x_3) + x_3
       \left[ 4(1-y_3) + 6 y_3 \right] \right) \right\}\right]\right)   \nonumber \\
      &=& (1 - x_1) + x_1
       \left( x_1
       \left[ 3 (1-x_1) + x_1
       \left\{ 2 (1-x_1) + x_1
       \left( 5(1-x_1) + x_1
       \left[ 4(1-x_1) + 6 x_1 \right] \right) \right\}\right]\right)   \nonumber \\
      &=& 1  + 5 x_1   \nonumber \\
    \Pi^Y &=&  x_1
       \left( 2(1-y_1)+y_1
       \left[ 1 (1-x_2) + x_2
       \left\{ 4 (1-y_2) + y_2
       \left( 3(1-x_3) + x_3
       \left[ 6(1-y_3) + 5 y_3 \right] \right) \right\}\right]\right). \nonumber \\
 &=&  x_1
       \left( 2(1-x_1)+x_1
       \left[ 1 (1-x_1) + x_1
       \left\{ 4 (1-x_1) + x_1
       \left( 3(1-x_1) + x_1
       \left[ 6(1-x_1) + 5 x_1 \right] \right) \right\}\right]\right). \nonumber \\
 &=&  5 x_1.
\end{eqnarray}
Here, it is immediately evident that player $X$ maximizes their
payoff by setting $x_1=1$ generating a sequence of play of
$(x_1,x_2,x_3,y_1,y_2,y_3)=(1,1,1,1,1,1)$ and payoffs of
$(\Pi^X,\Pi^Y)=(6,5)$.

A similar result is obtained from optimizing the expected
payoffs via
\begin{eqnarray}
    X: \max_{p_1}\;\; \langle\Pi^X\rangle
    &=& \sum_{x_1,y_1,x_2,y_2,x_3,y_3=0}^{1}
          P^{X}(x_1) \delta_{y_1x_1}
          \delta_{x_2y_1} \delta_{y_2x_2}
          \delta_{x_3y_2} \delta_{y_3x_3}  \Pi^X  \nonumber \\
    &=& \sum_{x_1=0}^{1}
          P^{X}(x_1) \Pi^X(x_1,x_1,x_1,x_1,x_1,x_1)  \nonumber \\
    &=& 1 + 5 p_1 \nonumber \\
    Y: \langle\Pi^Y\rangle
    &=& \sum_{x_1,y_1,x_2,y_2,x_3,y_3=0}^{1}
          P^{X}(x_1) \delta_{y_1x_1}
          \delta_{x_2y_1} \delta_{y_2x_2}
          \delta_{x_3y_2} \delta_{y_3x_3} \Pi^Y  \nonumber \\
    &=& \sum_{x_1=0}^{1}
          P^{X}(x_1)  \Pi^Y(x_1,x_1,x_1,x_1,x_1,x_1)  \nonumber \\
    &=& 5 p_1.
\end{eqnarray}
Here, player $Y$ has left themselves no choices in any stage.
As a result, the optimization is completed by
\begin{equation}
    \frac{\partial\langle\Pi^X\rangle}{\partial p_1} = 5 > 0,
\end{equation}
so $X$ sets $p_1=1$ to choose $x_1=1$ and plays across in stage
1. This choice is mimicked in every subsequent stage giving
$(x_1,y_1,x_2,y_2,x_3,y_3)=(1,1,1,1,1,1)$ to generate payoffs to
the players of $\left(\langle\Pi^X\rangle,\langle\Pi^Y\rangle
\right)=(6,5)$.

\subsection{The space ${\cal P}^X_B|_{x_2=y_1,x_3=y_2}\times{\cal P}^Y_B|_{y_2=x_2,y_3=x_3}$}

In the joint probability space ${\cal
P}^X_B|_{x_2=y_1,x_3=y_2}\times{\cal P}^Y_B|_{y_2=x_2,y_3=x_3}$
the variable assignment reduces to $y_3=x_3=y_2=x_2=y_1$ so the
payoff functions become
\begin{eqnarray}
    \Pi^X &=& (1 - x_1) + x_1
       \left( y_1
       \left[ 3 (1-x_2) + x_2
       \left\{ 2 (1-y_2) + y_2
       \left( 5(1-x_3) + x_3
       \left[ 4(1-y_3) + 6 y_3 \right] \right) \right\}\right]\right)   \nonumber \\
    &=& (1 - x_1) + x_1
       \left( y_1
       \left[ 3 (1-y_1) + y_1
       \left\{ 2 (1-y_1) + y_1
       \left( 5(1-y_1) + y_1
       \left[ 4(1-y_1) + 6 y_1 \right] \right) \right\}\right]\right)   \nonumber \\
    &=& 1 - x_1 + 6 x_1  y_1   \nonumber \\
 \Pi^Y &=&  x_1
       \left( 2(1-y_1)+y_1
       \left[ 1 (1-x_2) + x_2
       \left\{ 4 (1-y_2) + y_2
       \left( 3(1-x_3) + x_3
       \left[ 6(1-y_3) + 5 y_3 \right] \right) \right\}\right]\right) \nonumber \\
&=&  x_1 ( 2 + 3 y_1).
\end{eqnarray}
These payoff functions establish that player $Y$ maximizes their
payoff by setting $y_1=1$ while player $X$ maximizes their
income by setting $x_1=1$ generating a sequence of play of
$(x_1,x_2,x_3,y_1,y_2,y_3)=(1,1,1,1,1,1)$ and payoffs of
$(\Pi^X,\Pi^Y)=(6,5)$.

The expected payoff functions optimization task becomes
\begin{eqnarray}
    X: \max_{p_1}\;\; \langle\Pi^X\rangle
    &=& \sum_{x_1,y_1,x_2,y_2,x_3,y_3=0}^{1}
          P^{X}(x_1) P^Y(y_1|x_1)
          \delta_{x_2y_1} \delta_{y_2x_2}
          \delta_{x_3y_2} \delta_{y_3x_3}  \Pi^X  \nonumber \\
    &=& \sum_{x_1y_1=0}^{1}
          P^{X}(x_1) P^Y(y_1|x_1) \Pi^X(x_1,y_1,y_1,y_1,y_1,y_1)  \nonumber \\
    &=& 1 - p_1 + 6 p_1 q_{1|1} \nonumber \\
    Y: \max_{q_{1|1}}\;\; \langle\Pi^Y\rangle
    &=& \sum_{x_1,y_1,x_2,y_2,x_3,y_3=0}^{1}
          P^{X}(x_1) P^Y(y_1|x_1)
          \delta_{x_2y_1} \delta_{y_2x_2}
          \delta_{x_3y_2} \delta_{y_3x_3} \Pi^Y  \nonumber \\
    &=& \sum_{x_1y_1=0}^{1}
          P^{X}(x_1) P^Y(y_1|x_1) \Pi^Y(x_1,y_1,y_1,y_1,y_1,y_1)  \nonumber \\
    &=& p_1 \left[ 2 + 3 q_{1|1}\right].
\end{eqnarray}
In this case, the optimization is completed by
\begin{eqnarray}
    \frac{\partial\langle\Pi^X\rangle}{\partial p_1}
    &=&  -1 + 6 q_{1|1} \nonumber \\
    \frac{\partial\langle\Pi^Y\rangle}{\partial q_{1|1}}
    &=&  3 p_1.
\end{eqnarray}
Essentially then, player $Y$ notes their positive gradient and
so sets $q_{1|1}=1$ to give $y_1=1$.  In turn, player $X$
deduces this and sets $p_1=1$ to give $x_1=1$.  Together, in the
joint probability space ${\cal
P}^X_B|_{x_2=y_1,x_3=y_2}\times{\cal P}^Y_B|_{y_2=x_2,y_3=x_3}$,
the optimization generates the play choices
$(x_1,y_1,x_2,y_2,x_3,y_3)=(1,1,1,1,1,1)$ to generate payoffs to
the players of $\left(\langle\Pi^X\rangle,\langle\Pi^Y\rangle
\right)=(6,5)$.

\begin{figure}[htb]
\centering
\includegraphics[width=\columnwidth,clip]{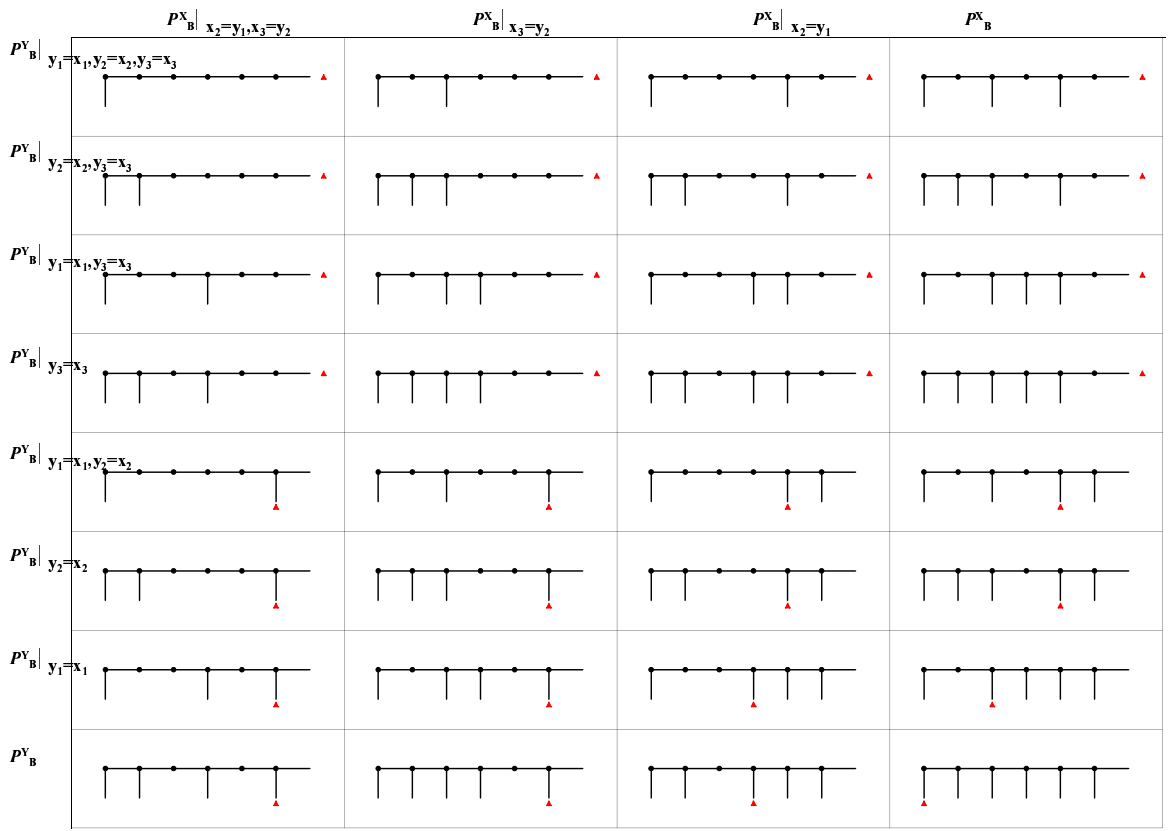}
\caption[The centipede game: Isomorphic probability spaces and decision
trees]{\em The 32 distinct trees and equilibrium pathways
(indicated by triangles) given that players $X$ and $Y$ adopt the
probability spaces shown. Dots indicate successive decision nodes,
where nodes with a descending vertical line are independent
decision points and nodes lacking a descending vertical line are
isomorphically constrained to equal the immediately preceding decision.
 \label{f_centipede_extended}}
\end{figure}

\subsection{Expected payoff comparison across multiple probability spaces}

Similar analysis to that above can be applied to evaluate the
expected payoffs in all the other combinations of joint
probability spaces to give the payoff combination table
\begin{equation}
    \begin{array}{c|cccc}
   \left(\langle\Pi^X\rangle,\langle\Pi^Y\rangle\right)
       & {\cal P}^X_B|_{x_2=y_1,x_3=y_2} & {\cal P}^X_B|_{x_3=y_2}  & {\cal P}^X_B|_{x_2=y_1}  & {\cal P}^X_B  \\ \hline
  {\cal P}^Y_B|_{y_1=x_1,y_2=x_2,y_3=x_3} &       (6,5)      &    (6,5)          &    (6,5)          &    (6,5)      \\
  {\cal P}^Y_B|_{y_1=x_1,y_3=x_3}         &       (6,5)      &    (6,5)          &    (6,5)          &    (6,5)      \\
  {\cal P}^Y_B|_{y_2=x_2,y_3=x_3}         &       (6,5)      &    (6,5)          &    (6,5)          &    (6,5)      \\
  {\cal P}^Y_B|_{y_3=x_3}                 &       (6,5)      &    (6,5)          &    (6,5)          &    (6,5)      \\
  {\cal P}^Y_B|_{y_1=x_1,y_2=x_2}         &       (4,6)      &    (4,6)          &    (5,3)          &    (5,3)      \\
  {\cal P}^Y_B|_{y_1=x_1}                 &       (4,6)      &    (4,6)          &    (5,3)          &    (5,3)      \\
  {\cal P}^Y_B|_{y_2=x_2}                 &       (4,6)      &    (4,6)          &    (2,4)          &    (3,1)      \\
  {\cal P}^Y_B                            &       (4,6)      &    (4,6)          &    (2,4)          &    (1,0).      \\
    \end{array}.
\end{equation}
The equivalent trees and equilibrium pathways are shown in Fig.
\ref{f_centipede_extended}.  Perusal of this table makes it
clear that players do not optimize their payoffs by choosing the
conventionally mandated joint probability space. Rather, it is
much more likely that $Y$ will choose any probability space in
which their last stage variable is isomorphically constrained.
In turn, this alters the payoffs for player $X$ in such a way as
to render them indifferent to any choice of probability space.
The net result will be that $X$ will find themselves playing
across in the first stage irrespective of which space they
adopt.

A more sophisticated analysis in a longer game would take into
account end-game effects where players might express some
preference for terminating the game slightly early.  Such
tendencies are similar to those seen in the finite iterated
prisoner's dilemma game, and as there, are not likely to make it
irrational for players to play across in the early stages of the
centipede game.

The extended analysis presented here produces game theoretic
predictions in substantial accord with observed human play in
the centipede game.  As noted above, this agreement contrasts
sharply with the manifest contradiction between the game
theoretic predictions of conventional analysis and observed play
tendencies. As such, we take these observations as evidence that
humans naturally take account of isomorphic constraints in
strategic play in game theory.

 \chapter{The Iterated Prisoner's Dilemma}
 \label{chap_iterated_prisoner_dilemma}

\section{Introduction}

Conventional game analysis holds that it is rational for players
in a finite iterated prisoner's dilemma to adopt the
noncooperative ``all defect" as the optimal solution under
common knowledge of rationality (CKR) even though human players
are commonly observed to increase payoffs by irrationally
adopting alternative strategies. There are many observations of
this mismatch between theoretical prediction and observed
behaviour
\cite{Cooper_96_18,Milinski_98_13,Davis_99_89,Croson_00_29}.
These mismatches have typically been explained by introducing
behavioral factors such as bounded rationality, incomplete
information, and other innate tendencies promoting cooperative
and altruistic behaviours.  In particular, these suggestions
include modifying definitions of rationality to include
reciprocity, fairness and altruism or to otherwise bound
rationality
\cite{Radner_80_13,Radner_86_38,Vegaredondo_94_18,Harborne_97_13,Anthonisen_99_14,Fehr_99_81,Fehr_03_785},
via modelling the evolution of cooperation
\cite{Axelrod_1984,Nowak_98_573}, by taking account of
incomplete information
\cite{Harsanyi_67_15,Kreps_82_24,Fudenberg_86_53,Sarin_99_10}
and uncertainty in the number of repeat stages
\cite{Neyman_99_45}, to bound the complexity of implementable
strategies \cite{Neyman_85_22,Rubinstein_86_83,Cho_99_93}, to
account for communication and coordination costs
\cite{Raff_00_10}, to incorporate reputation and experimentation
effects \cite{Evans_97_11} or secondary utility functions as in
benevolence theory \cite{Selten_78_12} or in moral discussions
\cite{Sheng_94_23}, to include adaptive learning
\cite{Groes_99_12} or fuzzy logic \cite{Song_99_63}, or more
directly, to employ comprehensive constructions of normal form
strategy tables incorporating belief strategies
\cite{Howard_71,Rapoport_67,Straffin_93}. Interestingly, quantum
correlations can be introduced to resolve the prisoner's dilemma
\cite{Eisert_99_30}.

\begin{figure}[htb]
\centering
\includegraphics[width=\columnwidth,clip]{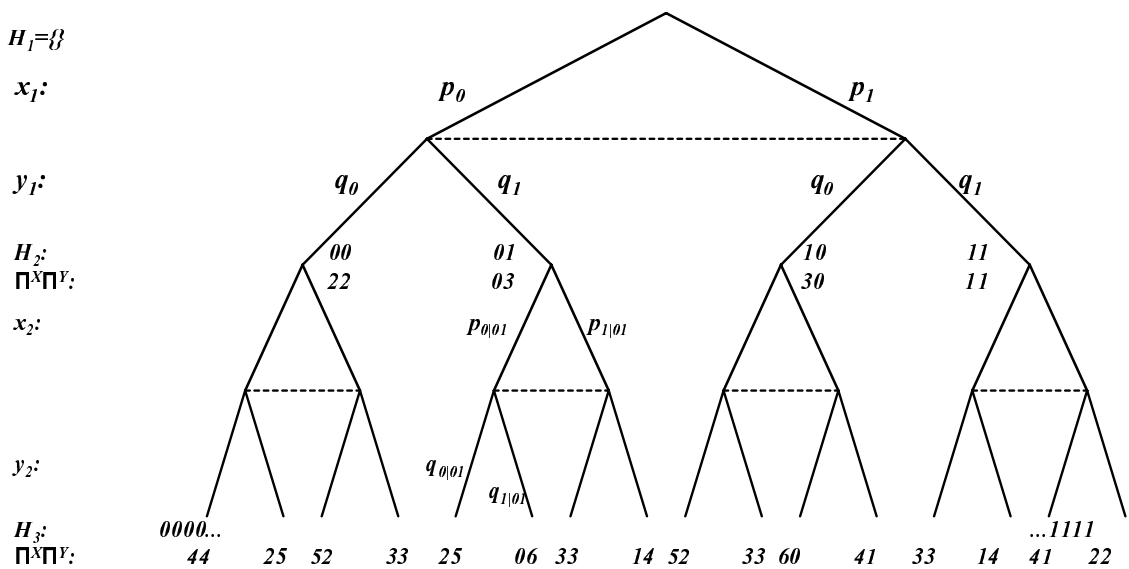}
\caption[A two stage iterated prisoner's dilemma]{\em A two stage
game decision tree where two non-communicating players
simultaneously choose moves $x_n$ or $y_n$ equal to ``0" or ``1"
at stage $n$ with respective probabilities $P^X(x_n|H_n)$ and
$P^Y(y_n|H_n)$ at every decision point.  At the beginning of each
stage, players know the history sets
$H_n=\{x_1,y_1,\dots,x_{n-1},y_{n-1}\}$ detailing the shared
information known to both players of all choices to that stage
(with $H_1=\{\}$).  Players also know their cumulative payoffs
$(\Pi^X,\Pi^Y)$ to that point. \label{f_two_stage_game_IPD}}
\end{figure}

\section{The finite Iterated Prisoner's Dilemma}
\label{sect_IPD}

In this chapter, we will examine the finite iterated prisoner's
dilemma while using the strong isomorphic mappings of
probability theory to construct our mixed and behavioural
strategy game spaces.  Our particular focus will be to examine
whether cooperation is rational in the finite iterated
prisoner's dilemma.  As usual, we assume our players are
rational and of unbounded capacity, and that they have adopted
common knowledge of rationality (CKR). An illustrative game tree
depicting a two stage iterated prisoner's dilemma is shown in
Fig. \ref{f_two_stage_game_IPD}.

The finite iterated prisoner's dilemma is defined here over a
finite number of $N$ stages, where at each stage $1\leq n\leq N$
two non-communicating players $X$ and $Y$ choose moves $x_n$ and
$y_n$ chosen to be either $0$ (cooperation) or $1$ (defection).
The payoffs gained in each stage are given by the payoff matrix
\begin{equation}\label{eq_2pIPD}
  \begin{array}{cc}
      & Y \\
    X &
    \begin{array}{c|cc}
      (\pi_x,\pi_y)  & 0 & 1 \\   \hline
      0 & (2,2) & (0,3) \\
      1 & (3,0) & (1,1), \\
    \end{array}
  \end{array}
\end{equation}
equivalent to the single stage payoff functions
\begin{eqnarray}        \label{eq_pi}
  \pi_x (x_n,y_n) & = &  2 + x_n - 2 y_n    \nonumber \\
  \pi_y (x_n,y_n) & = &  2 - 2 x_n + y_n.
\end{eqnarray}
For multiple stage games, total game payoffs of a finite $N$
stage game are simply the sum of single stage payoffs.  The
optimization problem for both players is then
\begin{eqnarray}     \label{eq_define_Pi}
    X: \max_{x_1,\dots,x_N}\;\;  \Pi^X(x_1,y_1,\dots x_N,y_N) &=& \sum_{n=1}^N (2+x_n-2y_n) \nonumber \\
    Y: \max_{y_1,\dots,y_N}\;\; \Pi^Y(x_1,y_1,\dots x_N,y_N) &=& \sum_{n=1}^N (2-2x_n+y_n).
\end{eqnarray}
Each player desires to maximize their respective endgame payoffs
by varying their respective move choices $x_n$ and $y_n$ over
every stage of the game. (The players know $N$ in advance.)

Yet more generally, players choose their moves probabilistically
to prevent their opponent predicting and exploiting
deterministic strategies.  The players will then adopt the joint
probability space ${\cal P}^X\times{\cal P}^Y$, and so seek to
maximize their respective expected payoffs
\begin{eqnarray}           \label{eq_expected_payff_functions}
    X: \max_{{\cal P}^X}\;\; \langle\Pi^X\rangle
       &=& \sum_{x_1\dots y_N=0}^1
          P^{XY}(x_1,y_1,\dots,x_N,y_N) \Pi^X(x_1,y_1,\dots,x_N,y_N)  \nonumber \\
       &=& \sum_{x_1\dots y_N=0}^1
          P^X(x_1)P^Y(y_1) P^X(x_2|H_2)P^Y(y_2|H_2)\dots \times  \nonumber \\
         && \hspace{2cm}   \dots  P^X(x_N|H_N)P^Y(y_N|H_N)
            \sum_{n=1}^N (2+x_n-2y_n)  \nonumber \\
       &=& 2N + \sum_{n=1}^N \sum_{\stackrel{x_1\dots x_n=0}{y_1\dots y_n=0}}^1
          P^X(x_1)P^Y(y_1) P^X(x_2|H_2)P^Y(y_2|H_2)\dots \times  \nonumber \\
       && \hspace{3cm}   \dots  P^X(x_n|H_n)P^Y(y_n|H_n)
        (x_n-2y_n)  \nonumber \\
    Y: \max_{{\cal P}^Y}\;\; \langle\Pi^Y\rangle
       &=& \sum_{x_1\dots y_N=0}^1
          P^{XY}(x_1,y_1,\dots,x_N,y_N)
          \Pi^Y(x_1,y_1,\dots,x_N,y_N) \nonumber \\
          &=& \sum_{x_1\dots y_N=0}^1
            P^X(x_1)P^Y(y_1)P^X(x_2|H_2)P^Y(y_2|H_2)\dots\times  \nonumber \\
       && \hspace{2cm}   \dots  P^X(x_N|H_N)P^Y(y_N|H_N)
            \sum_{n=1}^N (2-2x_n+y_n)  \nonumber \\
       &=& 2N+\sum_{n=1}^N \sum_{\stackrel{x_1\dots x_n=0}{y_1\dots y_n=0}}^1
          P^X(x_1)P^Y(y_1) P^X(x_2|H_2)P^Y(y_2|H_2)\dots \times  \nonumber \\
       && \hspace{3cm}   \dots  P^X(x_n|H_n)P^Y(y_n|H_n)
        (y_n-2x_n),
\end{eqnarray}
We have written $P^Z(z_n|H_n)$ as the conditioned probability
distribution at stage $n$ that player $Z$ chooses move $z_n$
(either $x_n$ or $y_n$) given history
$H_n=\{x_1,y_1,\dots,x_{n-1},y_{n-1}\}$ detailing the shared
information known to both players of all choices to that stage
(with $H_1=\{\}$).  We further write
$P^X(x_1|H_1)=P^X(x_1)=p_1$, $P^Y(y_1|H_1)=P^Y(y_1)=q_1$,
$P^X(x_n|H_n)=p_{x_n|H_n}$ and $P^Y(y_n|H_n)=q_{y_n|H_n}$.  The
expected payoffs are obtained by summing over every possible
path through the game tree specified by the move choices
$x_1,y_1,\dots,x_N,y_N$, with each path weighted by the joint
probability of that path being selected
$P^{XY}(x_1,y_1,\dots,x_N,y_N)$, and where each path generates a
payoff of $\Pi^Z(x_1,y_1,\dots,x_N,y_N)$ for player $Z$.

Here, as usual, the players $X$ and $Y$ vary their choice of
respective probability space ${\cal P}^X$ and ${\cal P}^Y$ so as
to maximize their expected payoff. That is, we hold that such
players will avail themselves of the strong isomorphic mappings
adopted by probability theory to construct their mixed or
behavioural strategy spaces.  Hence, each player will
sequentially analyze situations where both players adopt altered
joint probability spaces ${\cal P}^X_i\times{\cal P}^Y_j$ for
$i,j=0,1,2,\dots$. The infinity of possible alternatives
mandates that some limits be placed on the search space.

In the following analysis, we will first consider the $N=1$
single stage prisoner dilemma game. This will inform our
subsequent analysis of the $N=2$ stage prisoner's dilemma. We
will analyze the $N=2$ stage game by comparing three strategies
commonly found in the literature---conventional independent
play, a Tit-For-Tat strategy, and All Defect---with a
functionally correlated Markovian probability strategy space.
This analysis will then be generalized to consider a total of
256 alternate joint probability spaces.  Finally, we will
consider a multiple stage game with $N$ arbitrary and analyze a
number of alternate joint probability spaces.

\section{The $N=1$ stage Prisoner's dilemma}

The single stage prisoner's dilemma has the players seeking to
optimize the payoff functions
\begin{eqnarray}
    X: \max_{x_1}\;\;  \Pi^X(x_1,y_1) &=&  2 + x_1 - 2 y_1 \nonumber \\
    Y: \max_{y_1}\;\; \Pi^Y(x_1,y_1) &=&   2 - 2 x_1 + y_1.
\end{eqnarray}
We suppose that players adopt a joint behavioural probability
space ${\cal P}^X_B\times{\cal P}^X_B$.  Because of the lack of
communication, the choices of the $x_1$ and $y_1$ variables are
independent. One possible joint probability space defines the
expected payoff optimization problem for each player as
\begin{eqnarray}
    X: \max_{p_1}\;\; \langle\Pi^X\rangle
        &=& \sum_{x_1y_1=0}^1  P^{XY}(x_1,y_1)\Pi^X(x_1,y_1) \nonumber \\
        &=& \sum_{x_1y_1=0}^1  P^X(x_1)P^Y(y_1)(2+x_1-2y_1) \nonumber \\
        &=& 2+p_1-2q_1 \nonumber \\
    Y: \max_{q_1}\;\; \langle\Pi^Y\rangle
        &=& \sum_{x_1y_1=0}^1  P^{XY}(x_1,y_1)\Pi^Y(x_1,y_1) \nonumber \\
        &=& \sum_{x_1y_1=0}^1  P^X(x_1)P^Y(y_1)(2-2x_1+y_1) \nonumber \\
        &=& 2-2p_1+q_1,
\end{eqnarray}
where use has been made of the normalization conditions
$p_0+p_1=1$ and $q_0+q_1=1$.  In this two-player-single-stage
game, each expected payoff function is a function of the
independent parameters $p_1$ and $q_1$ and so are maximized by
the gradient operator
\begin{equation}
 \nabla=\left[ \frac{\partial}{\partial p_1},\frac{\partial}{\partial q_1} \right],
\end{equation}
giving the joint optimization conditions
\begin{eqnarray}
    \frac{\partial \langle\Pi^X\rangle}{\partial p_1}  &=& 1 \nonumber \\
    \frac{\partial \langle\Pi^Y\rangle}{\partial q_1}  &=& 1.
\end{eqnarray}
Together, these make it evident that each player optimizes their
expected payoff by maximizing their defection probability
(choosing $p_1=1$ and $q_1=1$) irrespective of their opponent's
choices.  That is, both players defect with certainty.  This is
the unique single-stage Nash equilibrium point \cite{Nash_51_28}
from which neither player can unilaterally alter their choice
without worsening their payoff.  Even so, payoffs are jointly
maximized when both players cooperate (via $x_1=y_1=0$) to yield
payoffs of $(\Pi^X,\Pi^Y)=(2,2)$. Herein lies the dilemma.

We now turn to consider the $N=2$ stage iterated prisoner's
dilemma.

\section{The $N=2$ stage prisoner's dilemma}

For the $N=2$ stage game, the optimization problem for both
players is
\begin{eqnarray}     \label{eq_define_Pi}
    X: \max_{x_1,x_2}\;\;  \Pi^X &=& \sum_{n=1}^2 (2+x_n-2y_n) \nonumber \\
    Y: \max_{y_1,y_2}\;\;  \Pi^Y &=& \sum_{n=1}^2 (2-2x_n+y_n).
\end{eqnarray}
The question which needs to be addressed by each player is how
to take account of all of the possible functional relationships
that might exist between the variables.  Of course, when the
variables are functionally related then this imposes constraints
onto the calculation of gradients which effects optimization
outcomes. Game theory presumes there exists a single space which
properly takes into account every possible functional
dependency.  Probability theory and optimization theory in
general hold that no such single space exists.  These fields
employ a multiplicity of distinct spaces in order to take
account of the different possible dependencies.  In what
follows, we will consider a small number of different possible
functional dependencies.

\subsection{The unconstrained space ${\cal P}^X_B\times{\cal P}^Y_B$}

Conventional game analysis assumes that rational players $X$ and
$Y$ will adopt a single specific joint probability space,
denoted here ${\cal P}^X_B\times{\cal P}^Y_B$. In this space,
the absence of isomorphism constraints means that all
behavioural strategies are independent allowing the game to be
decomposed into subgames in every history separating the last
stage from the preceding stage. Then, optimization in the last
stage is independent of both prior and non-existent future
events, so the last stage is identically a single stage game and
optimized in the prisoner's dilemma via the unique single stage
Nash equilibria of mutual defection.  This process can then be
iterated backwards through the game (backwards induction) to
locate the unique Nash equilibria for the entire game of mutual
defection in every stage. We now detail this analysis.

In the space ${\cal P}^X_B\times{\cal P}^Y_B$, players seek to
optimize their respective expected payoffs
\begin{eqnarray}     \label{eq_optimization_total}
    X: \max\;\; \langle\Pi^X\rangle &=&
          2N+\sum_{n=1}^2 \sum_{x_1\dots y_n=0}^1
          P^X(x_1)P^Y(y_1)\dots P^X(x_n|H_n)P^Y(y_n|H_n)
          (x_n-2y_n)  \nonumber \\
       &=& 4+p_1-2q_1+ \left[1-p_1\right] \left[1-q_1\right]
             \left[ p_{1|00}-2q_{1|00}\right] \nonumber \\
       &&      + \left[1-p_1\right]
             q_1 \left[ p_{1|01}-2q_{1|01}\right] \nonumber   \\
       &&      + p_1  \left[1-q_1\right]
             \left[ p_{1|10}-2q_{1|10}\right] +  p_1 q_1
             \left[ p_{1|11}-2q_{1|11}\right]  \nonumber \\
    Y: \max\;\; \langle\Pi^Y\rangle &=&
          2N+\sum_{n=1}^2 \sum_{x_1\dots y_n=0}^1
          P^X(x_1)P^Y(y_1)\dots P^X(x_n|H_n)P^Y(y_n|H_n)
          (y_n-2x_n)  \nonumber \\
       &=& 4-2p_1+q_1+ \left[1-p_1\right]
              \left[1-q_1\right]
               \left[ q_{1|00}-2p_{1|00}\right]\nonumber \\
       &&   +\left[1-p_1\right]
             q_1  \left[ q_{1|01}-2p_{1|01}\right]  \nonumber\\
          &&  + p_1
              \left[1-q_1\right]
              \left[ q_{1|10}-2p_{1|10}\right]+ p_1 q_1
              \left[ q_{1|11}-2p_{1|11}\right].
\end{eqnarray}
These expected payoff functions can take account of every
possible state of correlation between the second stage variables
$x_2$ and $y_2$ and the first stage variables $x_1$ and $y_1$.
The first stage probability variables $p_1,q_1$, together with
the second stage variables
$p_{1|00},p_{1|01},p_{1|10},p_{1|11}$, and
$q_{1|00},q_{1|01},q_{1|10},q_{1|11}$ are all freely varying
over the range $[0,1]$. As a result, the relevant gradient
operator used by both players to analyze this particular
probability space is
\begin{eqnarray}
    \nabla &=&
    \left[
    \frac{\partial}{\partial p_1},\frac{\partial}{\partial q_1},
    \frac{\partial}{\partial p_{1|00}},\frac{\partial}{\partial p_{1|01}},
    \frac{\partial}{\partial p_{1|10}},\frac{\partial}{\partial
    p_{1|11}}, \frac{\partial}{\partial q_{1|00}},\frac{\partial}{\partial q_{1|01}},
    \frac{\partial}{\partial q_{1|10}},\frac{\partial}{\partial q_{1|11}}
    \right].
\end{eqnarray}
Immediately then, optimization with respect to second stage
variables by player $X$ gives
\begin{eqnarray}      \label{eq_BI_two_stage_a}
    \frac{\partial\langle\Pi^X\rangle}{\partial p_{1|00}} &=&
      \left[1-p_1\right]  \left[1-q_1\right] \geq 0 \nonumber  \\
    \frac{\partial\langle\Pi^X\rangle}{\partial p_{1|01}} &=&
      \left[1-p_1\right]  q_1 \geq 0 \nonumber  \\
    \frac{\partial\langle\Pi^X\rangle}{\partial p_{1|10}} &=&
      p_1  \left[1-q_1\right] \geq 0 \nonumber  \\
    \frac{\partial\langle\Pi^X\rangle}{\partial p_{1|11}} &=&
      p_1  q_1 \geq 0,
\end{eqnarray}
with similar results applying for $Y$.  As the rate of change of
the expected payoff is essentially positive with increasing last
stage defection probability, each player maximizes their
expected payoff by defecting with certainty in the last stage.
That is, each player sets $p_{1|x_1y_1}=1$ and $q_{1|x_1y_1}=1$
on every pathway. Taking account of this last stage result
simplifies the optimization for the first stage probability
variables (backwards induction), giving
\begin{equation}      \label{eq_BI_two_stage_b}
   \frac{\partial\langle\Pi^X\rangle}{\partial p_1} = 1,
\end{equation}
with similar results applying for $Y$.  Again, players will
defect in the first stage by setting $p_1=1$ and $q_1=1$. Hence,
players conclude that, given the adoption of the joint
probability space ${\cal P}^X_B\times{\cal P}^Y_B$, they
maximize their expected payoffs by defecting in every stage of
the game $(x_1,y_1,x_2,y_2)=(1,1,1,1)$ to derive a joint
expected payoff of
$\left(\langle\Pi^X\rangle,\langle\Pi^Y\rangle
\right)=(N,N)=(2,2)$.  This is the unique Nash equilibrium
pathway for the finite iterated prisoner's dilemma, given the
adoption of the joint probability space ${\cal P}^X_B\times{\cal
P}^Y_B$.

\subsection{Alternate isomorphic probability spaces}

In this section we suppose that players $X$ and $Y$ consider
only a choice of four possible alternate probability spaces,
namely, the conventional independent behavioural strategy space,
a functionally correlated Markovian probability space, a
Tit-For-Tat strategy space, and an All Defect strategy space.

When adopting a Markovian space, each player functionally
correlates their second stage choices to their opponent's first
stage choices.  That is, player $X$ implements
\begin{eqnarray}   \label{eq_mkv_x2}
   x_2             &=& y_1 \nonumber \\
   p_{x_2|x_1y_1}  &=& \delta_{x_2y_1},
\end{eqnarray}
while player $Y$ chooses
\begin{eqnarray}     \label{eq_mkv_y2}
   y_2             &=& x_1 \nonumber \\
   q_{y_2|x_1y_1}  &=& \delta_{y_2x_1}.
\end{eqnarray}
We denote these spaces respectively as ${\cal P}^X_B|_{x_2=y_1}$
and ${\cal P}^Y_B|_{y_2=x_1}$.

When adopting Tit-For-Tat, each player chooses to cooperate in
the first stage and then functionally correlate their second
stage choice to the opponent's first stage choice.  Player $X$
implements Tit-For-Tat via
\begin{eqnarray}
   x_1             &=& 0 \nonumber \\
   p_{x_1}         &=& \delta_{x_10} \nonumber \\
   x_2             &=& y_1 \nonumber \\
   p_{x_2|x_1y_1}  &=& \delta_{x_2y_1},
\end{eqnarray}
while player $Y$ will implement
\begin{eqnarray}
   y_1             &=& 0 \nonumber \\
   q_{y_1}         &=& \delta_{y_10} \nonumber \\
   y_2             &=& x_1 \nonumber \\
   q_{y_2|x_1y_1}  &=& \delta_{y_2x_1}.
\end{eqnarray}
We denote these probability spaces respectively as ${\cal
P}^X_B|_{x_1=0,x_2=y_1}$ and ${\cal P}^Y_B|_{y_1=0,y_2=x_1}$.

Finally, by adopting the ALL DEFECT space, each player chooses
to defect in every stage. Player $X$ chooses
\begin{eqnarray}
   x_1             &=& 1 \nonumber \\
   p_{x_1}         &=& \delta_{x_11} \nonumber \\
   x_2             &=& 1 \nonumber \\
   p_{x_2|x_1y_1}  &=& \delta_{x_21},
\end{eqnarray}
and player $Y$ chooses
\begin{eqnarray}
   y_1             &=& 1 \nonumber \\
   q_{y_1}         &=& \delta_{y_11} \nonumber \\
   y_2             &=& 1 \nonumber \\
   q_{y_2|x_1y_1}  &=& \delta_{y_21}.
\end{eqnarray}
We denote these probability spaces respectively as ${\cal
P}^X_B|_{x_1=x_2=1}$ and ${\cal P}^Y_B|_{y_1=y_2=1}$.

Subsequently, players of unbounded rationality will then
sequentially examine the alternate isomorphic probability spaces
available to the players.  Within each possible space, they will
locate the constrained equilibria optimizing outcomes, and then
later compare these outcomes in a comparison table. We complete
this process now.

\begin{figure}[htb]
\centering
\includegraphics[width=\columnwidth,clip]{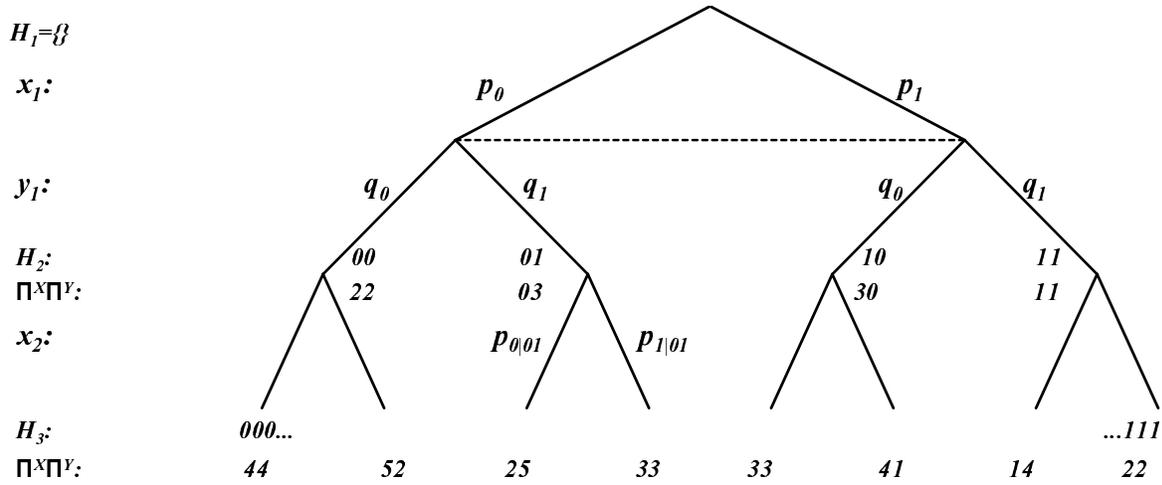}
\caption[Independent and Markovian strategies]{\em The case where players $(X,Y)$
adopt Independent versus Markovian strategies in the ${\cal P}^X_{B}\times{\cal P}^Y_B|_{y_2=x_1}$ joint
probability space. The second stage choices of player $Y$ are isomorphically
constrained and so are not freely varying parameters and
do not appear in the decision tree. \label{f_IND_MKV}}
\end{figure}

\subsection{$N=2$ stage: Independent versus Markovian strategies}

Supposing that the players examine the case where they adopt
Independent versus Markovian strategies and so jointly adopt the
${\cal P}^X_{B}\times{\cal P}^Y_B|_{y_2=x_1}$ probability space.
In this space, the players seek to optimize
(\ref{eq_optimization_total}) subject to the isomorphic
constraint $y_2=x_1$.  This constraint alters the expected
payoff optimization problems to be
\begin{eqnarray}   \label{eq_IND_MKV_expected_payoffs}
    X: \max_{p_1,p_{1|x_1y_1}}\;\; \langle\Pi^X\rangle
       &=& 4+p_1-2q_1+ \nonumber \\
        &&   \left[1-p_1\right]
             \left[1-q_1\right]
             p_{1|00} +       \nonumber \\
        &&     \left[1-p_1\right]
             q_1
              p_{1|01} + \nonumber \\
       &&    p_1
             \left[1-q_1\right]
             \left[ p_{1|10}-2\right] +        \nonumber \\
        &&   p_1
             q_1
             \left[ p_{1|11}-2\right]  \nonumber \\
    Y: \max_{q_1}\;\; \langle\Pi^Y\rangle
       &=& 4-2p_1+q_1+ \nonumber \\
          && -2\left[1-p_1\right]
              \left[1-q_1\right]
               p_{1|00}-      \nonumber \\
        &&      2 \left[1-p_1\right]
             q_1
             p_{1|01}+\nonumber   \\
          &&  p_1
              \left[1-q_1\right]
              \left[ 1-2p_{1|10}\right]+      \nonumber \\
        &&     p_1
              q_1
              \left[ 1-2p_{1|11}\right].
\end{eqnarray}
These expected payoffs are continuous multivariate functions
dependent only on the freely varying parameters
$[p_1,q_1,p_{1|00},p_{1|01},p_{1|10},p_{1|11}]$.  Consequently,
the relevant gradient operator used by both players to analyze
this particular probability space is
\begin{eqnarray}
    \nabla &=&
    \left[
    \frac{\partial}{\partial p_1},\frac{\partial}{\partial q_1},
    \frac{\partial}{\partial p_{1|00}},\frac{\partial}{\partial p_{1|01}},
    \frac{\partial}{\partial p_{1|10}},\frac{\partial}{\partial
    p_{1|11}}  \right] \nonumber \\
    &&
\end{eqnarray}
while the resulting game tree is shown in Fig. \ref{f_IND_MKV}.
Optimization then proceeds as usual via
\begin{eqnarray}      \label{eq_BI_IND_MKV_a}
    \frac{\partial\langle\Pi^X\rangle}{\partial p_{1|00}} &=&
      \left[1-p_1\right]  \left[1-q_1\right] \geq 0 \nonumber  \\
    \frac{\partial\langle\Pi^X\rangle}{\partial p_{1|01}} &=&
      \left[1-p_1\right]  q_1 \geq 0 \nonumber  \\
    \frac{\partial\langle\Pi^X\rangle}{\partial p_{1|10}} &=&
      p_1  \left[1-q_1\right] \geq 0 \nonumber  \\
    \frac{\partial\langle\Pi^X\rangle}{\partial p_{1|11}} &=&
      p_1  q_1 \geq 0,
\end{eqnarray}
ensuring that player $X$ defects with certainty in the last
stage by setting $p_{1|x_1y_1}=1$ on every pathway.  These
choices then allow evaluating
\begin{eqnarray}      \label{eq_BI_IND_MKV_b}
    X: \max_{p_1}\;\; \langle\Pi^X\rangle &=& 5 - p_1 - 2 q_1 \nonumber \\
    \frac{\partial\langle\Pi^X\rangle}{\partial p_1} &=&-1 \leq 0,
\end{eqnarray}
so player $X$ cooperates with certainty in the first stage by
setting $p_1=0$.  In contrast, the analysis by player $Y$ must
simply determine their first stage variable (taking account of
the optimized moves by player $X$) via
\begin{eqnarray}      \label{eq_BI_IND_MKV_c}
    Y: \max_{q_1}\;\; \langle\Pi^Y\rangle &=& 2 +2 q_1 \nonumber \\
    \frac{\partial\langle\Pi^Y\rangle}{\partial q_1} &=&1 \geq 0,
\end{eqnarray}
so player $Y$ defects in the first stage by setting $q_1=1$.
Altogether, when players $(X,Y)$ adopt the ${\cal
P}^X_{B}\times{\cal P}^Y_B|_{y_2=x_1}$ joint probability space,
they play the move combinations $(x_1,y_1,x_2,y_2)=(0,1,1,0)$ to
garner payoffs
$\left(\langle\Pi^X\rangle,\langle\Pi^Y\rangle\right)= (3,3)$.
Here, in this particular joint probability space, the player
adopting an Independent strategy must cooperate in the first
stage to ensure that their mimicking opponent playing a
Markovian will cooperate in the second stage so setting them up
for a sucker's payoff in that stage. However, this gains them
little as their opponent can still freely defect in the first
stage so in the end, players end up with equal payoffs.

\begin{figure}[htb]
\centering
\includegraphics[width=\columnwidth,clip]{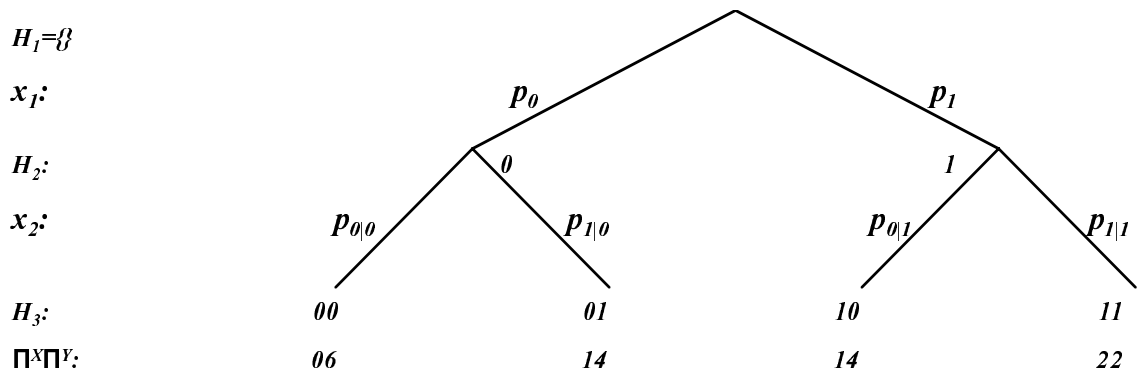}
\caption[Independent and mutual defection strategies]{\em The case where players $(X,Y)$
adopt Independent versus All Defect strategies in the ${\cal P}^X_{B}\times{\cal
P}^Y_B|_{y_1=y_2=1}$ joint
probability space. We write $p_{x_2|x_11}\rightarrow p_{x_2|x_1}$. Here,
neither first nor second stage choices of player $Y$ appear in the
game tree as they have been isomorphically constrained. \label{f_IND_DD}}
\end{figure}

\subsection{$N=2$ stage: Independent versus All Defect strategies}

Suppose now that players examine the situation where they
jointly adopt Independent versus All Defect strategies in the
${\cal P}^X_{B}\times{\cal P}^Y_B|_{y_1=y_2=1}$ probability
space. After resolution of the adopted isomorphic constraints,
the expected payoff optimization problems become
\begin{eqnarray}
    X: \max_{p_1,p_{1|01},p_{1|11}}\;\; \langle\Pi^X\rangle
       &=& 2+p_1+  \left[1-p_1\right] \left[ p_{1|01}-2\right] +
          p_1 \left[ p_{1|11}-2\right]  \nonumber \\
    Y: \langle\Pi^Y\rangle
       &=& 5-2p_1+ \left[1-p_1\right] \left[ 1-2p_{1|01}\right]+
           p_1 \left[ 1-2p_{1|11}\right].
\end{eqnarray}
Given the isomorphic constraints adopted by the players, these
expected payoff functions are dependent solely on the freely
varying parameters $[p_1,p_{1|01},p_{1|11}]$ so the relevant
gradient operator used by both players in their analysis is
\begin{eqnarray}
    \nabla&=&
    \left[
    \frac{\partial}{\partial p_1},
    \frac{\partial}{\partial p_{1|01}},
    \frac{\partial}{\partial p_{1|11}}  \right].
\end{eqnarray}
Consequently, optimization for player $X$ gives
\begin{eqnarray}      \label{eq_BI_IND_DD_a}
    \frac{\partial \langle\Pi^X\rangle}{\partial p_{1|01}} &=& \left[1-p_1\right] \geq 0 \nonumber \\
    \frac{\partial \langle\Pi^X\rangle}{\partial p_{1|11}} &=& p_1 \geq 0,
\end{eqnarray}
leading, essentially, to defection on all second stage histories
via $p_{1|x_11}=1$ and $q_{1|x_11}=1$ on every pathway.  Taking
account of these last stage results then gives
\begin{eqnarray}      \label{eq_BI_IND_DD_b}
    X: \max_{p_1}\;\; \langle\Pi^X\rangle &=& 1 +p_1 \nonumber \\
    \frac{\partial \langle\Pi^X\rangle}{\partial p_1} &=& 1,
\end{eqnarray}
so player $X$ also defects in the first stage with certainty
through the choice $p_1=1$.  Altogether, the ${\cal
P}^X_{B}\times{\cal P}^Y_B|_{y_1=y_2=1}$ joint probability space
leads both players to mutual defection in every stage to garner
expected payoffs of
$\left(\langle\Pi^X\rangle,\langle\Pi^Y\rangle\right)=(N,N)=(2,2)$.

\begin{figure}[htb]
\centering
\includegraphics[width=\columnwidth,clip]{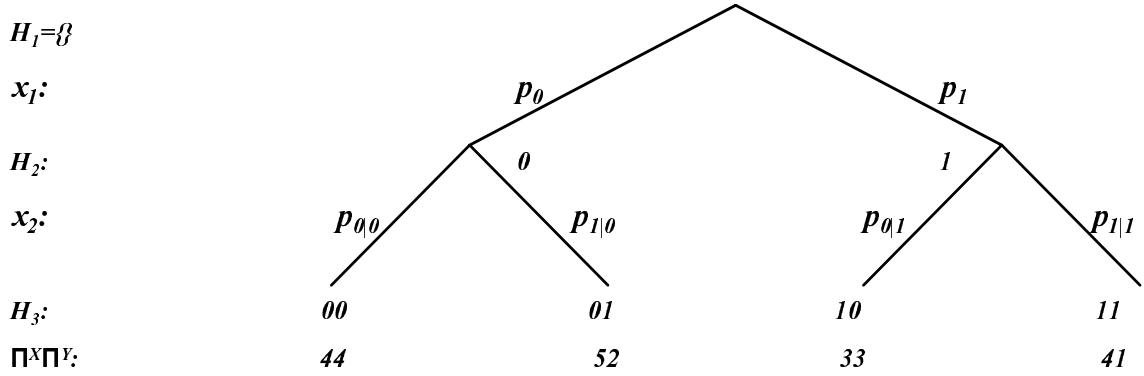}
\caption[Independent and Tit-For-Tat strategies]{\em The case where players $(X,Y)$
adopt Independent versus Tit-For-Tat strategies in the ${\cal P}^X_{B}\times{\cal
P}^Y_B|_{y_1=0,y_2=x_1}$ joint
probability space. We write $p_{x_2|x_10}=p_{x_2|x_1}$. Again,
neither first nor second stage choices of player $Y$ appear in the
game tree as they have been isomorphically constrained and so are not
freely varying parameters. \label{f_IND_TFT}}
\end{figure}

\subsection{$N=2$ stage: Independent versus Tit-For-Tat strategies}

If, on the other hand, players $(X,Y)$ suppose that together
they adopt the ${\cal P}^X_{B}\times{\cal
P}^Y_B|_{y_1=0,y_2=x_1}$ joint probability space, then the
expected payoff function optimization problem becomes
\begin{eqnarray}
    X: \max_{p_1,p_{1|00},p_{1|10}}\;\; \langle\Pi^X\rangle
       &=& 4+p_1+  \left[1-p_1\right] p_{1|00} +
        p_1 \left[ p_{1|10}-2\right]  \nonumber \\
    \langle\Pi^Y\rangle
       &=& 4-2p_1- 2\left[1-p_1\right] p_{1|00}+
         p_1 \left[ 1-2p_{1|10}\right].
\end{eqnarray}
As such, the expected payoff functions are dependent only on the
freely varying parameters $[p_1,p_{1|00},p_{1|10}]$ so the
relevant gradient operator used by both players in their
analysis is
\begin{eqnarray}
    \nabla &=&
    \left[
    \frac{\partial}{\partial p_1},
    \frac{\partial}{\partial p_{1|00}},
    \frac{\partial}{\partial p_{1|10}} \right].
\end{eqnarray}
Consequently, optimization for player $X$ gives
\begin{eqnarray}      \label{eq_BI_IND_TFT_a}
    \frac{\partial \langle\Pi^X\rangle}{\partial p_{1|00}} &=& \left[1-p_1\right] \geq 0 \nonumber \\
    \frac{\partial \langle\Pi^X\rangle}{\partial p_{1|10}} &=& p_1 \geq 0,
\end{eqnarray}
leading, essentially, to defection on all second stage histories
via $p_{1|x_10}=1$ on every pathway.  Taking account of these
last stage results then gives
\begin{eqnarray}      \label{eq_BI_IND_TFT_b}
    X: \max_{p_1}\;\; \langle\Pi^X\rangle &=& 5-p_1 \nonumber \\
    \frac{\partial \langle\Pi^X\rangle}{\partial p_1} &=& -1,
\end{eqnarray}
so player $X$ cooperates in the first stage with certainty
through the choice $p_1=0$.  Altogether, the ${\cal
P}^X_{B}\times{\cal P}^Y_B|_{y_1=0,y_2=x_1}$ joint probability
space leads players to the move combinations
$(x_1,y_1,x_2,y_2)=(0,0,1,0)$ to garner payoffs
$\left(\langle\Pi^X\rangle,\langle\Pi^Y\rangle\right)= (5,2)$.

\begin{figure}[htb]
\centering
\includegraphics[width=\columnwidth,clip]{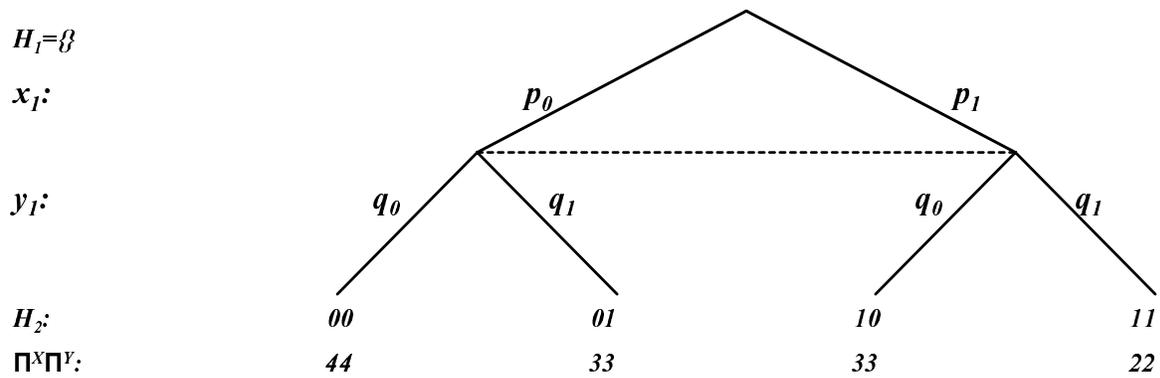}
\caption[Dual Markovian strategies]{\em The case where players $(X,Y)$ adopt
Markovian versus Markovian strategies in the
${\cal P}^X_B|_{x_2=y_1}\times{\cal P}^Y_B|_{y_2=x_1}$ joint probability
space. As both players functionally assign their second stage
choices, the only freely varying parameters are the first stage
choices of each player. \label{f_MKV_MKV}}
\end{figure}

\subsection{$N=2$ stage: Markovian versus Markovian strategies}

Suppose now that players $(X,Y)$ jointly assume they both adopt
the ${\cal P}^X_B|_{x_2=y_1}\times{\cal P}^Y_B|_{y_2=x_1}$
probability space. After resolution of the adopted isomorphic
constraints, the expected payoff function optimization problems
become
\begin{eqnarray}
    X: \max_{p_1}\;\; \langle\Pi^X\rangle &=& 4-p_1-q_1 \nonumber \\
    Y: \max_{q_1}\;\; \langle\Pi^Y\rangle &=& 4-p_1-q_1,
\end{eqnarray}
which are dependent only on the freely varying parameters
$[p_1,q_1]$, so immediately, the gradient operator used by each
player in their analysis is
\begin{eqnarray}
    \nabla &=&   \left[
    \frac{\partial}{\partial p_1},
    \frac{\partial}{\partial q_1} \right].
\end{eqnarray}
Optimization then proceeds straightforwardly giving respectively
for each player
\begin{eqnarray}      \label{eq_BI_MKV_MKV_a}
    \frac{\partial \langle\Pi^X\rangle}{\partial p_1} &=& -1
    \nonumber \\
    \frac{\partial \langle\Pi^Y\rangle}{\partial q_1} &=& -1,
\end{eqnarray}
ensuring that in this space, both players cooperate with
certainty in the first stage by setting $p_1=q_1=0$. Altogether,
when players $(X,Y)$ adopt the ${\cal
P}^X_B|_{x_2=y_1}\times{\cal P}^Y_B|_{y_2=x_1}$ joint
probability space, they cooperate via the move combinations
$(x_1,y_1,x_2,y_2)=(0,0,0,0)$ to garner payoffs
$\left(\langle\Pi^X\rangle,\langle\Pi^Y\rangle\right)= (4,4)$.
That is, under a joint constraint where each player mimics their
opponent's previous moves, a strategy of cooperation is rational
as it maximizes expected payoffs for both players.

\begin{figure}[htb]
\centering
\includegraphics[width=\columnwidth,clip]{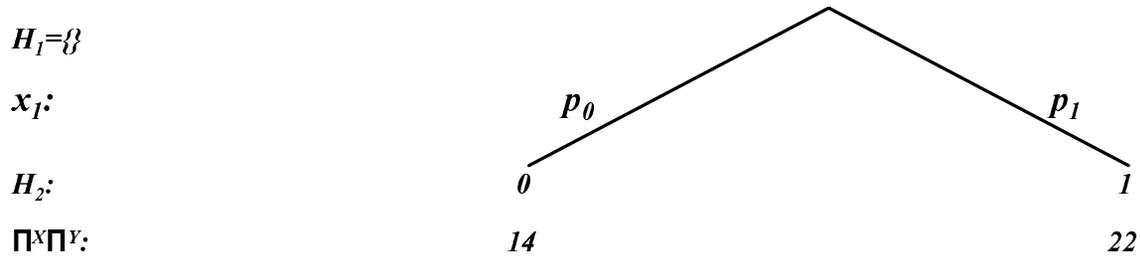}
\caption[Markovian and mutual defection strategies]{\em The case where players $(X,Y)$
adopt Markovian versus All Defect strategies in
the ${\cal P}^X_B|_{x_2=y_1}\times{\cal P}^Y_B|_{y_1=y_2=1}$ joint
probability space. As both players functionally assign all of
their second stage choices while player $Y$ defects with certainty
in the first stage, the only freely varying parameter left is the
first stage choice of player $X$ reducing the game to being a
single-player-single-stage situation as shown. \label{f_MKV_DD}}
\end{figure}

\subsection{$N=2$ stage: Markovian versus All Defect strategies}

Suppose now that players $(X,Y)$ analyze the case where they
jointly adopt the ${\cal P}^X_B|_{x_2=y_1}\times{\cal
P}^Y_B|_{y_1=y_2=1}$ probability space.  The resolution of the
adopted constraints means that the expected payoff function
optimization problem for the players becomes
\begin{eqnarray}
    X: \max_{p_1}\;\; \langle\Pi^X\rangle &=& 1+p_1 \nonumber \\
    \langle\Pi^Y\rangle &=& 4-2p_1,
\end{eqnarray}
which are dependent only on the freely varying parameter $p_1$,
so immediately, the gradient operator used by each player in
their analysis is
\begin{equation}
    \nabla = \frac{\partial}{\partial p_1}.
\end{equation}
Player $X$ then evaluates
\begin{equation}      \label{eq_BI_MKV_DD_a}
    \frac{\partial \langle\Pi^X\rangle}{\partial p_1} = 1,
\end{equation}
ensuring that this player defects with certainty in the first
stage by setting  $p_1=1$.  Altogether, when players $(X,Y)$
jointly adopt the ${\cal P}^X_B|_{x_2=y_1}\times{\cal
P}^Y_B|_{y_1=y_2=1}$ probability space, they generate the
optimal move combination $(x_1,y_1,x_2,y_2)=(1,1,1,1)$ to garner
payoffs
$\left(\langle\Pi^X\rangle,\langle\Pi^Y\rangle\right)=(2,2)$.

\begin{figure}[htb]
\centering
\includegraphics[width=\columnwidth,clip]{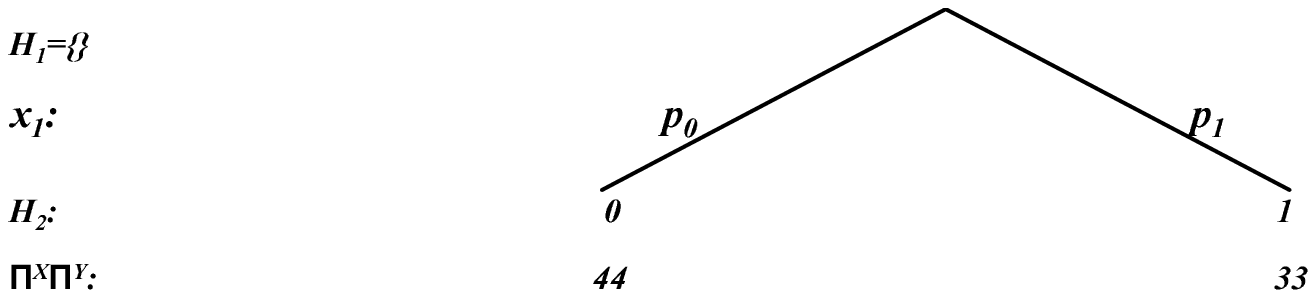}
\caption[Markovian and Tit-For-Tat strategies]{\em The case where players $(X,Y)$
adopt Markovian verses Tit-For-Tat strategies in
the ${\cal P}^X_B|_{x_2=y_1}\times{\cal P}^Y_B|_{y_1=0,y_2=x_1}$ joint
probability space. As both players functionally assign all of
their second stage choices while player $Y$ cooperates with
certainty in the first stage, the only freely varying parameter
left is the first stage choice of player $X$ reducing the game to
being a single-player-single-stage situation as shown.
\label{f_MKV_TFT}}
\end{figure}

\subsection{$N=2$ stage: Markovian verses Tit-For-Tat strategies}

Suppose now that players $(X,Y)$ jointly assume that together
they adopt the ${\cal P}^X_B|_{x_2=y_1}\times{\cal
P}^Y_B|_{y_1=0,y_2=x_1}$ probability space.  After resolution of
the isomorphic constraints, the expected payoff function
optimization problems become
\begin{eqnarray}
    X: \max_{p_1}\;\; \langle\Pi^X\rangle &=& 4-p_1 \nonumber \\
    \langle\Pi^Y\rangle &=& 4-p_1,
\end{eqnarray}
which are dependent only on the freely varying parameter $p_1$,
so immediately, the gradient operator used by each player in
their analysis is
\begin{equation}
    \nabla = \frac{\partial}{\partial p_1}.
\end{equation}
Player $X$ then evaluates
\begin{equation}      \label{eq_BI_MKV_TFT_a}
    \frac{\partial \langle\Pi^X\rangle}{\partial p_1} = -1,
\end{equation}
ensuring that this player cooperates with certainty in the first
stage by setting $p_1=0$.  Altogether, when players $(X,Y)$
jointly adopt the ${\cal P}^X_B|_{x_2=y_1}\times{\cal
P}^Y_B|_{y_1=0,y_2=x_1}$ probability space, they generate the
optimal move combination $(x_1,y_1,x_2,y_2)=(0,0,0,0)$ to garner
payoffs
$\left(\langle\Pi^X\rangle,\langle\Pi^Y\rangle\right)=(4,4)$.

\subsection{$N=2$ stage: Comparing payoffs}

The remainder of the possible probability spaces that the
players might analyze, Tit-For-Tat versus Tit-For-Tat (${\cal
P}^X_B|_{x_1=0,x_2=y_1}\times{\cal P}^Y_B|_{y_1=0,y_2=x_1}$),
Tit-For-Tat versus All Defect (${\cal
P}^X_B|_{x_1=0,x_2=y_1}\times{\cal P}^Y_B|_{y_1=y_2=1}$), and
All Defect versus All defect (${\cal
P}^X_B|_{x_1=x_2=1}\times{\cal P}^Y_B|_{y_1=y_2=1}$), possess no
free variables whatsoever and so merely involve an evaluation of
the expected payoffs in each case. Altogether, under the
assumption that either player might adopt any of the four
probability spaces considered here, then players must compare 16
possible isomorphically constrained optima to locate their
optimal choice of probability space.  The comparison table
showing every possible combination of adopted probability space
for either player is
\begin{equation}
    \begin{array}{c|cccc}
      \left(\langle\Pi^X\rangle,\langle\Pi^Y\rangle\right) & {\cal P}^Y_B|_{y_2=x_1}   & {\cal P}^Y_B   & {\cal P}^Y_B|_{y_1=0,y_2=x_1}   & {\cal P}^Y_B|_{y_1=y_2=1}     \\ \hline
                                                           &       &       &       &       \\
      {\cal P}^X_B|_{x_2=y_1}                              & (4,4) & (3,3) & (4,4) & (2,2) \\
                                                           &       &       &       &       \\
      {\cal P}^X_B                                         & (3,3) & (2,2) & (5,2) & (2,2) \\
                                                           &       &       &       &       \\
      {\cal P}^X_B|_{x_1=0,x_2=y_1}                        & (4,4) & (2,5) & (4,4) & (1,4) \\
                                                           &       &       &       &       \\
      {\cal P}^X_{x_1=x_2=1}                               & (2,2) & (2,2) & (4,1) & (2,2).\\
                                                           &       &       &       &       \\
    \end{array}
\end{equation}
This table of alternate expected payoffs makes it evident that
the Tit-For-Tat and All Defect probability spaces are weakly
dominated by the Markovian and Independent probability spaces.
Player's choices of optimal probability spaces then come down
effectively to a comparison of the Markovian or the Independent
probability spaces. Perusal of the table shows that adopting the
Markovian probability space offers the better returns to either
player.

Given this admittedly small set of possible strategy
constraints, rational players will maximize their expected
payoffs by adopting a Markovian strategy and rationally
cooperate in the finite iterated prisoner's dilemma. The
traditional result of conventional game analysis that mutual all
defection is the unique Nash equilibria for this game is an
incomplete analysis based on the unjustified restriction that
players can only employ a restricted set of independent
probability distributions.

\begin{figure}[htbp]
\centering
\includegraphics[width=\columnwidth,clip]{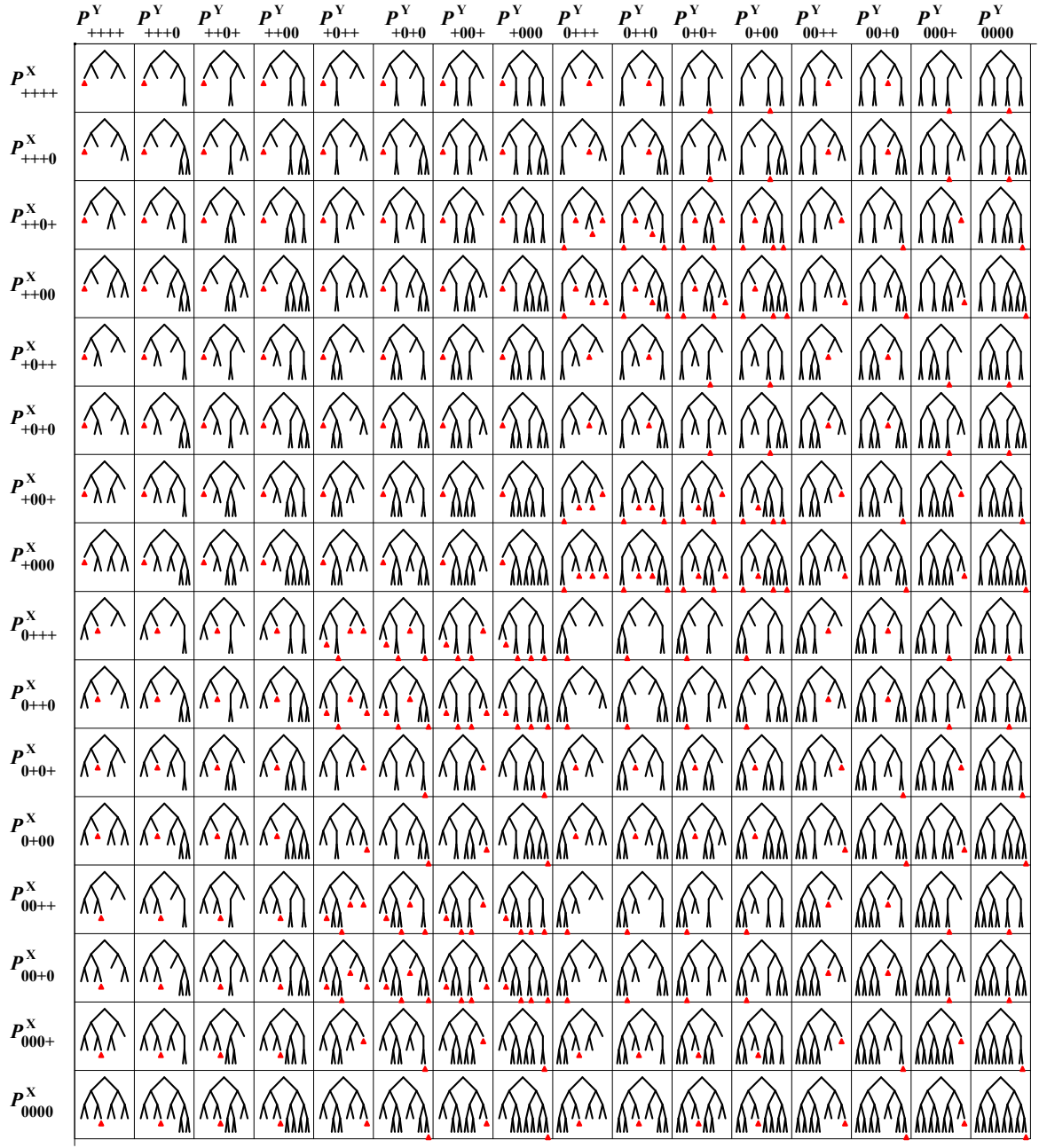}
\caption[Alternate probability spaces and decision trees]{\em The generated trees and
the equilibrium pathways (indicated by small dots, with multiple
dots indicating mixed equilibrium pathways) assuming that player
$X$ adopts the probability space shown on the vertical axis and
that player $Y$ adopts the probability space shown horizontally.
(When the $x_2$ choice is correlated and the $y_2$ choice is
independent, a vertical line is shown to maintain the relative
spacings of each tree.)  The expected payoffs under each strategy
combination are shown in Table \protect\ref{t_extended_analysis}.
\label{f_extended_analysis}}
\end{figure}

\subsection{$N=2$ stage: Extended isomorphic constraints}

An immediate question of interest is whether the conclusion that
cooperation is rational survives an extended analysis employing
a wider class of possible isomorphic constraints which we
investigate now. We here examine a total of 256 alternate
probability spaces for the $N=2$ stage iterated prisoner's
dilemma game.  The resulting game trees are shown in Fig.
\ref{f_extended_analysis} (appearing in exploded form), with
optimized expected payoffs derived in each joint probability
space shown in Table \ref{t_extended_analysis}.

We suppose that each of our players, denoted $Z\in\{X,Y\}$,
chooses whether each of their four second stage behavioural
strategies $P^Z(z_2|x_1y_1)$ are either independent, denoted
``$0$", or perfectly correlated to their opponent's previous
move, denoted ``$+$". (Perfect anti-correlations are also
possible, but these are not considered here.)  There are four
histories $(x_1,y_1)\in\{(0,0),(0,1),(1,0),(1,1)\}$.
Admittedly, it is unusual to specify whether a behavioural
strategy implemented at a single node of a game tree is either
independent of previous events or correlated with previous
events.  However, there is nothing preventing this from
occurring---it might not be an optimal choice but it is a
possible set of choices that a player might make when optimizing
their payoffs over a game tree.

Consequently, if player $Z$ chooses to make all of the second
stage behavioural strategy probability distributions
$P^Z(z_2|x_1y_1)$ independent then the adopted space is ${\cal
P}^Z_{0000}$.  This means that the randomized choices player $Z$
makes at every second stage node of the game tree are
independent of every other event (as is usually the case).
However, if $Z$ chooses to functionally correlate all of their
second stage behavioural strategy probability distributions
$P^Z(z_2|x_1y_1)$ then the adopted space is ${\cal P}^Z_{++++}$.
In this case, the dice roll that $Z$ uses to make their choice
of $y_2$ in the case $(x_1,y_1)=(0,0)$ will be perfectly
correlated to the previous event $x_1=0$.  As noted, this is an
unusual choice but nevertheless it is still a possible choice.
Intermediate cases include when, for instance, $Z$ decides to
make $P^Z(z_2|00)$ and $P^Z(z_2|10)$ independent, and to
functionally correlate $P^Z(z_2|01)$ and $P^Z(z_2|11)$, in which
case the adopted space is ${\cal P}^Z_{0+0+}$, and so on.
Altogether, there are 16 possible choices that player $Z$ might
make about their probability space, namely $\{{\cal
P}^Z_{0000},{\cal P}^Z_{000+},{\cal P}^Z_{00+0},{\cal
P}^Z_{00++},\dots,{\cal P}^Z_{++++}\}$. In combination, both
players can jointly adopt one of $16^2=256$ different joint
probability spaces, in each of which a potentially different
constrained equilibria exists, and all of these optima must be
compared so that players can decide which probability space they
can rationally choose.

\begin{table}
\centering
\begin{sideways}
\scriptsize
\begin{tabular}{c|cccccccccccccccc}
  $\left(\langle\Pi^X\rangle,\langle\Pi^Y\rangle\right)$
                          &             &             &             &             &             &             &             &             &             &             &             &             &             &             &             &           \\
    & ${\cal P}^Y_{++++}$ & ${\cal P}^Y_{+++0}$ & ${\cal P}^Y_{++0+}$ & ${\cal P}^Y_{++00}$ & ${\cal P}^Y_{+0++}$ & ${\cal P}^Y_{+0+0}$ & ${\cal P}^Y_{+00+}$ & ${\cal P}^Y_{+000}$ & ${\cal P}^Y_{0+++}$ & ${\cal P}^Y_{0++0}$ & ${\cal P}^Y_{0+0+}$ & ${\cal P}^Y_{0+00}$ & ${\cal P}^Y_{00++}$ & ${\cal P}^Y_{00+0}$ & ${\cal P}^Y_{000+}$ & ${\cal P}^Y_{0000}$ \\\hline
                          &             &             &             &             &             &             &             &             &             &             &             &             &             &             &             &           \\
  ${\cal P}^X_{++++}$     &    (4,4)    &    (4,4)    &    (4,4)    &    (4,4)    &    (4,4)    &    (4,4)    &    (4,4)    &    (4,4)    &    (3,3)    &    (3,3)    &    (3,3)    &    (3,3)    &    (3,3)    &    (3,3)    &    (3,3)    &    (3,3)  \\
                          &             &             &             &             &             &             &             &             &             &             &             &             &             &             &             &           \\
  ${\cal P}^X_{+++0}$     &    (4,4)    &    (4,4)    &    (4,4)    &    (4,4)    &    (4,4)    &    (4,4)    &    (4,4)    &    (4,4)    &    (3,3)    &    (3,3)    &    (3,3)    &    (3,3)    &    (3,3)    &    (3,3)    &    (3,3)    &    (3,3)  \\
                          &             &             &             &             &             &             &             &             &             &             &             &             &             &             &             &           \\
  ${\cal P}^X_{++0+}$     &    (4,4)    &    (4,4)    &    (4,4)    &    (4,4)    &    (4,4)    &    (4,4)    &    (4,4)    &    (4,4)    &    ($\frac{8}{3},\frac{7}{3}$)    &    ($\frac{8}{3},\frac{7}{3}$)    &    ($\frac{8}{3},\frac{7}{3}$)    &    ($\frac{8}{3},\frac{7}{3}$)    &    (2,2)    &    (2,2)    &    (2,2)    &    (2,2)  \\
                          &             &             &             &             &             &             &             &             &             &             &             &             &             &             &             &           \\
  ${\cal P}^X_{++00}$     &    (4,4)    &    (4,4)    &    (4,4)    &    (4,4)    &    (4,4)    &    (4,4)    &    (4,4)    &    (4,4)    &    ($\frac{8}{3},\frac{7}{3}$)    &    ($\frac{8}{3},\frac{7}{3}$)    &    ($\frac{8}{3},\frac{7}{3}$)    &    ($\frac{8}{3},\frac{7}{3}$)    &    (2,2)    &    (2,2)    &    (2,2)    &    (2,2)  \\
                          &             &             &             &             &             &             &             &             &             &             &             &             &             &             &             &           \\
  ${\cal P}^X_{+0++}$     &    (4,4)    &    (4,4)    &    (4,4)    &    (4,4)    &    (4,4)    &    (4,4)    &    (4,4)    &    (4,4)    &    (3,3)    &    (3,3)    &    (3,3)    &    (3,3)    &    (3,3)    &    (3,3)    &    (3,3)    &    (3,3)  \\
                          &             &             &             &             &             &             &             &             &             &             &             &             &             &             &             &           \\
  ${\cal P}^X_{+0+0}$     &    (4,4)    &    (4,4)    &    (4,4)    &    (4,4)    &    (4,4)    &    (4,4)    &    (4,4)    &    (4,4)    &    (3,3)    &    (3,3)    &    (3,3)    &    (3,3)    &    (3,3)    &    (3,3)    &    (3,3)    &    (3,3)  \\
                          &             &             &             &             &             &             &             &             &             &             &             &             &             &             &             &           \\
  ${\cal P}^X_{+00+}$     &    (4,4)    &    (4,4)    &    (4,4)    &    (4,4)    &    (4,4)    &    (4,4)    &    (4,4)    &    (4,4)    &    ($\frac{8}{3},\frac{7}{3}$)    &    ($\frac{8}{3},\frac{7}{3}$)    &    ($\frac{8}{3},\frac{7}{3}$)    &    ($\frac{8}{3},\frac{7}{3}$)    &    (2,2)    &    (2,2)    &    (2,2)    &    (2,2)  \\
                          &             &             &             &             &             &             &             &             &             &             &             &             &             &             &             &           \\
  ${\cal P}^X_{+000}$     &    (4,4)    &    (4,4)    &    (4,4)    &    (4,4)    &    (4,4)    &    (4,4)    &    (4,4)    &    (4,4)    &    ($\frac{8}{3},\frac{7}{3}$)    &    ($\frac{8}{3},\frac{7}{3}$)    &    ($\frac{8}{3},\frac{7}{3}$)    &    ($\frac{8}{3},\frac{7}{3}$)    &    (2,2)    &    (2,2)    &    (2,2)    &    (2,2)  \\
                          &             &             &             &             &             &             &             &             &             &             &             &             &             &             &             &           \\
  ${\cal P}^X_{0+++}$     &    (3,3)    &    (3,3)    &    (3,3)    &    (3,3)    &    ($\frac{7}{3},\frac{8}{3}$)    &    ($\frac{7}{3},\frac{8}{3}$)    &    ($\frac{7}{3},\frac{8}{3}$)    &    ($\frac{7}{3},\frac{8}{3}$)    &    (3,3)    &    (3,3)    &    (3,3)    &    (3,3)    &    (3,3)    &    (3,3)    &    (3,3)    &    (3,3)  \\
                          &             &             &             &             &             &             &             &             &             &             &             &             &             &             &             &           \\
  ${\cal P}^X_{0++0}$     &    (3,3)    &    (3,3)    &    (3,3)    &    (3,3)    &    ($\frac{7}{3},\frac{8}{3}$)    &    ($\frac{7}{3},\frac{8}{3}$)    &    ($\frac{7}{3},\frac{8}{3}$)    &    ($\frac{7}{3},\frac{8}{3}$)    &    (3,3)    &    (3,3)    &    (3,3)    &    (3,3)    &    (3,3)    &    (3,3)    &    (3,3)    &    (3,3)  \\
                          &             &             &             &             &             &             &             &             &             &             &             &             &             &             &             &           \\
  ${\cal P}^X_{0+0+}$     &    (3,3)    &    (3,3)    &    (3,3)    &    (3,3)    &    (2,2)    &    (2,2)    &    (2,2)    &    (2,2)    &    (3,3)    &    (3,3)    &    (3,3)    &    (3,3)    &    (2,2)    &    (2,2)    &    (2,2)    &    (2,2)  \\
                          &             &             &             &             &             &             &             &             &             &             &             &             &             &             &             &           \\
  ${\cal P}^X_{0+00}$     &    (3,3)    &    (3,3)    &    (3,3)    &    (3,3)    &    (2,2)    &    (2,2)    &    (2,2)    &    (2,2)    &    (3,3)    &    (3,3)    &    (3,3)    &    (3,3)    &    (2,2)    &    (2,2)    &    (2,2)    &    (2,2)  \\
                          &             &             &             &             &             &             &             &             &             &             &             &             &             &             &             &           \\
  ${\cal P}^X_{00++}$     &    (3,3)    &    (3,3)    &    (3,3)    &    (3,3)    &    ($\frac{7}{3},\frac{8}{3}$)    &    ($\frac{7}{3},\frac{8}{3}$)    &    ($\frac{7}{3},\frac{8}{3}$)    &    ($\frac{7}{3},\frac{8}{3}$)    &    (3,3)    &    (3,3)    &    (3,3)    &    (3,3)    &    (3,3)    &    (3,3)    &    (3,3)    &    (3,3)  \\
                          &             &             &             &             &             &             &             &             &             &             &             &             &             &             &             &           \\
  ${\cal P}^X_{00+0}$     &    (3,3)    &    (3,3)    &    (3,3)    &    (3,3)    &    ($\frac{7}{3},\frac{8}{3}$)    &    ($\frac{7}{3},\frac{8}{3}$)    &    ($\frac{7}{3},\frac{8}{3}$)    &    ($\frac{7}{3},\frac{8}{3}$)    &    (3,3)    &    (3,3)    &    (3,3)    &    (3,3)    &    (3,3)    &    (3,3)    &    (3,3)    &    (3,3)  \\
                          &             &             &             &             &             &             &             &             &             &             &             &             &             &             &             &           \\
  ${\cal P}^X_{000+}$     &    (3,3)    &    (3,3)    &    (3,3)    &    (3,3)    &    (2,2)    &    (2,2)    &    (2,2)    &    (2,2)    &    (3,3)    &    (3,3)    &    (3,3)    &    (3,3)    &    (2,2)    &    (2,2)    &    (2,2)    &    (2,2)  \\
                          &             &             &             &             &             &             &             &             &             &             &             &             &             &             &             &           \\
  ${\cal P}^X_{0000}$     &    (3,3)    &    (3,3)    &    (3,3)    &    (3,3)    &    (2,2)    &    (2,2)    &    (2,2)    &    (2,2)    &    (3,3)    &    (3,3)    &    (3,3)    &    (3,3)    &    (2,2)    &    (2,2)    &    (2,2)    &    (2,2)  \\
                          &             &             &             &             &             &             &             &             &             &             &             &             &             &             &             &           \\
\end{tabular}
\normalsize
\end{sideways}
\caption[The prisoner's dilemma: Extended payoff table]{\em
Table of expected payoffs for various isomorphically constrained
equilibria. The trees generated under each joint probability
space and their equilibrium pathways are shown in Fig.
\protect\ref{f_extended_analysis}. } \label{t_extended_analysis}
\end{table}

Here, without presenting the details of the calculations, we
show the results of comparing all 16 possible probability spaces
of each player against all 16 of their opponent's possible
probability spaces---see Fig. \ref{f_extended_analysis} and
Table \ref{t_extended_analysis}. (In cases where players are
indifferent to move choice, we arbitrarily choose cooperation.)
We also note that it turns out that there is only one
isomorphically constrained equilibria in each probability space
and some of these are in mixed strategies.

It is of course possible to use Table \ref{t_extended_analysis}
to locate globally optimal choices of probability space.
Examination of this table shows that many rows and columns are
identical. Numbering each row from top to bottom by $r_i$ and
each column from left to right by $c_j$ ($1\leq i,j\leq 16$), we
have $r_1=r_2=r_5=r_6$, $r_3=r_4=r_7=r_8$,
$r_9=r_{10}=r_{13}=r_{14}$, and $r_{11}=r_{12}=r_{15}=r_{16}$.
As well, we have $c_{1}=c_{2}=c_{3}=c_{4}$,
$c_{5}=c_{6}=c_{7}=c_{8}$, $c_{9}=c_{10}=c_{11}=c_{12}$, and
$c_{13}=c_{14}=c_{15}=c_{16}$. Removing all identical rows and
columns leaves the variational payoff table
\begin{equation}
    \begin{array}{c|cccc}
   \left(\langle\Pi^X\rangle,\langle\Pi^Y\rangle\right)
                      & {\cal P}^Y_{++++} & {\cal P}^Y_{+000}           & {\cal P}^Y_{0+++}               & {\cal P}^Y_{0000}  \\ \hline
                        &             &                                 &                                 &               \\
  {\cal P}^X_{++++}     &    (4,4)    &    (4,4)                        &    (3,3)                        &    (3,3)      \\
                        &             &                                 &                                 &               \\
  {\cal P}^X_{+000}     &    (4,4)    &    (4,4)                        &    (\frac{8}{3},\frac{7}{3})    &    (2,2)      \\
                        &             &                                 &                                 &               \\
  {\cal P}^X_{0+++}     &    (3,3)    &    (\frac{7}{3},\frac{8}{3})    &    (3,3)                        &    (3,3)      \\
                        &             &                                 &                                 &               \\
  {\cal P}^X_{0000}     &    (3,3)    &    (2,2)                        &    (3,3)                        &    (2,2)      \\
                        &             &                                 &                                 &               \\
    \end{array}.
\end{equation}
An inspection by eye (checked by numerical calculation) confirms
that the only ``equilibria" in this reduced table of constrained
equilibria are the uninteresting combinations in the bottom
right of $\left({\cal P}^X_{0000},{\cal P}^Y_{0+++}\right)$,
$\left({\cal P}^X_{0+++},{\cal P}^Y_{0000}\right)$, and
$\left({\cal P}^X_{0+++},{\cal P}^Y_{0+++}\right)$, and the more
interesting payoff maximizing equilibria in the top left of
$\left({\cal P}^X_{++++},{\cal P}^Y_{++++}\right)$, $\left({\cal
P}^X_{++++},{\cal P}^Y_{+000}\right)$, $\left({\cal
P}^X_{+000},{\cal P}^Y_{++++}\right)$, and $\left({\cal
P}^X_{+000},{\cal P}^Y_{+000}\right)$. In these latter
equilibria, as long as players functionally correlate their
behavioural strategies in the second stage following from the
history $(x_1,y_1)=(0,0)$, then they will conclude that it is
payoff maximizing to cooperate in this finite iterated
prisoner's dilemma to garner joint payoffs of
$\left(\langle\Pi^X\rangle,\langle\Pi^Y\rangle\right)=(4,4)$.
Any other choice is not rational.

Again, we conclude that players of unrestricted rationality will
cooperate in the finite iterated prisoner's dilemma.  As such,
our analysis reconciles game theoretic predictions and the
cooperative human behaviours observed in experimental tests
\cite{Cooper_96_18,Milinski_98_13}.

\section{$N>2$ stages: A limited investigation}

We now consider the case where the number of stages is known and
finite and greater than two.  We will consider how players might
vary their choice of probability space or of isomorphic
constraints so as to optimize the expected payoffs of Eq.
\ref{eq_expected_payff_functions} when the number of stages
$N>2$. Our analysis will be limited as with each additional
stage the number of possible joint probability spaces that might
be considered by the players increases exponentially. In the
present section, we suppose that players adopt either a
conventional independent behavioural space or a Markovian space
in which current stage choices are correlated to the immediately
preceding stage choices.  In more detail, the choices open to
the players include adopting either a conventional independent
behavioural strategy space ${\cal P}^X_B$ and ${\cal P}^Y_B$, or
a Markovian probability space ${\cal P}^X_B|_{x_n=y_{n-1}}$ and
${\cal P}^Y_B|_{y_n=x_{n-1}}$. In subsequent sections, we will
examine the various combinations of probability space that might
be adopted, and we will finally allow players to try preemptive
defection near the terminal stages of the game.  This will allow
us to check whether these defections propagate backwards as
required by a standard backwards induction analysis.

\subsection{$N\geq 2$: Independent strategies}

We first presume that players $X$ and $Y$ each examine the case
where they jointly adopt the space ${\cal P}^X_B\times{\cal
P}^Y_B$ in which all of their behavioural strategies on every
possible history set are independent of any other event.  The
players seek to optimize their respective expected payoff
functions in Eq. \ref{eq_expected_payff_functions}.

Every behavioural probability parameter (after normalization) is
independent so the gradient operator used by both players to
analyze optimal play are
\begin{equation}
    \nabla =
    \left[
    \frac{d}{d P^X(1)},\frac{d}{d P^Y(1)},
    \frac{d}{d P^X(1|H_1)},\frac{d}{d P^Y(1|H_1)}, \dots,
        \frac{d}{d P^X(1|H_N)},\frac{d}{d P^Y(1|H_N)}
    \right].
\end{equation}
where gradients are taken with respect to all possible history
sets $H_n$.  Also, gradients are taken via total derivatives
rather than partial derivatives to facilitate calculations---the
normalization constraint $P^X(0|H_n)=1-P^X(1|H_n)$ allows
writing the total rate of change of the expected payoff function
with respect to the changing probability parameters as
\begin{equation}
  \frac{d \langle \Pi^X \rangle}{d P^X(1|H_n)}
   = \frac{\partial \langle \Pi^X \rangle}{\partial P^X(1|H_n)}
    - \frac{\partial \langle \Pi^X \rangle}{\partial P^X(0|H_n)}.
\end{equation}

Each player can then straightforwardly use this gradient
operator defined within the joint probability space ${\cal
P}^X_B\times{\cal P}^Y_B$ to evaluate their optimal choices. In
particular, the shorthand notation $H_{n}=\{H_{n-1},x_n,y_n\}$
and some algebra allows writing the optimization conditions for
player $X$ as the set of simultaneous equations
\begin{eqnarray}
 \frac{d \langle \Pi^X \rangle}{d P^X(1)}&=& \dots  \nonumber \\
 &\vdots&   \nonumber \\
 \frac{d \langle \Pi^X \rangle}{d P^X(1|H_{n-1})}
 &=& 1 +  \nonumber \\
 && \sum_{\stackrel{x_1\dots x_{N-2}=0}{y_1\dots y_{N-2}=0}}^1
       P^X(x_1)P^Y(y_1) \dots P^X(x_{N-2}|H_{N-2})P^Y(y_{N-2}|H_{N-2}) \times \nonumber \\
 && \sum_{y_{N-1}=0}^1 P^Y(y_{N-1}|H_{N-1})
    \sum_{x_N y_N=0}^1 (x_N-2y_N) \times \nonumber \\
 && \hspace{1cm} \left[ \right. P^X(x_N|H_{N-1},1,y_{N-1})
        P^Y(y_N|H_{N-1},1,y_{N-1}) - \nonumber \\
 && \hspace{2cm} P^X(x_N|H_{N-1},0,y_{N-1})
           P^Y(y_N|H_{N-1},0,y_{N-1}) \left. \right] \nonumber \\
 \frac{d \langle \Pi^X \rangle}{d P^X(1|H_N)} &=& 1.
\end{eqnarray}
The equivalent simultaneous optimization conditions for player
$P_y$ are
\begin{eqnarray}
 \frac{d \langle \Pi^Y \rangle}{d P^Y(1)}&=& \dots  \nonumber \\
 &\vdots&   \nonumber \\
 \frac{d \langle \Pi^Y \rangle}{d P^Y(1|H_{N-1})}
 &=& 1 +  \nonumber \\
 && \sum_{\stackrel{x_1\dots x_{N-2}=0}{y_1\dots y_{N-2}=0}}^1
       P^X(x_1)P^Y(y_1) \dots P^X(x_{N-2}|H_{N-2})P^Y(y_{N-2}|H_{N-2}) \times \nonumber \\
 && \sum_{x_{N-1}=0}^1 P^X(x_{N-1}|H_{N-1})
    \sum_{x_N y_N=0}^1 (y_N-2x_N) \times \nonumber \\
 && \hspace{1cm} \left[ \right.
        P^X(x_N|H_{N-1},x_{N-1},1)
        P^Y(y_N|H_{N-1},x_{N-1},1) - \nonumber \\
 &&     \hspace{2cm}   P^X(x_N|H_{N-1},x_{N-1},0)
           P^Y(y_N|H_{N-1},x_{N-1},0) \left. \right] \nonumber \\
 \frac{d \langle \Pi^Y \rangle}{d P^Y(1|H_N)} &=& 1.
\end{eqnarray}
Subsequently each player solves their respective sets of
simultaneous equations to maximize their expected payoff in the
joint probability space ${\cal P}^X_B\times{\cal P}^Y_B$ by
setting $P^X(1|H_N)=1$ and $P^Y(1|H_N)=1$ for all history sets
$H_N$, and by setting $P^X(1|H_{N-1})=1$ and $P^Y(1|H_{N-1})=1$
for all history sets $H_{N-1}$, and so on. The final result is
that both players defect at every stage giving optimal choices
as $(x_n,y_n)=(1,1)\equiv(D,D)$ for all $n$. At this point,
payoffs are
$(\langle\Pi^X_B\rangle,\langle\Pi^Y_B\rangle)=(N,N)$.

\subsection{$N\geq 2$: Markovian versus Independent spaces}

Suppose now that players $X$ and $Y$ jointly examine the case
where $Y$ adopts the independent probability space while $X$
adopts isomorphic constraints to implement Markovian play. In
this case the joint probability space is ${\cal
P}^X_B|_{x_n=y_{n-1}}\times{\cal P}^Y_B$. Here, $X$ adopts the
isomorphic constraints
\begin{eqnarray}     \label{eq_Mark_x}
  x_n &=& y_{n-1}  \nonumber \\
  P^X(x_n|H_n) &=& \delta_{x_ny_{n-1}},
\end{eqnarray}
for $2\leq n\leq N$ and on every history $H_n$. As usual, these
isomorphic constraints must be resolved before the optimization
can proceed rendering the optimization problem for each player
as
\begin{eqnarray}    \label{eq_MKV_IND}
 X: \max_{P^X(1)}\;\; \langle \Pi^X \rangle
  &=& 2N + \left[ \sum_{x_1=0}^1 P^X(x_1) x_1 \right] + \nonumber \\
  && -  \sum_{n=1}^{N-1} \sum_{x_1y_1\dots y_n=0}^1
        P^X(x_1) P^Y(y_1) \dots  P^Y(y_n|H'_n) y_n + \nonumber \\
  &&  -2 \sum_{x_1y_1\dots y_N=0}^1 P^X(x_1)P^Y(y_1) \dots  P^Y(y_N|H'_N) y_N,  \nonumber \\
  Y: \max_{P^Y(1),P^Y(1|H_n)}\;\; \langle \Pi^Y \rangle
  &=& 2N -2 \left[ \sum_{x_1=0}^1 P^X(x_1) x_1 \right] +  \\
  &&   - \sum_{n=1}^{N-1} \sum_{x_1y_1\dots y_n=0}^1
      P^X(x_1)P^Y(y_1) \dots  P^Y(y_n|H'_n) y_n   \nonumber \\
  && + \sum_{x_1y_1\dots y_N=0}^1 P^X(x_1)P^Y(y_1) \dots  P^Y(y_N|H'_N) y_N.  \nonumber
\end{eqnarray}
Here, a modified history set appears due to the delta-function
constraints so that, for instance,
$H'_3=\{x_1,y_1,x_2,y_2\}=\{x_1,y_1,y_1,y_2\}$.  Hereinafter,
primes are dropped.

The shorthand notation $H_{n}=\{H_{n-1},y_n\}$ for $n\geq 2$ and
some algebra allows writing the optimization conditions for player
$Y$ as the set of simultaneous equations
\begin{eqnarray}
 \frac{d \langle \Pi^Y\rangle}{d P^Y(1)}&=& \dots,  \nonumber \\
          &\vdots&   \nonumber \\
 \frac{d \langle \Pi^Y \rangle}{d P^Y(1|H_{N-1})} &=& -1 + \nonumber \\
    && + \sum_{x_1y_1\dots y_{N-2}=0}^1 P^X(x_1)P^Y(y_1) \dots
           P^Y(y_{N-2}|H_{N-2}) \times \nonumber \\
    && \sum_{y_N=0}^1 y_N
            \left[ P^Y(y_N|H_{N-1},1)-
             P^Y(y_N|H_{N-1},0)\right],   \nonumber \\
 \frac{d \langle \Pi^Y \rangle}{d P(1|H_N)} &=& 1.
\end{eqnarray}
Hence, player $Y$ optimizes their payoff by setting $P^Y(1|H_N)=1$
for every history set $H_N$, and by setting $P^Y(1|H_{N-1})=0$ for
every history set $H_{N-1}$, and eventually by setting
$P^Y(1|H_n)=0$ for $1\leq n\leq (N-1)$.  That is, $Y$ maximizes
their expected payoff by cooperating in every stage but the last.

Player $X$ is well able to calculate the same optimal choices for
their opponent, and uses this knowledge to simplify their own
optimization problem to eventually give the condition
\begin{equation}
 \frac{d \langle \Pi^X \rangle}{d P^X(1)}  = 1.
\end{equation}
Consequently, $X$ optimizes their expected payoff by setting
$P^X(1)=1$ and so defects in this first stage.

In the joint probability space ${\cal
P}^X_B|_{x_n=y_{n-1}}\times{\cal P}^Y_B$, the players locate the
constrained equilibria at the point
$(x_1,y_1,\ldots,y_N)=(1,0,\dots,0,1)$ generating the play
sequence
\begin{eqnarray}
 (x_n,y_n) &=& (1,0),(0,0),\dots,(0,0),(0,1) \nonumber \\
  &=& (D,C) (C,C) \dots (C,C) (C,D),
\end{eqnarray}
to give expected payoffs
$(\langle\Pi^X\rangle,\langle\Pi^Y\rangle) = (2N-1,2N-1)$. Here,
$X$ defects in the first stage as their opponent cannot respond
without decreasing their payoff, while $Y$ can defect in the
last stage when $X$ can no longer respond.

\subsection{$N\geq 2$: Markovian versus Markovian strategies}

Each player might well then analyze the case where both players
adopt Markovian strategies and thereby implement the joint
probability space ${\cal P}^X_B|_{x_n=y_{n-1}}\times{\cal
P}^Y_B|_{y_n=x_{n-1}}$.  Here, $X$ adopts the isomorphic
constraints
\begin{eqnarray}
  x_n &=& y_{n-1}  \nonumber \\
  P^X(x_n|H_n) &=& \delta_{x_ny_{n-1}},
\end{eqnarray}
for $2\leq n\leq N$ and every history set $H_n$, while $Y$
adopts the isomorphic constraints
\begin{eqnarray} \label{eq_TFT_xy}
  y_n &=& x_{n-1}  \nonumber \\
  P^Y(y_n|H_n) &=& \delta_{y_nx_{n-1}},
\end{eqnarray}
for $2\leq n\leq N$  and every history set $H_n$. These
constraints must be resolved before the optimization can proceed
reducing the optimization problem for each player to
\begin{eqnarray}
 X: \max_{P^X(1)}\;\; \langle \Pi^X \rangle
   &=& \sum_{x_1,y_1=0}^1 P^X(x_1) P^Y(y_1)\Pi^X(x_1,y_1), \nonumber   \\
 Y: \max_{P^Y(1)}\;\; \langle \Pi^Y \rangle
   &=& \sum_{x_1,y_1=0}^1 P^X(x_1) P^Y(y_1)\Pi^Y(x_1,y_1),
\end{eqnarray}
where the payoffs for a given play sequence $(x_1,y_1)$ are
\begin{eqnarray}
  \Pi^X(x_1,y_1) & = &  \left\{
   \begin{array}{ll}
 2N - \frac{N}{2}x_1 - \frac{N}{2} y_1, & N \mbox{ even},\\
      &    \\
 2N - \frac{N-3}{2}x_1 - \frac{N+3}{2}y_1, & N \mbox{ odd},\\
   \end{array}
          \right.    \nonumber \\
          \nonumber   \\
  \Pi^Y(x_1,y_1) & = &  \left\{
   \begin{array}{lc}
 2N - \frac{N}{2} x_1 - \frac{N}{2} y_1, & N \mbox{ even},\\
      &    \\
 2N - \frac{N+3}{2}x_1 - \frac{N-3}{2}y_1, & N \mbox{ odd}.\\
   \end{array}
          \right.
\end{eqnarray}
The adoption of the joint probability space ${\cal
P}^X_B|_{x_n=y_{n-1}}\times{\cal P}^Y_B|_{y_n=x_{n-1}}$ has
effectively reduced the $N$ stage supergame to a single stage
game with variables $x_1$ and $y_1$ and payoff matrices, for $N$
even of
\begin{equation}
  \begin{array}{cc}
      & Y \\
    X &
    \begin{array}{c|cc}
      (\Pi^X,\Pi^Y) & C                           &  D                          \\   \hline
            C       & (2N,2N)                     & (\frac{3}{2}N,\frac{3}{2}N) \\
            D       & (\frac{3}{2}N,\frac{3}{2}N) & (N,N),                      \\
    \end{array}
  \end{array}
\end{equation}
and for odd $N$ of
\begin{equation}
  \begin{array}{cc}
      & Y \\
    X &
    \begin{array}{c|cc}
      (\Pi^X,\Pi^Y) & C                    &  D                   \\   \hline
            C       & (2N,2N)              & \frac{3}{2}[N-1,N+1] \\
            D       & \frac{3}{2}[N+1,N-1] & (N,N).               \\
    \end{array}
  \end{array}
\end{equation}
That is, in the joint probability space ${\cal
P}^X_B|_{x_n=y_{n-1}}\times{\cal P}^Y_B|_{y_n=x_{n-1}}$, the
normal form game (and equivalent game tree) is described by an
effective payoff matrix with altered off-diagonal elements which
naturally modify equilibria.

As usual, the constrained equilibria in the joint space ${\cal
P}^X_B|_{x_n=y_{n-1}}\times{\cal P}^Y_B|_{y_n=x_{n-1}}$ are now
located via
\begin{eqnarray} \label{dx1_dy1}
  \frac{d\langle\Pi^X\rangle}{d P^X(1)} & = &  \left\{
   \begin{array}{ll}
 - \frac{N}{2}, & N \mbox{ even},\\
      &    \\
 - \frac{1}{2}(N-3), & N \mbox{ odd},\\
   \end{array}
          \right.    \nonumber \\
    &&      \nonumber   \\
    &&      \nonumber   \\
 \frac{d\langle\Pi^Y\rangle}{d P^Y(1)} & = &  \left\{
   \begin{array}{lc}
 - \frac{N}{2}, & N \mbox{ even},\\
      &    \\
  - \frac{1}{2}(N-3), & N \mbox{ odd}.\\
   \end{array}
          \right.
\end{eqnarray}
Thus, for either $N$ even or for $N$ odd and greater than 3 we
have the equilibrium points $P^X(1)=0$ and $P^Y(1)=0$ or
$(x_1,y_1)=(0,0)\equiv(C,C)$.  Alternatively, for $N=1$ the
equilibria is $P^X(1)=1$ and $P^Y(1)=1$ or
$(x_1,y_1)=(1,1)\equiv(D,D)$. When $N=3$ these conditions are
satisfied for any values of $(x_1,y_1)$ requiring examination of
actual payoffs motivating the selection
$(x_1,y_1)=(0,0)\equiv(C,C)$. The generated sequences of play
are
\begin{equation} \label{sequence}
\begin{array}{c|c|l|c}
 N       &  (x_1,y_1) &               & (\langle\Pi^X\rangle,\langle\Pi^Y\rangle) \\  \hline
         &            &               &        \\
 1       &  (1,1)     & (DD)          & (1,1)  \\
         &            &               &        \\
 N\geq 2 &  (0,0)     & (CC)\dots(CC) & (2N,2N). \\
\end{array}
\end{equation}

\subsection{$N\geq 2$: Comparing payoffs}

Each player must then compare the payoffs they expect given that
together they jointly adopt the probability space combinations
examined above.  A table of all possible outcomes for an $N\geq
2$ stage game given the probability spaces under consideration
takes the form
\begin{equation}
    \begin{array}{c|cccc}
   \left(\langle\Pi^X\rangle,\langle\Pi^Y\rangle\right)
                             & {\cal P}^Y_B|_{y_n=x_{n-1}} & {\cal P}^Y_B \\ \hline
                             &                            &                               \\
  {\cal P}^X_B|_{x_n=y_{n-1}} &    (2N,2N)                 &    (2N-1,2N-1)                      \\
                             &                            &                               \\
  {\cal P}^X_B               &    (2N-1,2N-1)             &    (N,N)                      \\
    \end{array}.
\end{equation}
This table makes it clear that in all the games considered here
with two or more stages, players of unbounded rationality
maximize their payoffs by each adopting the joint probability
space ${\cal P}^X_B|_{x_n=y_{n-1}}\times{\cal
P}^Y_B|_{y_n=x_{n-1}}$ in which they adopt isomorphic
constraints to correlate all of their choices in every stage
after the first with their opponents. Once each player has
adopted this particular probability space, this means that they
have adopted a ``roulette" randomization device which allows
them no further choices in any stage after the first, and they
have done this as it maximizes their expected payoff.

As in the $N=2$ stage game, we conclude that while players of
bounded rationality implementing a conventional analysis will
defect in the multiple stage game, players of unrestricted
rationality will cooperate in the finite iterated prisoner's
dilemma.  Again, our analysis is consistent with observed human
behaviours \cite{Cooper_96_18,Milinski_98_13}.

\subsection{$N\geq 2$: Endgame analysis}

The simplified analysis of the previous section does not allow
consideration of ``endgame" strategies where players seek to
defect in the final stages of a multiple stage game to preempt
the defection of their opponent.  It is these preemptive
defections in backwards induction which conventionally require
players of bounded rationality to defect in every stage of the
finite iterated prisoner's dilemma.  The question now is, does
such mutual preemption apply in an unbounded rational analysis
where players consider a wider range of possible alternate
probability spaces.  To this end, we suppose that player $X$
adopts a probability space ${\cal P}^X_k$ where they
functionally correlate their moves for stage $2\leq n\leq (N-k)$
to their opponent's previous choices via
\begin{eqnarray}
 x_n &=& y_{n-1}  \nonumber \\
 P^X(x_n|H_n) &=& \delta_{x_ny_{n-1}},
\end{eqnarray}
for $2\leq n \leq N-k$ and for every history $H_n$, but chooses to
make their choices in subsequent stages independently so that for
$(N-k+1)\leq n\leq N$, all distributions $P^X(x_n|H_n)$ for all
histories $H_n$ represent independent behavioural random
variables. Similarly, we suppose that player $Y$ adopts a
probability space ${\cal P}^Y_j$ where they functionally correlate
their moves for stage $2\leq n\leq (N-j)$ to their opponent's
previous choices $N$ via
\begin{eqnarray}
 y_n &=& x_{n-1}  \nonumber \\
 P^Y(y_n|H_n) &=& \delta_{y_nx_{n-1}},
\end{eqnarray}
for $2\leq n \leq N-j$ and for every history $H_n$, but chooses to
make their choices in subsequent stages independently so that for
$(N-j+1)\leq n\leq N$, all distributions $P^Y(y_n|H_n)$ for all
histories $H_n$ represent independent behavioural random
variables.

For either player, the probability space ${\cal P}^Z_k$ subsumes
a number of other possible probability spaces of interest. For
instance, setting either $k=N-1$ or $k=N$ makes all of player
$Z$'s behavioural variables throughout the entire game
independent, so ${\cal P}^Z_N={\cal P}^Z_{N-1}={\cal P}^Z_B$.
More interestingly, this probability space subsumes certain
deterministic alternatives. To see this, suppose that player $Z$
considers a probability space enforcing defection with certainty
in the last $k$ stages. However, it is not difficult to see that
this probability space is weakly dominated by space ${\cal
P}^Z_k$---this latter space allows players to either defect
whenever that is payoff maximizing so they will do as well as
defecting with certainty, or to cooperate whenever that is
payoff maximizing so they will do as well as cooperating with
certainty. That is, the motivation to preemptively defect in the
endgame for a larger payoff is taken into account when
considering the probability space ${\cal P}^Z_k$. Exactly
similar considerations establish that ${\cal P}^Z_k$ weakly
dominates spaces enforcing a deterministic play of Tit-For-Tat
which specify cooperation in the first stage.

We now suppose that players $X$ and $Y$ together adopt the joint
probability spaces ${\cal P}^X_k\times{\cal P}^Y_j$ to examine
rational choices for the cessation of cooperative play and the
onset of preemptive defections.  In this particular joint
probability space, the optimization problem for each player
becomes
\begin{eqnarray}
 X: \max_{p_1,P^X(1|H_{N-k+1}),\dots,P^X(1|H_N)}\;\; \langle \Pi^X_{kj} \rangle
   &=&  \nonumber \\
   && \hspace{-6cm} \sum_{\stackrel{x_1,x_{N-k+1},\dots,x_N=0}{y_1,y_{N-j+1},\dots y_N=0}}^1
        P^X(x_1)P^Y(y_1)
        P^X(x_{N-k+1}|H'_{N-k+1})P^Y(y_{N-j+1}|H'_{N-j+1}) \times \dots \nonumber \\
 &&   \hspace{-6cm} \dots \times P^X(x_N|H'_N)P^Y(y_N|H'_N)
      \Pi^X_{kj}(x_1,x_{N-k+1},\dots,x_N,y_1,y_{N-j+1},\dots, y_N)\nonumber \\
   & & \nonumber \\
 Y: \max_{q_1,P^Y(1|H_{N-j+1}),\dots,P^Y(1|H_N)}\;\; \langle \Pi^Y_{kj} \rangle
   &=& \\
   &&  \hspace{-6cm}  \sum_{\stackrel{x_1,x_{N-k+1},\dots,x_N=0}{y_1,y_{N-j+1},\dots y_N=0}}^1
        P^X(x_1)P^Y(y_1)
        P^X(x_{N-k+1}|H'_{N-k+1})P^Y(y_{N-j+1}|H'_{N-j+1}) \times \dots \nonumber \\
 &&   \hspace{-6cm} \dots \times P^X(x_N|H'_N)P^Y(y_N|H'_N)
      \Pi^Y_{kj}(x_1,x_{N-k+1},\dots,x_N,y_1,y_{N-j+1},\dots y_N), \nonumber
\end{eqnarray}
where again, care must be taken in writing the delta-function
modified history sets $H'_n$.

In this equation, the attained payoffs for any given play sequence
$(x_1,x_{N-k+1},\dots,x_N,y_1,y_{N-j+1}\dots,y_N)$, assuming for
simplicity that $N\geq 3$, are variously:
\begin{eqnarray}
   && \nonumber  \\
   && \hspace{-2cm} 1\leq k\leq (N-1), j=0:
      \mbox{independent variables: }
      x_1,x_{N-k+1},\dots,x_{N},y_1 \\
      &&  \nonumber \\
  \Pi^X_{kj} & = &  \left\{
   \begin{array}{lc}
 2N  + \frac{k-N}{2} x_1 + \frac{k-4-N}{2}y_1
          - \sum_{n=N-k+1}^{N-1} x_n + x_N, & (N-k) \mbox{ even}\\
      &    \\
 2N  + \frac{k-1-N}{2} x_1 + \frac{k-3-N}{2}y_1
          - \sum_{n=N-k+1}^{N-1} x_n + x_N, & (N-k) \mbox{ odd}.\\
   \end{array}
          \right.  \nonumber   \\
      &&  \nonumber \\
      &&  \nonumber \\
  \Pi^Y_{kj} & = &  \left\{
   \begin{array}{ll}
 2N + \frac{k-N}{2} x_1 + \frac{2+k-N}{2}y_1
          - \sum_{n=N-k+1}^{N-1} x_n - 2 x_N, & (N-k) \mbox{ even}\\
      &    \\
 2N + \frac{k-1-N}{2} x_1 + \frac{3+k-N}{2}y_1
          - \sum_{n=N-k+1}^{N-1} x_n - 2 x_N, & (N-k) \mbox{ odd}.\\
   \end{array}
          \right.  \nonumber \\
      &&  \nonumber \\
   && \nonumber  \\
   && \hspace{-2cm} 1\leq k\leq (N-1), j=(N-1):
      \mbox{independent variables: }
      x_1,x_{N-k+1},\dots,x_{N},y_1,\dots,y_N \nonumber \\
      &&  \nonumber \\
  \Pi^X_{kj} & = &  2N + x_1 + \sum_{n=N-k+1}^{N} x_n
        - \sum_{n=1}^{N-k-1} y_n - 2 \sum_{n=N-k}^{N} y_n , \nonumber \\
      &&  \nonumber \\
      &&  \nonumber \\
  \Pi^Y_{kj} & = &  2N - 2 x_1 - 2 \sum_{n=N-k+1}^{N} x_n
        - \sum_{n=1}^{N-k-1} y_n + \sum_{n=N-k}^{N} y_n \nonumber  \\
      &&  \nonumber \\
   && \nonumber  \\
   && \hspace{-2cm} k=j, 1\leq k\leq (N-1):
      \mbox{independent variables: }
      x_1,x_{N-k+1},\dots,x_{N},y_1,y_{N-k+1},\dots,y_{N} \nonumber \\
      &&  \nonumber \\
  \Pi^X_{kj} & = &  \left\{
   \begin{array}{ll}
 2N + \frac{k-N}{2}x_1+ \frac{k-N}{2} y_1
          + \sum_{n=N-k+1}^{N} x_n - 2 \sum_{n=N-k+1}^{N} y_n, & (N-k) \mbox{ even}\\
      &    \\
 2N + \frac{3+k-N}{2}x_1+ \frac{k-3-N}{2} y_1
          + \sum_{n=N-k+1}^{N} x_n - 2 \sum_{n=N-k+1}^{N} y_n, & (N-k) \mbox{ odd}\\
   \end{array}
          \right.    \nonumber \\
      &&  \nonumber   \\
      &&  \nonumber \\
  \Pi^Y_{kj} & = &  \left\{
   \begin{array}{lc}
 2N + \frac{k-N}{2}x_1+ \frac{k-N}{2} y_1
          - 2 \sum_{n=N-k+1}^{N} x_n + \sum_{n=N-k+1}^{N} y_n, & (N-k) \mbox{ even}\\
      &    \\
 2N + \frac{k-3-N}{2}x_1+ \frac{3+k-N}{2} y_1
          - 2 \sum_{n=N-k+1}^{N} x_n + \sum_{n=N-k+1}^{N} y_n, & (N-k) \mbox{ odd}.\\
   \end{array}
          \right.  \nonumber \\
      &&  \nonumber \\
   && \nonumber \\
   && \hspace{-2cm} k>j, 1\leq k,j\leq (N-1):
      \mbox{independent variables: }
      x_1,x_{N-k+1},\dots,x_{N},y_1,y_{N-j+1},\dots,y_{N} \nonumber \\
      &&  \nonumber \\
   \Pi^X_{kj} & = &  \left\{
   \begin{array}{l}
 2N + \frac{k-N}{2}x_1+ \frac{k-4-N}{2} y_1
          - \sum_{n=N-k+1}^{N-j-1} x_n
          + \sum_{n=N-j}^{N} x_n
          - 2 \sum_{n=N-j+1}^{N} y_n, \\
      \hspace{11cm} (N-k) \mbox{ even}    \\
      \\
 2N + \frac{k-1-N}{2}x_1+ \frac{k-3-N}{2} y_1
          - \sum_{n=N-k+1}^{N-j-1} x_n
          + \sum_{n=N-j}^{N} x_n
          - 2 \sum_{n=N-j+1}^{N} y_n, \\
      \hspace{11cm}  (N-k) \mbox{ odd} \\
   \end{array}
          \right.    \nonumber \\
      &&    \nonumber   \\
      &&  \nonumber \\
  \Pi^Y_{kj} & = &  \left\{
   \begin{array}{l}
 2N + \frac{k-N}{2}x_1+ \frac{2+k-N}{2} y_1
          - \sum_{n=N-k+1}^{N-j-1} x_n
          - 2 \sum_{n=N-j}^{N} x_n
          + \sum_{n=N-j+1}^{N} y_n, \\
       \hspace{11cm} (N-k) \mbox{ even}    \\
       \\
 2N + \frac{k-1-N}{2}x_1+ \frac{3+k-N}{2} y_1
          - \sum_{n=N-k+1}^{N-j-1} x_n
          - 2 \sum_{n=N-j}^{N} x_n
          + \sum_{n=N-j+1}^{N} y_n,  \\
          \hspace{11cm} (N-k) \mbox{ odd}. \\
   \end{array}
          \right.  \nonumber \\
   && \nonumber
\end{eqnarray}

\begin{table}
\centering
\begin{sideways}
\scriptsize
\begin{tabular}{c|cccccccc}
  $(\langle\Pi^X_{kj}\rangle,\langle\Pi^Y_{kj}\rangle)$ & $j=0$   & $1$  & $2$               & $3$                      & $4$                       &  $\cdots$ & $N-2$                    & $N-1$        \\ \hline
                &                         &                       &                          &                          &                           &           &                          &              \\
      $k=0$     & $2N,2N$                 & $2N-2,2N+1$           &                 =        &                =         &                =          &  $\cdots$ & $2N-2,\frac{2N+1}{2N-2}$ & $2N-1,2N-1$  \\
                &                         &                       &                          &                          &                           &           &                          &              \\
      $1$       & $2N+1,2N-2$             & $2N-1,2N-1$           & $2N-3,2N$                &                =         &                =          &  $\cdots$ & $2N-3,\frac{2N}{2N-3}$   & $2N-2,2N-2$  \\
                &                         &                       &                          &                          &                           &           &                          &              \\
      $2$       &       "                 & $2N,2N-3$             & $2N-2,2N-2$              & $2N-4,2N-1$              &                =          &  $\cdots$ & $2N-4,\frac{2N-1}{2N-4}$ & $2N-3,2N-3$  \\
                &                         &                       &                          &                          &                           &           &                          &              \\
      $3$       &       "                 &       "               & $2N-1,2N-4$              & $2N-3,2N-3$              &  $2N-5,2N-2$              &  $\cdots$ & $2N-5,\frac{2N-2}{2N-5}$ & $2N-4,2N-4$  \\
                &                         &                       &                          &                          &                           &           &                          &              \\
      $4$       &       "                 &       "               &             "            & $2N-2,2N-5$              &  $2N-4,2N-4$              &  $\cdots$ & $2N-6,\frac{2N-3}{2N-6}$ & $2N-5,2N-5$  \\
                &                         &                       &                          &                          &                           &           &                          &              \\
      \vdots    & \vdots                  & \vdots                & \vdots                   &  \vdots                  &  \vdots                   &  $\cdots$ & \vdots                   & \vdots       \\
                &                         &                       &                          &                          &                           &           &                          &              \\
      $N-4$     &       "                 &       "               &             "            &            "             &  $2N-3,2N-6$              &  $\cdots$ & $N+2,\frac{N+5}{N+2}$    & $N+3,N+3$    \\
                &                         &                       &                          &                          &                           &           &                          &              \\
      $N-3$     & $\frac{2N+1}{2N-2},2N-2$& $\frac{2N}{2N-3},2N-3$& $\frac{2N-1}{2N-4},2N-4$ & $\frac{2N-2}{2N-5},2N-5$ &  $\frac{2N-3}{2N-6},2N-6$ &  $\cdots$ & $N+1,\frac{N+4}{N+1}$    & $N+2,N+2$    \\
                &                         &                       &                          &                          &                           &           &                          &              \\
      $N-2$     &                      "  &              "        &            "             &            "             &             "             &  $\cdots$ & $N+2,N+2$                & $N+1,N+1$    \\
                &                         &                       &                          &                          &                           &           &                          &              \\
      $N-1$     & $2N-1,2N-1$             & $2N-2,2N-2$           & $2N-3,2N-3$              & $2N-4,2N-4$              &  $2N-5,2N-5$              &  $\cdots$ & $N+1,N+1$                & $N,N$.       \\
                &                         &                       &                          &                          &                           &           &                          &              \\
\end{tabular}  \normalsize
\normalsize
\end{sideways}
\caption[The prisoner's dilemma: Alternate isomorphic
equilibria]{A partial listing of isomorphic equilibria when
players $X$ and $Y$ jointly adopt the probability space ${\cal
P}^X_k\times{\cal P}^Y_j$.  In this space, $X$ functionally
correlates their moves for stage $2\leq n\leq (N-k)$ to their
opponent's previous choices but adopts independent behavioural
strategies in stages $(N-k+1)$ to $N$, while player $Y$
functionally correlates their moves for stage $2\leq n\leq
(N-j)$ to their opponent's previous choices but adopts
independent behavioural strategies in stages $(N-j+1)$ to $N$.
Here, every shown payoff pair is a isomorphic equilibrium point
making selection of a single best payoff maximization strategy
difficult. Fractions indicate alternate equilibria with distinct
payoffs shown in the numerator and denominator.  Ditto signs (")
and equal signs (=) copy values downwards and to the right
respectively. \label{tab_equil} }
\end{table}

The respective constrained equilibria with the optimized payoffs
as shown in Table \ref{tab_equil} for all combinations of $k$
and $j$. Every listed payoff pair in Table \ref{tab_equil} is an
isomorphically constrained equilibrium point optimizing payoffs
given imposed constraints. As noted previously, there is no
generally accepted method to choose between alternate
equilibria. However, it is tempting to use the rules of game
theory to try to select an optimal choice of play.  In Table
\ref{tab_equil}, each alternate probability space becomes a
strategy choice, and each equilibrium point becomes a pair of
payoffs. Standard techniques can then be applied to determine
global equilibria among the located constrained equilibria.
However, we note that in general we have to take care to deal
with multiple equilibria generated by particular joint
probability spaces. By applying the Nash equilibrium definition
to Table \ref{tab_equil}, we obtain global equilibria at ${\cal
P}^X_k\times{\cal P}^Y_j$ for either $k=0$ and $3\leq j\leq
(N-2)$, or $j=0$ and $3\leq k\leq (N-2)$.

These global equilibria can be considered rational for the
iterated prisoner's dilemma in this restricted class of joint
probability spaces, and there is no established way to select a
particular one among these.  The more important feature given
from this analysis is that cooperation still naturally arises
from these equilibria.  The pathways produced by these
equilibria are dominated by cooperation apart from some
different choices at the last stage. This cooperative behaviour
results when players of unbounded rationality examine alterative
probability spaces to optimize their payoffs, in contrast to the
conventionally mandated analysis wherein players are able to
examine only a single probability space and are thus of bounded
rationality.

 \chapter{Conclusion}
 \label{chap_conclusion}

\section{The foundations of strategic analysis}

Strategic game analysis begins by defining the set of players
\begin{equation}
I=\{1,2,\dots,n\}
\end{equation}
with $n\geq 2$.  The choice $n=1$ corresponds to decision
theory.  This immediately begs the question as to whether $n$ is
fixed or variable, and what effect this might have on the
structure of the game analysis space.  The number of players $n$
acts as an index denoting the size of all subsequent spaces, and
$n$ would normally be considered as a constant taking different
values.  Suppose however, that a player wanted to construct a
single space which ``contained" all the possible spaces defined
by each value of $n$.  Would this single space adopt isomorphic
mappings or allow uncertainty in the number of players to
influence strategic decisions?

Subsequently, each player $i$ has a set of pure strategies
$S_i=\{1,2,\dots,m_i\}$ which combine together to give a set of
pure strategy profiles $S=S_1\times \dots\times S_n$.  It is
commonly assumed that an unconstrained rational player must
consider every one of their moves with some (possibly
infinitesimal) probability and thus that the structure of the
strategy set specifies the structure of the game.  In contrast,
we have shown that different probability spaces can be applied
to the set of all possible strategies.  Hence, it is a mistake
to assume that the dimensionality of the strategy set somehow
determines the dimensionality of the game space.

A payoff function $\Pi:S\rightarrow \Re^n$ with
$\Pi(s)=[\Pi_1(s),\dots,\Pi_n(s)]$ then defines the payoff that
player $i$ receives when strategy profile $s\in S$ is played.
Subsequently, a player $i$'s mixed strategy is defined as a
probability distribution over the pure strategy set $S_i$ to
locate a point in an $(m_i-1)-$dimensional standard simplex
\begin{equation}
  \Delta_i = \left\{ x_i\in R^{m_i}: \forall j=1\dots m_i:
      x_{ij}\geq 0: \sum_{j=1}^{m_i} x_{ij}=1 \right\}.
\end{equation}
The mixed strategy profile is then a vector
$x=\{x_1,\dots,x_n\}$.  The mixed strategy space is a
multi-simplex $\Delta=\Delta_1\times\dots\times\Delta_n$. This
simplex is held to be ``complete" and to contain every possible
probability distribution that might describe a game.  It
certainly contains every possible value of every possible
probability distribution, but optimization requires it to
contain every possible value and gradient of each probability
distribution at a minimum.  (Situations requiring greater
generality could well be envisaged.)

Finally, following Von Neumann and Morgenstern, it is
universally held that every player's randomizations are
independent and hence that there are no constraints acting on
the probability distributions of the mixed strategy space.
Thus, the probability of a pure strategy profile $s$ given $x$
is
\begin{equation}
  x(s) = \prod_{i=1}^{n} x_{is_i}
\end{equation}
and the expected payoff to player $i$ is
\begin{equation}
  u_i(x) = \sum_{s\in S} x_i(s) \Pi_i(s).
\end{equation}
This payoff definition acts to limit the scope of possible games
considered in game theory.  There is no reason why games have to
be restricted to consider only poly-linear expected payoff
functions, and we argue here that these restrictions have
limited the ability to analyze games. Payoffs can be assigned to
players based on the probability distributions that they adopt,
or on the gradients of the adopted probability distributions, or
on their ability to maximize entropy or uncertainty or mutual
information or Fisher information. Game probability
distributions can be actualized by having players adjust the
probability of light transmission through painted glass, or by
altering the placement and number of pins effecting the fall of
balls or of water streams.  More mundanely, players can instruct
agents allowing referees to repeat games many times to deduce
adopted probability distributions to assign payoffs.  Further,
in the absence of a complete theory of games, we simply do not
know if players of unbounded rationality would optimize their
outcomes by calculating the Fisher Information of a game, or by
maximizing the Log Likelihood function.  No limits should be
placed on rationality in formulating a complete theory of games.

Present practice in game theory discards isomorphism constraints
allowing the mixed strategy space to take the form of a compact
convex polyhedron in which expected payoff functions are
quasiconcave and continuous polylinear functions of the mixed
strategies of each player. This, in turn, allows the use of
fixed point theorems to locate Nash equilibria, points at which
no player can unilaterally improve their expected payoffs by
changing their mixed strategy \cite{Nash_50_48,Nash_51_28}.
However, no rationale has ever been offered for why the tangent
spaces of the embedded source probability distributions need to
be overwritten. That is, the strength of the isomorphisms
underlying the construction of mixed strategy spaces has never
been considered. Whenever analysis is transferred from one space
to another, then the strength of the isomorphism underlying the
transfer mapping must be established. Von Neumann did precisely
this when he provided the mathematical foundations of quantum
mechanics.  In its early stages, quantum mechanics appeared in
two seemingly distinct forms, matrix mechanics and wave
mechanics. Von Nuemann unified these approaches by establishing
an exact isomorphism between the space of states in matrix
mechanics and the space of wave functions including all relevant
derivatives using theorems from functional analysis
\cite{vonNeumann_1955}. From that point on, the proven existence
of this isomorphic mapping allowed quantum analysis to use
either matrix or wave approaches as desired.  In game theory,
the strength of the isomorphic mapping underlying the embedding
of probability spaces within mixed strategy spaces has not yet
been established.

If, following probability theory, the original tangent spaces of
the source probability distributions describing a game are
retained within the mixed strategy space, then this impacts on
the boundaries, shape, dimensionality, and geometry of the mixed
strategy space. In turn, this alters the strategic analysis. For
example, different tangent spaces can change the convexity and
polylinearity properties of expected payoff functions---one
tangent space might ensure expected payoff functions are convex
and polylinear so established existence theorems can define Nash
equilibria, while a different tangent space might support
nonconvex and non-polylinear expected payoff functions.  In such
spaces established existence theorems cannot be used to define
Nash equilibria.

Probability theory models two perfectly correlated variables as
necessarily possessing perfectly correlated trembles, and
accomplishes this by using a one-dimensional tangent space. In
contrast, in the mixed strategy space two perfectly correlated
variables can exhibit independent trembles because the mixed
strategy tangent space permits this. Similarly, probability
theory models independent variables as necessarily possessing
independent trembles in a two dimensional tangent space.  In
contrast, independent variables in the mixed strategy space must
exhibit correlated trembles if they are to remain independent in
the enlarged tangent space of the mixed strategy space. (They
must fluctuate together to maintain the separability of their
joint distribution.) The different tangent spaces adopted by
probability theory and game theory impact on which probabilities
can be trembled and on the possibility of equilibrium
refinements. As trembles are the differential variations of
probability parameters within the adopted tangent space, so
different tangent spaces modify both possible trembles and
defined gradient operators. Altering the differential
fluctuations and gradients of a probability space correspond to
altering which moves can occur at each stage of a game and even
of the number of stages in a game. In turn, these altered move
trees impact on the implementation of optimization algorithms
such as ``backwards induction". In general, the adopted tangent
space underlies all optimization algorithms in both game and
probability theory. Game theory imposes the tangent space of the
mixed strategy simplex on all the probability distributions
modelling a game, while probability theory associates different
tangent spaces with each probability distribution.  It is
natural to expect that these different adopted tangent spaces
will lead to different optimization outcomes.

In this work, we have shown that we can define and employ
probability distributions possessing properties which differ
from any ``contained" within the mixed strategy simplex. These
probability distributions possess a different differential
geometry to that of the simplex.  This has not generally been
considered as probability spaces are not supposed to possess a
geometrical interpretation. However, optimizing random functions
within probability spaces often takes advantage of the
geometrical properties of those spaces, and when those spaces
are isomorphically embedded within enlarged probability spaces,
then those geometrical properties must be preserved.

We further note that mixed strategy spaces are supposed to
contain all cases of deterministic dependencies.  Every
deterministic dependency equates to every possible functional
dependency, and there are standard techniques for dealing with
these functional dependencies.  Players can embed their decision
making processes within deterministic functional spaces of
arbitrary dimension and scope.  The resulting analysis must be
consistent with multi-variate calculus and differential
geometry.  Should probability distributions be applied to these
analytical structures, then the analysis should be consistent
with probability theory.

There are essentially no limits to the scope of the analysis
that can be brought to bear by a rational optimizing agent in a
game.  And game theory needs to provide a treatment consistent
with these other approaches. If a player, following the rules of
game theory, cannot accurately calculate properties of a game,
then they have bounded rationality.  In order to properly
calculate game properties, players must use isomorphic
probability spaces.  Isomorphic mappings are necessary in order
to exhibit unbounded rationality.

In this paper, we hold that game theory must be fully consistent
with both probability theory and optimization theory in general.
Further, we hold that rational players must be able to reproduce
any result from probability theory or optimization theory when
analyzing a game or a decision tree. Indeed, a rational player
should, if they chose, be able to exclusively use techniques
from probability theory and find perfect accord with the results
of game theory. Probability theory mandates that appropriate
constraints designed to preserve tangent spaces must be used
whenever probability distributions are embedded within an
enlarged space in order to preserve all properties. Game theory
has eschewed use of any constraints when embedding distributions
within the mixed strategy probability space, and this leads to
contradictions with probability theory. These discrepancies stem
from the different tangent spaces adopted by probability theory
and game theory, and an examination of these issues promises to
cast light on some of the paradoxes of game theory.  At the very
least, these issues require examination even if only to
establish their irrelevance.

In this work, we consider how to locate the best possible optima
from many different functions defined over different
incommensurate spaces. One way to approach this problem is to
sequentially select each space, and then each function within
that space, and then to locate each of the optima of that
function, and finally to compare all optima to locate the best
outcome. An alternative approach is to embed every possible
function from each space into a single enlarged function, and
then to apply standard techniques to locate the optima of that
function. This approach is in common use in decision theory,
game theory, and in artificial intelligence where multistage
search and decision problems are concatenated together into a
single, enlarged, multivariate mapping from choices to outcomes.
However, the typical embeddings used in these fields do not
preserve gradient information specific to the source function.
That is, an embedding of a source function $f(x)$ within a
surface $g(x,y)$ can be via either $lim_{y\rightarrow y_0}
g(x,y)=f(x)$ or $g(x,y)|_{y=y_0}=f(x)$.  The first of these
methods does not necessarily preserve gradient information as
$\lim_{y\rightarrow y_0}\nabla g(x,y)\neq \nabla g(x,y)|_{y=y_0}
= \nabla f(x)$. In other words, the surface gradient generally
does not replicate the line gradient of the function embedded
within it. This means that a single surface containing many
embedded functions can't reproduce gradient information and
hence can't be used to locate optima of those embedded
functions.

 \mbox{} \newpage
\addcontentsline{toc}{chapter}{BIBLIOGRAPHY}

\begin{thebibliography}{100}

\bibitem{vonNeumann_44} J.~{von Neumann} and O.~Morgenstern.
\newblock {\em Theory of Games and Economic Behavior}.
\newblock Princeton University Press, Princeton, 1944.

\bibitem{Nash_50_48} J.~F. Nash.
\newblock Equilibrium points in $n$-person games.
\newblock {\em Proceedings of the National Academy of Sciences of the United
  States of America}, 36(1):48--49, 1950.

\bibitem{Nash_51_28} J.~Nash.
\newblock Non-cooperative games.
\newblock {\em Annals of Mathematics}, 54(2):286--295, 1951.

\bibitem{Kuhn_1953} H.~W. Kuhn.
\newblock Extensive games and the problem of information.
\newblock In H.~W. Kuhn and A.~W. Tucker, editors, {\em Contributions to the
  Theory of Games, Volume II}, Princeton Annals of Mathematical Studies, No.
  28, Princeton, 1953. Princeton University Press.

\bibitem{Hart_92_19} S.~Hart.
\newblock Games in extensive and strategic forms.
\newblock In R.~J. Aumann and S.~Hart, editors, {\em Handbook of Game Theory
  with Economic Applications}, pages 19--40, Amsterdam, 1992. North Holland.

\bibitem{Selton_1975} R.~Selten.
\newblock A reexamination of the perfectness concept for equilibrium points in
  extensive games.
\newblock {\em International Journal of Game Theory}, 4:25--55, 1975.

\bibitem{Chatterjee_2005} D.~Chatterjee.
\newblock {\em Abstract Algebra}.
\newblock Prentice-Hall, New Delhi, 2005.

\bibitem{Ito_1984} K.~Ito.
\newblock {\em Introduction to Probability Theory}.
\newblock Cambridge University Press, Cambridge, 1984.

\bibitem{Gray_2009} R.~M. Gray.
\newblock {\em Probability, Random Processes and Ergodic Processes}.
\newblock Springer, Dordrecht, 2009.

\bibitem{Walters_1982} P.~Walters.
\newblock {\em An Introduction to Ergodic Theory}.
\newblock Springer-Verlag, New York, 1982.

\bibitem{Sernesi_1993} E.~Sernesi.
\newblock {\em Linear Algebra: A Geometric Approach}.
\newblock Chapman and Hall, Boca Raton, 1993.

\bibitem{Georgii_2008} H.-O. Georgii.
\newblock {\em Stochastics: Introduction to Probability and Statistics}.
\newblock de Gruyter, Berlin, 2008.

\bibitem{Burk_1998} F.~Burk.
\newblock {\em Lebesgue Measure and Integration: An Introduction}.
\newblock New York, Wiley, 1998.

\bibitem{Hayashi_2006} M.~Hayashi.
\newblock {\em Quantum Information: An Introduction}.
\newblock Springer, Berlin, 2006.

\bibitem{Pfeiffer_1990} P.~E. Pfeiffer.
\newblock {\em Probability for Application}.
\newblock Springer, New York, 1990.

\bibitem{Gersho_1991} A.~Gersho and R.~M. Gray.
\newblock {\em Vector Quantization and Signal Compression}.
\newblock Springer, London, 1991.

\bibitem{Insall_2009} M.~Insall, T.~Rowland, and E.~W.
    Weisstein.
\newblock Embedding.
\newblock From MathWorld---A Wolfram Web Resource.
  http://mathworld.wolfram.com/Embedding.html, 2009.

\bibitem{Ramsey_1928_543} F.~P. Ramsey.
\newblock A mathematical theory of savings.
\newblock {\em Economic Journal}, 38(152):543--559, 1928.

\bibitem{Kelly_94} D.~G. Kelly.
\newblock {\em Introduction to Probability}.
\newblock Macmillan, New York, 1994.

\bibitem{Cover_91} T.~M. Cover and J.~A. Thomas.
\newblock {\em Elements of Information Theory}.
\newblock Wiley, New York, 1991.

\bibitem{vanDamme_92_41} E.~van Damme.
\newblock Strategic equilibrium.
\newblock In R.~J. Aumann and S.~Hart, editors, {\em Handbook of Game Theory
  with Economic Applications}, pages 1521--1596, Amsterdam, 1992. North
  Holland.

\bibitem{Pinter_2012} J.~Pinter.
\newblock {\em Global Optimization}.
\newblock From MathWorld--A Wolfram Web Resource, created by Eric W. Weisstein.
  http://mathworld.wolfram.com/GlobalOptimization.html.

\bibitem{Aumann_74_67} R.~J. Aumann.
\newblock Subjectivity and correlation in randomized strategies.
\newblock {\em Journal of Mathematical Economics}, 1:67--96, 1974.

\bibitem{Milgrom_1982_280} P.~Milgrom and J.~Roberts.
\newblock Predation, reputation, and entry deterrence.
\newblock {\em Journal of Economic Theory}, 27:280--312, 1982.

\bibitem{Selten_78_12} R.~Selten.
\newblock The chain store paradox.
\newblock {\em Theory and Decision}, 9:127--159, 1978.

\bibitem{Rosenthal_1981_92} R.~W. Rosenthal.
\newblock Games of perfect information, predatory pricing and the chain-store
  paradox.
\newblock {\em Journal of Economic Theory}, 25:92--100, 1981.

\bibitem{Kreps_1982_253} D.~M. Kreps and R.~Wilson.
\newblock Reputation, and imperfect information.
\newblock {\em Journal of Economic Theory}, 27:253--279, 1982.

\bibitem{Davis_85_13} L.~H. Davis.
\newblock No chain store paradox.
\newblock {\em Theory and Decision}, 18(2):139--144, 1985.

\bibitem{Trockel_86_16} W.~Trockel.
\newblock The chain-store paradox revisited.
\newblock {\em Theory and Decision}, 21(2):163--179, 1986.

\bibitem{Wilson_1992_305} R.~Wilson.
\newblock Strategic models of entry deterrence.
\newblock In R.~J. Aumann and S.~Hart, editors, {\em Handbook of Game Theory
  with Economic Applications}, pages 305--329, Amsterdam, 1992. North Holland.

\bibitem{Kreps_1990_1} D.~M. Kreps.
\newblock Corporate culture and economic theory.
\newblock In J.~E. Alt and K.~A. Shepsle, editors, {\em Perspectives on
  Positive Political Economy}, page~1, Cambridge, UK, 1990. Cambridge
  University.

\bibitem{Berg_1995_122} J.~Berg.
\newblock Turst, reciprocity, and social history.
\newblock {\em Games and Economic Behaviour}, 10:122--142, 1995.

\bibitem{King-Casas_2005_78} B.~King-Casas, D.~Tomlin, C.~Anen,
    C.~F. Camerer, S.~R. Quartz, and P.~R.
  Montague.
\newblock Getting to know you: Reputation and trust in a two-person economic
  exchange.
\newblock {\em Science}, 308:78--83, 2005.

\bibitem{Guth_82_36} W.~G\"{u}th, R.~Schmittberger, and
    B.~Schwarze.
\newblock An experimental analysis of ultimatum bargaining.
\newblock {\em Journal of Economic Behavior and Organization}, 3(4):367--388,
  1982.

\bibitem{Stahl_72} I.~Stahl.
\newblock {\em Bargaining Theory}.
\newblock Economic Research Institute, Stockholm, 1972.

\bibitem{Rubinstein_82_97} A.~Rubinstein.
\newblock Perfect equilibrium in a bargaining model.
\newblock {\em Econometrica}, 50:97--109, 1982.

\bibitem{Binmore_85_11} K.~Binmore, A.~Shaked, and J.~Sutton.
\newblock Testing noncooperative bargaining theory: A preliminary study.
\newblock {\em American Economic Review}, 75:1178--1180, 1985.

\bibitem{Ochs_89_35} J.~Ochs and A.~E. Roth.
\newblock An experimental study of sequential bargaining.
\newblock {\em American Economic Review}, 79(3):355--384, 1989.

\bibitem{Bolton_91_10} G.~E. Bolton.
\newblock A comparative model of bargaining: Theory and evidence.
\newblock {\em American Economic Review}, 81(5):1096--1136, 1991.

\bibitem{Oosterbeek_2004_171} H.~Oosterbeek, R.~Sloof, and
    G.~{van de Kuilen}.
\newblock Cultural differences in ultimatum game experiments: Evidence from a
  meta-analysis.
\newblock {\em Experimental Economics}, 7:171--188, 2004.

\bibitem{Hoffman_96_28} E.~Hoffman, K.~A. Mc{C}abe, and V.~L.
    Smith.
\newblock On expectations and the monetary stakes in ultimatum games.
\newblock {\em International Journal of Game Theory}, 25(3):289--302, 1996.

\bibitem{Slonim_98_56} R.~Slonim and A.~E. Roth.
\newblock Learning in high stakes ultimatum games: An experiment in the
  {S}lovak republic.
\newblock {\em Econometrica}, 66(3):569--596, 1998.

\bibitem{Cameron_99_47} L.~A. Cameron.
\newblock Raising the stakes in the ultimatum game: Experimental evidence from
  {I}ndonesia.
\newblock {\em Economic Inquiry}, 37(1):47--59, 1999.

\bibitem{Roth_91_10} A.~E. Roth, V.~Prasnikar,
    M.~Okuno-Fujiwara, and S.~Zamir.
\newblock Bargaining and market behavior in {J}erusalem, {L}jubljana,
  {P}ittsburgh, and {T}okyo.
\newblock {\em American Economic Review}, 81(5):1068--1095, 1991.

\bibitem{Henrich_00_97} J.~Henrich.
\newblock Does culture matter in economic behavior? {U}ltimatum game bargaining
  among the {M}achiguenga of the {P}eruvian {A}mazon.
\newblock {\em American Economic Review}, 90(4):973--979, 2000.

\bibitem{Thaler_88_19} R.~H. Thaler.
\newblock The ultimatum game.
\newblock {\em Journal of Economic Perspectives}, 2(4):195--206, 1988.

\bibitem{Roth_95_12} A.~E. Roth.
\newblock Bargaining experiments.
\newblock In J.~Kagel and A.~E. Roth, editors, {\em Handbook of Experimental
  Economics}, Princeton, NJ, 1995. Princeton University Press.

\bibitem{Camerer_95_20} C.~Camerer and R.~H. Thaler.
\newblock Anomalies - ultimatums, dictators and manners.
\newblock {\em Journal of Economic Perspectives}, 9(2):209--219, 1995.

\bibitem{Zamir_00} S.~Zamir.
\newblock Rationality and emotions in ultimatum bargaining.
\newblock mimeo, {L}ecture, {C}onf\'{e}rence {D}es {A}nnales, {J}une 19, 2000.

\bibitem{Winter_97} E.~Winter and S.~Zamir.
\newblock An experiment with ultimatum bargaining in a changing environment.
\newblock {T}he {H}ebrew {U}niversity, {C}enter for {R}ationality and
  {I}nteractive {D}ecision {T}heory, {DP No. 159}, 1997.

\bibitem{Ruffle_1998_247} B.~J. Ruffle.
\newblock More is better, but fair is fair: Tipping in dictator and ultimatum
  games.
\newblock {\em Games and Economic Behavior}, 23(2):247--265, 1998.

\bibitem{Prasnikar_92_86} V.~Prasnikar and A.~E. Roth.
\newblock Considerations of fairness and strategy: Experimental data from
  sequential games.
\newblock {\em Quarterly Journal of Economics}, 107(3):865--888, 1992.

\bibitem{Rabin_93_12} M.~Rabin.
\newblock Incorporating fairness into game theory and economics.
\newblock {\em American Economic Review}, 83(5):1281--1302, 1993.

\bibitem{Loewenstein_93_13} G.~Loewenstein, I.~Samuel,
    C.~Camerer, and L.~Babcock.
\newblock Self-serving assessments of fairness and pretrial bargaining.
\newblock {\em Journal of Legal Studies}, 22:135--159, 1993.

\bibitem{Forsythe_94_34} R.~Forsythe, J.~L. Horowitz, N.~E.
    Savin, and M.~Sefton.
\newblock Fairness in simple bargaining experiments.
\newblock {\em Games and Economic Behavior}, 6(3):347--369, 1994.

\bibitem{Blount_95_13} S.~Blount.
\newblock When social outcomes aren't fair: The effect of causal attributions
  on preferences.
\newblock {\em Organizational Behavior and Human Decision Processes},
  63(2):131--144, 1995.

\bibitem{Bolton_00_16} G.~E. Bolton and A.~Ockenfels.
\newblock {ERC}: A theory of equity, reciprocity and competition.
\newblock {\em American Economic Review}, 90(1):166--193, 2000.

\bibitem{Kagel_96_10} J.~H. Kagel, C.~Kim, and D.~Moser.
\newblock Fairness in ultimatum games with asymmetric information and
  asymmetric payoffs.
\newblock {\em Games and Economic Behavior}, 13(1):100--110, 1996.

\bibitem{Burnell_99_22} S.~J. Burnell, L.~Evans, and S.~Yao.
\newblock The ultimatum game: Optimal strategies without fairness.
\newblock {\em Games and Economic Behavior}, 26:221--252, 1999.

\bibitem{Dufwenberg_98_0} M.~Dufwenberg and G.~Kirchsteiger.
\newblock A theory of sequential reciprocity.
\newblock mimeo, {C}ent{ER} for {E}conomic {R}esearch, Tilberg, 1998.

\bibitem{Falk_00_0} A.~Falk and U.~Fischbacher.
\newblock A theory of reciprocity.
\newblock Institute for Empirical Research in Economics: Working Paper Series,
  Working Paper No. 6, See http://www.unizh.ch/iew/wp/, 2000.

\bibitem{Kirchsteiger_94_37} G.~Kirchsteiger.
\newblock The role of envy in ultimatum games.
\newblock {\em Journal of Economic Behavior and Organization}, 25(3):373--390,
  1994.

\bibitem{Bolton_95_95} G.~E. Bolton and R.~Zwick.
\newblock Anonymity versus punishment in ultimatum bargaining.
\newblock {\em Games and Economic Behavior}, 10:95--121, 1995.

\bibitem{Fehr_99_81} E.~Fehr and K.~M. Schmidt.
\newblock A theory of fairness, competition and cooperation.
\newblock {\em Quarterly Journal of Economics}, 114:817--868, 1999.

\bibitem{Levine_98_59} D.~Levine.
\newblock Modeling altruism and spitefulness in experiments.
\newblock {\em Review of Economic Dynamics}, 1:593--622, 1998.

\bibitem{Hoffman_94_34} E.~Hoffman, K.~Mc{C}abe, K.~Shachat, and
    V.~Smith.
\newblock Preferences, property rights and anonymity in bargaining games.
\newblock {\em Games and Economic Behavior}, 7:346--380, 1994.

\bibitem{Roth_95_16} A.~E. Roth and I.~Erev.
\newblock Learning in extensive form games: Experimental data and simple
  dynamic models in the intermediate term.
\newblock {\em Games and Economic Behavior}, 8(1):164--212, 1995.

\bibitem{Gale_95_56} J.~Gale, K.~G. Binmore, and L.~Samuelson.
\newblock Learning to be imperfect: The ultimatum game.
\newblock {\em Games and Economic Behavior}, 8(1):56--90, 1995.

\bibitem{Vriend_97_9} N.~J. Vriend.
\newblock Will reasoning improve learning?
\newblock {\em Economics Letters}, 55(1):9--18, 1997.

\bibitem{Duffy_99_13} J.~Duffy and N.~Feltovich.
\newblock Does observation of others affect learning in strategic environments?
  {A}n experimental study.
\newblock {\em International Journal of Game Theory}, 28(1):131--152, 1999.

\bibitem{Nowak_00_17} M.~A. Nowak, K.~M. Page, and K.~Sigmund.
\newblock Fairness versus reason in the ultimatum game.
\newblock {\em Science}, 289:1773--1775, 2000.

\bibitem{Huck_99_13} S.~Huck and J.~Oechssler.
\newblock The indirect evolutionary approach to explaining fair allocations.
\newblock {\em Games and Economic Behavior}, 28:13--24, 1999.

\bibitem{Guth_92_23} W.~G\"{u}th and M.~Yaari.
\newblock An evolutionary approach to explain reciprocal behavior in a simple
  strategic game.
\newblock In U.~Witt, editor, {\em Explaining Process and Change: Appproaches
  to Evolutionary Economics}, pages 23--34, Ann Arbor, 1992.

\bibitem{Peters_00_31} R.~Peters.
\newblock Evolutionary stability in the ultimatum game.
\newblock {\em Group Decision and Negotiation}, 9(4):315--324, 2000.

\bibitem{Hardin_1968_1243} G.~Hardin.
\newblock The tragedy of the commons.
\newblock {\em Science}, 162:1243--1248, 1968.

\bibitem{Fehr_2000_980} E.~Fehr and S.~Gachter.
\newblock Cooperation and punishment in public goods experiments.
\newblock {\em American Economic Review}, 90:980--994, 2000.

\bibitem{Nowak_98_573} M.~A. Nowak and K.~Sigmund.
\newblock Evolution of indirect reciprocity by image scoring.
\newblock {\em Nature}, 393:573--577, 1998.

\bibitem{Wedekind_2000_850} C.~Wedekind and M.~Milinski.
\newblock Cooperation through image scoring in humans.
\newblock {\em Science}, 288:850--852, 2000.

\bibitem{Fehr_04_185} E.~Fehr and U.~Fischbacher.
\newblock Social norms and human cooperation.
\newblock {\em Trends in Cognitive Sciences}, 8(4):185--190, 2004.

\bibitem{Nowak_2005_1291} M.~A. Nowak and K.~Sigmund.
\newblock Evolution of indirect reciprocity.
\newblock {\em Nature}, 437:1291--1298, 2005.

\bibitem{Chen_03_0301013} R.~Beausoleil K.-Y.~Chen, T.~Hogg.
\newblock A practical quantum mechanism for the public goods game.
\newblock Eprint Archive:quant-phys/0301013 (See
  http://arxiv.org/abs/quant-ph/0301013), 2003.

\bibitem{Kreps_1990} D.~M. Kreps.
\newblock {\em A Course in Microeconomic Theory}.
\newblock Harvester Wheatsheaf, New York, 1990.

\bibitem{McKelvey_1992_803} R.~McKelvey and T.~Palfrey.
\newblock An experimental study of the centipede game.
\newblock {\em Econometrica}, 60(4):803--836, 1992.

\bibitem{Nagel_1998_356} R.~Nagel and F.~F. Tang.
\newblock An experimental study on the centipede game in normal form: An
  investigation on learning.
\newblock {\em Journal of Mathematical Psychology}, 42:356--384, 1998.

\bibitem{Binmore_87_17} K.~Binmore.
\newblock Modeling rational players: Part {I}.
\newblock {\em Economics and Philosophy}, 3:179--214, 1987.

\bibitem{Binmore_88_9} K.~Binmore.
\newblock Modeling rational players: Part {II}.
\newblock {\em Economics and Philosophy}, 4:9--55, 1988.

\bibitem{Binmore_1994_150} K.~Binmore.
\newblock Rationality in the centipede.
\newblock In R.~Fagin, editor, {\em Theoretical Aspects Of Rationality And
  Knowledge (TARK 1994): Proceedings of the 5th Conference on Theoretical
  Aspects of Reasoning about Knowledge}, pages 150--159, Pacific Grove,
  California, 1994. Morgan Kaufmann.

\bibitem{Aumann_1995_6} R.~J. Aumann.
\newblock Backward induction and common knowledge of rationality.
\newblock {\em Games and Economic Behavior}, 8:6--19, 1995.

\bibitem{Binmore_1996_135} K.~Binmore.
\newblock A note on backward induction.
\newblock {\em Games and Economic Behavior}, 17:135--137, 1996.

\bibitem{Aumann_1996_138} R.~J. Aumann.
\newblock Reply to {B}inmore.
\newblock {\em Games and Economic Behavior}, 17:138--146, 1996.

\bibitem{Aumann_1998_97} R.~J. Aumann.
\newblock Note on the centipede game.
\newblock {\em Games and Economic Behavior}, 23:97--105, 1998.

\bibitem{Pettit_99_17} P.~Pettit and R.~Sugden.
\newblock The backward induction paradox.
\newblock {\em The Journal of Philosophy}, 136(4):169--182, 1999.

\bibitem{Broome_1999_237} J.~Broome and W.~Rabinowicz.
\newblock Backwards induction in the centipede game.
\newblock {\em Analysis}, 59(4):237--242, 1999.

\bibitem{Sobel_1993_114} J.~H. Sobel.
\newblock Backward-induction arguments: A paradox regained.
\newblock {\em Philosophy of Science}, 60(1):114--133, 1993.

\bibitem{Sigmund_2000_949} K.~Sigmund and M.~A. Nowak.
\newblock A tale of two selves.
\newblock {\em Science}, 290:949--950, 2000.

\bibitem{Cooper_96_18} R.~Cooper, D.~V.~De Jong, R.~Forsythe,
    and T.~W. Ross.
\newblock Cooperation without reputation: Experimental evidence from prisoner's
  dilemma games.
\newblock {\em Games and Economic Behavior}, 12(2):187--218, 1996.

\bibitem{Milinski_98_13} M.~Milinski and C.~Wedekind.
\newblock Working memory constrains human cooperation in the prisoner's
  dilemma.
\newblock {\em Proceedings of the National Academy of Sciences of the United
  States of America}, 95(23):13755--13758, 1998.

\bibitem{Davis_99_89} D.~D. Davis and C.~A. Holt.
\newblock Equilibrium cooperation in two-stage games: Experimental evidence.
\newblock {\em International Journal of Game Theory}, 28(1):89--109, 1999.

\bibitem{Croson_00_29} R.~T.~A. Croson.
\newblock Thinking like a game theorist: Factors affecting the frequency of
  equilibrium play.
\newblock {\em Journal of Economic Behavior and Organization}, 41(3):299--314,
  2000.

\bibitem{Radner_80_13} R.~Radner.
\newblock Collusive behaviour in non-cooperative epsilon-equilibria in
  oligopolies with long but finite lives.
\newblock {\em Journal of Economic Theory}, 22:136--154, 1980.

\bibitem{Radner_86_38} R.~Radner.
\newblock Can bounded rationality resolve the prisoner's dilemma.
\newblock In A.~Mas-Colell and W.~Hildenbrand, editors, {\em Essays in Honor of
  Gerard Debreu}, pages 387--399, Amsterdam, 1986. North-Holland.

\bibitem{Vegaredondo_94_18} F.~Vegaredondo.
\newblock Bayesian boundedly rational agents play the finitely repeated
  prisoner's dilemma.
\newblock {\em Theory and Decision}, 36(2):187--206, 1994.

\bibitem{Harborne_97_13} S.~W. {Harborne Jr.}
\newblock Common belief of rationality in the finitely repeated prisoners'
  dilemma.
\newblock {\em Games and Economic Behavior}, 19(1):133--143, 1997.

\bibitem{Anthonisen_99_14} N.~Anthonisen.
\newblock Strong rationalizability for two-player noncooperative games.
\newblock {\em Economic Theory}, 13:143--169, 1999.

\bibitem{Fehr_03_785} E.~Fehr and U.~Fischbacher.
\newblock The nature of human altruism.
\newblock {\em Nature}, 425:785--791, 2003.

\bibitem{Axelrod_1984} R.~Axelrod.
\newblock {\em The Evolution of Cooperation}.
\newblock Basic Books, New York, 1984.

\bibitem{Harsanyi_67_15} J.~C. Harsanyi.
\newblock Games with incomplete information played by ``{B}ayesian" players.
\newblock {\em Management Science}, 14(3):159--182, 1967.

\bibitem{Kreps_82_24} D.~M. Kreps, P.~Milgrom, J.~Roberts, and
    R.~Wilson.
\newblock Rational cooperation in the finitely repeated prisoner's dilemma.
\newblock {\em Journal of Economic Theory}, 27:245--252, 1982.

\bibitem{Fudenberg_86_53} D.~Fudenberg and E.~Maskin.
\newblock The {F}olk {T}heorem in repeated games with discounting and
  incomplete information.
\newblock {\em Econometrica}, 54:533--554, 1986.

\bibitem{Sarin_99_10} R.~Sarin.
\newblock Simple play in the prisoner's dilemma.
\newblock {\em Journal of Economic Behavior and Organization}, 40(1):105--113,
  1999.

\bibitem{Neyman_99_45} A.~Neyman.
\newblock Cooperation in repeated games when the number of stages is not known.
\newblock {\em Econometrica}, 67(1):45--64, 1999.

\bibitem{Neyman_85_22} A.~Neyman.
\newblock Bounded complexity justifies cooperation in the finitely repeated
  prisoner's dilemma.
\newblock {\em Economics Letters}, 19:227--229, 1985.

\bibitem{Rubinstein_86_83} A.~Rubinstein.
\newblock Finite automata play the repeated prisoner's dilemma.
\newblock {\em Journal of Economic Theory}, 39:83--96, 1986.

\bibitem{Cho_99_93} I.-K. Cho and H.~Li.
\newblock How complex are networks playing repeated games.
\newblock {\em Economic Theory}, 13:93--123, 1999.

\bibitem{Raff_00_10} H.~Raff and D.~Schmidt.
\newblock Cumbersome coordination in repeated games.
\newblock {\em International Journal of Game Theory}, 29(1):101--118, 2000.

\bibitem{Evans_97_11} R.~Evans and J.~P. Thomas.
\newblock Reputation and experimentation in repeated games with two long-run
  players.
\newblock {\em Econometrica}, 65(5):1153--1173, 1997.

\bibitem{Sheng_94_23} C.~L. Sheng.
\newblock A note on the prisoner dilemma.
\newblock {\em Theory and Decision}, 36(3):233--246, 1994.

\bibitem{Groes_99_12} E.~Groes, H.~J. Jacobsen, and B.~Sloth.
\newblock Adaptive learning in extensive form games and sequential equilibria.
\newblock {\em Economic Theory}, 13:125--142, 1999.

\bibitem{Song_99_63} Q.~A. Song and A.~Kandel.
\newblock A fuzzy approach to strategic games.
\newblock {\em {IEEE} Transactions on Fuzzy Systems}, 7(6):634--642, 1999.

\bibitem{Howard_71} N.~Howard.
\newblock {\em Paradoxes of Rationality: Theory of Metagames and Political
  Behavior}.
\newblock MIT Press, Cambridge, Mass, 1971.

\bibitem{Rapoport_67} A.~Rapoport.
\newblock Escape from paradox.
\newblock {\em Scientific American}, 217:50--56, July 1967.

\bibitem{Straffin_93} P.~D. Straffin.
\newblock {\em Game Theory and Strategy}.
\newblock Mathematical Association of America, Washington, 1993.

\bibitem{Eisert_99_30} J.~Eisert, M.~Wilkens, and M.~Lewenstein.
\newblock Quantum games and quantum strategies.
\newblock {\em Physical Review Letters}, 83(15):3077--3080, 1999.

\bibitem{vonNeumann_1955} J.~{von Neumann}.
\newblock {\em Mathematical Foundations of Quantum Mechanics}.
\newblock Princeton University Press, Princeton, 1955.
\newblock First published in 1932.

\end{thebibliography}

\end{document}